\documentclass[12pt, a4paper, twoside, onecolumn, openright]{book} 		

\usepackage[latin1]{inputenc}			
\usepackage[T1]{fontenc}		    		
\usepackage[english]{babel}	
\usepackage{amsfonts}  					
\usepackage{amsmath}					
\usepackage{amssymb}					
\usepackage{fancyhdr}			
\usepackage[top=2.5cm, bottom=2.5cm, inner=2.5cm, outer=2.5cm]{geometry}	
\usepackage{setspace}	
\usepackage[dvipsnames]{xcolor}
\usepackage{color}
\usepackage[squaren,Gray,cdot]{SIunits}  
\usepackage{enumerate}
\usepackage{graphicx}
\usepackage{subfigure}
\usepackage{multirow}			
\usepackage[labelsep=endash]{caption}   
\usepackage{todonotes}
\usepackage[rightcaption]{sidecap}       
\usepackage{xspace} 
\usepackage{caption}
\usepackage[a4paper]{meta-donnees}
\usepackage{eurosym}
\usepackage{comment}
\usepackage{cleveref}
\usepackage{subfigmat}
\usepackage{etoolbox}
\usepackage{lmodern}
\usepackage{geometry}

\usepackage[a4paper]{meta-donnees}
\usepackage{eso-pic} 

\usepackage{silence}
\usepackage{relsize}

\usepackage{array,multirow,makecell}


\hypersetup{
  colorlinks = true,
  linkcolor  = black
}

\begin{document}

\setcounter{secnumdepth}{4}							
\setcounter{tocdepth}{3}	
\renewcommand\figurename{\small\textbf{Figure}}

\pagestyle{empty}
\fancyhf{}
			





\Specialite{Physique de la Mati\`ere Condens\'ee et du Rayonnement}
\Arrete{23 Avril 2009}
\Auteur{Guillaume Lang}
\Directeur{Anna Minguzzi}
\CoDirecteur{Frank Hekking}    
\Laboratoire{du laboratoire LPMMC}
\EcoleDoctorale{l'école doctorale de physique}         
\Titre{Correlations in low-dimensional quantum gases}
\Depot{27 octobre 2017}




\Jury{
\UGTPresident{M. Jean-Christian ANGLES D'AURIAC}{Directeur de recherche CNRS, Institut N\'eel, Grenoble}
\UGTRapporteur{M. Grigori ASTRAKHARCHIK}{Professeur Assistant, Universitat Polit\`ecnica de Catalunya, Barcelone, Espagne}
\UGTRapporteur{M. Dimitri GANGARDT}{Professeur, University of Birminghan, Birminghan, Royaume-Uni}
\UGTExaminateur{M. Jean-S\'ebastien CAUX}{Professeur, University of Amsterdam, Amsterdam, Pays Bas}

\UGTDirecteur{Mme Anna MINGUZZI}{Directeur de recherche CNRS, LPMMC, Grenoble}

}

\MakeUGthesePDG    




\tableofcontents 

\pagestyle{fancy}
\fancyhf{}
\renewcommand{\headrulewidth}{1pt}
\rfoot[\thepage]{\thepage}
\onehalfspacing

\newpage

\Huge{Acknowledgments}
\\

\normalsize{

This manuscript being an advanced milestone of a long adventure, there are so many people I would like to thank for their direct or indirect contribution to my own modest achievements that I will not even try and list them all. May they feel sure that in my heart at least, I forget none of them.

First of all, I am grateful to the hundreds of science teachers who contributed to my education, and in particular to Serge Gibert, Thierry Schwartz, Jean-Pierre Demange and Jean-Dominique Mosser, who managed to trigger my sense of rigour, and to Alice Sinatra for her precious help in crucial steps of my orientation. Thanks to those who have devoted their precious time to help my comrades and I get firm enough basis in physics and chemistry to pass the selective 'agr\'egation' exam, most peculiarly Jean Cviklinski, Jean-Baptiste Desmoulins, Pierre Villain and Isabelle Ledoux.

I would like to thank my internship advisors for their patience and enthusiasm. Nicolas Leroy, who gave me a first flavor of the fascinating topic of gravitational waves, Emilia Witkowska, who gave me the opportunity to discover the wonders of Warsaw and cold atoms, for her infinite kindness, Anna Minguzzi and Frank Hekking who taught me the basics of low-dimensional physics and helped me make my first steps into the thrilling topic of integrable models.
Thank to their network, I have had the opportunity to closely collaborate with Patrizia Vignolo, Vanja Dunjko and Maxim Olshanii. I would like to thank the latter for his interest in my ideas and his enthusiasm. Our encounter remains one of the best memories of my life as a researcher. I also appreciated to take part to the Superring project and discuss with the members of H\'el\`ene Perrin's team.

A few collaborations have not been as rewarding in terms of concrete achievements but taught me much as well, I would like to thank in particular Eiji Kawasaki for taking time to think together about spin-orbit coupling in low dimensions, Luigi Amico for private lectures on the XYZ model and his kind offers of collaboration, and Davide Rossini for his kind welcome in Pisa.

Thanks and greetings to all former and current LPMMC members, permanent researchers, visitors, interns, PhD students and Postdocs. I really enjoyed these months or years spent together, your presence to internal seminars and challenging questions, your availability and friendship. In particular, I would like to thank those to whom I felt a little closer: Marco Cominotti first, then Eiji Kawasaki with whom I shared the office and spent good moments out of the lab, Malo Tarpin with whom it always was a pleasure to discuss, and Katharina Rojan who has been so kind to all of us.

Thanks to all of those who took time to discuss during conferences and summer schools, in particular Bruno Naylor, David Cl\'ement and Thierry Giamarchi, they who made interesting remarks helping improve my works or shared ideas on theirs, in particular Fabio Franchini, Mart\'on Kormos, Maxim Olshanii, Sylvain Prolhac, Zoran Ristivojevic and Giulia de Rosi. Thanks also to my jury members for accepting to attend my defence, and for their useful comments and kind remarks.
I also have a thought for my advisor Frank Hekking, who would have been glad to attend my defence as well. I always admired him for his skills as a researcher, a teacher and for his qualities as a human being in general, and would like to devote this work to his memory, and more generally, to all talented people who devoted (part of) their life to the common adventure of science. May their example keep inspiring new generations.

Aside my research activity, I have devoted time and energy to teaching as well, in this respect I would like to thank Jean-Pierre Demange for allowing me to replace him a couple of times, Michel Faye for giving me the opportunity to teach at Louis le Grand for a few months, then my colleagues at Universit\'e Grenoble Alpes, in peculiar Sylvie Zanier and, most of all, Christophe Rambaud with whom it was a pleasure to collaborate.
Now that I have moved to full-time teaching, thanks to all of my colleagues at Auxerre for their kind welcome, in peculiar to the CPGE teachers with whom I interact most, Cl\'ement Dunand for his charism and friendship, and Fanny Bavouzet who is always willing to help me and give me good advice.

To finish with, my way to this point would not have been the same without my family and their support, nor without wonderful people whom I met on the road, among others Lor\`ene, Michel, Jo\"elle, Jean-Guillaume and Marie-Anne, without whom I would have stopped way before, then Charles-Arthur, Guillaume, Thibault, C\'ecile, S\'ebastien, Pierre, Nicolas, Vincent, Cl\'elia, Delphine, Am\'elie, Iulia, \'Elodie, Cynthia, F\'elix and Ariane. Special thanks to Paul and Marc who have been my best friends all along this sneaky and tortuous way to the present.

Most of all, my thoughts go to my sun and stars, pillar and joy of my life, friend and soulmate. Thank you so much, Claire!

With love,

G. Lang.

}

\newpage

This thesis consists of an introductory text, followed by a summary of my research.

A significant proportion of the original results presented has been published in the following articles:\\ \\
(i) \underline{Guillaume Lang}, Frank Hekking and Anna Minguzzi, \textit{Dynamic structure factor and drag force in a one-dimensional Bose gas at finite temperature}, Phys. Rev. A \textbf{91}, 063619 (2015), Ref.~\cite{LangHekkingMinguzzi2015}
\\ \\
(ii) \underline{Guillaume Lang}, Frank Hekking and Anna Minguzzi, \textit{Dimensional crossover in a Fermi gas and a cross-dimensional Tomonaga-Luttinger model}, Phys. Rev. A \textbf{93}, 013603 (2016), Ref.~\cite{LangHekkingMinguzzi2016}
\\ \\
(iii) \underline{Guillaume Lang}, Frank Hekking and Anna Minguzzi, \textit{Ground-state energy and excitation spectrum of the Lieb-Liniger model : accurate analytical results and conjectures about the exact solution}, SciPost Phys. \textbf{3}, 003 (2017), Ref.~\cite{LangHekkingMinguzzi2017}
\\ \\
(iv) \underline{Guillaume Lang}, Patrizia Vignolo and Anna Minguzzi, \textit{Tan's contact of a harmonically trapped one-dimensional Bose gas: strong-coupling expansion and conjectural approach at arbitrary interactions}, Eur. Phys. J. Special Topics \textbf{226}, 1583-1591 (2017), Ref.~\cite{LangVignoloMinguzzi2017}
\\ \\
(v) Maxim Olshanii, Vanja Dunjko, Anna Minguzzi and \underline{Guillaume Lang}, \textit{Connection between nonlocal one-body and local three-body correlations of the Lieb-Liniger model}, Phys. Rev. A \textbf{96}, 033624 (2017), Ref.~\cite{OlshaniiDunjkoMinguzziLang}
\\ \\
\\
Other publication by the author, not presented in this thesis:
\\ \\
(vi) \underline{Guillaume Lang} and Emilia Witkowska, \textit{Thermodynamics of a spin-1 Bose gas with fixed magnetization}, Phys. Rev. A \textbf{90}, 043609 (2014), Ref.~\cite{LangWitkowska}

\newpage
\textit{New topics constantly appear, that bring researchers away from old problems. Mastering the latter, precisely because they have been so much studied, requires an ever increasing effort of understanding, and this is unpleasing. It turns out that most researchers prefer considering new, less developed problems, that require less knowledge, even if they are not challenging. Nothing can be done against it, but formatting old topics with good references$\dots$ so that later developments may follow, if destiny decides so.} 
\\
\\
Felix Klein, translation from German by the author
\newpage

\chapter{Introduction: this thesis/cette th\`ese}

This theoretical thesis summarizes the research activity I have performed during the three years of my PhD studies at Laboratoire de Physique et Mod\'elisation des Milieux Condens\'es (LPMMC) in Grenoble, as a student of the \'Ecole doctorale de Physique of Universit\'e Grenoble Alpes, under the supervision of Dr. Anna Minguzzi, and late Prof. Frank Hekking.

My work deals with ultracold atom physics \cite{BlochDalibardZwerger2008, GiorginiPitaevskiiStringari2008}, where the high versatility of current experiments allows to probe phase diagrams of various systems in detail. I put the emphasis on low-dimensional setups, in particular degenerate quantum gases kinetically confined to one spatial dimension (1D gases), that became available in the early years of the twenty-first century \cite{CazalillaCitroGiamarchiOrignacRigol2011}, but had already been studied as toy models since the early days of quantum physics.

I have focused on analytical methods and techniques, sometimes at the verge of mathematical physics, and left aside advanced numerical tools in spite of their increasing importance in modern theoretical physics. Experimental aspects are secondary in this manuscript, but have been a guideline to my investigations, as I have taken part to a joint programm with an experimental group at Laboratoire de Physique et des Lasers (LPL) in Villetaneuse, the SuperRing project.

The key notion of this thesis is the one of strongly-correlated systems, that can not be described in terms of weakly-interacting parts. Solving models that feature strong correlations is among the most challenging problems encountered in theoretical physics, since the strong-coupling regime is not amenable to perturbative techniques. In this respect, reduction of dimensionality is of great help as it makes some problems analytically amenable, thanks to powerful tools such as Bethe Ansatz (BA), bosonization \cite{GiamarchiBook} or conformal field theory (CFT) \cite{FrancescoMathieuSenechal}. Another interesting point is that parallels between high-energy, condensed-matter and statistical physics are especially strong nowadays, since the theoretical tools involved are of the same nature \cite{GogolinNersesyanTsvelik, Witten2016, Wilczek2016}. I focus on the low-energy sector and use a condensed-matter language, but readers from other communities may find interest in the techniques all the same.

I tackle various aspects of the many-body problem, with auto-correlation functions of the many-body wavefunction as a common denominator, and a means to characterize low-dimensional ultracold atoms. The manuscript is composed of four main parts, whose outline follows:

\textbf{Chapter \ref{secII}} is a general introduction to various experimental and theoretical aspects of the many-body problem in reduced dimension. I give a brief account of the main specificities of one-dimensional gases, and introduce correlation functions as a suitable observable to characterize such fluids. Experimental and theoretical studies that allowed this reduction of dimensionality are summarized. I present powerful theoretical tools that are commonly used to solve integrable models, such as Bethe Ansatz, bosonization in the framework of Luttinger liquid theory and Conformal Field Theory. Their common features are put into light and simple illustrations are given. To finish with, I present the main known methods to increase the effective dimension of a system, as an introduction to the vast topic of dimensional crossovers.

\textbf{Chapter \ref{secIII}} deals with local and non-local equal-time equilibrium correlations of the Lieb-Liniger model. The latter is the paradigmatic model to describe 1D Bose gases, and has the property of being integrable. It has a long history, and this chapter may serve as an introduction to the topic, but deals with advanced aspects as well. In particular, I have made a few contributions towards the analytical exact ground-state energy, based on the analysis of mathematical structures that emerge in weak- and strong-coupling expansions. Then, I delve into the issue of correlation functions and the means to construct them from integrability in a systematic way. I introduce the notion of connection, that binds together in a single formalism a wide variety of relationships between correlation functions and integrals of motion. Keeping in mind that most experiments involve a trap that confines the atoms, I then show how the Bethe Ansatz formalism can be combined to the local density approximation (LDA) to describe trapped interacting gases in the non-integrable regime of inhomogeneous density, through the so-called BALDA (Bethe Ansatz LDA) formalism.

\textbf{Chapter \ref{secIV}} is devoted to the dynamical correlations of the Lieb-Liniger model. They are investigated in order to discuss the notion of superfluidity, through the concept of drag force induced by a potential barrier stirred in the fluid. The drag force criterion states that a superfluid flow is linked to the absence of a drag force under a critical velocity, generaling Landau's criterion for superfluidity. Computing the drag force in linear response theory requires a good knowledge of the dynamical structure factor, an observable worth studying for itself as well since it is experimentally accessible by Bragg scattering and quite sensitive to interactions and dimensionality. This gives me an opportunity to investigate the validity range of the Tomonaga-Luttinger liquid theory in the dynamical regime, and tackle a few finite-temperature aspects. I also study the effect of a finite width of the barrier on the drag force, putting into light a decrease of the drag force, hinting at a quasi-superfluid regime at supersonic flows.

In \textbf{chapter \ref{secV}}, I study the dimensional crossover from 1D to higher dimensions. A simple case, provided by noninteracting fermions in a box trap, is treated exactly and in detail. The effect of dimensionality on the dynamical structure factor and drag force is investigated directly and through multi-mode structures, the effect of a harmonic trap is treated in the local density approximation.

After a general conclusion, a few Appendices provide details of calculations and introduce transverse issues. I did not reproduce all the derivations published in my articles, the interested reader can find them there and in references therein.
\\
\\
\\
\\
\\
Cette th\`ese th\'eorique r\'esume les principaux r\'esultats que j'ai obtenus au cours de mes trois ann\'ees de doctorat au LPMMC, \`a Grenoble, sous la direction d'Anna Minguzzi et de feu Frank Hekking.

Elle s'inscrit dans le cadre de la physique de la mati\`ere condens\'ee, et plus particuli\`erement des atomes ultrafroids, qui suscite l'int\'er\^et de par la possibilit\'e qu'offrent ces syst\`emes de simuler toutes sortes de mod\`eles et d'\'etudier en d\'etail les diagrammes de phase qui leurs sont associ\'es. Je m'int\'eresse plus particuli\`erement \`a des gaz d\'eg\'en\'er\'es dont les degr\'es de libert\'e spatiaux transversaux sont entrav\'es par des pi\`eges au point que leur dynamique est strictement unidimensionnelle. Bien qu'ils aient fait l'objet d'\'etudes th\'eoriques depuis des d\'ecennies, de tels syst\`emes ont \'et\'e r\'ealis\'es pour la premi\`ere fois au tournant du XXI-\`eme si\`ecle, ravivant l'int\'er\^et pour ces derniers.

Parmi les nombreuses m\'ethodes disponibles pour d\'ecrire les gaz quantiques unidimensionnels, j'ai plus particuli\`erement port\'e mon attention sur les techniques analytiques, d\'elaissant volontairement les aspects num\'eriques pour lesquels, en d\'epit de leur importance croissante et de leur int\'er\^et ind\'eniable, je n'ai pas d'affinit\'e particuli\`ere. Je n'insiste pas non plus outre mesure sur les aspects exp\'erimentaux, dont je suis loin d'\^etre expert, mais ils restent pr\'esents en toile de fond comme source d'inspiration. En particulier, certaines th\'ematiques que j'ai abord\'ees l'ont \'et\'e dans le cadre du projet SuperRing, conjoint avec des exp\'erimentateurs du LPL \`a Villetaneuse.

La notion de syst\`eme fortement corr\'el\'e joue un r\^ole essentiel dans mon projet de recherche. De tels syst\`emes ne peuvent \^etre appr\'ehend\'es en toute g\'en\'eralit\'e par les m\'ethodes perturbatives usuelles, qui ne s'appliquent pas dans le r\'egime de couplage fort. De ce fait, ils constituent un formidable d\'efi pour la physique th\'eorique actuelle. La r\'eduction de dimension le rend abordable, mais pas trivial pour autant, loin s'en faut. Les outils phares qui en permettent l'\'etude analytique sont connus sous les noms d'Ansatz de Bethe, de bosonisation et de th\'eorie des champs conforme. Une particularit\'e qui me tient particuli\`erement \`a c\oe ur est le parall\`ele fort qui existe actuellement entre la physique des hautes \'energies, de la mati\`ere condens\'ee et la physique statistique du fait de leurs emprunts mutuels de formalisme et de techniques. Bien que je m'int\'eresse ici plus sp\'ecifiquement \`a la physique de basse \'energie, des chercheurs d'autres communaut\'es sont susceptibles de trouver un int\'er\^et pour les techniques et le formalisme employ\'es.

J'aborde divers aspects du probl\`eme \`a N corps, centr\'es autour des multiples fonctions de corr\'elation qu'on peut d\'efinir \`a partir de la seule fonction d'onde, qui constituent un formidable outil pour caract\'eriser les syst\`emes d'atomes froids en basse dimension. J'ai d\'ecid\'e de les pr\'esenter dans quatre parties distinctes, qui constituent chacune un chapitre de ce manuscrit.

Le Chapitre II consistue une introduction g\'en\'erale au probl\`eme \`a N corps quantique en dimension r\'eduite. J'y pr\'esente quelques caract\'eristiques sp\'ecifiques aux gaz unidimensionnels, puis explique comment les efforts conjoints des th\'eoriciens et exp\'erimentateurs ont permis leur r\'ealisation. Certains mod\`eles phares des basses dimensions s'av\`erent \^etre int\'egrables, aussi je pr\'esente les m\'ethodes analytiques qui permettent d'en \'etudier de mani\`ere exacte les propri\'et\'es thermodynamiques et les fonctions de corr\'elation, \`a savoir l'Ansatz de Bethe, la bosonisation appliqu\'ee aux liquides de Tomonaga-Luttinger et la th\'eorie conforme des champs. Ces techniques sont en partie compl\'ementaires, mais j'insiste \'egalement sur leurs similarit\'es. Enfin, \`a rebours de la d\'emarche qui consiste \`a chercher \`a r\'eduire la dimension d'un syst\`eme, je m'int\'eresse au probl\`eme oppos\'e, qui consiste \`a augmenter la dimension effective de mani\`ere graduelle, et pr\'esente les quelques m\'ethodes \'eprouv\'ees \`a ce jour.

Le Chapitre III traite des corr\'elations locales et non-locales dans l'espace mais \`a temps \'egaux et \`a l'\'equilibre du mod\`ele de Lieb et Liniger. Il s'agit l\`a d'un paradigme couramment appliqu\'e pour d\'ecrire les gaz de Bose unidimensionnels, et les techniques pr\'esent\'ees au chapitre pr\'ec\'edent s'y appliquent car ce mod\`ele est int\'egrable. De par sa longue histoire et le bon millier d'articles qui lui ont \'et\'e consacr\'es, il constitue \`a lui seul un vaste sujet dont ma pr\'esentation peut faire guise d'introduction. J'y aborde \'egalement des aspects techniques avanc\'es concernant l'\'energie exacte du gaz dans son \'etat fondamental. J'ai notamment am\'elior\'e les estimations analytiques de cette derni\`ere par une \'etude fine des structures math\'ematiques apparaissant dans les d\'eveloppements en couplage fort et faible. Cette \'etude pr\'eliminaire d\'ebouche sur celle des fonctions de corr\'elation, et notamment la fonction de corr\'elation \`a un corps que je m'emploie \`a construire de fa\c con syst\'ematique en me fondant sur l'int\'egrabilit\'e du mod\`ele de Lieb et Liniger. En explicitant les premi\`eres \'etapes de cette construction, j'ai \'et\'e amen\'e \`a introduire la notion de connexion, qui englobe dans un formalisme unique l'ensemble des formules connues actuellement qui lient les fonctions de corr\'elations et les int\'egrales du mouvement. En fait, la plupart des exp\'eriences actuelles font intervenir un pi\`ege pour confiner les atomes, ce qui rend le gaz inhomog\`ene et prive le mod\`ele de sa propri\'et\'e d'int\'egrabilit\'e. Toutefois, une astucieuse combinaison de l'approximation de la densit\'e locale et de l'Ansatz de Bethe permet d'acc\'eder quand m\^eme \`a la solution exacte moyennant des calculs plus \'elabor\'es.

Dans le Chapitre IV, je m'int\'eresse aux corr\'elations dynamiques du mod\`ele de Lieb et Liniger, qui apportent des informations sur les propri\'et\'es de superfluidit\'e \`a travers le concept de force de tra\^in\'ee induite par une barri\`ere de potentiel mobile. Le crit\`ere de superfluidit\'e associ\'e \`a la force de tra\^in\'ee stipule qu'un \'ecoulement superfluide est associ\'e \`a une force de tra\^in\'ee rigoureusement nulle. Cette derni\`ere peut \^etre \'evalu\'ee dans le formalisme de la r\'eponse lin\'eaire, \`a condition de conna\^itre le facteur de structure dynamique du gaz, une autre observable traditionnellement mesur\'ee par diffusion de Bragg, et tr\`es sensible \`a l'intensit\'e des interactions ainsi qu'\`a la dimensionnalit\'e. Cette \'etude me donne une opportunit\'e de discuter du domaine de validit\'e de la th\'eorie des liquides de Tomonaga-Luttinger dans le r\'egime dynamique, et de m'int\'eresser \`a quelques aspects thermiques. Enfin, en \'etudiant plus sp\'ecifiquement l'effet de l'\'epaisseur de la barri\`ere de potentiel sur la force de tra\^in\'ee, je mets en \'evidence la possibilit\'e d'un r\'egime supersonique particulier, qu'on pourrait qualifier de quasi-superfluide.

Dans le Chapitre V, j'\'etudie la transition progressive d'un gaz unidimensionnel vers un gaz de dimension sup\'erieur \`a travers l'exemple, conceptuellement simple, de fermions sans interaction plac\'es dans un pi\`ege parall\'el\'epip\'edique. La simplicit\'e du mod\`ele autorise un traitement analytique exact de bout en bout, qui met en \'evidence les effets dimensionnels sur les observables d\'ej\`a \'etudi\'ees dans le chapitre pr\'ec\'edent, le facteur de structure dynamique et la force de tra\^in\'ee, tant de fa\c con directe que par la prise en compte d'une structure multimodale en \'energie obtenue par ouverture graduelle du pi\`ege. L'effet d'un pi\`ege harmonique est tra\^it\'e ult\'erieurement, toujours \`a travers l'approximation de la densit\'e locale.

Apr\`es une conclusion globale, quelques appendices compl\`etent cette vision d'ensemble en proposant des digressions vers des sujets transverses ou en approfondissant quelques d\'etails techniques in\'edits.

\newpage

\chapter{From 3D to 1D and back to 2D}
\label{secII}

\section{Introduction}

We perceive the world as what mathematicians call three-dimensional (3D) Euclidian space, providing a firm natural framework for geometry and physics until the modern times. Higher-dimensional real and abstract spaces have pervaded physics in the course of the twentieth century, through statistical physics where the number of degrees of freedom considered is comparable to the Avogadro number, quantum physics where huge Hilbert spaces are often involved, general relativity where in addition to a fourth space-time dimension one considers curvature of a Riemannian manifold, or string theory where more dimensions are considered before compactification.

Visualizing a higher-dimensional space requires huge efforts of imagination, for a pedagogical illustration the reader is encouraged to read the visionary novel \textit{Flatland} \cite{Flatland}. As a general rule, adding dimensions has dramatic effects due to the addition of degrees of freedom, that we do not necessarily apprehend intuitively. The unit ball has maximum volume in 5D, for instance.

This is not the point I would like to emphasize however, but rather ask this seemingly innocent, less debated question: we are obviously able to figure out lower-dimensional spaces, ranging from 0D to 3D, but do we really have a good intuition of them and of the qualitative differences involved? As an example, a random walker comes back to its starting point in finite time in 1D and 2D, but in 3D this is not always the case. One of the aims of this thesis is to point out such qualitative differences in ultracold gases, that will manifest themselves in their correlation functions. To put specific phenomena into light, I will come back and forth from the three-dimensional Euclidian space, to a one-dimensional line-world.

As far as dimension is concerned, there is a deep dichotomy between the experimental point of view, where reaching a low-dimensional regime is quite challenging, and the theoretical side, where 1D models are far easier to deal with, while powerful techniques are scarce in 3D. Actually, current convergence of experimental and theoretical physics in this field concerns multi-mode quasi-one dimensional systems and dimensional crossovers from 1D to 2D or vice-versa.

This introductory, general chapter is organized as follows: first, I present a few peculiarities of 1D quantum systems and introduce the concept of correlation functions as an appropriate tool to characterize them, then I present a few experimental breakthroughs involving low-dimensional gases, and the main analytical tools I have used during my thesis to investigate such systems. To finish with, I present a few approaches to the issue of dimensional crossovers to higher dimension.
\\
\\
\\
\\
\\
La dimension d'un espace correspond au nombre de directions ind\'ependantes qui le caract\'erisent. En ce sens, la fac\c on dont nous percevons le monde par le biais de nos sens am\`ene naturellement \`a le mod\'eliser par un espace euclidien de dimension trois. La possibilit\'e d'envisager des espaces (r\'eels ou abstraits) de dimension plus \'elev\'ee a fait son chemin des math\'ematiques vers la physique, o\`u cette id\'ee est d\'esormais courante dans les th\'eories modernes. En m\'ecanique hamiltonienne et en physique statistique, le nombre de degr\'es de libert\'e envisag\'es est de l'ordre du nombre d'Avogadro, la physique quantique fait appel \`a des espaces de Hilbert de grande dimension, tandis que la relativit\'e g\'en\'erale consid\`ere un espace-temps quadridimensionnel o\`u la courbure locale joue un r\^ole primordiale, et la th\'eorie des cordes envisage encore plus de dimensions spatiales avant l'\'etape finale de compactification.

Ces espaces de dimension sup\'erieure soul\`event la probl\'ematique de leur visualisation, qui n'a rien de simple. Je recommande \`a ce sujet la lecture d'un roman visionnaire intitul\'e \textit{Flatland}, qui invite \`a y m\'editer. Pour les lecteurs francophones int\'eress\'es par une approche plus formelle, je conseille \'egalement la lecture de la r\'ef\'erence \cite{Espacephy}. On retiendra qu'en r\`egle g\'en\'erale, une augmentation de la dimension de l'espace s'accompagne d'effets importants et pas n\'ecessairement triviaux du fait de l'accroissement concomitant du nombre de degr\'es de libert\'e. Certains de ces effets ne s'appr\'ehendent pas intuitivement, un exemple qui me pla\^it est le fait qu'un d\'eplacement al\'eatoire ram\`ene au point de d\'epart en temps fini m\^eme si l'espace est infini en une et deux dimensions, ce qui n'est pas n\'ecessairement le cas en trois dimensions. Un des objectifs de cette th\`ese est de mettre en \'evidence des effets dimensionnels non-triviaux dans le domaine des gaz d'atomes ultrafroids, notamment en ce qui concerne les fonctions d'auto-corr\'elation associ\'ees \`a la fonction d'onde. Pour les comprendre, il sera n\'ecessaire d'envisager \`a la fois un monde lin\'eaire, unidimensionnel, et des espaces euclidiens de dimension sup\'erieure.

La probl\'ematique de la dimension d'un syst\`eme s'appr\'ehende de mani\`ere relativement diff\'erente selon qu'on est exp\'erimentateur ou th\'eoricien. Dans les exp\'eriences, il est difficile de diminuer la dimension d'un syst\`eme, tandis que du c\^ot\'e th\'eorique, les mod\`eles unidimensionnels sont bien plus faciles \`a traiter que les mod\`eles 3D du fait du nombre restreint de m\'ethodes efficaces dans ce dernier cas. On assiste aujourd'hui \`a une convergence des probl\'ematiques th\'eoriques et exp\'erimentales au passage de 1D \`a 2D et vice-versa, \`a travers la notion de syst\`eme multi-mode quasi-1D.

Ce chapitre est organis\'e de la mani\`ere suivante: dans un premier temps, je pr\'esente quelques particularit\'es des syst\`emes quantiques 1D et explique que leurs corr\'elations les caract\'erisent, puis je r\'ecapitule les principales avanc\'ees th\'eoriques et exp\'erimentales dans ce domaine, apr\`es quoi j'introduis dans les grandes lignes les techniques th\'eoriques que j'ai utilis\'ees dans ma th\`ese. Enfin, j'aborde la probl\'ematique de l'augmentation de la dimension d'un syst\`eme \`a travers les quelques techniques connues \`a ce jour.

\section{Welcome to Lineland}
\label{First dive into 1D}
\subsection{Generalities on one-dimensional systems}

It is quite intuitive that many-particle physics in one dimension must be qualitatively different from any higher dimension whatsoever, since particles do not have the possibility of passing each other without colliding. This topological constraint has exceptionally strong effects on systems of non-ideal particles, however weakly they may interact, and the resulting collectivization of motion holds in both classical and quantum theories.

An additional effect of this crossing constraint is specific to the degenerate regime and concerns quantum statistics. While in three dimensions particles are either bosons or fermions, in lower dimension the situation is more intricate. To understand why, we shall bear in mind that statistics is defined through the symmetry of the many-body wavefunction under two-particle exchange: it is symmetric for bosons and antisymmetric for fermions. Such a characterization at the most elementary level is experimentally challenging \cite{Roos2017}, but quite appropriate for a Gedankenexperiment. In order to directly probe the symmetry of the many-body wavefunction, one shall engineer a physical process responsible for the interchange of two particles, that would not otherwise disturb the system. A necessary condition is that the particles be kept apart enough to avoid the influence of interaction effects.

In two dimensions, this operation is possible provided that interactions are short-ranged, although performing the exchange clockwise or counter-clockwise is not equivalent, leading to the (theoretical) possibility of intermediate particle statistics \cite{LeinaasMyrrheim, Wilczek1982, Haldane1991}. The corresponding particles are called anyons, as they can have \textit{any} statistics between fermionic and bosonic, and are defined through the symmetry of their many-body wavefunction under exchange as 
\begin{eqnarray}
\psi(\dots x_i,\dots, x_j,\dots)=e^{i\chi}\psi(\dots x_j,\dots, x_i,\dots),
\end{eqnarray}
where $\chi$ is real.

In one dimension, such an exchange process is utterly forbidden by the crossing constraint, making particle statistics and interactions deeply intertwined: the phase shifts due to scattering and statistics merge, arguably removing all meaning from the very concept of statistics. I will nonetheless, in what follows, consider particles as fermions or bosons, retaining by convention the name they would be given if they were free in 3D (for instance, $^{87}$Rb atoms are bosons and $^{40}$K atoms are fermions). The concept of 1D anyons is more tricky and at the core of recent theoretical investigations \cite{Kundu1999, BatchelorGuanHe2007, Patu2007, CalabreseMintchev2007, PiroliCalabrese2017}, but I leave this issue aside.

The origin of the conceptual difficulty associated with statistics in 1D is the fact that we are too accustomed to noninteracting particles in 3D. Many properties that are fully equivalent in the three-dimensional Euclidian space, and may unconsciously be put on equal footing, are not equivalent anymore in lower dimension. For instance, in 3D, bosons and fermions are associated to Bose-Einstein and Fermi-Dirac statistics respectively. Fermions obey the Pauli principle (stating that two or more identical fermions can not occupy the same quantum state simultaneously), the spin-statistics theorem implies that bosons have integer spin and fermions an half-integer one \cite{Pauli1940}, and any of these properties looks as fundamental as any other. In one dimension however, strongly-interacting bosons possess a Fermi sea structure and can experience a kind of Pauli principle due to interactions. These manifestations of a statistical transmutation compel us to revise, or at least revisit, our conception of statistics in arbitrary dimension.

For fermions with spin, the collision constraint has an even more dramatic effect. A single fermionic excitation has to split into a collective excitation carrying charge (a 'chargon', the analog of a sound wave) and another one carrying spin (called spin wave, or 'spinon'). They have different velocities, meaning that electrons, that are fundamental objects in 3D, break into two elementary excitations. As a consequence, in one dimension there is a complete separation between charge and spin degrees of freedom. Stated more formally, the Hilbert space is represented as a product of charge and spin sectors, whose parameters are different. This phenomenon is known as 'spin-charge separation' \cite{Kollath2005}, and is expected in bosonic systems as well \cite{Kollath2008}. 

These basic facts should be sufficient to get a feeling that 1D is special. We will see many other concrete illustrations in the following in much more details, but to make physical predictions that illustrate peculiarities of 1D systems and characterize them, it is first necessary to select a framework and a set of observables. Actually, the intertwined effect of interactions and reduced dimensionality is especially manifest on correlation functions.

\subsection{Correlation functions as a universal probe for many-body quantum systems}

Theoretical study of condensed-matter physics in three dimensions took off after the laws of many-body quantum mechanics were established on firm enough grounds to give birth to powerful paradigms. A major achievement in this respect is Landau's theory of phase transitions. In this framework, information on a system is encoded in its phase diagram, obtained by identifying order parameters that take a zero value in one phase and are finite in the other phase, and studying their response to variations of external parameters such as temperature or a magnetic field in the thermodynamic limit. Laudau's theory is a versatile paradigm, that has been revisited over the years to encompass notions linked to symmetry described through the theory of linear Lie groups. It turns out that symmetry breaking is the key notion underneath, as in particle physics, where the Higgs mechanism plays a significant role.

In one dimension, however, far fewer finite-temperature phase transitions are expected, and none in systems with short-range interactions. This is a consequence of the celebrated Mermin-Wagner-Hohenberg theorem, that states the impossibility of spontaneous breakdown of a continuous symmetry in 1D quantum systems with short-range interactions at finite temperature \cite{MerminWagner}, thus forbidding formation of off-diagonal long-range order.

In particular, according to the definition proposed by Yang \cite{Yang1962}, this prevents Bose-Einstein condensation in uniform systems, while this phenomenon is stable to weak interactions in higher dimensions. This example hints at the fact that Landau's theory of phase transitions may not be adapted in most cases of interest involving low-dimensional systems, and a shift of paradigm should be operated to characterize them efficiently.

An interesting, complementary viewpoint suggested by the remark above relies on the study of correlation functions of the many-body wavefunction in space-time and momentum-energy space. In mathematics, the notion of correlation appears in the field of statistics and probabilities as a tool to characterize stochastic processes. It comes as no surprise that correlations have become central in physics as well, since quantum processes are random, and extremely huge numbers of particles are dealt with in statistical physics.

The paradigm of correlation functions first pervaded astrophysics with the Hanbury Brown and Twiss experiment \cite{HanburyBrown1956}, and has taken a central position in optics, with Michelson, Mach-Zehnder and Sagnac interferometers as fundamental setups, where typically electric field or intensity temporal correlations are probed, to quantify the coherence between two light-beams and probe the statistics of intensity fluctuations respectively.

In parallel, this formalism has been successfully transposed and developed to characterize condensed-matter systems, where its modern form partly relies on the formalism of linear response theory, whose underlying idea is the following: in many experimental configurations, the system is probed with light or neutrons, that put it slightly out of equilibrium. Through the response to such external excitations, one can reconstruct equilibrium correlations \cite{Pottier}.

Actually, the paradigm of correlation functions allows a full and efficient characterization of 1D quantum gases. In particular, it is quite usual to probe how the many-body wavefunction is correlated with itself. For instance, one may be interested in density-density correlations, or their Fourier transform known as the dynamical structure factor.
It is natural to figure out, and calculations confirm it, that the structure of correlation functions in 1D is actually much different from what one would expect in higher dimensions. At zero temperature, in critical systems correlation functions decay algebraically in space instead of tending to a finite value or even of decaying exponentially, while in energy-momentum space low-energy regions can be kinetically forbidden, and power-law divergences can occur at their thresholds. These hallmarks of 1D systems are an efficient way to probe their effective dimension, and will be investigated much in detail throughout this thesis. However, recent developments such as far from equilibrium dynamics \cite{Kinoshita2006}, thermalization or its absence after a quench \cite{CalabreseCardy2006, RigolDunjkoYurovskyOlshanii2007, CauxEssler2013, DeNardis2014, Atas2016, NardisPanfil2016} or periodic driving to a non-equilibrium steady state \cite{Eckardt2017} are beyond its scope. More recent paradigms, such as topological matter and information theory (with entanglement entropy as a central notion \cite{Eisert2010}), will not be tackled neither.

I proceed to describe dimensional reduction in ultracold atom systems and the possibilities offered by the crossover from 3D to 1D.

\section{From 3D to 1D in experiments}

While low-dimensional models have had the status of toy models in the early decades of quantum physics, they are currently realized to a good approximation in a wide variety of condensed-matter experimental setups. The main classes of known 1D systems are spin chains, some electronic wires, ultracold atoms in a tight waveguide, edge states (for instance in the Quantum Hall Effect), and magnetic insulators. Their first representatives have been experimentally investigated in the 1980's, when the so-called Bechgaard salts have provided examples of one-dimensional conductors and superconductors \cite{Bechgaard1980}. As far as 2D materials are concerned, the most remarkable realizations are high-temperature superconductors \cite{Bednorz1986}, graphene \cite{GeimNovoselov2004} and topological insulators \cite{Moore2007}.

A revolution came later on from the field of ultracold atoms, starting in the 1990's. The main advantage of ultracold atom gases over traditional condensed-matter systems is that, thanks to an exceptional control over all parameters of the gaseous state, they offer new observables and tunable parameters, allowing for exhaustive exploration of phase diagrams, to investigate macroscopic manifestations of quantum effects such as superfluidity, and clean realizations of quantum phase transitions (such transitions between quantum phases occur at zero temperature by varying a parameter in the Hamiltonian, and are driven by quantum fluctuations, contrary to 'thermal' ones where thermal fluctuations play a major role \cite{Sachdev1999}). Ultracold gases are a wonderful platform for the simulation of condensed-matter systems \cite{Lewenstein2007} and theoretical toy-models, opening the field of quantum simulation \cite{BlochDalibardNascimbene2012}, where experiments are designed to realize models, thus reversing the standard hierarchy between theory and experiment \cite{Feynman1982}.

With ultracold atoms, the number of particles and density are under control, allowing for instance to construct a Fermi sea atom per atom \cite{Wenz2013}. The strength and type of interactions can be modified as well: tuning the power of the lasers gives direct control over the hopping parameters in each direction of an optical lattice, whereas Feshbach resonance allows to tune the interaction strength \cite{Courteille1998, Chin2010}. Neutral atoms interact through a short-ranged potential, while dipolar atoms and molecules feature long-range interactions \cite{Griesmaier2005}.

Particles are either bosons or fermions, but any mixture of different species is a priory feasible. Recently, a mixture of degenerate bosons and fermions has been realized using the lithium-6 and lithium-7 isotopes of Li atoms \cite{Sushi}, and in lower dimensions, anyons may become experimentally relevant. Internal atomic degrees of freedom can be used to produce multicomponent systems in optical traps, the so-called spinor gases, where a total spin $F$ leads to $2F\!+\!1$ components \cite{Ho1998, StamperKurn1998}.

Current trapping techniques allow to modify the geometry of the gas through lattices, i.e. artificial periodic potentials similar to the ones created by ions in real solids, or rings and nearly-square boxes that reproduce ideal theoretical situations and create periodic and open boundary conditions respectively \cite{Gaunt2013}. Although (nearly) harmonic traps prevail, double-well potentials and more exotic configurations yield all kinds of inhomogeneous density profiles. On top of that, disorder can be taylored, from a single impurity \cite{Kohl2010} to many ones \cite{Palzer2009}, to explore Anderson localization \cite{Anderson1958, Billy2008} or many-body localization \cite{BAA, Choi2016}.

As far as thermal effects are concerned, in condensed-matter systems room temperature is usually one or two orders of magnitude lower than the Fermi temperature, so one can consider $T\!=\!0$ as a very good approximation. In ultracold atom systems, however, temperature scales are much lower and span several decades, so that one can either probe thermal fluctuations, or nearly suppress them at will to investigate purely quantum fluctuations \cite{Dettmer2001, Esteve2006}.

Recently, artificial gauge fields similar to real magnetic fields for electrons could be applied to these systems \cite{Lin2009, Dalibard2011}, giving access to the physics of ladders \cite{Atala2014}, quantum Hall effect \cite{Price2015} and spin-orbit coupling \cite{Galitski2013}.

The most famous experimental breakthrough in the field of ultracold atoms is the demonstration of Bose-Einstein condensation, a phenomenon linked to Bose statistics where the lowest energy state is macroscopically occupied \cite{CornellBEC, KetterleBEC, BradleyBEC}, 70 years after its prediction \cite{Einstein1924, Einstein1925}. This tour de force has been allowed by continuous progress in cooling techniques (essentially by laser and evaporation \cite{Ketterle1996}) and confinement. Other significant advances are the observation of the superfluid-Mott insulator transition in an optical lattice \cite{Greiner2002}, degenerate fermions \cite{deMarcoJin1998}, the BEC-BCS crossover \cite{Regal2004}, and of topological defects such as quantized vortices \cite{Matthews1999, Dalibard2000} or solitons \cite{Fleischer2003}.

Interesting correlated phases appear both in two-dimensional and in one-dimensional systems, where the most celebrated achievements are the observation of the Berezinskii-Kosterlitz-Thouless (BKT) transition \cite{Hadzibabic2006}, an unconventional phase transition in 2D that does not break any continuous symmetry \cite{KosterlitzThouless}, and the realization of the fermionized, strongly-correlated regime of impenetrable bosons in one dimension \cite{Paredes2004, Kinoshita2004}, the so-called Tonks-Girardeau gas \cite{Girardeau1960}.

Such low-dimensional gases are obtained by a strong confinement along one (2D gas) or two (1D gas) directions, in situations where all energy scales of the problem are smaller than the transverse confinement energy, limiting the transverse motion of atoms to zero point oscillations. This tight confinement is experimentally realized through very anisotropic trapping potentials.

The crossover from a 2D trapped gas to a 1D one has been theoretically investigated in \cite{Olshanii98}, under the following assumptions: the waveguide potential is replaced by an axially symmetric two-dimensional harmonic potential of frequency $\omega_{\perp}$, and the forces created by the potential act along the $x\!-\!y$ plane.
The atomic motion along the $z$-axis is free, in other words no longitudinal trapping is considered. As usual with ultracold atoms, interactions between the atoms are modeled by Huang's pseudopotential \cite{Huang1957}
 \begin{eqnarray}
 U(r)=g\,\delta(r)\frac{\partial}{\partial r}(r\cdot),
 \end{eqnarray}
where $g\!=\!4\pi\hbar^2a_s/m$, $a_s$ being the $s$-wave scattering length for the true interaction potential, $\delta$ the dirac function and $m$ the mass of an atom. The regularization operator $\frac{\partial}{\partial r}(r\cdot)$, that removes the $1/r$ divergence from the scattered wave, plays an important role in the derivation.
The atomic motion is cooled down below the transverse vibrational energy $\hbar \omega_{\perp}$. Then, at low velocities the atoms collide in the presence of the waveguide and the system is equivalent to a 1D gas subject to the interaction potential $U_{\mbox{\scriptsize{1D}}}(z)=g_{\mbox{\scriptsize{1D}}}\delta(z)$, whose interaction strength is given by \cite{Olshanii98}

\begin{eqnarray}
g_{\mbox{\scriptsize{1D}}}=\frac{2\hbar^2}{ma_{\perp}}\frac{a_s/a_{\perp}}{1-C a_s/a_{\perp}}.
\end{eqnarray}
In this equation, $a_{\perp}\!=\!\sqrt{\frac{\hbar}{m\omega_{\perp}}}$ represents the size of the ground state of the transverse Hamiltonian and $C\!=\!-\zeta(1/2)\simeq 1.46$, where $\zeta$ is the Riemann zeta function.

In subsequent studies, the more technical issue of the crossover from 3D to 1D for a trapped Bose gas has also been discussed \cite{LiebSeiringerYngvason2004, LiebSolovejSeiringerYngvason2005}. Recently, the dimensional crossover from 3D to 2D in a bosonic gas through strengthening of the transverse confinement, has been studied by renormalization group techniques \cite{Lammers2016}.

The experimental realization of the necessary strongly-anisotropic confinement potentials is most commonly achieved via two schemes. In the first one, atoms are trapped in 2D optical lattices that are created by two orthogonal standing waves of light, each of them obtained by superimposing two counter-propagating laser beams. The dipole force acting on the atoms localizes them in the intensity extrema of the light wave, yielding an array of tightly-confining 1D potential tubes \cite{Bloch2005}.

In the second scheme, atoms are magnetically trapped on an atom chip \cite{Amerongen2008}, where magnetic fields are created via a current flowing in microscopic wires and electrodes, that are micro-fabricated on a carrier substrate. The precision in the fabrication of such structures allows for a very good control of the generated magnetic field, that designs the potential landscape via the Zeeman force acting on the atoms. In this configuration, a single 1D sample is produced, instead of an array of several copies as in the case of an optical lattice.

Both techniques are used all around the world. The wire configuration thereby obtained corresponds to open boundary conditions, but there is also current interest and huge progress in the ring geometry, associated to periodic boundary conditions. This difference can have a dramatic impact on observables at the mesoscopic scale, especially if there are only a few particles. The effect of boundary conditions is expected to vanish in the thermodynamic limit.

The ring geometry has already attracted interest in condensed-matter physics in the course of last decades: supercurrents in superconducting coils are used on a daily basis to produce strong magnetic fields (reaching several teslas), and superconducting quantum interference devices (SQUIDs), based on two parallel Josephson junctions in a loop, allow to measure magnetic flux quanta \cite{Doll1961}. In normal (as opposed to superconducting) systems, mesoscopic rings have been used to demonstrate the Aharonov-Bohm effect \cite{Webb1985} (a charged particle is affected by an electromagnetic potential despite being confined to a space region where both magnetic and electric fields are zero, as predicted by quantum physics \cite{AharonovBohm1959}), and persistent currents \cite{Bluhm2009}.

Ring geometries are now investigated in ultracold gases as well. Construction of ring-shaped traps and study of the superfluid properties of an annular gas is receiving increasing attention from various groups worldwide. The driving force behind this development is its potential for future applications in the fields of quantum metrology and quantum information technology, with the goal of realising high-precision atom interferometry \cite{Arnold2006} and quantum simulators based on quantum engineering of persistent current states \cite{Cominotti2014}, opening the field of 'atomtronics' \cite{Seaman2007}.

Among the ring traps proposed or realized so far, two main categories can be distinguished. In a first kind of setup, a cloud of atoms is trapped in a circular magnetic guide of a few centimeters \cite{Sauer2001}, or millimeters in diameter \cite{StamperKurn2005, Pritchard2012}. Such large rings can be described as annular wave-guides. They can be used as storage rings, and are preferred when it comes to developing guided-atom, large-area interferometers designed to measure rotations.

The second kind of ring traps, designed to study quantum fluid dynamics, has a more recent history, and associated experiments started with the first observation of a persistent atomic flow \cite{Ryu2007}. To maintain well-defined phase coherence over the whole cloud, the explored radii are much smaller than in the previous configuration. A magnetic trap is pierced by a laser beam, resulting in a radius of typically $10$ to $20\mu m$ \cite{Phillips2011, Boshier2013, Kumar2016}. The most advanced experiments of this category rely mostly on purely optical traps, combining a vertical confinement due to a first laser beam, independent of the radial confinement realized with another beam propagating in the vertical direction, in a hollow mode \cite{Moulder2012, StamperKurn2015}.

\begin{figure}
\includegraphics[width=4cm, keepaspectratio, angle=0]{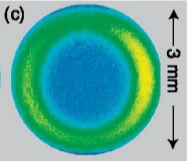}
\includegraphics[width=4cm, keepaspectratio, angle=0]{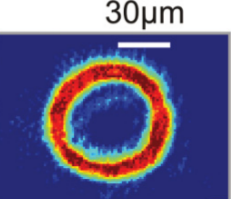}

\includegraphics[width=13cm, keepaspectratio, angle=0]{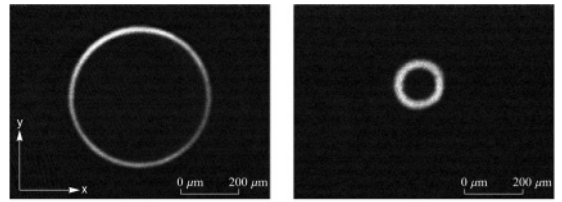}
\caption{In-situ experimental pictures of ring-trapped atomic gases, illustrated the variety of radii obtained with a magnetic trap (top left, from Ref.~\cite{StamperKurn2005}), an optical trap (top right, from Ref.~\cite{Moulder2012}), and with the combined techniques and a radio-frequency field, allowing to tune the radius (bottom, from Ref.~\cite{Foot2011})}
\label{figrings}
\end{figure}

Other traps make use of a combination of magnetic, optical and radio-frequency fields \cite{Morizot2006, Foot2010, Foot2011, Bell2016, Chakraborty2016, VonKlitzing2016}. They can explore radii between $20$ and $500\mu m$, bridging the gap between optical traps and circular waveguides. As an illustration, Fig.~\ref{figrings} shows in-situ images of trapped gases obtained by the techniques presented above. The tunable parameters in radio-frequency traps are the ring radius and its ellipticity. Moreover, vertical and radial trapping frequencies can be adjusted independently, allowing to explore both the 2D and 1D regime.

In the following, I shall always consider a ring geometry, though in most cases it will be of no importance whatsoever once the thermodynamic limit is taken. In the next section, I present the main analytical tools I have used to study 1D gases on a ring.

\section{Analytical methods to solve 1D quantum models}
\label{General theoretical tools for 1D quantum systems}

Condensed-matter theorists are confronted to the tremendously challenging issue of the description of many-body interacting systems. In three dimensions, in some cases one may eliminate the main complicated terms in a many-electron problem and merely incorporate the effect of interactions into parameters (such as the mass) of new excitations called quasiparticles, that are otherwise similar noninteracting fermions. This adiabatic mapping is the essence of Landau's Fermi liquid theory \cite{Landau1957, Landau1957bis, LandauLifschitzVol9}, that has been the cornerstone of theoretical solid-state physics for a good part of the $20^{th}$ century. This approach provides
the basis for understanding metals in terms of weakly-interacting electron-like particles, describes superconductivity and superfluidity, but is restricted to fermions and breaks down in 1D \cite{Voit1995}. For these reasons, other tools are needed to study low-dimensional strongly-correlated gases, a fortiori bosonic ones.

Actually, there are three main theoretical approaches to one-dimensional strongly-correlated systems. Either one tries and find exact solutions of many-body theories, typically using Bethe Ansatz techniques, or reformulate complicated interacting models in such a way that they become weakly-interacting, which is the idea at the basis of bosonization. These techniques are complementary, both will be used throughout this thesis. The third approach is the use of powerful numerical tools and will not be tackled here. Let me only mention that a major breakthrough in this field over the recent years has been the spectacular development of the density matrix renormalization group (DMRG) method \cite{White}. It is an iterative, variational method within the space of matrix product states, that reduces effective degrees of freedom to those most important for a target state, giving access to the ground state of 1D models, see e.g. \cite{Schollwock2005}. To study finite-temperature properties and large systems, quantum Monte Carlo (QMC) remains at the forefront of available numerical methods.

In this section, I give an introduction to the notion of (quantum) integrability, a feature shared by many low-dimensional models, including some spin chains and quantum field theories in the continuum in 1D, as well as classical statistical physics models in 2D. There are basically two levels of understanding, corresponding to coordinate Bethe Ansatz and algebraic Bethe Ansatz, that yield the exact thermodynamics and correlation functions respectively. Then, I consider noninteracting systems separately, as trivial examples of integrable systems. They are especially relevant in 1D due to an exact mapping between the Bose gas with infinitely strong repulsive interactions and a gas of noninteracting fermions. I also give a short introduction to the non-perturbative theory of Tomonaga-Luttinger liquids. It is an integrable effective field theory that yields the universal asymptotics of correlation functions of gapless models at large distances and low energies. To finish with, I present conformal field theory as another generic class of integrable models, providing a complementary viewpoint to the Tomonaga-Luttinger liquid theory, and put the emphasis on parallels between these formalisms.

\subsection{Quantum integrability and Bethe Ansatz techniques}

\textit{One can ask, what is good in 1\!+\!1-dimensional models, when our spacetime is 3\!+\!1-dimensional. There are several particular answers to this question.}

\textit{(a) The toy models in 1\!+\!1 dimension can teach us about the realistic field-theoretical models in a nonperturbative way. Indeed such phenomena as renormalisation, asymptotic freedom, dimensional transmutation (i.e. the appearance of mass via the regularisation parameters) hold in integrable models and can be described exactly.}

\textit{(b) There are numerous physical applications of the 1\!+\!1 dimensional models in condensed-matter physics.}

\textit{(c) [\dots] conformal field theory models are special massless limits of integrable models.}

\textit{(d) The theory of integrable models teaches us about new phenomena, which were not appreciated in the previous developments of Quantum Field Theory, especially in connection with the mass spectrum.}

\textit{(e) [\dots] working with the integrable models is a delightful pastime. They proved also to be very successful tool for educational purposes.}
\\
\\
Ludwig Fadeev
\\
\\
\\
\\
\\

Quantum field theory (QFT) is a generic denomination for theories based on the application of quantum mechanics to fields, and is a cornerstone of modern particle and condensed-matter physics. Such theories describe systems of several particles and possess a huge (often infinite) number of degrees of freedom. For this reason, in general they can not be treated exactly, but are amenable to perturbative methods, based on expansions in the coupling constant. Paradigmatic examples are provided by quantum electrodynamics, the relativistic quantum field theory of electrodynamics that describes how light and matter interact, where expansions are made in the fine structure constant, and quantum chromodynamics, the theory of strong interaction, a fundamental force describing the interactions between quarks and gluons, where high-energy asymptotics are obtained by expansions in the strong coupling constant.

One of the main challenges offered by QFT is the quest of exact, thus non-perturbative, methods, to circumvent the limitations of perturbation theory, such as difficulty to obtain high-order corrections (renormalization tools are needed beyond the lowest orders, and the number of processes to take into account increases dramatically) or to control approximations, restricting its validity range. In this respect, the concept of integrability turns out to be extremely powerful. If a model is integrable, then it is possible to calculate exactly quantities like the energy spectrum, the scattering matrix that relates initial and final states of a scattering process, or the partition function and critical exponents in the case of a statistical model.

The theoretical tool allowing to solve quantum integrable models is called Bethe Ansatz, that could be translated as 'Bethe's educated guess'. Its discovery coincides with the early days of quantum field theory, when Bethe found the exact eigenspectrum of the 1D Heisenberg model (the isotropic quantum spin-$1/2$ chain with nearest-neighbor interactions, a.k.a. the XXX spin chain), using an Ansatz for the wavefunction \cite{Bethe1931}. This solution, provided analytically in closed form, is highly impressive if one bears in mind that the Hamiltonian of the system is a $2^N\!\times\!2^N$ matrix, technically impossible to diagonalize by brute force for long chains. Bethe's breakthrough was followed by a multitude of exact solutions to other 1D models, especially flourishing in the 1960's. Most of them enter the three main categories: quantum 1D spin chains, low-dimensional QFTs in the continuum or on a lattice, and classical 2D statistical models.

The typical form for the Hamiltonian of spin chains with nearest-neighbor interactions is
\begin{eqnarray}
\label{HSC}
\hat{H}^{SC}=-\sum_{i=1}^N\left(J_x\hat{S}_i^x\hat{S}_{i+1}^x+J_y\hat{S}_i^y\hat{S}_{i+1}^y+J_z\hat{S}_i^z\hat{S}_{i+1}^z\right),
\end{eqnarray}
where the spin operators satisfy local commutations
\begin{eqnarray}
[\hat{S}_k^a,\hat{S}_l^b]=i\hbar \delta_{k,l}\epsilon_{a,b,c}\hat{S}_k^c,
\end{eqnarray}
with $\delta$ and $\epsilon$ the Kronecker and Levi-Civita symbols respectively ($\epsilon_{a,b,c}$ takes the value $0$ if there are repeated indices, $1$ if $(a,b,c)$ is obtained by an even permutation of $(1,2,3)$ and $-1$ if the permutation is odd).

In the case of a spin-$1/2$ chain, spin operators are usually represented by the Pauli matrices.
The XXX spin chain solved by Bethe corresponds to the special case where $J_x\!=\!J_y\!=\!J_z$ in Eq.~(\ref{HSC}), and the anisotropic XXZ model, solved later on by Yang and Yang \cite{YangYang1966I, YangYang1966II}, to $J_x\!=\!J_y$. A separate thread of development began with Onsager's solution of the two-dimensional, square-lattice Ising model \cite{Onsager1944}. Actually, this solution consists of a Jordan-Wigner transformation to convert Pauli matrices into fermionic operators, followed by a Bogoliubov rotation to diagonalize the quadratic form thereby obtained \cite{SchultzMattisLieb1964}. Similar techniques allow to diagonalize the XY spin chain Hamiltonian, where $J_z\!=\!0$ \cite{LiebSchultzMattis1961}.

As far as QFT models in the continuum are concerned, the most general Hamiltonian for spinless bosons interacting through a two-body potential is
\begin{eqnarray}
\hat{H}^{SB}=\sum_{i=1}^N\left[\frac{\hat{p}_i^2}{2m}+V_{ext}(\hat{x}_i)\right]+\sum_{\{i\neq j\}}V_{int}(\hat{x}_i-\hat{x}_j),
\end{eqnarray}
where $\hat{p}_i$ and $\hat{x}_i$ are the momentum and position operators, $V_{ext}$ is an external potential, while $V_{int}$ represents inter-particle interactions.
A few integrable cases have been given special names. The most famous ones are perhaps the Lieb-Liniger model \cite{LiebLiniger1963}, defined by 
\begin{eqnarray}
V^{LL}_{ext}(x)=0,\quad V^{LL}_{int}(x)=g_{\mathrm{1D}}\,\delta(x), 
\end{eqnarray}
with $\delta$ the dirac function and $g_{\mathrm{1D}}$ the interaction strength, and the Calogero-Moser model \cite{Calogero1969, Calogero1971}, associated to the problem of particles interacting pairwise through inverse cube forces ('centrifugal potential') in addition to linear forces ('harmonic potential'), i.e. such that 
\begin{eqnarray}
V^{CM}_{ext}(x)=\frac{1}{2}m\omega^2 x^2,\quad V^{CM}_{int}(x)=\frac{g}{x^2}.
\end{eqnarray}

The Lieb-Liniger model has been further investigated soon after by McGuire \cite{McGuire1964} and Berezin et al. \cite{Berezin1964}. Its spin-$1/2$ fermionic analog has been studied in terms of the number $M$ of spins flipped from the ferromagnetic state, in which they would all be aligned. The case $M\!=\!1$ was solved by McGuire \cite{McGuire1965}, $M\!=\!2$ by Flicker and Lieb \cite{Flicker1967}, and the arbitrary $M$ case by Gaudin \cite{Gaudin1967} and Yang \cite{Yang1967, Yang1968}, which is the reason why spin-$1/2$ fermions with contact interactions in 1D are known as the Yang-Gaudin model.
Higher-spin Fermi gases have been investigated by Sutherland \cite{Sutherland1968}.

The models presented so far in the continuum are Galilean-invariant, but Bethe Ansatz can be adapted to model with Lorentz symmetry as well, as shown by its use to treat certain relativistic field theories, such as the massive Thirring model \cite{Bergknoff1979}, and the equivalent quantum sine-Gordon model \cite{Coleman1975}, as well as the Gross-Neveu \cite{Andrei1979} (a toy-model for quantum chromodynamics) and SU(2)-Thirring models \cite{Belavin1979}. A recent study of the non-relativistic limit of such models shows the ubiquity of the Lieb-Liniger like models for non-relativistic particles with local interactions \cite{Bastianello2016, Bastianello2017}.

\begin{figure}
\includegraphics[width=9cm, keepaspectratio, angle=0]{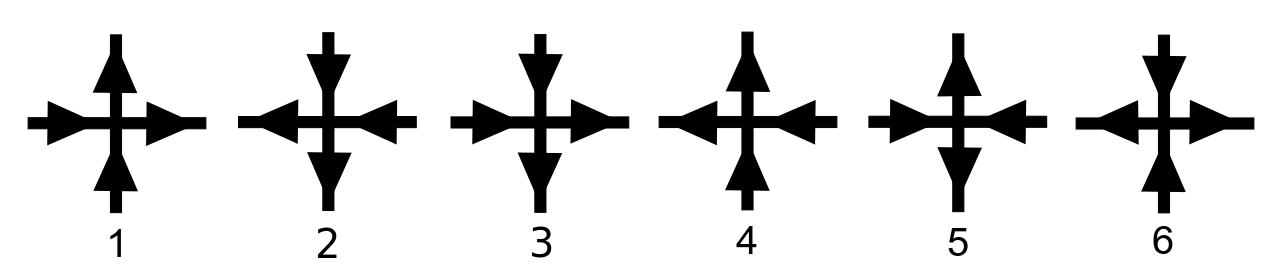}
\caption{The six configurations allowed by the ice rule in the 6-vertex model}
\label{icerule}
\end{figure}

The last category, i.e. classical statistical physics models in 2D, is essentially composed of classical 2D spin chains, and of ice-type models. When water freezes, each oxygen atom is surrounded by four hydrogen ions. Each of them is closer to one of its neighboring oxygens, but always in such a way that each oxygen has two hydrogens closer to it and two further away. This condition is known as the ice rule, and allows to model the system as a 2D square lattice, where each vertex hosts an oxygen atom and each bond between two vertices is depicted with an arrow, indicating to which of the two oxygens the hydrogen ion is closer, as illustrated in Fig.~\ref{icerule}.

Due to the ice rule, each vertex is surrounded by two arrows pointing towards it, and two away: this constraint limits the number of possible vertex configurations to six, thus the model is known as the 6-vertex model. Its solution has been obtained stepwise \cite{Lieb1967, CPYang1967}.
Baxter's solution of the 8-vertex model includes most of these results \cite{Baxter1971} and also solves the XYZ spin chain, that belongs to the first category.

The general approach introduced by Hans Bethe and refined in the many works cited above is known as coordinate Bethe Ansatz. It provides the excitation spectrum, as well as some elements of thermodynamics. The non-trivial fact that Bethe Ansatz provides solutions to both 1D quantum and 2D classical models is due to an exact, general mapping between dD quantum models at zero temperature and (d+1)D classical models with infinitely many sites, since the imaginary time $\tau$ in the path integral description of quantum systems plays the role of an extra dimension \cite{Suzuki1976}. This quantum-classical mapping implies that studying quantum fluctuations in 1D quantum systems amounts to studying thermal fluctuations in 2D classical ones, and is especially useful as it allows to solve quantum models with numerical methods designed for classical ones.

Computing the exact correlation functions of quantum-integrable models is a fundamental problem in order to enlarge their possibilities of application, and the next step towards solving them completely. Unfortunately, coordinate Bethe Ansatz does not provide a simple answer to this question, as the many-body wavefunction becomes extremely complicated when the number of particles increases, due to summations over all possible permutations.

The problem of the construction of correlation functions from integrability actually opened a new area in the field in the 1970's, based on algebraic Bethe Ansatz, that is essentially a second-quantized form of the coordinate one. A major step in the mathematical discussion of quantum integrability was the introduction of the quantum inverse scattering method (QISM) \cite{KorepinBogoliubovIzergin} by the Leningrad group of Fadeev \cite{Sklyanin1982}. Roughly, this method relies on a spectrum-generating algebra, i.e. operators that generate the eigenvectors of the Hamiltonian by successive action on a pseudo-vacuum state, and provides an algebraic framework for quantum-integrable models. Its development has been fostered by an advanced understanding of classical integrability (for an introduction to this topic, see e.g. \cite{Torrielli}), soliton theory (see e.g. \cite{Dauxois}), and a will to transpose them to quantum systems.

The original work of Gardner, Greene, Kruskal and Miura \cite{GardnerGreeneKruskalMiura1967} has shown that the initial value problem for the nonlinear Korteweg-de Vries equation of hydrodynamics (describing a wave profile in shallow water) can be reduced to a sequence of linear problems. The relationship between integrability, conservation laws, and soliton behavior was clearly exhibited by this technique. Subsequent works revealed that the inverse scattering method is applicable to a variety of non-linear equations, including the classical versions of the non-linear Schr\"odinger \cite{ZakharovShabat1972} and sine-Gordon \cite{Ablowitz1973} equations. The fact that the quantum non-linear Schr\"odinger equation could also be exactly solved by Bethe Ansatz suggested a deep connection between inverse scattering and Bethe Ansatz. This domain of research soon became an extraordinary arena of interactions between various branches of theoretical physics, and has strong links with several fields of mathematics as well, such as knot invariants \cite{Wu1992}, topology of low-dimensional manifolds, quantum groups \cite{Drinfeld1986} and non-commutative geometry.

I will only try and give a glimpse of this incredibly vast and complicated topic, without entering into technical details. To do so, following \cite{Wittenintegrability}, I will focus on integrable models that belong to the class of continuum quantum field theories in 1D.

\begin{figure}
\includegraphics[width=6cm, keepaspectratio, angle=0]{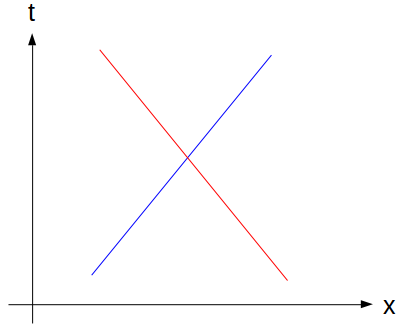}
\caption{A space-time picture of a two-body elastic scattering in 1D}
\label{Fig1}
\end{figure}

Figure \ref{Fig1} shows a spacetime diagram, where a particle of constant velocity is represented by a straight line. It shows the immediate vicinity of a collision process involving two particles. Due to energy and momentum conservation, after scattering, the outgoing particles go off at the same velocities as the incoming ones. In a typical relativistic quantum field theory (such theories are sometimes relevant to condensed matter), particle production processes may be allowed by these symmetries. In a $N\!=\!2\!\to\!N\!=\!3$ scattering event (where $N$ represents the number of particles), the incoming and outgoing lines can be assumed to all end or begin at a common point in spacetime. However, integrable models have extra conserved quantities that commute with the velocity, but move a particle in space by an amount that depends on its velocity.

Starting with a spacetime history in which the incoming and outgoing lines meet at a common point in space-time, a symmetry that moves the incoming and outgoing lines by a velocity-dependent amount creates an history in which the outgoing particles could have had no common origin in spacetime, leading to a contradiction. This means that particle production is not allowed in integrable models. By contrast, two-particle scattering events happen even in integrable systems, but are purely elastic, in the sense that the initial and final particles have the same masses. Otherwise, the initial and final velocities would be different, and considering a symmetry that moves particles in a velocity-dependent way would again lead to a contradiction. In other words, the nature of particles is also unchanged during scattering processes in integrable models.

\begin{figure}
\includegraphics[width=10cm, keepaspectratio, angle=0]{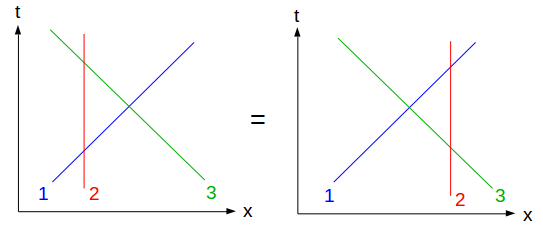}
\caption{There are two ways to express a $N\!=\!3\!\to\!N\!=\!3$ scattering event as a succession of three $N\!=\!2\!\to\!N\!=\!2$ scattering events. The scattering matrix of the full process is the product of the two-body scattering matrices, and both situations are equivalent, according to the Yang-Baxter equation (\ref{YBS}).}
\label{Fig2}
\end{figure}

The situation becomes more interesting when one considers three particles in both the initial and final state. Since we can be moved relative to each other, leaving their slopes fixed, the scattering process is only composed of pairwise collisions. There are two ways to do this, as shown in Fig.~\ref{Fig2}, and both must yield the same result. More formally, the equivalence of these pictures is encoded in the celebrated Yang-Baxter equation \cite{Baxter1971}, that schematically reads
\begin{eqnarray}
\label{YBS}
S(1,2,3)=S(1,2)S(1,3)S(2,3)=S(2,3)S(1,3)S(1,2),
\end{eqnarray}
in terms of scattering matrices, where $S(1,2,\dots)$ is the coefficient relating the final- and inital-state wavefunctions in the collision process involving particles $1$, $2\dots$

The Yang-Baxter equation (\ref{YBS}) guarantees that a multi-body scattering process can be factorized as a product of two-body scattering events, in other words, that scattering is not diffractive. Two-body reducible dynamics (i.e. absence of diffraction for models in the continuum) is a key point to quantum integrability, and may actually be the most appropriate definition of this concept \cite{CauxMossel2011}.




To sum up with, a $N$-particle model is quantum-integrable if the number and nature of particles are unchanged after a scattering event, i.e. if its $S$-matrix can be factorized into a product of $\binom{N}{2}$ two-body scattering matrices, and satisfies the Yang-Baxter equation (\ref{YBS}).

I proceed to consider the most trivial example of integrable model: a gas of noninterating particles, whose relevance in 1D stems from an exact mapping involving a strongly-interacting gas.

\subsection{Exact solution of the Tonks-Girardeau model and Bose-Fermi mapping}
In the introduction to this section devoted to analytical tools, I mentioned that a possible strategy to solve a strongly-interacting model is to try and transform it into a noninteracting problem. Actually, there is a case where such a transformation is exact, known as the Bose-Fermi mapping \cite{Girardeau1960}. It was put into light by Girardeau in the case of a one-dimensional gas of hard-core bosons, the so-called Tonks-Girardeau gas (prior to Girardeau, Lewi Tonks had studied the classical gas of hard spheres \cite{Tonks1936}). Hard-core bosons can not pass each other in 1D, and a fortiori can not exchange places. The resulting motion can be compared to a traffic jam, or rather to a set of 1D adjacent billiards whose sizes vary with time, containing one boson each.

The infinitely strong contact repulsion between the bosons imposes a constraint to the many-body wave function of the Tonks-Girardeau gas, that must vanish whenever two particles meet. As pointed out by Girardeau, this constraint can be implemented by writing the many-body wavefunction as follows:
\begin{eqnarray}
\label{BFmap}
\psi^{TG}(x_1,\dots,x_N)=A(x_1,\dots,x_N)\psi^F(x_1,\dots,x_N),
\end{eqnarray}
where
\begin{eqnarray}
\label{antisymm}
A(x_1,\dots,x_N)=\prod_{\{i>j\}}\mathrm{sign}(x_i\!-\!x_j),
\end{eqnarray}
where $\psi^F$ is the many-body wavefunction of a fictitious gas of noninteracting, spinless fermions. The antisymmetric function $A$ takes values in $\{-1,1\}$ and compensates the sign change of $\psi^F$ whenever two particles are exchanged, yielding a wavefunction that obeys Bose permutation symmetry, as expected. Furthermore, eigenstates of the Tonks-Girardeau Hamiltonian must satisfy the same Schr\"odinger equation as the ones of a noninteracting spinless Fermi gas when all coordinates are different. The ground-state wavefunction of the free Fermi gas is a Slater determinant of plane waves, leading to a Vandermonde determinant in 1D, hence the pair-product, Bijl-Jastrow form \cite{Girardeau1960}
\begin{eqnarray}
\psi^{TG}(x_1,\dots,x_N)=\sqrt{\frac{2^{N(N-1)}}{N!L^N}}\prod_{\{i>j\}}\left|\sin\!\left[\frac{\pi}{L}(x_i\!-\!x_j)\right]\right|.
\end{eqnarray}
This form is actually generic of various 1D models in the limit of infinitely strong repulsion, such as the Lieb-Liniger model \cite{LiebLiniger1963}.

The ground-state energy of the Tonks-Girardeau gas in the thermodynamic limit is then \cite{Girardeau1960}
\begin{eqnarray}
\label{energyTG}
E_0^{TG}\!=\!N\frac{(\pi\hbar n_0)^2}{6m}\!=\!E_0^F,
\end{eqnarray}
thus it coincides with the one of $N$ noninteracting spinless fermions, which is another important feature of the Bose-Fermi mapping. More generally, their thermodynamics are utterly equivalent. Even the excitation spectrum of the Tonks-Girardeau gas, i.e. the set of its excitations above the ground state, coincides with the one of a noninteracting spinless Fermi gas. The total momentum $P$ and energy $E$ of the model are given by $P=\hbar\sum_{j=1}^N k_j$ and $E=\frac{\hbar^2}{2m}\sum_{j=1}^N k_j^2$ respectively, where the set of quasi-momenta $\{k_j\}_{j=1,\dots,N}$ satisfies
\begin{eqnarray}
k_j=\frac{2\pi}{L}I_j.
\end{eqnarray}
The Bethe numbers $\{I_j\}_{j=1,\dots,N}$ are integer for odd values of $N$ and half-odd if $N$ is even. The quasi-momenta can be ordered in such a way that $k_1<k_2<\dots <k_N$, or equivalently $I_1<I_2<\dots<I_N$. The ground state corresponds to $I_j=-\frac{N+1}{2}+j$, its total momentum $P_{GS}=0$. I use the notations $p\!=\!P\!-\!P_{GS}$ and $\epsilon\!=\!E\!-\!E_{GS}$ to denote the total momentum and energy of an excitation with respect to the ground state, so that the excitation spectrum is given by $\epsilon(p)$. For symmetry reasons, I only consider excitations such that $p\geq 0$, those with $-p$ having the same energy.

The Tonks-Girardeau gas features two extremal excitation branches, traditionally called type I and type II. Type-I excitations occur when the highest-energy particle with $j\!=\!(N\!-\!1)/2$ gains a momentum $p_n\!=\!\hbar 2\pi n/L$ and an energy $\epsilon_n^I=\frac{\hbar^2\pi^2}{2mL^2}[(N-1+2n)^2-(N-1)^2]$. The corresponding continuous dispersion relation is \cite{Brand2004}
\begin{eqnarray}
\label{EETG}
 \epsilon^I(p)=\frac{1}{2m}\left[2p_F p\left(1-\frac{1}{N}\right)+p^2\right],
\end{eqnarray}
where $p_F\!=\!\pi \hbar N/L$ is the Fermi momentum.

\begin{figure}
\includegraphics[width=9cm, keepaspectratio, angle=0]{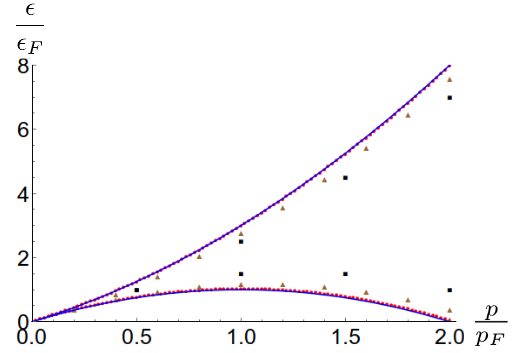}
\caption{Excitation energy of the Tonks-Girardeau gas in units of the Fermi energy in the thermodynamic limit, as a function of the excitation momentum in units of the Fermi momentum, for $N\!=\!4$ (black squares), $N\!=\!10$ (brown triangles) and $N\!=\!100$ hard-core bosons (red dots). The last case is quasi-indistinguishable from the thermodynamic limit (solid, blue).}
\label{Fig41}
\end{figure}

Type-II excitations correspond to the case where a particle inside the Fermi sphere is excited to occupy the lowest energy state available, carrying a momentum $p_n\!=\!2\pi \hbar n/L$. This type of excitation amounts to shifting all the quasi-momenta with $j\!>\!n$ by $2\pi\hbar/L$, thus leaving a hole in the Fermi sea. This corresponds to an excitation energy $\epsilon_n^{II}\!=\!\frac{\hbar^2\pi^2}{2mL^2}[(N+1)^2-(N+1-2n)^2]$, yielding the type-II excitation branch \cite{Brand2004}
\begin{eqnarray}
\epsilon^{II}(p)=\frac{1}{2m}\left[2p_F p\left(1+\frac{1}{N}\right)-p^2\right],
\end{eqnarray}
that acquires the symmetry $p \leftrightarrow 2p_F\!-\!p$ at large number of bosons. Any combination of one-particle and one-hole excitations is also possible, giving rise to intermediate excitation energies between $\epsilon^I(p)$ and $\epsilon^{II}(p)$, that form a continuum in the thermodynamic limit, known as the particle-hole continuum. Figure~\ref{Fig41} shows the type-I and type-II excitation spectra of the Tonks-Girardeau gas. Below $\epsilon^{II}$, excitations are kinematically forbidden, which is another peculiarity of dimension one.

The Bose-Fermi mapping offers a possibility of investigating exactly and relatively easily a peculiar point of the phase diagram of 1D models, and in particular of calculating even-order auto-correlation functions of the wavefunction. I illustrate this point on the example of the density-density correlation function of the Tonks-Girardeau gas, using the mapping onto noninteracting fermions in the form
\begin{eqnarray}
\psi^{TG}=|\psi^F|.
\end{eqnarray}
In particular,
\begin{eqnarray}
(n^{TG})^k=|\psi^{TG}|^{2k}=|\psi^F|^{2k}=(n^F)^k,
\end{eqnarray}
where $n$ is the density. As a consequence of Wick's theorem \cite{Wick}, the quantum-statistical average of the equal-time density correlations of a Tonks-Girardeau gas at zero temperature is
\begin{eqnarray}
\label{discretenn}
 \langle n(x)n(0)\rangle^{TG}=n_0^2+\frac{1}{L^2}\sum_{k,k'}e^{-i(k-k')x}\Theta(k_F-|k|)\Theta(|k'|-k_F),
\end{eqnarray}
where $x$ is the distance between the two probed points, $k$ and $k'$ are the quantized momenta, integer multiples of $2\pi/L$, $\Theta$ is the Heaviside step function, $n_0\!=\!N/L$ is the density of the homogeneous gas, and $k_F\!=\!\pi n_0$ the norm of the Fermi wavevector.

In the thermodynamic limit, Eq.~(\ref{discretenn}) transforms into
\begin{eqnarray}
\label{nTG}
 \frac{\langle n(x)n(0)\rangle^{TG}}{n_0^2}\!=\!1\!+\!\frac{1}{(2k_F)^2}\int_{-k_F}^{k_F}dk\, e^{-ikx}\left[\int_{-\infty}^{-k_F}+\int_{k_F}^{+\infty}\right]dk'\, e^{ik'x}\!=1\!-\!\frac{\sin^2(k_F x)}{(k_F x)^2}.
\end{eqnarray}
This quantity represents the probability of observing simultaneously two atoms separated by a distance $x$. The fact that it vanishes at $x\!=\!0$ is a consequence of Pauli's principle, known as the Pauli hole, and the oscillating structure is typical of Friedel oscillations.

Actually, one can even go a step further and treat time-dependent correlations, since the Bose-Fermi mapping remains exact even in the time-dependent problem \cite{YukalovGirardeau2005, MinguzziGangardt2005}. It yields
\begin{eqnarray}
\langle n(x,t)n(0,0)\rangle^{TG}=n_0^2+\frac{1}{L^2}\sum_{k,k'}e^{-i(k-k')x}e^{i\frac{\hbar}{2m}t(k^2-k'^2)}\Theta(k_F-|k|)\Theta(|k'|-k_F),
\end{eqnarray}
and in the thermodynamic limit I obtain
\begin{eqnarray}
&&\frac{\langle n(x,t)n(0,0)\rangle^{TG}}{n_0^2}\!=\nonumber\\
&&1\!+\!\frac{1}{4k_F^2}\!\int_{-k_F}^{k_F}\!\!\!dk\, e^{i\left(\frac{\hbar k^2 t}{2m}-kx\right)}\!\left[\int_{-\infty}^{+\infty}\!\!\!dk'e^{-i\left(\frac{\hbar k'^2t}{2m}-k'x \right)}\!-\!\int_{-k_F}^{k_F}\!\!\!dk'e^{-i\left(\frac{\hbar k'^2t}{2m}-k'x\right)}\!\right]\!\!.
\end{eqnarray}
To evaluate it, I define
\begin{eqnarray}
I(x,t)\!=\!\int_{-\infty}^{+\infty}dk\, e^{-i\left(\frac{\hbar^2k^2}{2m}t-kx\right)},\,J(x,t)=\int_{-k_F}^{k_F}dk\, e^{i\left(\frac{\hbar^2k^2}{2m}t-kx\right)}.
\end{eqnarray}
Then, doing natural changes of variables and using the property
\begin{eqnarray}
\int_0^{+\infty}dx \sin(x^2)=\int_0^{+\infty}dx\cos(x^2)=\frac{1}{2}\sqrt{\frac{\pi}{2}},
\end{eqnarray}
I find
\begin{eqnarray}
I(x,t)=e^{\frac{imx^2}{2\hbar t}}\sqrt{\frac{2m\pi}{\hbar t}}e^{-i\pi/4}.
\end{eqnarray}
This term represents a decaying wave packet, and is equal to $2\pi$ times the propagator of free fermions.

The total correlation function can be split into two parts, one 'regular' and real-valued, the other complex and associated to the wave packet, such that
\begin{eqnarray}
\label{wpreg}
\langle n(x,t)n(0,0)\rangle^{TG}=\langle n(x,t)n(0,0)\rangle^{TG}_{reg}+\langle n(x,t)n(0,0)\rangle^{TG}_{wp}.
\end{eqnarray}
Then, defining the Fresnel integrals as
\begin{eqnarray}
S(x)=\frac{2}{\sqrt{2\pi}}\int_0^x dt\sin(t^2), \,C(x)=\frac{2}{\sqrt{2\pi}}\int_0^x dt\cos(t^2),
\end{eqnarray}
focusing on the regular part I find
\begin{eqnarray}
\label{predTG}
&&\!\!\!\!\frac{\langle n(x,t)n(0,0)\rangle^{TG}_{reg}}{n_0^2}=1-\frac{1}{4k_F^2}\left|J(x,t)\right|^2\nonumber\\
&&\!\!\!\!\!=\!1\!-\!\frac{\pi}{8}\frac{1}{\omega_F t}\!\left\{C\!\left[\sqrt{\!\frac{m}{2 \hbar t}}(x\!+\!v_F t)\right]\!-\!C\!\left[\sqrt{\!\frac{m}{2 \hbar t}}(x\!-\!v_F t)\right]\right\}^2\nonumber\\
&&\,\,\,\,-\frac{\pi}{8}\frac{1}{\omega_F t}\!\left\{S\!\left[\sqrt{\!\frac{m}{2 \hbar t}}(x\!+\!v_F t)\right]\!-\!S\!\left[\sqrt{\!\frac{m}{2 \hbar t}}(x\!-\!v_F t)\right]\right\}^2
\end{eqnarray}
where $v_f\!=\!\frac{\hbar k_F}{m}$ is the Fermi velocity, and $\omega_F\!=\!\frac{\hbar k_F^2}{2m}$.

The Tonks-Girardeau case will serve as a comparison point several times in the following, as the only case where an exact closed-form solution is available. One should keep in mind, however, that any observable involving the phase of the wavefunction, although it remains considerably less involved than the general finitely-interacting case, is not as easy to obtain.

Another advantage of the Bose-Fermi mapping is that it holds even in the presence of any kind of trapping, in particular if the hard-core bosons are harmonically trapped, a situation that will be encountered below, in chapters \ref{secIII} and \ref{secV}. It has also been extended to spinor fermions \cite{Girardeau2011} and Bose-Fermi mixtures \cite{GirardeauMinguzzi2007}, and another generalization maps interacting bosons with a two-body $\delta$-function interaction onto interacting fermions, except that the roles of strong and weak couplings are reversed \cite{CheonShigehara1999}. This theorem is peculiarly important, in the sense that any result obtained for bosons is also valid for fermions. An extension to anyons has also been considered \cite{Girardeau2006}.

To sum up with, the hard-core Bose gas is known as the Tonks-Girardeau gas, and according to the Bose-Fermi mapping, it is partially equivalent to the noninteracting spinless Fermi gas, in the sense that their ground-state wavefunctions differ only by a multiplicative function that assumes two values, $\pm 1$. Their energies, excitation spectra and density correlation functions are identical as well.

Along with this exact mapping, another technique exists, where the mapping from an interacting to a noninteracting problem is only approximate, and is called bosonization. I proceed to study its application to interacting fermions and bosons in 1D, yielding the formalism of Tomonaga-Luttinger liquids.

\subsection{Bosonization and Tomonaga-Luttinger liquids}

The first attempts to solve many-body, strongly-correlated problems in one dimension have focused on fermions. It turns out that a non-perturbative solution can be obtained by summing an infinite number of diverging Feynman diagrams, that correspond to particle-hole and particle-particle scattering \cite{DzyaloshinskiiLarkin1973}, in the so-called Parquet approximation. This tour de force, supplemented by renormalization group techniques \cite{Shankar1994}, is known as the Dzyaloshinskii-Larkin solution.

There is, actually, a much simpler approach to this problem. It is based on a procedure called bosonization, introduced independently in condensed-matter physics \cite{LutherPeschel1974} and particle physics \cite{Mandelstam1975} in the 1970's. In a nutshell, bosonization consists in a reformulation of the Hamiltonian in a more convenient basis involving free bosonic operators (hence the name of the method), that keeps a completely equivalent physical content. To understand the utility of bosonization, one should bear in mind that interaction terms in the fermionic problem are difficult to treat as they involve four fermionic operators. The product of two fermions being a boson, it seems interesting to expand fermions on a bosonic basis to obtain a quadratic, and thus diagonalizable, Hamiltonian.

The main reason for bosonization's popularity is that some problems that look intractable when formulated in terms of fermions become easy, and sometimes even trivial, when formulated in terms of bosonic fields. The importance and depth of this change of viewpoint is such, that it has been compared to the Copernician revolution \cite{GogolinNersesyanTsvelik}, and to date bosonization remains one of the most powerful non-perturbative approches to many-body quantum systems.

Contrary to the exact methods discussed above, bosonization is only an effective theory, but has non-negligible advantages as a bosonized model is often far easier to solve than the original one when the latter is integrable. Moreover, the bosonization technique yields valuable complementary information to Bethe Ansatz, about its universal features (i.e., those that do not depend on microscopic details), and allows to describe a wide class of non-integrable models as well.

Tomonaga was the first to identify boson-like behavior of certain elementary excitations in a 1D theory of interacting fermions \cite{Tomonaga1950}. A precise definition of these bosonic excitations in terms of bare fermions was given by Mattis and Lieb \cite{MattisLieb1965}, who took the first step towards the full solution of a model of interacting 1D fermions proposed by Luttinger \cite{Luttinger1963}. The completion of this formalism was realized later on by Haldane \cite{Haldane1981}, who coined the expression 'Luttinger liquid' to describe the model introduced by Luttinger, exactly solved by bosonization thanks to its linear dispersion relation. 
 
Actually, 1D systems with a Luttinger liquid structure range from interacting spinless fermions to fermions with spin, and interacting Bose fluids. Condensed-matter experiments have proved its wide range of applicability, from Bechgaard salts \cite{Schwartz1998} to atomic chains and spin ladders \cite{Klanjsek2008, Bouillot2011, Jeong2016}, edge states in the fractional quantum Hall effect \cite{Wen90, Milliken1996}, carbon \cite{Bockrath1999} and metallic \cite{Gao2004} nanotubes, nanowires \cite{Levy2012}, organic conductors \cite{Dardel1993, Lebed2008} and magnetic insulators \cite{Lake2005}.

The harmonic fluid approach to bosonic Luttinger liquids, following Cazalilla \cite{Cazalilla2004}, is adapted to any statistics at the price of tiny modifications in the fermionic case. This approach operates a change of viewpoint compared to the historical development, as it \textit{defines} a Tomonaga-Luttinger liquid as a 1D model described by the generic Hamiltonian
\begin{equation}
\label{HTLL}
H^{TL}=\frac{\hbar v_s}{2\pi}\int_0^L dx \left[K\left(\partial_x \Phi\right)^2+\frac{1}{K}\left(\partial_x\theta\right)^2\right],
\end{equation}
where $\theta$ and $\Phi$ are scalar fields that satisfy the commutation relation
\begin{eqnarray}
[\partial_x\theta(x),\Phi(x')]=i\pi \delta(x\!-\!x').
\end{eqnarray}
Note that this commutation relation is anomalous, as it involves a partial derivative of one of the fields. The motivation behind this definition is that the Tomonaga-Luttinger Hamiltonian is the simplest that can be obtained by expanding an interaction energy in the deviations from constant density and zero current. The cross-term $\partial_x\theta\partial_x\phi$ does not arise, as it could be removed by Galilean transformation. Usually, the fields $\partial_x\theta$ and $\Phi$ correspond to local density fluctuations around the mean value,
\begin{eqnarray}
\frac{1}{\pi}\partial_x\theta(x)\!=\!n(x)\!-\!n_0,
\end{eqnarray}
and to phase respectively. The positive coefficients $K$ and $v_s$ in Eq.~(\ref{HTLL}) are phenomenological, model-dependent parameters. To be more specific, $K$ is a dimensionless stiffness and $v_s$ represents the sound velocity.

Qualitatively, two limiting regimes are expected. If $K$ is large, density fluctuations are important and phase fluctuations are reduced. It corresponds to a classical regime that looks like a BEC phase, but can not be so due to the impossibility of symmetry breaking \cite{MerminWagner}.
If $K$ is small, the system looks like a crystal to some extent. Note also the $\Phi\leftrightarrow\theta$ and $K\leftrightarrow 1/K$ duality, suggesting that the value $K\!=\!1$ has a special meaning. Actually, it corresponds to noninteracting fermions, as will be shown below.

Since the Tomonaga-Luttinger Hamiltonian (\ref{HTLL}) is a bilinear form, it can be diagonalized. A convenient basis is provided by bosonic creation and annihilation operators with standard commutation relations, $[b_q,b_{q'}^{\dagger}]=\delta_{q,q'}$. Neglecting topological terms that are crucial at the mesoscopic scale but irrelevant in the thermodynamic limit \cite{Cazalilla2004} and using periodic boundary conditions, the original fields are expressed in this basis as
\begin{equation}
\label{thetastatic}
\theta(x)=\frac{1}{2}\sum_{q\neq 0}\left|\frac{2\pi K}{qL}\right|^{1/2}(e^{iqx}b_q+e^{-iqx}b_q^{\dagger})
\end{equation}
and
\begin{equation}
\Phi(x)=\frac{1}{2}\sum_{q\neq 0}\left|\frac{2 \pi}{qLK}\right|^{1/2}\mathrm{sign}(q)(e^{iqx}b_q+e^{-iqx}b_q^{\dagger}).
\end{equation}

Inserting these normal mode expansions into Eq.~(\ref{HTLL}) and using the bosonic commutation relations yields the diagonalized form of the Hamiltonian,
\begin{equation}
\label{diagH}
H^{TL}=\sum_{q\neq 0} \hbar \omega(q)b_q^\dagger b_q,
\end{equation}
where
\begin{eqnarray}
\label{lindesp}
\omega(q)=v_s|q|
\end{eqnarray}
is a sound-like spectrum associated to density waves, justifying a posteriori the notation $v_s$. Equations (\ref{diagH}) and (\ref{lindesp}) describe gapless, linear-dispersing (i.e.~collective phonon-like) excitations, sealing the absence of individual (i.e.~particle-like) excitation in the low-energy spectrum, and the breakdown of the Fermi liquid picture in 1D. Actually, Eqs.~(\ref{diagH}) and (\ref{lindesp}) could as well serve as definition of a homogeneous Tomonaga-Luttinger liquid. Then, it is obvious that Luttinger liquid theory is based on linearization of the dispersion relation. This is indeed the procedure used to construct the field theory from a given microscopic model.

Another striking point in Eq.~(\ref{lindesp}) is that it does not explictly depend on the parameter $K$, suggesting that $v_s$ and $K$ are linked together. This property can be shown using the density-phase approach. Writing the wavefunction as
\begin{eqnarray}
\label{densityphase}
\psi(x)=\sqrt{n(x)}e^{i\Phi(x)},
\end{eqnarray}
and inserting this identity into the kinetic part of the microscopic Hamiltonian, that reads
\begin{eqnarray}
H_{kin}=\frac{\hbar^2}{2m} \int dx\,\partial_x\psi^{\dagger}\partial_x\psi,
\end{eqnarray}
yields by identification with Eq.~(\ref{HTLL}) the relation 
\begin{eqnarray}
\label{linkGal}
v_sK=v_F,
\end{eqnarray}
as derivatives of the operator $\psi$ are absent from the interaction term. Equation (\ref{linkGal}) is valid for translation-invariant models. The Luttinger parameters are also linked together by the relation, valid for any model,
\begin{eqnarray}
\frac{K}{v_s}=\pi n_0^2\kappa,
\end{eqnarray}
where
\begin{eqnarray}
\kappa=-\frac{1}{L}\left(\frac{\partial L}{\partial P}\right)_N
\end{eqnarray}
is the compressibility, accessible to experiments.
The density-phase picture of Eq.~(\ref{densityphase}), pioneered by Schotte and Schotte \cite{SchotteSchotte1969}, is another key ingredient of bosonization. In principle, it allows to justify the form of Eq.~(\ref{HTLL}) for quantum field theories in the continuum, starting from the microscopic level. 

To illustrate the predictive power of the Tomonaga-Luttinger liquid formalism on a concrete example, I proceed to study its density correlations.
It is well-known (I refer to Appendix \ref{dernn} for a detailed derivation) that the density-density correlations of a Tomonaga-Luttinger liquid have the following structure in the thermodynamic limit \cite{EfetovLarkin1976, Haldane1981bis}:
\begin{eqnarray}
\label{nnTLL}
 \frac{\langle n(x)n(0)\rangle^{TL}}{n_0^2}=1-\frac{K}{2}\frac{1}{(k_Fx)^2}+\sum_{m=1}^{+\infty}A_m(K)\frac{\cos(2mk_F x)}{(k_Fx)^{2Km^2}},
\end{eqnarray}
where $\{A_m(K)\}_{m>0}$ is an infinite set of model-dependent functions of $K$ called form factors. Equation (\ref{nnTLL}) is one of the greatest successes of the Tomonaga-Luttinger liquid theory, as it explicitly yields the large-distance structure of a non-local correlation function, that would otherwise be difficult to obtain by Bethe Ansatz techniques.

Since Eq.~(\ref{HTLL}) is an effective Hamiltonian, its validity range is not clear when it is used to describe a given model, such as the Lieb-Liniger model. A good starting point to investigate this crucial issue is to check whether Eq.~(\ref{nnTLL}) is compatible with the known exact result in the Tonks-Girardeau regime. Although Eq.~(\ref{nnTLL}) looks far more complicated than Eq.~(\ref{nTG}), setting $K\!=\!1$, $A_1(1)\!=\!\frac{1}{2}$, and $A_{m}(1)\!=\!0$, $\forall m\!>\!1$, yields
\begin{eqnarray}
 \frac{\langle n(x)n(0)\rangle^{TL}_{K=1}}{n_0^2}\!=\!1\!-\!\frac{1}{k_F^2x^2}\left[\frac{1\!-\!\cos(2k_Fx)}{2}\right]=1-\frac{\sin^2(k_F x)}{(k_F x)^2}= \frac{\langle n(x)n(0)\rangle^{TG}}{n_0^2}.
\end{eqnarray}
Therefore, the Tomonaga-Luttinger liquid theory is able to reproduce the exact static density correlations of a Tonks-Girardeau gas, or equivalently a gas of noninteracting fermions, albeit at the price of fine-tuning an infinite set of coefficients.
Note that one could have found the value of the Luttinger parameter $K$ associated to noninteracting fermions by deriving the Luttinger Hamiltonian for the latter. Another possibility is to use Eq.~(\ref{linkGal}) and bear in mind that $v_s\!=\!v_F$, which is even more straightforward. None of these approaches, however, yields the values of $\{A_m\}$.

This fine-tuning is assuredly a considerable shortcoming, unless it is possible to find the whole, infinite set of unknown coefficients in non-trivial cases (i.e. at finite interaction strength), by a systematic procedure that does not require thorough knowledge of the exact solution. Fortunately, the large-distance decay of the power law contributions in Eq.~(\ref{nnTLL}) becomes faster with increasing order $m$, and coefficients $\{A_m(K)\}_{m>1}$ are known to be negligible compared to $A_1(K)$ in the thermodynamic limit for the Lieb-Liniger model with repulsive interactions \cite{Shashi2011, Shashi2012}. Thus, only two coefficients are needed in practice to describe the statics at large distances: the Luttinger parameter $K$, and the first form factor $A_1(K)$.

The explicit expression of $K$ in the effective Hamiltonian in terms of the microscopic parameters of the model it is derived from is sometimes found constructively, e.g. for noninteracting fermions \cite{Haldane1981} or interacting fermions in the g-ology context \cite{Solyom1979}, where Eq.~(\ref{HTLL}) can be obtained from a more fundamental analysis, starting from the microscopic Hamiltonian. It has also been derived in peculiar from the hydrodynamic Hamiltonian of a one-dimensional liquid in the weakly-interacting case \cite{LandauLifschitzVol9, Bovo2016}. In most other contexts, a constructive derivation is lacking, but is not required to make quantitative predictions, as long as the two necessary parameters can be obtained from outside considerations, stemming from Bethe Ansatz, DMRG or experiments. As an example, for the Lieb-Liniger model, $K$ can be extracted by coordinate Bethe Ansatz using thermodynamic relations \cite{LiebLiniger1963}. For this model, it varies between $K\!=\!1$ in the infinitely-interacting regime and $K\!\to\!+\infty$ for vanishing interactions. The form factor $A_1(K)$ has been obtained in the repulsive regime, based on algebraic Bethe Ansatz \cite{Shashi2011, Shashi2012}.

I have shown above that the Tomonaga-Luttinger liquid formalism reproduces exactly the \textit{static} density correlations of the Tonks-Girardeau gas. However, for many purposes, one may be interested in time-dependent correlations as well. Time dependence is taken into account in the Schr\"odinger picture by $A(x,t)=e^{iHt}A(x)e^{-iHt}$, where $A$ is any observable and $H$ is the Hamiltonian. Using the equations of motion or the Baker-Campbell-Haussdorff lemma, from Eqs.~(\ref{thetastatic}), (\ref{diagH}) and (\ref{lindesp}), one finds
\begin{equation}
\theta(x,t)\!=\!\frac{1}{2}\sum_{q\neq 0}\left|\frac{2\pi K}{qL}\right|^{1/2}\left(e^{i[qx-\omega(q)t]}b_q+e^{-i[qx-\omega(q)t]}b_q^{\dagger}\right),
\label{thet}
\end{equation}
and after some algebra (details can be found again in Appendix \ref{dernn}),
\begin{equation}
\label{nn}
\!\!\!\frac{\langle n(x,t) n(0,0)\rangle^{TL}}{n_0^2}\!=\!1\!-\!\frac{K}{4k_F^2}\!\left[\frac{1}{(x\!-\!v_s t)^2}\!+\!\frac{1}{(x\!+\!v_s t)^2}\right]\!+\!\!\sum_{m=1}^{+\infty}\!A_m(K)\frac{\cos(2mk_F x)}{k_F^{2Km^2}(x^2\!-\!v_s^2 t^2)^{Km^2}}.
\end{equation}
When truncated to any finite order, Eq.~(\ref{nn}) is divergent on the mass shell, defined by $x\!=\!\pm v_s t$, and is usually regularized as $x\!=\!\pm(v_s t-i\epsilon)$ where $\epsilon$ is a short-distance cut-off, that mimics a lattice regularization. Sometimes, the term light-cone is also used instead of mass shell, in analogy with special relativity. Indeed, the bosons describing the dispersion are massless, since they verify the relativistic dispersion $\epsilon(p)\!=\!c\sqrt{M^2c^2\!+\!p^2}$, where $\epsilon(q)\!=\!\hbar \omega(q)$, with a mass term $M\!=\!0$, and $v_s$ plays the same role as $c$, the speed of light.

In the Tonks-Girardeau regime, corresponding to $K\!=\!1$, the whole set of coefficients $\{A_m\}_{m\geq 1}$ has already been obtained from the static treatment. Equation (\ref{nn}) then reduces to
\begin{eqnarray}
\label{nnfft}
\frac{\langle n(x,t) n(0,0)\rangle^{TL}_{K=1}}{n_0^2}=1-\frac{1}{4k_F^2}\left[\frac{1}{(x-v_F t)^2}+\frac{1}{(x+v_F t)^2}\right]+ \frac{1}{2}\frac{\cos(2k_F x)}{k_F^{2}(x^2-v_F^2 t^2)}.
\end{eqnarray}
A natural question is to what extent this expression captures the exact dynamics of the Tonks-Girardeau gas. Since Eq.~(\ref{predTG}) involves special functions, it is not obvious whether it is consistent with Eq.~(\ref{nnfft}). However, considering a specific point far from the mass shell, i.e. such that $k_F|x\pm v_F t|\gg 1$, using the expansions $S(z)\simeq_{z\gg 1}\frac{1}{2}-\frac{1}{\sqrt{2\pi}z}\cos(z^2)$
and $C(z)\simeq_{z\gg 1}\frac{1}{2}+\frac{1}{\sqrt{2\pi}z}\sin(z^2)$, keeping only the lowest orders in $|x\pm v_Ft|$, I finally recover Eq.~(\ref{nnfft}). Had I not previously used the Galilean invariance argument, the condition $v_s\!=\!v_F$ would have been imposed at this stage for consistency.

The conclusion of this first-order study is that the Tomonaga-Luttinger liquid theory captures the short-time, long-distance dynamics of the Tonks-Girardeau gas, i.e. far from the 'light-cone', except for the contribution of the decaying wave packet, that vanishes at long times. A quantitative validity criterion is thus $k_F|x| \gg \omega_F|t| \gg 1$, and in a relativistic language Eq.~(\ref{nnfft}) holds deep in the space-like region. Recalling the splitting of the Tonks-Girardeau correlation function into regular and wavepacket part in Eq.~(\ref{wpreg}), I find that the Tomonaga-Luttinger liquid fails to describe the regular part of the time-dependent density-density correlations of the Tonks-Girardeau gas at larger time scales, as can be seen at next order already. Using expansions of the Fresnel integrals $S$ and $C$ to higher orders around $t\!=\!0$ in Eq.~(\ref{predTG}), keeping terms in the light-cone variables $u\!=\!x\!-\!v_Ft$ and $v\!=\!x\!+\!v_Ft$ to same order, I find
\begin{eqnarray}
\label{more}
&&\frac{\langle n(x,t)n(0,0)\rangle^{TG}_{reg}}{n_0^2}=\frac{\langle n(x,t)n(0,0)\rangle^{TL,TG}}{n_0^2}+\frac{\omega_Ft}{k_F^4}\frac{\sin(2k_Fx)}{x^2-v_F^2t^2}\left[\frac{1}{v^2}-\frac{1}{u^2}\right]\nonumber\\
&&+\frac{2(\omega_Ft)^2}{k_F^6}\!\left\{\frac{5}{2}\left[\frac{1}{v^6}+\frac{1}{u^6}\right]\!+\frac{\cos(2k_Fx)}{(x^2-v_F^2t^2)^3}-\frac{3\cos(2k_Fx)}{x^2-v_F^2t^2}\left[\frac{1}{v^4}+\frac{1}{u^4}\right]\right\}\nonumber\\
&&+\frac{12(\omega_Ft)^3}{k_F^8}\left\{-\frac{5\sin(2k_Fx)}{x^2-v_F^2t^2}\left[\frac{1}{v^6}-\frac{1}{u^6}\right]+\frac{\sin(2k_Fx)}{(x^2-v_F^2t^2)^3}\left[\frac{1}{v^2}-\frac{1}{u^2}\right]\right\}+\dots
\end{eqnarray}
I have checked that Eq.~(\ref{more}) is equivalent to the series expansion obtained in \cite{Korepin1990}. The new terms in the density-density correlations compared to the first-order expansion described by the Tomonaga-Luttinger theory are all proportional to a power of $\omega_Ft$ and, as such, vanish at equal times, as expected. None of them is reproduced by the effective field theory.

To obtain better agreement with the exact expansion, a generalized effective theory should predict higher-order terms as well. To do so, it would be natural to include non-linearities in the Tomonaga-Luttinger Hamiltonian, that correspond to curvature of the dispersion relation. Note, however, that the expression Eq.~(\ref{more}) diverges at all orders on the mass shell, except when they are all resummed, plaguing this approach at the perturbative level in the vicinity of the mass shell \cite{Aristov2007}. Similar conclusions have been drawn from the comparison of the Tomonaga-Luttinger liquid theory and exact Tonks-Girardeau results, focusing on other observables, such as the Green function \cite{Pereira2012}.

The Tomonaga-Luttinger result Eq.~(\ref{nnfft}) also misses the part of the exact density-density correlation function associated to the wave packet, whose expansion reads
\begin{eqnarray}
&&\frac{\langle n(x,t)n(0,0)\rangle^{TG}_{wp}}{n_0^2}\nonumber\\
&&\!=\!\frac{\pi}{4}\frac{e^{-i\pi/4}}{\sqrt{2}}\frac{1}{\omega_Ft}\left\{C\left(\!\sqrt{\frac{m}{2\hbar t}}v\right)\!-\!C\left(\!\sqrt{\frac{m}{2\hbar t}}u\right)\!+\!i\!\left[S\left(\!\sqrt{\frac{m}{2\hbar t}}v\right)\!-\!S\left(\!\sqrt{\frac{m}{2\hbar t}}u\right)\right]\right\}\nonumber\\
&&\simeq i\frac{\sqrt{\pi}}{4}e^{-i\pi/4}\frac{1}{\sqrt{\omega_Ft}}e^{-i\frac{(k_Fx)^2}{4\omega_Ft}}\left[\frac{e^{i(k_Fx-\omega_Ft)}}{k_Fu}-\frac{e^{-i(k_Fx+\omega_Ft)}}{k_Fv}\right]+\dots
\end{eqnarray}
It has been shown in \cite{Kozlowski2011} that in the general case (i.e. for arbitrary interaction strengths in the microscopic model), the wave packet term coincides with the saddle point.

The unpleasant conclusion is that, as far as dynamics of density-density correlations is concerned, the standard Tomonaga-Luttinger liquid approach presented here misses on the one hand an infinite number of regular terms, and on the other hand the wave packet term. Thus, it is not adapted to investigate long-time dynamics.

To sum up with, in this section I have presented the Tomonaga-Luttinger Hamiltonian, its diagonalization via the bosonization procedure, and given the structure of its density-density correlation function. Even at zero temperature, correlators decay as power laws, that indicate the absence of a characteristic length scale.
The tendency towards certain ordering is defined by the most weakly-decaying correlation and this, in turn, is determined by the sole Luttinger parameter $K$, renormalized by interactions as would be the case for a Fermi liquid. However, the Tomonaga-Luttinger liquid is a paradigmatic example of a non-Fermi liquid \cite{Schofield1999}, and unlike the latter applies to bosons and insulating magnetic materials as well.

The main conundrums of bosonization are the built-in ultraviolet cut-off, that calls for external form-factor calculations, and its limitation to low energy due to the linear spectrum assumption. These points will be investigated in chapter \ref{secIV}, partly devoted to the dynamical correlations of Tomonaga-Luttinger liquids in momentum-energy space.

To circumvent limitation to low energies, an intuitive approach would be to try and include terms describing curvature of the dispersion relation. However, upon closer inspection, such terms would break Lorentzian invariance and doom this technique at the perturbative level. The extended Tomonaga-Luttinger model that has emerged as the mainstream paradigm in the first decade of the twenty-first century is the Imambekov-Glazman formalism of 'beyond Luttinger liquids' (see \cite{ImambekovSchmidtGlazman2012} for a review), that is based on a multiband structure and an impurity formalism, instead of including curvature. In particle physics, bosonization has also been extended to new formalisms where the bosons of the new basis are interacting, and non-abelian bosonization has been developed \cite{Witten1984}.

Another major problem of the Tomonaga-Luttinger liquid theory in its standard form is that proving the validity of the bosonization formalism in explicit detail and ironing out its subtleties is considerably harder than merely applying it. This issue has been widely ignored and may look even more obsolete nowadays regarding the success of the Imambekov-Glazman paradigm, but has not been investigated deeply enough, in my opinion. In the next chapters, I shall come back to this issue regularly and try and fill a few gaps in the previous literature.

The lack of obvious generalization to higher dimensions is also often deplored. Despite numerous efforts and reflexions in this direction \cite{Wen1990, Bartosch1999}, a general construction of an efficient Tomonaga-Luttinger liquid formalism in higher dimensions is still lacking. I shall come back to this issue in chapter \ref{secV}, where I construct a higher-dimensional Tomonaga-Luttinger model in a peculiar case.

\subsection{Conformal Field Theory}

To conclude this section on theoretical tools, I give the basics of conformal field theory (CFT). This research topic is extremely wide and active, so I will not attempt to introduce all of its (even elementary) aspects, but rather select those, that are useful to my purpose, i.e. essentially the ones linked to finite-size and temperature thermodynamics and correlation functions.

My first motivation is that conformal invariant systems are a subclass of integrable models, the second is that CFT provides an alternative to bosonization when it comes to evaluate finite-size and finite-temperature effects on correlation functions. Conformal field theory has also become essential in its role of a complementary tool to numerical methods, as it enables to extrapolate results obtained at finite particle number (typically from exact diagonalization) to the thermodynamic limit.

As a starting point, let me introduce the notion of conformal transformation. Since it has a geometric nature, the tensorial approach to differential geometry, also used in general relativity, provides compact notations for a general discussion \cite{FrancescoMathieuSenechal}. In arbitrary dimension, the space-time interval is written in terms of the metric tensor $g_{\mu \nu}$ as
\begin{eqnarray}
ds^2=g_{\mu \nu}dx^{\mu}dx^{\nu},
\end{eqnarray}
where I use Einstein's convention that a pair of identical covariant and contravariant indices represents a summation over all values of the index. The metric tensor is assumed to be symmetric, $g_{\mu \nu}\!=\!g_{\nu \mu}$, and non-degenerate, det$(g_{\mu \nu})\!\neq\!0$, thus the pointwise metric tensor has an inverse, $g^{\mu \nu}(x)$, such that
\begin{eqnarray}
g^{\mu \nu}(x)g_{\nu \lambda}(x)=\delta^{\mu}_{\lambda},
\end{eqnarray}
where $\delta^{\mu}_{\lambda}$ represents the identity tensor. A coordinate transformation $x\to x'$ yields a covariant transformation of the metric tensor,
\begin{eqnarray}
g'_{\mu\nu}(x')=\frac{\partial x^{\alpha}}{\partial x'^{\mu}}\frac{\partial x^{\beta}}{\partial x'^{\nu}}g_{\alpha\beta}(x).
\end{eqnarray}
An infinitesimal transformation of the coordinates $x^{\mu}\to x'^{\mu}=x^{\mu}+\epsilon^{\mu}(x)$ can be inverted as $x^{\mu}=x'^{\mu}-\epsilon^{\mu}(x')+O(\epsilon^2)$, hence
\begin{eqnarray}
\frac{\partial x^{\rho}}{\partial x'^{\mu}}=\delta^{\rho}_{\mu}-\partial_{\mu}\epsilon^{\rho},
\end{eqnarray}
transforming the metric according to
\begin{eqnarray}
\label{transmet}
g_{\mu\nu}\to g'_{\mu\nu}=g_{\mu\nu}+\delta g_{\mu\nu}=g_{\mu\nu}-(\partial_{\mu}\epsilon_{\nu}+\partial_{\nu}\epsilon_{\mu}).
\end{eqnarray}
By definition, a conformal transformation preserves the angle between two vectors, thus it must leave the metric invariant up to a local scale factor:
\begin{eqnarray}
\label{eqconstraint}
g'_{\mu\nu}(x')=\Omega(x)g_{\mu\nu}(x).
\end{eqnarray}
For this condition to be realized, the transformation described by Eq.~(\ref{transmet}) must be such, that the variation of the metric is proportional to the original metric itself. A more explicit expression of this constraint is obtained by taking the trace, that corresponds to a contraction in the tensorial formalism:
\begin{eqnarray}
g^{\mu \nu}g_{\mu \nu}=D,
\end{eqnarray}
where $D\!=\!d\!+\!1$ is the space-time dimension, yielding
\begin{eqnarray}
g^{\mu \nu}(\partial_{\mu}\epsilon_{\nu}+\partial_{\nu}\epsilon_{\mu})=2\partial^{\nu}\epsilon_{\nu}.
\end{eqnarray}
In the end, the constraint Eq.~(\ref{eqconstraint}) has been transformed into
\begin{eqnarray}
\partial_{\mu}\epsilon_{\nu}+\partial_{\nu}\epsilon_{\mu}=\frac{2}{D}\partial_{\rho}\epsilon^{\rho}g_{\mu\nu},
\end{eqnarray}
the so-called conformal Killing equation. Its solutions in Euclidian space, the conformal Killing vectors, are of the form
\begin{eqnarray}
\label{solKilling}
\epsilon_{\mu}=a_{\mu}+\omega_{\mu \nu}x^{\nu}+\lambda x_{\mu}+b_{\mu}\vec{x}^2-2(\vec{b}\cdot \vec{\mu})x_{\mu},
\end{eqnarray}
where $\omega_{\mu \nu}$ is antisymmetric. It can be shown that in space-time dimension strictly larger than two, the allowed conformal transformations, found by exponentiation of infinitesimal ones described by Eq.~(\ref{solKilling}), are of four types.

They correspond to translations, such that
\begin{eqnarray}
x'^{\mu}=x^{\mu}+a^{\mu},
\end{eqnarray}
dilations such that
\begin{eqnarray}
x'^{\mu}=\lambda x^{\mu},
\end{eqnarray}
where $\lambda$ is a non-negative number, rotations
\begin{eqnarray}
x'^{\mu}=({\delta^{\mu}}_{\nu}+{\omega^{\mu}}_{\nu})x^{\nu}={M^{\mu}}_{\nu}x^{\nu}
\end{eqnarray}
and the less intuitive 'special conformal transformations' that correspond to a concatenation of inversion, translation and inversion:
\begin{eqnarray}
x'^{\mu}=\frac{x^{\mu}-b^{\mu}\vec{x}^2}{1-2\vec{b}\cdot \vec{x}+b^2\vec{x}^2}.
\end{eqnarray}
The conclusion is that the group of conformal transformations is finite in this case.

Space-time dimension two, however, appears to be special, as the constaint (\ref{eqconstraint}) reduces to
\begin{eqnarray}
\label{CReq}
&&\partial_1\epsilon_1=\partial_2\epsilon_2, \,\partial_1\epsilon_2=-\partial_2\epsilon_1.
\end{eqnarray}
Equation (\ref{CReq}) is nothing else than the well-known Cauchy-Riemann condition that appears in complex analysis, and characterizes holomorphic functions. In other words, since any holomorphic function generates a conformal transformation in a (1+1)D QFT, the dimension of the conformal group is infinite. Actually, this property is of considerable help to solve models that feature conformal invariance.

In field theory, the interest started in 1984 when Belavin, Polyakov and Zamolodchikov introduced CFT as a unified approach to models featuring gapless linear spectrum in (1+1)D \cite{BelavinPolyakovZamolodchikov1984}. This property implies that this formalism shares its validity range with the Tomonaga-Luttinger liquid theory, hinting at an intimate link between CFT and bosonization, as first noticed in \cite{DotsenkoFateev1984}. The infinite-dimensional conformal symmetry actually stems from the spectrum linearity. From the point of view of integrability, the most important result of CFT is that correlation functions of critical systems obey an infinite-dimensional number of so-called Ward identities. Their solution uniquely determines all correlation functions, and in this respect CFT is a substitute to the Hamiltonian formalism to exactly solve a gapless model.

Let us follow for a while the analogy with bosonization. Within the CFT formalism, correlation functions are represented in terms of correlators of bosonic fields. The two-point correlation function is defined from the action $S$ in the path-integral formalism as
\begin{eqnarray}
\label{pathint}
\langle \phi_1(x_1)\phi_2(x_2)\rangle=\frac{1}{Z}\int D[\phi]\, \phi_1(x_1)\phi_2(x_2)e^{-S[\phi]}
\end{eqnarray}
where $Z\!=\!\int D[\phi]\, e^{-S[\phi]}$ is the partition function of the model. Equation (\ref{pathint}) is then simplified, using the properties of the conformal transformations listed above \cite{Polyakov1970}.

Enforcing rotational and translational invariance imposes that the two-point correlation function depends only on $|x_1-x_2|$. Scale invariance in turn yields
\begin{eqnarray}
\langle \phi_1(x_1)\phi_2(x_2)\rangle=\lambda^{\Delta_1+\Delta_2}\langle \phi_1(\lambda x_1)\phi_2(\lambda x_2)\rangle,
\end{eqnarray}
where $\Delta_{1,2}$ are the dimensions of the fields $\phi_{1,2}$, and combining these invariances yields
\begin{eqnarray}
\langle \phi_1(x_1)\phi_2(x_2)\rangle= \frac{C_{12}}{|x_1\!-\!x_2|^{\Delta_1+\Delta_2}}.
\end{eqnarray}
Applying also conformal invariance, one obtains
\begin{eqnarray}
\label{fineqCFT}
\langle \phi_1(x_1)\phi_2(x_2)\rangle= \delta(\Delta_1-\Delta_2)\frac{C_{12}}{|x_1\!-\!x_2|^{2\Delta_1}}.
\end{eqnarray}
More generally, all correlation functions of a model described by CFT decay like power law at large distance, as in the Tomonaga-Luttinger framework, as a consequence of the operator product expansion.

Equation (\ref{fineqCFT}) would be of little importance, however, if it were not supplemented by an extremely useful result. There is a connection between finite-size scaling effects and conformal invariance \cite{Cardy1984, BloteCardyNightingale1986}, allowing to investigate mesoscopic effects from the knowledge of the thermodynamic limit.
For instance, the first-order finite-size correction to the energy with respect to the thermodynamic limit is
\begin{eqnarray}
\delta E=-\frac{\pi c v_s}{6L},
\end{eqnarray}
where $c$ is another key concept of CFT known as the conformal charge, interpreted in this context as the model-dependent proportionality constant in the Casimir effect.

Actually, conformal field theories are classified through the conformal dimensions of their primary fields, $\{\Delta_i\}$, and their conformal charge. When $0<c<1$, critical exponents of the correlation functions are known exactly, and due to unitarity conformal charge can only take quantized, rational values \cite{Friedan1984},
\begin{eqnarray}
c=1-\frac{6}{m(m+1)},\, m\geq 3.
\end{eqnarray}
When $c\geq 1$ exponents of the large-distance asymptotics of the correlation functions may depend on the parameters of the model. This implies that Tomonaga-Luttinger liquids enter this category. As central charge also corresponds to the effective number of gapless degrees of freedom, Tomonaga-Luttinger liquids have a central charge $c\!=\!1$, and lie in the universality class of free fermions and bosons.

As far as correlation functions are concerned, primary fields are defined by their transformation 
\begin{eqnarray}
\phi(z,\overline{z})\to \left(\frac{\partial w}{\partial z}\right)^{\Delta}\left(\frac{\partial \overline{w}}{\partial \overline{z}}\right)^{\Delta}\phi'\left[w(z),\overline{w}(\overline{z})\right],
\end{eqnarray}
under conformal transformations of the complex variable $w\!=\!v_s\tau+ix$. For instance, finite-size effects are obtained through the transformation from the infinite punctured $z$-plane to the $w$-cylinder:
\begin{eqnarray}
w(z)=\frac{L}{2\pi}\ln(z)\leftrightarrow z=e^{\frac{2\pi w}{L}}.
\end{eqnarray}
Mesoscopic physics is, however, not always far from macroscopic one. More interestingly, so, this correspondence also yields finite-temperature corrections. In particular, CFT allows to evaluate finite-size and finite-temperature correlations of a Tomonaga-Luttinger liquid. The most relevant terms read \cite{Cazalilla2004}
\begin{eqnarray}
\label{nnfftL}
\frac{\langle n(x,t) n(0,0)\rangle^{TL}_{L<+\infty}}{n_0^2}=1-\frac{K}{4}\left(\frac{\pi}{k_FL}\right)^2\left\{\frac{1}{\sin^2\left[\frac{\pi(x-v_st)}{L}\right]}+\frac{1}{\sin^2\left[\frac{\pi(x+v_st)}{L}\right]}\right\}\nonumber\\
+A_1(K)\left(\frac{\pi}{k_FL}\right)^{2K}\frac{\cos(2k_Fx)}{\sin^K\left[\frac{\pi(x-v_st)}{L}\right]\sin^K\left[\frac{\pi(x+v_st)}{L}\right]}
\end{eqnarray}
at finite size, and at finite temperature, \cite{GiamarchiBook}
\begin{eqnarray}
\label{nnfftT}
\frac{\langle n(x,t) n(0,0)\rangle^{TL}_{T>0}}{n_0^2}=1-\frac{K}{4}\left(\frac{\pi}{k_FL_T}\right)^2\left\{\frac{1}{\sinh^2\left[\frac{\pi(x-v_s t)}{L_T}\right]}+\frac{1}{\sinh^2\left[\frac{\pi(x+v_s t)}{L_T}\right]}\right\}\nonumber\\
+A_1(K)\left(\frac{\pi}{k_FL_T}\right)^{2K}\frac{\cos(2k_Fx)}{\sinh^K\left[\frac{\pi(x-v_st)}{L_T}\right]\sinh^K\left[\frac{\pi(x+v_st)}{L_T}\right]},
\end{eqnarray}
where $L_T\!=\!\beta \hbar v_s$ plays the role of a thermal length, with $\beta=1/k_BT$ the inverse temperature.

Equations (\ref{nnfftL}) and (\ref{nnfftT}) are written in a way that puts the emphasis on their similar structure, with the correspondence $L\leftrightarrow L_T$ and $\sin \leftrightarrow \sinh$, as expected from the mirror principle. In both cases a stripe in the complex plane is mapped onto a cylinder, the introduction of an imaginary time and the property $\sin(ix)=i\sinh(x)$ are the reasons for the similitudes and the slight differences between the two expressions. Note that I have not specified the dependence of $A_1$ on $L$ and $T$. Actually, the theory does not predict whether one should write $A_1(K,T)$, $A_1[K(T)]$ or $A_1[K(T),T]$ at finite temperature, for instance. 

\begin{figure}
\includegraphics[width=8cm, keepaspectratio, angle=0]{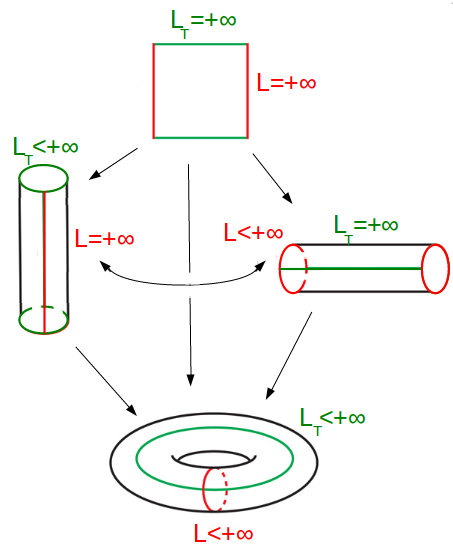}
\caption{Illustration of the complex coordinate conformal transformations to obtain the finite-size, finite-temperature (cylinder geometries), as well as finite size and temperature correlations (torus geometry) from the zero temperature correlations in the thermodynamic limit (plane)}
\label{FigCFT}
\end{figure}

I have also recovered Eqs.~(\ref{nnfftL}) and (\ref{nnfftT}) by generalizations of the bosonization procedure to finite system size and temperature \cite{LangHekkingMinguzzi2015} (elements of derivation are given in Appendix \ref{dernnT}). These equations are valid in the scaling limit, i.e. for $x\gg \epsilon$, $L_{(T)}-x\gg \epsilon$, where $\epsilon$ is the short-distance cut-off, of the order of $1/n_0$, and $L_{(T)} \gg x$. Far from the light-cone, Eq.~(\ref{nnfftT}), scales exponentially, but in both cases, the thermodynamic limit result is recovered at short distance and time. Actually, CFT even allows to go a step further, and investigate the intertwined effects of finite temperature and system size. This topic is far more advanced, however. The underlying idea consists in folding one of the cylinders at finite size or temperature into a torus, as illustrated in Fig.~\ref{FigCFT}. One can anticipate that the structure of the density-density correlation function will be similar to Eq.~(\ref{nnfftL}), with the sine function replaced by a doubly-periodic function. This double periodicity is at the heart of the field of elliptic functions, a wide subclass of special functions.

The final result reads (cf Appendix \ref{Demo2} for elements of derivation)
\begin{eqnarray}
\label{nnfftTL}
&&\frac{\langle n(x,t) n(0,0)\rangle^{TL}_{L<+\infty,T>0}}{n_0^2}\!=\!1\!-\!\frac{K}{4}\!\left(\frac{\pi}{k_FL}\right)^2\!\left\{\!\left[\frac{\theta_1''(\frac{\pi u}{L},e^{-\frac{\pi L_T}{L}})}{\theta_1(\frac{\pi u}{L},e^{-\frac{\pi L_T}{L}})}-\frac{\theta_1'(\frac{\pi u}{L},e^{-\frac{\pi L_T}{L}})^2}{\theta_1(\frac{\pi u}{L},e^{-\frac{\pi L_T}{L}})^2}\right]\!+\!u\!\leftrightarrow\!v\!\right\}\nonumber\\
&&+A_1(K)\left(\frac{\pi}{k_FL}\right)^{2K}\left[\theta_1'\left(0,e^{-\pi\frac{L_T}{L}}\right)\right]^{2K}\frac{\cos(2k_Fx)}{\left[\theta_1\left(\frac{\pi u}{L},e^{-\pi\frac{L_T}{L}}\right)\theta_1\left(\frac{\pi v}{L},e^{-\pi\frac{L_T}{L}}\right)\right]^{K}},
\end{eqnarray}
where $u\!=\!x\!-\!v_st$ and $v\!=\!x\!+\!v_st$ are the light-cone coordinates,
\begin{eqnarray}
\theta_1(z,q)=2q^{1/4}\sum_{k=0}^{+\infty}(-1)^kq^{k(k+1)}\sin[(2k\!+\!1)z],\, |q|<1,
\end{eqnarray}
is the first elliptic theta function, and $'$ denotes derivation with respect to the variable $z$. The first two terms in Eq.~(\ref{nnfftTL}) agree with the result of \cite{Affleck2011} as can be checked by easy algebraic manipulations. Note that all the expressions above have been obtained assuming periodic boundary conditions, i.e. a ring geometry, but space-time correlations depend on boundary conditions at finite size \cite{Cazalilla2004}.

To conclude, in this section I have presented conformal field theory as a formalism that allows to deal with critical 2D classical or 1D quantum models, where physics is scale-invariant. It can be viewed as an alternative to the bosonization procedure to derive the correlation functions of Luttinger liquids at finite size and temperature. Beyond this basic aspect, CFT remains central to the current understanding of fundamental physics. In recent years, attention has switched to conformal field theories in higher dimensions. The conformal bootstrap provides the most accurate evaluations of the critical exponents of the 3D Ising model \cite{Vichi2014}, and the discovery of the AdS/CFT duality has provoked a revolution in theoretical high-energy physics \cite{Maldacena1998, Witten1998} that may propagate to condensed-matter physics as well \cite{Zaanen}.

\section{From Lineland to Flatland: multi-component systems and dimensional crossovers}

To finish the dimensional roundabout, I shall come back to higher dimensions. Actually, in the aforementioned issue of bridging 1D and 2D, theory and experiment face opposite difficulties. In an experimental context, reducing the dimension of a system of ultracold atoms by lowering temperature and strengthening confinement is challenging. In theoretical physics, both analytically and numerically, one-dimensional models are considerably easier to deal with, as the number of available tools is far greater. These techniques can often be adapted to multi-component systems, obtained by adding a degree of freedom, at the price of much more efforts and cumbersome summation indices in the equations. Actually, multi-component models are often better suited to describe experiments than strictly-1D models.

Two deep questions are in order: first, can this degree of freedom realize or simulate an additional space dimension? Second, is it possible to deal with the limit where the parameter associated to this degree of freedom allows to approximate a higher-dimensional system? This phenomenon is generically called dimensional crossover, and from a theoretical perspective, it would be highly appreciated if it would give access to a regime where no efficient analytical tool is available to tackle the problem in a direct way.

One of the aims of studying dimensional crossovers is to gain insight into dimensional-dependent phenomena. The two most relevant examples I have in mind are the following: one-dimensional waveguides are not a priviledged place to observe Bose-Einstein condensation, but the latter is allowed in 3D. If one-dimensional atomic wires are created so that they are dense enough in space, then Bose-Einstein condensation occurs at a critical density \cite{Vogler2014, Pelster2017}.
As a second illustration, in two and three dimensions, interacting fermions are often described by the Fermi liquid theory. The latter fails in 1D, where low-energy physics is described by Luttinger liquid theory, that fails in higher dimensions. Dimensional crossovers may shed light on the subtleties of the Luttinger-Fermi liquid crossover. I will give a simple, qualitative explanation in chapter \ref{secV}.

There are actually many approaches and tools to treat the problem of dimensional crossovers, and not all of them are equivalent. 
One can think of gradually changing the dimension of the system, in such a way that its dimension $d$ takes non-integer values. This tomographic approach is taken formally within renormalization group techniques, but in this context non-integer values do not really have a physical meaning. This approach has been used to investigate the Luttinger-Fermi liquid transition \cite{Castellani1994, Bellucci2001}.
One can also imagine that the fractal dimension of the system is tunable. Although fractal systems are being devoted attention to \cite{Tsomokos2010, Aidelsburger2013}, this possibility does not sound realistic with current experimental techniques.

More promising is the idea of coupling one-dimensional gases, and paving space to obtain a higher-dimensional system in the limit where the number of components becomes infinite. This seductive idea, perhaps the most suited to explore the Luttinger to Fermi liquid crossover \cite{Guinea1993, Arrigoni1999, Biermann2001}, is actually quite difficult to formalize, but this interaction-driven scenario has been implemented experimentally \cite{Armijo2011, Hulet2016}.

The most attractive trend in recent years is to use an internal degree of freedom, such as spin or orbital angular momentum, to simulate higher-dimensional systems through the so-called synthetic dimensions \cite{Lewenstein2012, Lewenstein2014, Zeng2015, Luo2015, Barbarino2016}. A variant relies on the use a dynamical process to simulate an additional dimension \cite{Goldman2017}.

The approach I will use in chapter \ref{secV} is different, and consists in releasing a transverse confinement to generate new degrees of freedom \cite{Cheng2016}, associated to a multi-mode structure in energy space \cite{LangHekkingMinguzzi2016}.

\section{Summary of this chapter/r\'esum\'e du chapitre}

In this introductory chapter, I have tried to give a glimpse of the conceptual difficulties associated to the notion of space dimension in physics. While dimensions larger than three are obviously difficult to apprehend, the lower number of degrees of freedom in a 1D world also leads to fascinating effects on ultracold gases, such as collectivization of motion, fermionization of bosons, or spin-charge separation. The Mermin-Wagner theorem prevents thermal phase transitions to occur in 1D, urging for a shift of paradigm to describe low-dimensional gases, better characterized through their correlation functions.

Ultracold atoms are a versatile tool to simulate condensed-matter physics, as virtually every parameter is tunable, from particle number and density, to the strength and type of interactions, geometry of the gas and internal degrees of freedom. Moreover, temperature scales span several decades, allowing to probe thermal or purely quantum fluctuations and their intertwined effects. Their effects are enhanced in low-dimensional gases, obtained through highly-anisotropic trapping potentials, and the ring geometry is at the core of current attention.

I have presented the main analytical tools that allow to study this experimental configuration. A fair number of low-dimension models are integrable, from quantum one-dimensional spin chains and quantum field theories, to two-dimensional classical spin chains and ice-type models, allowing to obtain their exact thermodynamics and excitation spectrum by coordinate Bethe Ansatz, or their correlation functions by algebraic Bethe Ansatz.

A trivial peculiar case of integrable model is provided by noninteracting spinless fermions. They are in a one-to-one correspondence to the Tonks-Girardeau gas of hard-core bosons according to Girardeau's Bose-Fermi mapping, allowing to obtain the exact energy, even-order correlation functions and excitation spectrum of this strongly-interacting model, even in the non-integrable case of a harmonic trap.

At finite interaction strength, bosonization provides a means to obtain the exact large-distance asymptotics of correlation functions of gapless models. I have applied it to the Tomonaga-Luttinger liquid, an effective model that universally describes many gapless microscopic models in 1D, and have discussed its validity range on the example of time-dependent density-density correlations of the Tonks-Girardeau gas. For this observable, the Tomonaga-Luttinger prediction is exact in the static case, and correct to first order far from the mass shell in the dynamical case.

I have also given a brief introduction to conformal field theory, as a peculiar case of integrable theory, and an alternative formalism to obtain the finite-size and finite-temperature correlation functions of Tomonaga-Luttinger liquids.

To finish with, I have introduced the problematics of dimensional crossovers to higher dimensions, and the different ways to realize them in ultracold atom setups, often based on multi-mode structures.
\\
\\
\\
\\
\\
Dans ce chapitre introductif, j'ai laiss\'e entrevoir certaines des difficult\'es conceptuelles soulev\'ees par la notion de dimension en physique. Le nombre restreint de degr\'es de libert\'e dans un espace unidimensionnel donne lieu \`a de fascinants effets, parfois non-triviaux, dans les syst\`emes de gaz ultrafroids, parmi lesquels l'\'emergence d'un comportement dynamique collectif, la fermionisation des bosons et la s\'eparation des excitations de densit\'e et de spin. Le th\'eor\`eme de Mermin-Wagner emp\^eche les transitions de phase traditionnelles dans grand nombre de syst\`emes unidimensionnels, appelant \`a changer de paradigme pour d\'ecrire au mieux les gaz de basse dimension, qui sont caract\'eris\'es bien plus efficacement, dans l'ensemble, par leurs fonctions de corr\'elation que par leur diagramme de phase.

Les gaz ultrafroids constituent un support polyvalent pour la simulation de syst\`emes issus de la physique de la mati\`ere condens\'ee, dans la mesure o\`u chaque param\`etre y est ajustable, du nombre de particules \`a la densit\'e, en passant par le type et l'intensit\'e des interactions, ainsi que la g\'eom\'etrie du gaz et ses degr\'es de libert\'e internes. Qui plus est, les \'echelles de temp\'erature mises en jeu varient sur plusieurs d\'ecades, ce qui permet de sonder aussi bien les fluctuations thermiques que quantiques, en modifiant leur rapport. Ces fluctuations sont encore plus intenses dans les gaz de basse dimension, obtenus par un confinement hautement anisotrope, qui permet d\'esormais d'acc\'eder \`a des g\'eom\'etries annulaires.

Un certain nombre de mod\`eles sont int\'egrables en basse dimension, des cha\^ines de spins quantiques aux th\'eories quantiques des champs unidimensionnelles, mais aussi des cha\^ines de spins classiques et des mod\`eles de glace issus de la physique statistique, ce qui permet d'obtenir leurs propri\'et\'es thermodynamiques, ainsi que leur spectre d'excitation, par Ansatz de Bethe. La version alg\'ebrique de cet outil donne m\^eme acc\`es aux fonctions de corr\'elation.

Le gaz de fermions libres est un exemple trivial de mod\`ele int\'egrable. Il est en bijection avec le gaz de bosons de c\oe ur dur, dit de Tonks-Girardeau, pour un certain nombre d'observables sujettes \`a la correspondance bosons-fermions, qui donne facilement acc\`es \`a l'\'energie, au spectre d'excitations et aux fonctions de corr\'elation en densit\'e de ce mod\`ele fortement corr\'el\'e, et ce m\^eme en pr\'esence d'un pi\`ege harmonique.

Quand l'intensit\'e des interactions est finie dans un mod\`ele sans gap, on peut obtenir la forme exacte de ses corr\'elations asymptotiques par bosonisation. J'ai appliqu\'e cette m\'ethode au mod\`ele de Tomonaga-Luttinger, qui est un mod\`ele effectif universel, et discut\'e son domaine de validit\'e concernant les corr\'elations en densit\'e. Le mod\`ele de Tomonaga-Luttinger s'av\`ere exact pour les corr\'elations statiques, et correct au premier ordre loin de la couche de masse pour les corr\'elations dynamiques. J'ai aussi donn\'e une br\`eve introduction \`a la th\'eorie conforme des champs, que j'envisage en tant qu'exemple de mod\`ele int\'egrable et de formalisme alternatif pour obtenir les fonctions de corr\'elation des liquides de Tomonaga-Luttinger \`a taille et temp\'erature finies. Pour finir, j'ai introduit la probl\'ematique de l'augmentation progressive de la dimension d'un gaz ultrafroid, et expos\'e les diff\'erentes mani\`eres d'y parvenir, qui se fondent le plus souvent sur une structure multi-mode.

\newpage

\chapter{Ground-state static correlation functions of the Lieb-Liniger model}
\label{secIII}

\section{Introduction}

In this chapter, I characterize a strongly-correlated, ultracold one-dimensional Bose gas on a ring through its equilibrium, static correlation functions. The gas is described by the Lieb-Liniger model, that corresponds to contact interactions. This model is arguably the most conceptually simple in the class of continuum quantum field theories, and the most studied. It is integrable, equivalent to the Tonks-Girardeau gas in the strongly-interacting regime, its low-energy sector lies in the universality class of Tomonaga-Luttinger liquids, and it can be seen as a conformal field theory with unit central charge. These properties allow for a quite thorough theoretical investigation, involving all the analytical tools presented in chapter \ref{secII}.

This chapter is organized as follows: first, I present the Lieb-Liniger model, and explain the main steps of its solution by the coordinate Bethe Ansatz technique at finite number of bosons. This method yields the exact many-body wavefunction, and a set of coupled equations (called Bethe Ansatz equations), whose solution yields the exact ground-state energy. The Bethe Ansatz equations can be solved numerically up to a few hundreds of bosons. In the thermodynamic limit, the infinite set of coupled equations can be recast in closed form as a set of three integral equations. Not only are they amenable to numerical techniques, but approximate analytical solutions can be obtained in a systematic way in the weak- and strong-coupling regimes. However, finding the exact ground-state energy at arbitrary interaction strength is a long-standing open problem. More pragmatically, a reasonable aim was to bridge the weak- and strong-coupling expansions at intermediate interaction strengths, with an accuracy that would compete with state-of-the-art numerical methods. I summarize the main historical breakthroughs in both regimes, and my own contributions to the problem. Once the energy is known with satisfying accuracy, various thermodyamic quantities can be extracted through thermodynamic relations.

Then in a second time, I delve into the issue of correlation functions. The most simple ones are the local auto-correlations of the many-body wavefunction. Actually, one does not need to know the many-body wavefunction explicitly to evaluate them, as they are related to the moments of the density of pseudo-momenta, a quantity already evaluated to obtain the ground-state energy. This allows me to investigate the local first-, second- and third-order correlation functions, that are experimentally relevant with current methods.

Adding one level of complexity, I address the issue of non-local correlations at short and large distance. I focus on the one-body correlation function, whose asymptotics are known to relatively high orders in the Tonks-Girardeau regime. In the general case of finite interaction strength, the Tomonaga-Luttinger liquid theory allows to tackle the large-distance regime. I construct short-distance expansions using Bethe Ansatz techniques, through relations that I have called 'connections'.

The Fourier transform of the one-body correlation function, known as the momentum distribution, is also amenable to ultracold atom experiments, through ballistic expansion. Once again, its asymptotics can be calculated exactly, and the dominant term of the large-momentum tail is universal as it always correspond to an inverse quartic power law at finite interaction strength. Its numerical coefficient, however, depends on the interaction strength and is known as Tan's contact. I use this observable to illustrate an extension of the Bethe Ansatz technique to the inhomogeneous, harmonically-trapped system, whose integrability is broken, by combining it to the local-density approximation scheme.

All along the discussion, several technical details, transverse issues and interesting alternative approaches are left aside, but a few of them are evoked in a series of appendices.
\\
\\
\\
\\
\\
Dans ce chapitre, je caract\'erise un gaz de Bose ultrafroid fortement corr\'el\'e, confin\'e sur un anneau unidimensionnel, \`a travers l'analyse de ses fonctions de corr\'elation statiques \`a l'\'equilibre. Un tel gaz est bien d\'ecrit par le mod\`ele de Lieb et Liniger, qui correspond \`a des interactions de contact et des conditions aux limites p\'eriodiques. Ce mod\`ele est sans doute le plus simple parmi les th\'eories quantiques des champs de basse dimension, et l'un des plus \'etudi\'es. Il est \`a la fois int\'egrable, \'equivalent au gaz de Tonks-Girardeau dans son r\'egime de fortes interactions, membre de la classe d'universalit\'e des liquides de Tomonaga-Luttinger \`a basse \'energie, et peut y \^etre d\'ecrit par une th\'eorie conforme des champs de charge centrale unit\'e. Ces propri\'et\'es, qui mettent en jeu l'ensemble des m\'ethodes analytiques pr\'esent\'ees au chapitre pr\'ec\'edent, autorisent une \'etude th\'eorique en profondeur.

Le pr\'esent chapitre s'articule comme suit: dans un premier temps, je pr\'esente le mod\`ele de Lieb et Liniger, et d\'etaille les principales \'etapes de sa r\'esolution par Ansatz de Bethe. Cette m\'ethode conduit \`a un syst\`eme d'\'equations transcendantes coupl\'ees, en nombre \'egal \`a celui de bosons, dont la solution donne l'\'energie exacte de l'\'etat fondamental, et qui peuvent \^etre r\'esolues num\'eriquement. Par passage \`a la limite thermodynamique, ce syst\`eme d'\'equations se ram\`ene \`a un ensemble de trois \'equations int\'egrales. Ces derni\`eres peuvent \^etre r\'esolues num\'eriquement, mais de surcro\^it, des m\'ethodes analytiques permettent d'en construire des solutions approch\'ees de mani\`ere syst\'ematique sans les r\'egimes de couplage faible et fort. La question de la solution analytique exacte reste enti\`ere, mais plus pragmatiquement, j'ai consacr\'e beaucoup de temps et d'\'energie \`a la recherche et au d\'eveloppement de solutions approch\'ees, jusqu'\`a atteindre un degr\'e de pr\'ecision comparable \`a celui des m\'ethodes num\'eriques les plus avanc\'ees. Une fois l'\'energie connue avec une pr\'ecision satisfaisante, d'autres observables peuvent en \^etre d\'eduites, au travers d'identit\'es thermodynamiques.

Dans un second temps, je me plonge dans la probl\'ematique des fonctions de corr\'elation. Les plus simples d'entre elles sont les fonctions locales d'autocorr\'elation de la fonction d'onde. Nul besoin de conna\^itre explicitement cette derni\`ere pour les \'evaluer, car elles s'obtiennent directement \`a partir des moments de la distribution des pseudo-impulsions, d\'ej\`a \'evalu\'ee lorsqu'il s'agissait d'en d\'eduire l'\'energie de l'\'etat fondamental. Cela me permet d'\'evaluer les fonctions de corr\'elation locales \`a un, deux ou trois corps, que les m\'ethodes exp\'erimentales actuelles permettent de mesurer.

J'ajoute ensuite un degr\'e de complexit\'e au probl\`eme, et m'int\'eresse aux fonctions de corr\'elation non-locales. Je me concentre sur la fonction de corr\'elation \`a un corps, dont le comportement asymptotique \`a courte et longue distance sont connues \`a des ordres \'elev\'es dans le r\'egime de Tonks-Girardeau. Dans le cas g\'en\'eral, \`a interaction finie, la th\'eorie des liquides de Tomonaga-Luttinger permet d'obtenir le comportement asymptotique \`a longue distance. \`A courte distance, je construis le d\'eveloppement limit\'e en m'appuyant sur l'Ansatz de Bethe et des identit\'es que j'ai baptis\'ees connexions. La transform\'ee de Fourier de la fonction de corr\'elation \`a un point, appel\'ee distribution en impulsion, est mesur\'ee par expansion balistique dans les exp\'eriences d'atomes ultrafroids. Une fois n'est pas coutume, ses d\'eveloppements limit\'e et asymptotique peuvent \^etre calcul\'es de mani\`ere exacte, et le terme dominant \`a haute impulsion s'av\`ere universel, car il d\'ecro\^it comme l'inverse de la puissance quatri\`eme de l'impulsion. Son pr\'efacteur num\'erique, en revanche, n'est pas universel et d\'epend de l'intensit\'e des interactions. Il s'agit du contact de Tan, sur lequel je m'appuie pour illustrer une extension de l'Ansatz de Bethe au cas d'un gaz inhomog\`ene dans un pi\`ege harmonique longitudinal. Bien que non-int\'egrable, je le r\'esous par une astucieuse combinaison avec l'approximation de la densit\'e locale.

Tout au long de la discussion, je passe sous silence des d\'etails techniques, probl\`emes transversaux et approches alternatives. Certains sont tout de m\^eme \'evoqu\'es dans une s\'erie d'appendices rattach\'es \`a ce chapitre.

\section{Exact ground-state energy of the Lieb-Liniger model}

I start by reviewing a few known results concerning the Lieb-Liniger model. For introductory texts and reviews, I refer to \cite{Panfilthesis, ZhuYangYangWen2015, Franchini2016}.

\subsection{Ground-state energy in the finite-$N$ problem}

The Lieb-Liniger model describes a given number $N$ of identical bosons, confined to one spatial dimension. It assumes that they are point-like and interact through a two-body, zero-range, time- and velocity-independent potential. If $m$ denotes the mass of each boson, and $\{x_i\}_{i=1,\dots,N}$ label their positions, then the dynamics of the system is given by the Lieb-Liniger Hamiltonian, that reads \cite{LiebLiniger1963}
\begin{eqnarray}
\label{HLL}
H^{LL}=\sum_{i=1}^N\left[-\frac{\hbar^2}{2m}\frac{\partial^2}{\partial x_i^2}+g_{\mbox{\scriptsize{1D}}}\sum_{\{j\neq i\}}\delta(x_i\!-\!x_j)\right].
\end{eqnarray}
The first term is associated to the kinetic energy, the second one represents the contact interactions, where $g_{\mbox{\scriptsize{1D}}}$ is the interaction strength or coupling constant, whose sign is positive if interactions are repulsive, as in the case considered by Lieb and Liniger, and negative otherwise.

I will not consider this opportunity in the following, so let me give a brief account of the main known results. The attractive regime is unstable, owing to its negative ground-state energy, and does not possess a proper thermodynamic limit \cite{McGuire1964}. However, the first excited state, known as the super Tonks-Girardeau (sTG) gas \cite{AstrakharchikBoronatCasullerasGiorgini2005, Batchelor2005}, has attracted enough attention to be be realized experimentally \cite{Haller2009}. In the cold atom context, this metastable state is mainly studied in quench protocols, where thermalization is the question at stake. The sTG gas also maps onto the ground state of attractive fermions, that is stable \cite{Chen2010}, and signatures of a sTG regime are expected in dipolar gases \cite{AstrakharchikLozovik2008, GirardeauAstrakharchik2012}. More generally, the attractive regime of the Lieb-Liniger model is the seat of a variety of mappings, onto a Bardeen-Cooper-Schrieffer (BCS) model \cite{Fuchs2004, Batchelor2004, Wadati2005}, the Kardar-Parisi-Zhang (KPZ) model \cite{LeDoussal2014}, directed polymers \cite{Luca2015} or three-dimensional black holes \cite{Panchenko2015}. Moreover, in a peculiar limit, the attractive Bose gas becomes stable and features the Douglas-Kazakov, third-order phase transition \cite{Flassig2016, PiroliCalabrese2016}.

Let us come back to the case of repulsive interactions. I use units where $\hbar^2/(2m)=1$, and the mathematical physics notation $c$ for the interaction strength, instead of $g_{\mathrm{1D}}$. The Schr\"odinger time-independent equation associated to the Hamiltonian Eq.~(\ref{HLL}) is
\begin{eqnarray}
\label{Schrod}
H^{LL}\psi_N(\textbf{x})=E_0\,\psi_N(\textbf{x}),
\end{eqnarray}
where $E_0$ is the ground-state energy, and $\psi_N$ the many-body wavefunction for coordinates $\textbf{x}\!=\!(x_1,\dots,x_N)$. As an eigenvalue problem, Eq.~(\ref{Schrod}) can be solved exactly by explicit construction of $\psi_N$. To do so, Lieb and Liniger applied, for the first time to a model defined in the continuum, the coordinate Bethe Ansatz \cite{LiebLiniger1963}.

According to Eq.~(\ref{HLL}), interactions only occur when two bosons are at contact. Outside this case, one can split the support of the $N$-body wavefunction into $N!$ sectors, that correspond to all possible spatial orderings of $N$ particles along a line. Since the wavefunction is symmetric with respect to any permutation of the bosons, let us arbitrarily consider the fundamental simplex $R$, such that $0<x_1< x_2<\dots <x_N<L$, where $L$ is the length of the atomic waveguide. 

In $R$, the original Schr\"odinger equation (\ref{Schrod}) is replaced by an Helmoltz equation for the wavefunction $\psi_N|_R$ restricted to the fundamental simplex, namely
\begin{eqnarray}
\label{Helmoltzeq}
-\sum_{i=1}^N\frac{\partial^2\psi_N|_R}{\partial x_i^2}=E_0\psi_N|_R,
\end{eqnarray}
together with the Bethe-Peierls boundary conditions,
\begin{eqnarray}
\label{cusp}
\left(\frac{\partial}{\partial x_{j+1}}-\frac{\partial}{\partial x_j}\right)\psi_N|_{x_{j+1}=x_j}=c\,\psi_N|_{x_{j+1}=x_j}.
\end{eqnarray}
The latter are mixed boundary conditions, whose role is to keep track of the interactions at the internal boundaries of $R$. They are obtained by integration of Eq.~(\ref{Schrod}) over an infinitesimal interval around a contact \cite{GaudinBethe}.

Boundary conditions, assumed to be periodic here due to the ring geometry, are taken into account through
\begin{eqnarray}
\label{contboundary}
\psi_N(0,x_2,\dots,x_N)=\psi_N(L,x_2,\dots,x_N)=\psi_N(x_2,\dots,x_N,L),
\end{eqnarray}
where the exchange of coordinates is performed to stay in the simplex $R$. There is also a continuity condition on the derivatives:
\begin{eqnarray}
\label{contboundaryder}
\frac{\partial}{\partial x}\psi_N(x,x_2,\dots,x_N)|_{x=0}=\frac{\partial}{\partial x}\psi_N(x_2,\dots,x_N,x)|_{x=L}.
\end{eqnarray}
Equations (\ref{cusp}, \ref{contboundary}, \ref{contboundaryder}) represent the full set of boundary conditions associated to the differential equation (\ref{Helmoltzeq}), so that the problem is well defined by now, and simpler than the original Schr\"odinger equation (\ref{Schrod}).

To solve it, the starting point (Ansatz) consists in guessing the structure of the wavefunction inside the fundamental simplex:
\begin{eqnarray}
\label{Ansatz}
\psi_N|_{R}=\sum_{P\in S_N}\!a(P)e^{i\sum_{j=1}^N\!k_{P(j)}x_j},
\end{eqnarray}
where $P$ are elements of the symmetric group $S_N$, i.e. permutations of $N$ elements, $\{k_i\}_{i=1,\dots,N}$ are the pseudo-momenta carried by the individual bosons (called so because they are not observable and should not be confused with the physical momentum), and $a(P)$ are scalar coefficients that takes interactions into account. In other words, one postulates that the wavefunction can be written as a weighted sum of plane waves (in analogy to the noninteracting problem), and $\{a(P)\}$ and $\{k_i\}$ are then determined so as to verify Eqs.~(\ref{Helmoltzeq}), (\ref{cusp}), (\ref{contboundary}) and (\ref{contboundaryder}).

The two-body scattering matrix $S$ is defined through
\begin{eqnarray}
a(P')=Sa(P),
\end{eqnarray}
where $P'$ is a permutation obtained by exchanging $P(j)$ and $P(j\!+\!1)$, i.e.
\begin{eqnarray}
P'=\{P(1),\dots, P(j\!-\!1),P(j\!+\!1),P(j),P(j\!+\!2),\dots,P(N)\}.
\end{eqnarray}
The cusp condition Eq.~(\ref{cusp}) is satisfied provided
\begin{eqnarray}
\label{PprimeP}
a(P')=\frac{k_{P(j)}-k_{P(j+1)}+ic}{k_{P(j)}-k_{P(j+1)}-ic}\,a(P).
\end{eqnarray}
Thus, this peculiar scattering process leads to an antisymmetric phase shift, and the scattering matrix, that has unit modulus according to Eq.~(\ref{PprimeP}), can be written as a pure phase. It reads
\begin{eqnarray}
\label{Sphase}
S=e^{-i\theta[k_{P(j)}-k_{P(j+1)}; c]},
\end{eqnarray}
where
\begin{eqnarray}
\label{phaseshiftLL}
\theta(k;c)=2\arctan\left(\frac{k}{c}\right)
\end{eqnarray}
is the function associated to the phase shift due to a contact interaction. In the limit $c\!\to\!+\infty$, the scattering phase is the same as the one of noninteracting fermions, which is a signature of the Bose-Fermi mapping, and of a Tonks-Girardeau regime. Furthermore, as consequence of Eq.~(\ref{Sphase}), the Yang-Baxter equation (\ref{YBS}) is satisfied. Thus, the Lieb-Liniger model is integrable.

Actually, the pseudo-momenta $\{k_i\}_{i=1,\dots,N}$ satisfy the following set of equations \cite{Cazalilla2004}:
\begin{eqnarray}
\label{eqk}
e^{ik_iL}=\prod_{\{j\neq i\}}\frac{k_i-k_j+ic}{k_i-k_j-ic}=-\prod_{j=1}^N\frac{k_i-k_j+ic}{k_i-k_j-ic},
\end{eqnarray}
where the global minus sign in the right-hand side is a signature of the periodic boundary condition, and would become a plus for an anti-periodic one. Using the property $\arctan(x)\!=\!\frac{i}{2}\ln\left(\frac{i+x}{i-x}\right)$ and a few algrebraic transformations, Eq.~(\ref{eqk}) is then recast in logarithmic form in terms of the phase-shift function $\theta$ as
\begin{eqnarray}
\label{logBAE}
\frac{2\pi}{L}I_i=k_i+\frac{1}{L}\sum_{j=1}^N\theta\!\left(k_i-k_j;c\right).
\end{eqnarray}
The $N$ coupled equations (\ref{logBAE}), where the unknowns are the pseudo-momenta, are the Bethe Ansatz equations. The Bethe numbers $\{I_i\}_{i=1,\dots,N}$ are integers if the number of bosons is odd and half-odd integers if $N$ is even. They play the role of quantum numbers, and characterize the state uniquely.

The Bethe Ansatz equations (\ref{logBAE}) are physically interpreted as follows \cite{Eliens}: a particle $i$ moving along the circle of circumference $L$ to which the gas is confined acquires, during one turn, a phase determined by its momentum $k_i$, as well as a scattering phase from interactions with the $N-1$ other bosons on the ring. Since scattering is diffractionless, as a consequence of the Yang-Baxter equation, the whole scattering phase is a sum of two-body phase shifts.
Rephrased once more, in order to satisfy the periodicity condition, the phase associated to the momentum plus the total scattering phase shall add up to $2\pi$ times an (half-odd) integer.

In the limit $c\!\to\!+\infty$, Eq.~(\ref{logBAE}) simplifies dramatically and it becomes obvious that if two quantum numbers are equal, say $I_i\!=\!I_j$, then their corresponding quasi-momenta coincide as well, i.e. $k_i\!=\!k_j$. Since in such case the Bethe wavefunction vanishes, the Bethe numbers must be distinct to avoid it. As a consequence, the ground state, that minimizes energy and momentum, corresponds to a symmetric distribution of quantum numbers without holes, i.e. a Fermi sea distribution, and then
\begin{eqnarray}
\label{Bethenumber}
I_i=-\frac{N+1}{2}+i,
\end{eqnarray}
as already obtained in the previous chapter for the Tonks-Girardeau gas.

If the coupling $c$ becomes finite, a scattering phase is slowly turned on so that, for fixed $I_i$, the solution $\{k_i\}_{i=1,\dots,N}$ to the Bethe Ansatz equations (\ref{logBAE}) moves away from the regular distribution. However, since level crossings are forbidden (there is no symmetry to protect a degeneracy and accidental ones can not happen
in an integrable model), the state defined by (\ref{Bethenumber}) remains the lowest-energy one at arbitrary interaction strength. Changing $c$ modifies the quasi-momenta, but has no effect on the quantum numbers, that are quantized. Each choice of quantum numbers yields an eigenstate, provided that all Bethe numbers are different.
This rule confers a fermionic nature to the Bethe Ansatz solution in quasi-momentum space whenever $c\!>\!0$, although the system is purely bosonic in real space.

The momentum and energy of the Lieb-Liniger model in the ground state are obtained by summing over pseudo-momenta, or equivalently over Bethe numbers:
\begin{eqnarray}
P_0=\sum_{i=1}^Nk_i=\frac{2\pi}{L}\sum_{i=1}^NI_i=0.
\end{eqnarray}
The second equality follows from the Bethe Ansatz Equations (\ref{logBAE}) and the property $\theta(-k)\!=\!-\theta(k)$, showing that momentum is quantized, and independent of the interaction strength. The last equality is a direct consequence of Eq.~(\ref{Bethenumber}). Analogously, according to Eq.~(\ref{Helmoltzeq}) the ground-state energy is given by
\begin{eqnarray}
\label{EGS}
E_0=\sum_{i=1}^N k_i^2,
\end{eqnarray}
and the eigenvalue problem is solved, after extension of the wavefunction to the full domain $\textbf{x}\in [0,L]^N$, obtained by symmetrization of $\psi_N|_R$:
\begin{eqnarray}
\psi_N(\textbf{x})=\frac{\prod_{j>i}(k_j-k_i)}{\sqrt{N!\prod_{j>i}[(k_j-k_i)^2+c^2]}}\sum_{P\in S_N}\prod_{j>i}\left[1-\frac{i\,c\, \mathrm{sign}(x_j\!-\!x_i)}{k_{P(j)}\!-\!k_{P(i)}}\right]\prod_{j=1}^Ne^{ik_jx_j}.
\end{eqnarray}
Note that the derivation, as presented here, does not \textit{prove} that the Bethe Ansatz form of the wavefunction, Eq.~(\ref{Ansatz}), minimizes the energy.
This important point has been checked in \cite{Dorlas}, where the construction of Lieb and Liniger has been rigorously justified.

Equations (\ref{logBAE}) and (\ref{EGS}) yield the exact ground state properties of the finite $N$ problem. The equations are transcendent (i.e., not equivalent to the problem of finding roots of polynomials with integer coefficients), but can be solved numerically at arbitrary interaction strength up to the order of a hundred bosons \cite{Cazalilla2004, Sakmann2005}. Actually, the thermodynamic limit is directly amenable to Bethe Ansatz. The construction, also due to Lieb and Liniger, is the object of next section.

Before proceeding, I shall summarize a few arguments and interpret them in the more general context of integrable systems.
In a one-dimensional setting, when two particles scatter, conservation of energy and momentum constrain the outgoing momenta to be
equal to the incoming ones. Thus, the effect of interaction is reduced to adding a phase shift to the wavefunction.

The first step of the resolution consists in identifying the two-particle phase-shift, given here by Eq.~(\ref{phaseshiftLL}). Having determined the
two-particle scattering phase, one checks that the Yang-Baxter equation holds, by verifying that an ansatz wavefunction constructed
as a superposition of plane-wave modes with unknown quasi-momenta as in Eq.~(\ref{Ansatz}), is an eigenstate of the
Hamiltonian. The Yang-Baxter equation constrains the coefficients of the superposition, so that the eigenstate depends uniquely on the quasi-momenta.

One also needs to specify boundary conditions. For a system of $N$ particles, the choice of periodic boundary conditions generates a series of
consistency conditions for the quasi-momenta of the eigenstate, known as the Bethe Ansatz equations (\ref{logBAE}). This set
of $N$ algebraic equations depends on as many quantum numbers, that specify uniquely the quantum
state of the system. For each choice of these quantum numbers, one solves the set of Bethe Ansatz equations (being algebraic, they constitute a much lighter task than solving the original
partial derivative Schr\"odinger equation) to obtain the quasi-momenta, and thus the eigenstate wavefunction. These states have a fermionic nature, in that
all quantum numbers have to be distinct. This is a general feature of the Bethe Ansatz solution.

Further simplifications are obtained by considering the thermodynamic limit. Then, one is interested in the density of quasi-momenta and the set of algebraic equations (\ref{logBAE}) can be recast into the form of an integral equation for this distribution. The problem looks deceiptively simpler, then, as I will show in the next paragraph.

\subsection{Ground-state energy in the thermodynamic limit}
\label{GSTL}

In second-quantized form, more appropriate to deal with the thermodynamic limit, the Lieb-Liniger Hamiltonian Eq.~(\ref{HLL}) becomes:
\begin{eqnarray}
\label{HLLbis}
 H^{LL}[\hat{\psi}]=\frac{\hbar^2}{2m}\int_0^L dx\, \frac{\partial\hat{\psi}^{\dagger}}{\partial x}\frac{\partial\hat{\psi}}{\partial x}+\frac{g_{\mbox{\scriptsize{1D}}}}{2}\int_0^L dx\, \hat{\psi}^{\dagger}\hat{\psi}^{\dagger}\hat{\psi}\hat{\psi},
\end{eqnarray}
where $\hat{\psi}$ is a bosonic field operator that satisfies the canonical equal-time commutation relations with its Hermitian conjugate: 
\begin{eqnarray}
[\hat{\psi}(x),\hat{\psi}^\dagger(x')]=\delta(x\!-\!x'),\,[\hat{\psi}(x),\hat{\psi}(x')]=[\hat{\psi}^{\dagger}(x),\hat{\psi}^{\dagger}(x')]=0.
\end{eqnarray}

The ground-state properties of the Lieb-Liniger model depend on a unique dimensionless parameter, measuring the interaction strength. It is usual, following Lieb and Liniger, to define this coupling as
\begin{eqnarray}
\label{defgamma}
\gamma=\frac{mg_{\mbox{\scriptsize{1D}}}}{\hbar^2n_0},
\end{eqnarray}
where $n_0$ represents the linear density of the homogeneous gas. It appears at the denominator, which is rather counter-intuitive compared to bosons in higher dimensions. In particular, this means that diluting the gas increases the coupling, which is a key aspect to approach the Tonks-Girardeau regime, that corresponds to the limit $\gamma\!\to\!+\infty$.

In the regime of weak interactions, the bosons do not undergo Bose-Einstein condensation, since long-range order is prevented by fluctuations. Nonetheless, one can expect that a large proportion of them is in the zero momentum state, and forms a quasi-condensate. Under this assumption, the problem can be treated semi-classically. The operator $\hat{\psi}$ is replaced by a complex scalar field $\psi(x)$, and the Euler-Lagrange equation stemming from Eq.~(\ref{HLLbis}) is the 1D Gross-Pitaevskii equation \cite{Gross1961, Pitaevskii1961},
\begin{eqnarray}
i\frac{\partial\psi}{\partial t}\!=\!-\frac{\partial^2\psi}{\partial x^2}\!+\!2c\,\psi^{*}\psi\psi.
\end{eqnarray}
However, this method being of mean field type, one can expect that its validity range is quite limited in low dimension. The exact solution by Bethe Ansatz shall provide a rare opportunity to study this validity range quantitatively.

To obtain the exact solution, let us consider the Bethe Ansatz equations (\ref{logBAE}), and take the thermodynamic limit, i.e. $N\!\to\!+\infty$, $L\!\to\!+\infty$, while keeping $n_0\!=\!N/L$ fixed and finite. After ordering the Bethe numbers $\{I_i\}$ (or equivalently the pseudo-momenta $\{k_i\}$, that are real if $\gamma>0$ \cite{LiebLiniger1963}), one can rewrite the Bethe Ansatz equations as
\begin{eqnarray}
\label{BAEbis}
k_i-\frac{1}{L}\sum_{j=1}^N\theta(k_i-k_j)=y(k_i)
\end{eqnarray}
where $y$ is a 'counting function', constrained by two properties: it is strictly increasing and satisfies the Bethe Ansatz equations at any of the quasi-momenta, i.e., according to Eq.~(\ref{logBAE}), such that
\begin{eqnarray}
y(k_i)=\frac{2\pi}{L}I_i.
\end{eqnarray}
The aim is then to go from the discrete to the continuum, defining a density of pseudo-momenta such that
\begin{eqnarray}
\label{definerho}
\rho(k_i)\!=\!\lim_{N,L\to+\infty, N/L=n_0}\frac{1}{L(k_{i+1}\!-\!k_i)}.
\end{eqnarray}
It is strictly positive as expected, thanks to the ordering convention. In the thermodynamic limit, the sum in Eq.~(\ref{BAEbis}) becomes an integral over $k$, 
\begin{eqnarray}
\sum_{j=1}^N\to L\int\!dk\,\rho(k),
\end{eqnarray}
and the derivative of $y$ with respect to $k$,
\begin{eqnarray}
y'(k_i)=\lim_{N,L\to +\infty, N/L=n_0} \frac{y(k_{i+1})-y(k_{i})}{k_{i+1}-k_{i}}=2\pi \rho(k_i),
\end{eqnarray}
so that
\begin{eqnarray}
\label{inty}
\frac{1}{2\pi}y(k)=\!\int^k\!dk'\rho(k').
\end{eqnarray}
With these definitions, the set of Bethe Ansatz equations (\ref{logBAE}) becomes a single integral equation, relating the counting function to the distribution of quasi-momenta:
\begin{eqnarray}
\label{eqy}
 y(k)=k-\int_{k_{min}}^{k_{max}}\!dk'\,\theta(k\!-\!k')\rho(k'),
\end{eqnarray}
where $k_{min}$ and $k_{max}$ represent the lowest and highest quasi-momenta allowed by the Fermi sea structure. They are finite in the ground state, and the limits of integration are symmetric as a consequence of Eq.~(\ref{Bethenumber}): $k_{min}\!=\!-k_{max}$.
Differentiating Eq.~(\ref{eqy}) with respect to $k$ yields, by combination with Eq.~(\ref{inty}),
\begin{eqnarray}
\label{rhoeq}
\rho(k)=\frac{1}{2\pi}-\frac{1}{2\pi}\int_{-k_{max}}^{k_{max}}\!dk'K(k-k')\rho(k'),
\end{eqnarray}
where $K(k)\!=\!\theta'(k)\!=\!-\frac{2c}{c^2+k^2}$.

In view of a mathematical treatment of Eq.~(\ref{rhoeq}), it is convenient to perform the following rescalings:
\begin{eqnarray}
k\!=\!k_{max}z,\,\,\,\,c\!=\!k_{max}\alpha,\,\,\,\,\rho(k)\!=\!g(z;\alpha),
\end{eqnarray}
where $z$ is the pseudo-momentum in reduced units such that its maximal value is $1$, $\alpha$ is a non-negative parameter, and $g(z;\alpha)$ denotes the distribution of quasi-momenta expressed in these reduced units.
Finally, in the thermodynamic limit, the set of Bethe Ansatz equations (\ref{logBAE}) boils down to a set of three equations only, namely
\begin{equation}
\label{Fredholm}
g(z;\alpha)-\frac{1}{2\pi}\int_{-1}^{1}dy\frac{2\alpha g(y;\alpha)}{\alpha^2+(y-z)^2}=\frac{1}{2\pi},
\end{equation}
where $\alpha$ is in one-to-one correspondence with the Lieb parameter $\gamma$ introduced in Eq.~(\ref{defgamma}) via a second equation,
\begin{equation}
\label{alphagamma}
\gamma \int_{-1}^1\!dy\, g(y;\alpha)=\alpha.
\end{equation}
The third equation yields the dimensionless average ground-state energy per particle $e$, linked to the total ground-state energy $E_0$ expressed in the original units, and to the reduced density of pseudo-momenta $g$ by
\begin{eqnarray}
 \label{energy}
e(\gamma)=\frac{2m}{\hbar^2}\frac{E_0(\gamma)}{Nn_0^2}=\frac{\int_{-1}^1dy g[y;\alpha(\gamma)]y^2}{\{\int_{-1}^1 dy g[y;\alpha(\gamma)]\}^3}.
\end{eqnarray}
Interestingly, Eq.~(\ref{Fredholm}) is decoupled from Eqs.~(\ref{alphagamma}) and (\ref{energy}), which is specific to the ground state \cite{YangYang1969}. It is a homogeneous type II Fredholm integral equation with Lorentzian kernel, whose closed-form, exact solution is unknown but amenable to various approximation methods.

Before solving these equations, it is convenient to recall a few general properties of density of pseudo-momenta.

(i) The function $g$ is unique \cite{LiebLiniger1963}.

(ii) It is an even function of $z$, in agreement with the particle-hole symmetry noticed above. To see this, it is convenient to rewrite Eq.~(\ref{Fredholm}) as
\begin{eqnarray}
g(z;\alpha)=\frac{1}{2\pi}\left[1+2\alpha\left(\int_{0}^1dy \frac{g(y;\alpha)}{\alpha^2+(y-z)^2}+\int_{0}^1dy \frac{g(-y;\alpha)}{\alpha^2+(y+z)^2}\right)\right].
\end{eqnarray}
Then, introducing
\begin{eqnarray}
g_s(z;\alpha)=\frac{g(z;\alpha)+g(-z;\alpha)}{2},
\end{eqnarray}
it is easy to check that $g_s(z;\alpha)$ and $g(z;\alpha)$ are both solution to the Lieb equation. However, according to (i) the solution is unique, imposing $g(-z;\alpha)=g(z;\alpha)$.

(iii) The function $g$ is infinitely differentiable (analytic) in $z$ if $\alpha>0$ \cite{LiebLiniger1963}. This implies in particular that it has an extremum (which turns out to be a minimum) at $z\!=\!1$. The non-analyticity in the interaction strength at $\alpha\!=\!0$ is a signature of the absence of adiabatic continuation in 1D between ideal bosons and interacting ones.

(iv) $\forall z \in [-1,1]$, $g(z;\alpha)>0$, as a consequence of Eq.~(\ref{definerho}), as expected for a density. Moreover, $\forall \alpha>0$, $\forall z\in [-1,1]$, $g(z;\alpha)>g^{TG}(z)=\frac{1}{2\pi}$. This property directly follows from the discussion below Eq.~(\ref{iter}) and the mapping between Love's equation (\ref{Loveeq}) and the Lieb equation (\ref{Fredholm}).

(v) $\forall z \in [-1,1]$, $g$ is bounded from above if $\alpha>0$ \cite{LiebLiniger1963}, as expected from the Mermin-Wagner-Hohenberg theorem, that prevents true condensation.

In order to determine the ground-state energy in the thermodynamic limit, the most crucial step is to solve the Lieb equation (\ref{Fredholm}). This was done numerically by Lieb and Liniger for a few values of $\alpha$, spanning several decades. The procedure relies on the following steps:
an arbitrary (positive) value is fixed for $\alpha$, and Eq.~(\ref{Fredholm}) is solved, i.e. the reduced density of pseudo-momenta $g(z;\alpha)$ is evaluated with the required accuracy as a function of $z$ in the interval $[-1,1]$.
Then, Eq. (\ref{alphagamma}) yields $\gamma(\alpha)$, subsequently inverted to obtain $\alpha(\gamma)$. In doing so, one notices that $\gamma(\alpha)$ is an increasing function, thus interaction regimes are defined the same way for both variables.
The ground-state energy is then obtained from Eq. (\ref{energy}), as well as many interesting observables, that are combinations of its derivatives. They all depend on the sole Lieb parameter $\gamma$, which is the key of the conceptual simplicity of the Lieb-Liniger model.

\subsection{Weak-coupling regime}

Analytical breakthroughs towards the exact solution of the Bethe Ansatz equations have been quite scarce, since Lieb and Liniger derived them. I figure out three possible explanations: 
first, the Bethe Ansatz equations (\ref{logBAE}) or (\ref{Fredholm}) are easily amenable to numerical calculations in a wide range of interaction strengths.
Furthermore, simple approximate expressions reach a global $10\%$ accuracy \cite{Cazalilla2004}, comparable to error bars in the first generation of ultracold atom experiments.
Finally, the set of Lieb equations is actually especially difficult to tackle analytically in a unified way. Indeed, one should keep in mind that, although it consists in only three equations, the latter are just a convenient and compact way to rewrite an infinite set of coupled ones.

In the weakly-interacting regime, finding accurate approximate solutions of Eq.~(\ref{Fredholm}) at very small values of the parameter $\alpha$ is quite an involved task, both numerically and analytically. This is a consequence of the singularity of the function $g$ at $\alpha\!=\!0$, whose physical interpretation is that noninteracting bosons are not stable in 1D.

A guess function was proposed by Lieb and Liniger \cite{LiebLiniger1963}, namely
\begin{eqnarray}
\label{guessLieb}
 g(z;\alpha)\simeq_{\alpha\ll 1} \frac{\sqrt{1-z^2}}{2\pi\alpha}.
\end{eqnarray}
It is a semi-circle law, rigorously justified in \cite{Hutson}, and suggests a link with random matrices theory \cite{Mehta}. Lieb and Liniger have also shown, increasing the list of constraints on the density of pseudo-momenta $g$, that the semi-circular law Eq.~(\ref{guessLieb}) is a strict lower bound for the latter \cite{LiebLiniger1963}.

Heuristic arguments have suggested the following correction far from the edges in the variable $z$ \cite{Hutson, Popov}:
\begin{eqnarray}
\label{gcorr}
 g(z;\alpha)\simeq_{\alpha\ll 1,\alpha \ll |1\pm z|}\frac{\sqrt{1-z^2}}{2\pi\alpha}+\frac{1}{4\pi^2\sqrt{1-z^2}}\left[z\ln\left(\frac{1-z}{1+z}\right)+\ln\left(\frac{16\pi}{\alpha}\right)+1\right],
\end{eqnarray}
rigorously derived much later in \cite{Wadati}. To my knowledge, no further correction is explicitly known to date in this regime. I will not delve into the inversion step $\gamma(\alpha)\leftrightarrow \alpha(\gamma)$ in the main text, but refer to Appendix \ref{capa}, where a link with a classical problem is discussed, namely the calculation of the exact capacitance of a circular plate capacitor.

As far as the ground-state energy is concerned, in their seminal article Lieb and Liniger showed that $e(\gamma)\leq \gamma$, and obtained the correction
\begin{eqnarray}
\label{eBogo}
e(\gamma)\simeq_{\gamma \ll 1}\gamma-\frac{4}{3\pi}\gamma^{3/2}
\end{eqnarray}
from a Bogoliubov expansion \cite{Bogoliubov1947}. This approximation predicts negative energies at high coupling, and must be discarded then, but works surprisingly well at very small interaction strengths ($\gamma\!\lesssim\!1$), given the fact that there is no Bose-Einstein condensation. Actually, the Bogoliubov expansion Eq.~(\ref{eBogo}) coincides with the approximate result obtained by inserting Eq.~(\ref{gcorr}) in the Lieb equation, as confirmed later in \cite{Gaudin}, and detailed in \cite{Wadati, IidaWadati}.

It was then inferred on numerical grounds that the next order is such that \cite{Takahashi}
\begin{eqnarray}
\label{nextstep}
e(\gamma)=\gamma-\frac{4}{3\pi}\gamma^{3/2}+\left[\frac{1}{6}-\frac{1}{\pi^2}\right]\gamma^2+o(\gamma^2),
\end{eqnarray}
a result that agrees with later indirect (where by 'indirect', I mean that the technique involved does not rely on the Lieb equation) numerical calculations performed in \cite{Lee}, and \cite{Leebis} where the value $0.06535$ is found for the coefficient of $\gamma^2$. Equation (\ref{nextstep}) was derived quasi-rigorously much later in \cite{Widom}, also by indirect means. Actually, no fully analytical calculation based on Bethe Ansatz has confirmed this term yet, as the quite technical derivation in Ref.~\cite{Wadatibis} apparently contains a non-identified mistake.

Next step is
\begin{eqnarray}
\label{eqE}
e(\gamma)=\gamma-\frac{4}{3\pi}\gamma^{3/2}+\left[\frac{1}{6}-\frac{1}{\pi^2}\right]\gamma^2+a_3\gamma^{5/2}+O(\gamma^3),
\end{eqnarray}
where the exact fourth term, derived in closed form as multiple integrals by indirect means, was numerically evaluated to $a_3 \simeq -0.001597$ in \cite{Leebis}. A similar value was then recovered by fitting accurate numerical data \cite{Kardar}: $a_3\simeq -0.001588$, and another had been obtained previously in \cite{Takahashi}: $a_3\simeq -0.0018$.

The general structure of the weak-coupling series is very likely to be \cite{Widom}
\begin{eqnarray}
\label{conjWidom}
e(\gamma)=\sum_{k=0}^{+\infty}a_k\gamma^{1+k/2},
\end{eqnarray}
but until quite recently it was doubtful that the exact value of the coefficient $a_3$ would be identified in a close future. Ground-breaking numerical results have been obtained in \cite{Prolhac}, where the few next unknown coefficients $a_{k\geq 3}$ have been evaluated with high accuracy, such as
\begin{eqnarray}
\label{accu}
&&a_3\simeq - 0.00158769986550594498929,\nonumber\\
&&a_4 \simeq - 0.00016846018782773903545,\nonumber\\
&&a_5 \simeq - 0.0000208649733584017408,
\end{eqnarray}
up to $a_{10}$ included, by an appropriate sampling in numerical integration of the Lieb equation, and a method that accelerates the convergence. In particular, $a_4$ is in relatively good agreement with the approximate value $a_4\simeq -0.000171$ previously obtained in \cite{Kardar}. The fabulous accuracy of (\ref{accu}) has allowed for and has been increased by an advance from experimental analytical number theory along the three subsequent arXiv versions of Ref.~\cite{Prolhac}.

I guessed the value
\begin{eqnarray}
\label{conjLang}
a_3=\left(\frac{3}{8}\zeta(3)-\frac{1}{2}\right)\frac{1}{\pi^3},
\end{eqnarray}
based on the following heuristic grounds: in \cite{Leebis}, an overall factor $1/\pi^3$ is found, and $a_3$ is written as a sum of two integrals, possibly corresponding to a sum of \textit{two} types of terms.
Also, in the first arXiv version of \cite{Prolhac}, Prolhac wrote that he could not identify $a_3$ as \textit{a low-order polynomial in $1/\pi$ with rational coefficients}.
Combining the two previous items, and in view of the relative simplicity of $a_0$, $a_1$ and $a_2$, one can legitimely infer that the factor of $1/\pi^3$ in $a_3$ should be irrational. One can then think of $\zeta(3)$, where $\zeta$ is the Riemann zeta function. Indeed, the previous coefficient, $a_2$, can be written as $\left(\zeta(2)-1\right)\frac{1}{\pi^2}$, and $\zeta(3)$ is irrational \cite{Apery}.

My conjecture Eq.~(\ref{conjLang}) has inspired Prolhac, who guessed
\begin{eqnarray}
a_4=\frac{a_3}{3\pi}=\left(\frac{1}{8}\zeta(3)-\frac{1}{6}\right)\frac{1}{\pi^4}
\end{eqnarray}
and even

\begin{eqnarray}
a_5=\left(-\frac{45}{1024}\zeta(5)+\frac{15}{256}\zeta(3)-\frac{1}{32}\right)\frac{1}{\pi^5},
\end{eqnarray}
using a code that identifies the rational coefficients of a linear combination of peculiar values of the zeta function when given a target value \cite{Prolhac}. In principle, numerical values of the next coefficients could be obtained by iterating the procedure further. However, guessing other exact numbers without further insight seems difficult, as the relative accuracy of their numerical values decreases at each step, while the required precision is expected to increase, in view of the apparently increasing number of terms involved in the linear combination. Although the generating function of the exact coefficients of the weakly-interacting expansion still remains quite obscure, it seems reasonable to guess that $a_k$ naturally contains a factor $1/\pi^k$ at all orders $k$, so that
\begin{eqnarray}
\label{myconjweake}
e(\gamma)=\sum_{k=0}^{+\infty}\frac{\tilde{a}_k}{\pi^k}\gamma^{1+k/2}.
\end{eqnarray}

\subsection{Strong- to intermediate coupling regime}
\label{GSenergy}

While the weak-coupling regime is tremendously difficult to tackle in the close vicinity of the singularity, at strong coupling the problem, though far from trivial, is much easier to deal with in comparison.
In the Tonks-Girardeau regime ($\gamma\!\to\!+\infty$), the reduced dimensionless energy is
\begin{eqnarray}
e^{TG}\!=\!\frac{\pi^2}{3},
\end{eqnarray}
and coincides with the well-known result for spinless noninteracting fermions, Eq.~(\ref{energyTG}), due to the Bose-Fermi mapping. It corresponds to a uniform distribution of pseudo-momenta, 
\begin{eqnarray}
\label{gTG}
g^{TG}(z)=\frac{1}{2\pi}\Theta(1-|z|).
\end{eqnarray}
For $z\in[-1,1]$, i.e. inside the pseudo-Fermi sea, finite-interaction corrections to Eq.~(\ref{gTG}) can be expressed as
\begin{eqnarray}
\label{gpoltrunc}
g(z;\alpha)\simeq_{\alpha \gg 1} \sum_{k=0}^{k_{m}}\frac{P_k(z)}{\alpha^k},
\end{eqnarray}
where $\{P_k\}_{k=0,\dots,k_m}$ are polynomials and $k_m$ is a cut-off. Then, this truncated expansion can be used to approximate $\alpha(\gamma)$, inverted in $\gamma(\alpha)$, and yields the corresponding expansion of the ground-state energy:
\begin{eqnarray}
\label{strongcouplinge}
e(\gamma)\simeq_{\gamma\gg 1}\sum_{k=0}^{k_{m}}\frac{e_k}{\gamma^k},
\end{eqnarray}
where $\{e_k\}_{k=0,\dots,k_m}$ are real coefficients.
Surprisingly, few non-trivial corrections to the Tonks-Girardeau limit are available in the literature, in view of the relative simplicity of the first few steps. In \cite{Zvonarev}, this procedure has been pushed to sixth order.

A systematic method was proposed by Ristivojevic in \cite{Ristivojevic}, where it was used to generate corrections to the Tonks-Girardeau regime up to order $8$. In \cite{LangHekkingMinguzzi2017}, I have studied this method in detail, and used it to obtain analytical expressions up to order $20$ in $1/\gamma$. In a few words, the method, detailed in Appendix \ref{Ristimethods}, yields an approximation to the density of pseudo-momenta of the form
\begin{eqnarray}
\label{dev}
g(z;\alpha,M)=\!\sum_{k=0}^{2M+2}\sum_{j=0}^{M}g_{jk}\frac{z^{2j}}{\alpha^{k}},
\end{eqnarray}
where the matrix coefficients $g_{jk}$ are, by construction, polynomials in $1/\pi$ with rational coefficients, and $M$ is an integer cut-off such that the truncated density of pseudo-momenta $g(z;\alpha,M)$ converges to $g(z;\alpha)$ as $M\!\to\!+\infty$.

The original article \cite{Ristivojevic} and ours \cite{LangHekkingMinguzzi2017}, together, give a faithful account of the strengths and weaknesses of this method: a major positive trait is that it yields two orders of perturbation theory in $1/\alpha$ at each step, and is automatically consistent to all orders.
The lowest interaction strength attainable within this expansion is $\alpha\!=\!2$, since the procedure relies on a peculiar expansion of the Lorentzian kernel in Eq.~(\ref{Fredholm}). This corresponds to a Lieb parameter $\gamma(\alpha\!=\!2)\!\simeq\!4.527$, which is an intermediate interaction strength. A priori, this value is small enough to recombine with the available expansions in the weakly-interacting regime, and thus obtain accurate estimates of the ground-state energy over the whole range of repulsive interactions. However, it could as well be seen as a strong limitation of the method, all the more so as, since convergence with $M$ to the exact solution is slow, one can not reasonably expect to obtain reliable results below $\gamma\!\simeq\!5$ even with a huge number of corrections.

This drawback stems from the fact that capturing the correct behavior of the density of pseudo-momenta $g$ as a function of $z$ in the whole interval $[-1,1]$ is crucial to obtain accurate expressions of the ground-state energy, whereas the approximation Eq.~(\ref{dev}) converges slowly to the exact value close to $z\!=\!\pm 1$ since the Taylor expansion is performed at the origin. This reflects in the fact that the maximum exponent of $z^2$ varies more slowly with $M$ than the one of $1/\alpha$. 
What is more, if one is interested in explicit analytical results, at increasing $M$ the method quickly yields too unhandy expressions for the function $g$, as it generates $1+(M+1)(M+2)(M+3)/3$ terms.
To finish with, it is difficult to evaluate the accuracy of a given approximation in a rigorous and systematic way.

\begin{figure}
\includegraphics[width=9cm, keepaspectratio, angle=0]{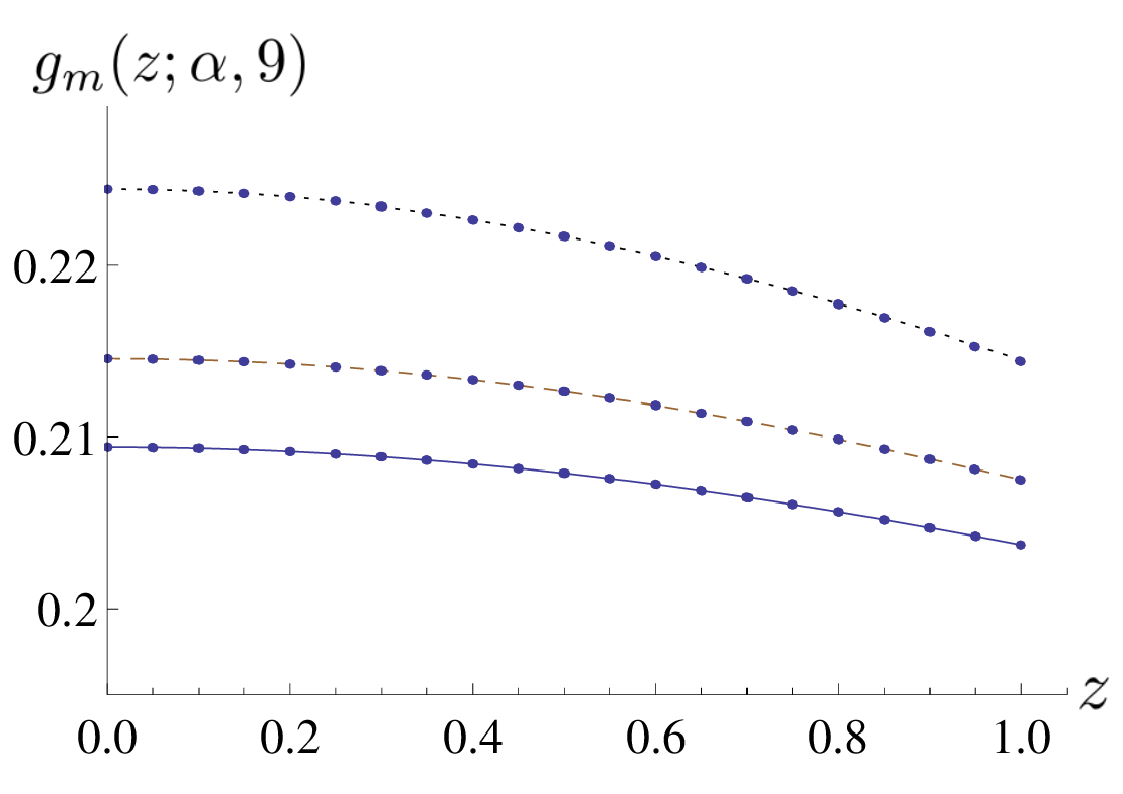}
\caption{Dimensionless function $g_{m}(z;\alpha,9)$, mean of the $18^{th}$ and $20^{th}$ order expansions in $1/\alpha$ of the density of pseudo-momenta $g(z;\alpha)$, as a function of the dimensionless variable $z$, at dimensionless parameters $\alpha\!=\!2.5$ (solid, blue), $\alpha\!=\!2.3$ (dashed, brown) and $\alpha\!=\!2$ (dotted, black) from bottom to top, compared to the corresponding numerically exact solutions (blue dots). Only a few numerical values are shown to improve the visibility, and numerical error is within the size of the dots.}
\label{Figaccu}
\end{figure}

Consistency at all orders is obviously the main quality of this method, the other points are rather drawbacks. After putting them into light, I have developed various methods to circumvent them during my thesis, but could not fix them simultaneously. The main improvements I have proposed are the following:

a) the huge number of corrections needed to reach $\alpha\simeq 2$ with good accuracy close to the Fermi surface $z\!=\!\pm 1$ seems redhibitory at first, but I have noticed that the arithmetic average of two consecutive orders in $M$, denoted by 
\begin{eqnarray}
\label{gmeanM}
 g_m(z;\alpha,M)=\frac{g(z;\alpha,M)+g(z;\alpha,M\!-\!1)}{2},
\end{eqnarray}
dramatically increases the precision. Figure~\ref{Figaccu} illustrates the excellent agreement, at $M\!=\!9$, with numerical calculations for $\alpha\!\in\![2,+\infty[$. Another approach consists in truncating expansions to their highest odd order in $1/\alpha$, more accurate than the even one, as already pointed out in \cite{Rao}.

b) I have also found a way to avoid expanding the Lorentzian kernel in Eq.~(\ref{Fredholm}), and adapt Ristivojevic's method to the whole range of repulsive interaction strengths $]0,+\infty[$, as detailed in Appendix \ref{Fabrikantmethod}. However, self-consistency at all orders, that was a quality of the method, is lost. Furthermore, the analytical expressions obtained are quite complicated, urging at a numerical evaluation at the very last step of the procedure. I have pushed the method to order $50$ in $z$, yielding the ground-state energy with machine precision over a very wide range of strong and intermediate interaction strengths and an interesting comparison point for other analytical and numerical approaches.

c) Last but not least, I have looked for compact analytical expressions, by identifying structures in the bare result of Ristivojevic's method. As far as the density of pseudo-momenta $g$ is concerned, I refer to \cite{LangHekkingMinguzzi2017} for a detailed account of these compact notations. My approach opens a new line of research, but I have not investigated it deeply enough to obtain fully satisfying expressions. Comparatively, the same method revealed quite powerful when applied to the ground-state energy.

Once again, the idea is based on experimental number theory. This time, the aim is not to guess numerical values of unknown coefficients as in the weakly-interacting regime, but rather to put regular patterns into light by scrutinizing the first few terms, and guess subsequent ones without actually computing them. To do so, I have considered all operations that yield the strong-coupling expansion of $e(\gamma)$, Eq.~(\ref{strongcouplinge}), as a black box, and focused on the result. A bit of reflexion hints at writing
\begin{eqnarray}
\label{struce}
 \frac{e(\gamma)}{e^{TG}}=\sum_{n=0}^{+\infty}\tilde{e}_n(\gamma),
\end{eqnarray}
where the index $n$ denotes a somewhat elusive notion of complexity, that corresponds to the level of difficulty to identify the pattern that defines $\tilde{e}_n$.

Focusing on the strong-coupling expansion Eq.~(\ref{strongcouplinge}), I have identified a first sequence of terms, conjectured that they appear at all higher orders as well, and resummed the series. I obtained
\begin{eqnarray}
\label{Liebe0}
\tilde{e}_0(\gamma)=\sum_{k=0}^{+\infty}\binom{k+1}{1}\left(-\frac{2}{\gamma}\right)^k\!=\frac{\gamma^2}{(2+\gamma)^2},
\end{eqnarray}
and noticed that the final expression for $\tilde{e}_0$ corresponds to Lieb and Liniger's approximate solution, that assumes a uniform density of pseudo-momenta in Eq.~(\ref{Fredholm}) \cite{LiebLiniger1963}. I have also found that the intermediate step in Eq.~(\ref{Liebe0}) appears in an appendix of Ref.~\cite{Astrakharchik2010}, but writing $k\!+\!1$ instead of the equivalent binomial interpretation, which is my main step forward, as will be seen below.

Using the strong-coupling expansion of $e(\gamma)$ up to $20^{th}$ order in $1/\gamma$, and guided by the property 
\begin{eqnarray}
 \sum_{k=0}^{+\infty}\binom{k\!+\!3n\!+\!1}{3n\!+\!1}\left(-\frac{2}{\gamma}\right)^k=\left(\frac{\gamma}{\gamma+2}\right)^{3n+2},
\end{eqnarray}
I have conjectured that the structure of the terms of complexity $n\!\geq\!1$ is
\begin{eqnarray}
\label{exp}
\tilde{e}_n(\gamma)=\frac{\pi^{2n}\gamma^2\mathcal{L}_n(\gamma)}{(2+\gamma)^{3n+2}},
\end{eqnarray}
where $\mathcal{L}_n$ is a polynomial of degree $n\!-\!1$ with non-zero, rational coefficients of alternate signs. The complexity turns out to be naturally related to the index $n$ in the right-hand side of this equation, and can be re-defined, a posteriori, from the latter. I have identified the first few polynomials as
\begin{eqnarray}
\label{Lpol}
\mathcal{L}_1(\gamma)\!\!\!\!&=&\!\!\!\!\frac{32}{15}\nonumber,\\
\mathcal{L}_2(\gamma)\!\!\!\!&=&\!\!\!\!-\frac{96}{35}\gamma+\frac{848}{315},\nonumber\\
\mathcal{L}_3(\gamma)\!\!\!\!&=&\!\!\!\!\frac{512}{105}\gamma^2-\frac{4352}{525}\gamma+\frac{13184}{4725},\nonumber\\
\mathcal{L}_4(\gamma)\!\!\!\!&=&\!\!\!\!-\frac{1024}{99}\gamma^3+\frac{131584}{5775}\gamma^2-\frac{4096}{275}\gamma+\frac{11776}{3465},\nonumber\\
\mathcal{L}_5(\gamma)\!\!\!\!&=&\!\!\!\!\frac{24576}{1001}\gamma^4-\frac{296050688}{4729725}\gamma^3+\frac{453367808}{7882875}\gamma^2-\frac{227944448}{7882875}\gamma+\frac{533377024}{212837625},\nonumber\\
\mathcal{L}_6(\gamma)\!\!\!\!&=&\!\!\!\!-\frac{4096}{65}\gamma^5\!+\!\frac{6140928}{35035}\gamma^4\!-\!\frac{4695891968}{23648625}\gamma^3\!+\!\frac{3710763008}{23648625}\gamma^2\nonumber\\
&&\!\!\!-\frac{152281088}{4729725}\gamma\!+\!\frac{134336512}{42567525}.
\end{eqnarray}
I have also conjectured that the coefficient of the highest-degree monomial of $\mathcal{L}_n$, written as
\begin{eqnarray}
\mathcal{L}_n(X)=\sum_{k=0}^{n-1}l_kX^k,
\end{eqnarray}
is
\begin{eqnarray}
\label{ConjHigh}
l_{n-1}=\frac{3\times(-1)^{n+1}\times2^{2n+3}}{(n+2)(2n+1)(2n+3)}.
\end{eqnarray}
Consulting the literature once more at this stage, it appears that the first correction $\tilde{e}_1$ had been rigorously predicted in \cite{IidaWadati}, supporting my conjecture on the structure of $e(\gamma)$ in the strong-coupling regime, Eq.~(\ref{exp}). Later on, Prolhac checked numerically that all coefficients in Eq.~(\ref{Lpol}) are correct, and that Eq.~(\ref{ConjHigh}) is still valid at larger values of $n$ \cite{Prolhacprivate}.

Innocent as it may look (after all, it is just another way of writing the strong-coupling expansion), the structure provided by Eq.~(\ref{exp}) has huge advantages.
Although the structure Eq.~(\ref{exp}) was not obvious at first, now that it has been found, identifying the polynomials $\mathcal{L}_n$ from Eq.~(\ref{strongcouplinge}) to all accessible orders becomes a trivial task. The expressions thereby obtained are more compact than the strong-coupling expansion, Eq.~(\ref{strongcouplinge}), and correspond to a partial resummation of the asymptotic series.
Last but not least, contrary to the $1/\gamma$ expansion, the combination of Eqs.~(\ref{struce}) and (\ref{exp}), truncated to the maximal order to which the polynomials are known in Eq.~(\ref{Lpol}), does not diverge at low $\gamma$. This fact considerably widens the validity range of the expansion.

Nonetheless, a few aspects are not satisfying so far. Progressively higher order expansions in $1/\gamma$ are needed to identify the polynomials in Eq.~(\ref{Lpol}). The expansion Eq.~(\ref{exp}) remains conjectural, and although a proof could be given using the techniques of \cite{IidaWadati}, this direct approach looks tremendously complicated. To finish with, I did not manage to identify the generating function of the polynomials in Eq.~(\ref{Lpol}), except for the first coefficient, Eq.~(\ref{ConjHigh}), preventing from infering higher-order polynomials in Eq.~(\ref{Lpol}) without relying on the $1/\gamma$ expansion, Eq.~(\ref{strongcouplinge}). Identifying this generating function may allow for a full resummation of the series, and thus to explicitly obtain the exact ground-state energy of the Lieb-Liniger model.

\subsection{Illustrations}

\begin{figure}
\includegraphics[width=10cm, keepaspectratio, angle=0]{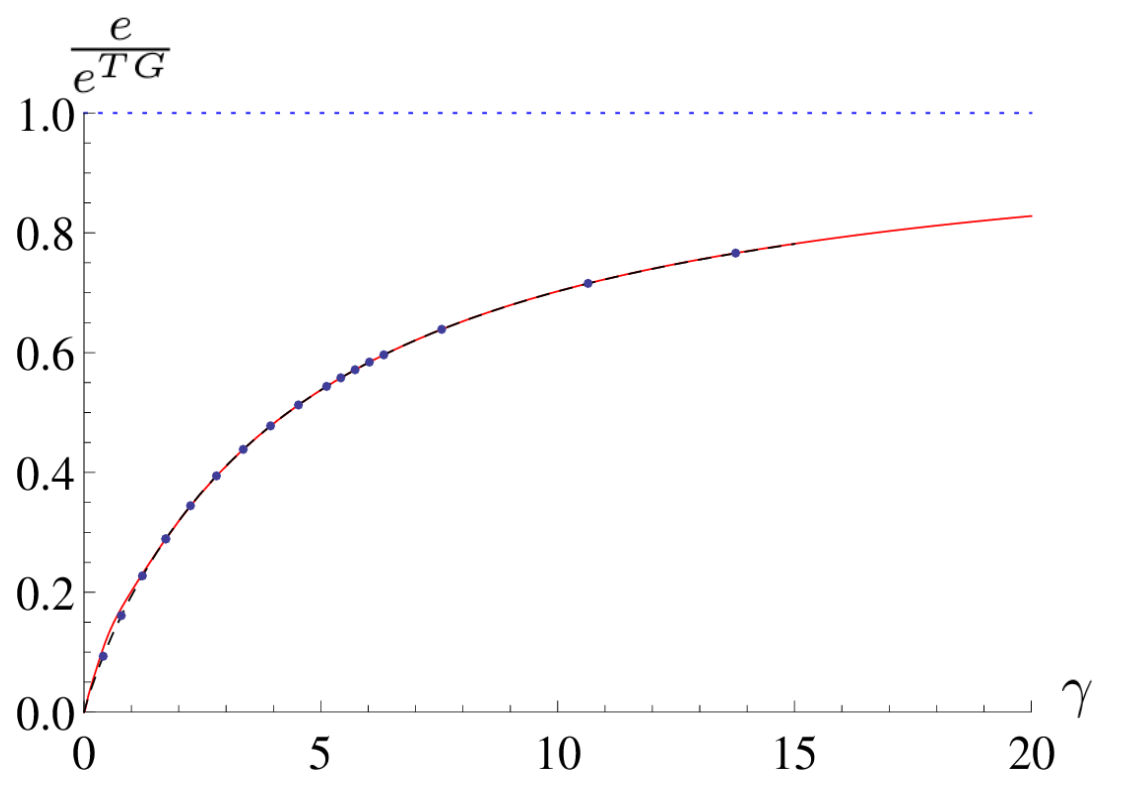}
\caption{Dimensionless ground state energy per particle $e$ normalized to its value in the Tonks-Girardeau limit $e^{TG}=\pi^2/3$ (dotted, blue), as a function of the dimensionless interaction strength $\gamma$: conjectural expansion at large $\gamma$ (solid, red) as given by Eq.~(\ref{exp}) to sixth order, small $\gamma$ expansion (dashed, black) and numerics (blue points).}
\label{Fig4}
\end{figure}

Bridging weak- and strong-coupling expansions, I have obtained the ground-state energy of the Lieb-Liniger model with good accuracy over the whole range of repulsive interactions, as illustrated in Fig.~\ref{Fig4}.

\begin{figure}
\includegraphics[width=10cm, keepaspectratio, angle=0]{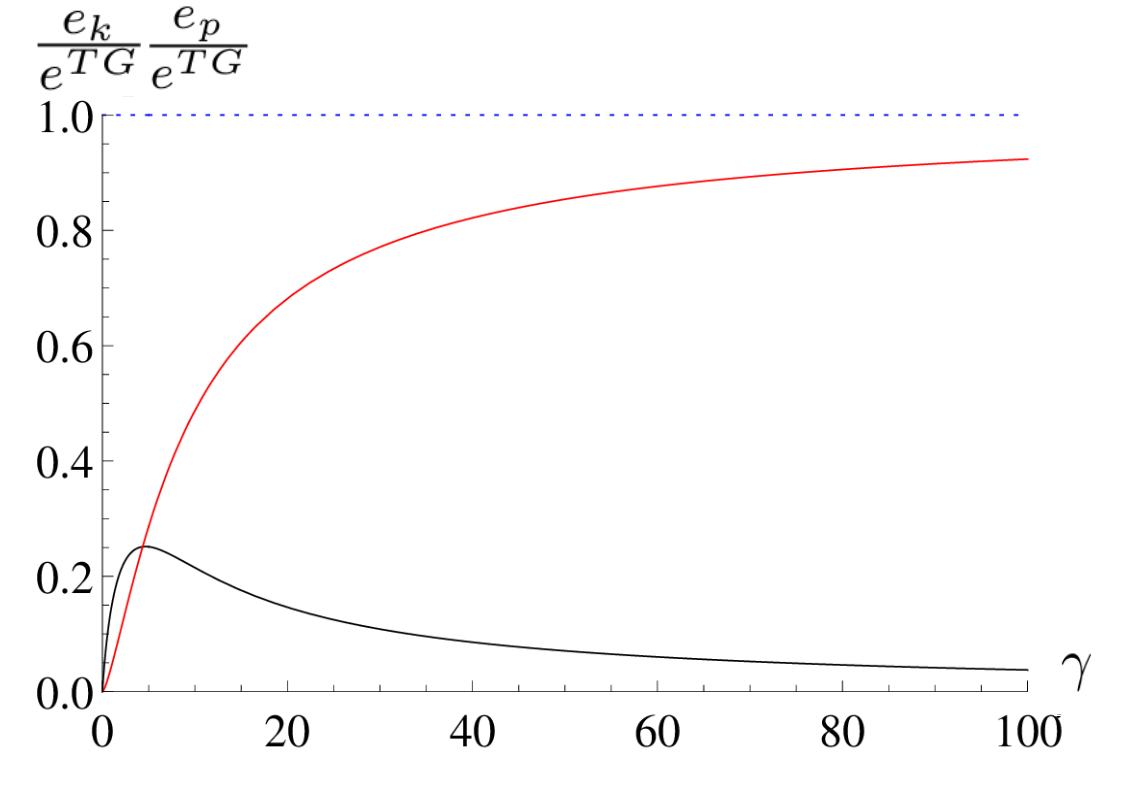}
\caption{Dimensionless ground-state kinetic energy per particle (red) and interaction energy per particle (black), normalized to the total energy per particle in the Tonks-Girardeau limit $e^{TG}$, as a function of the dimensionless interaction strength $\gamma$. The horizontal line (dotted, blue) is a guide to the eye. The results are indistinguishable from the numerical estimation of Ref.~\cite{XuRigol2015}.}
\label{Fig5}
\end{figure}

It can be split into two parts, that correspond to kinetic and interaction energy respectively, as \cite{LiebLiniger1963}
\begin{eqnarray}
e(\gamma)=[e(\gamma)-\gamma e'(\gamma)]+\gamma e'(\gamma)=e_{kin}(\gamma)+e_{int}(\gamma).
\end{eqnarray}
As illustrated in Fig.~\ref{Fig5}, the dimensionless kinetic energy per particle, $e_{kin}$, is a monotonic function of $\gamma$, while the dimensionless interaction energy $e_{int}$ (positive since interactions are repulsive), reaches a global maximum at an intermediate interaction strength.

This fact can be qualitatively understood on physical grounds. When $\gamma\!=\!0$, i.e. for a noninteracting Bose gas, the density of quasi-momenta is a Dirac-delta function, bosons are individually at rest, and the kinetic energy of the gas is zero. The interaction energy is null too, by definition. Switching on interactions adiabatically, the interaction energy increases abruptly, but can be treated perturbatively.
In the opposite, Tonks-Girardeau limit, in $k$-space the gas is equivalent to noninteracting fermions due to the Bose-Fermi mapping, thus its interaction energy is also zero. While decreasing interactions, the gas is equivalent to weakly-interacting fermions \cite{CheonShigehara1999}, thus the interaction energy increases, but at a slow pace due to the remnant artificial Pauli principle. By continuity, at intermediate $\gamma$ the interaction energy must reach a maximum, that corresponds to a subtle interplay of statistics in $k$-space and interactions in real space. Since the interaction energy is at its apex, this must be the regime where perturbative approaches are not adapted, explaining the counter-intuitive fact that intermediate interactions are the least amenable to analytical methods.

\begin{figure}
\includegraphics[width=12cm, keepaspectratio, angle=0]{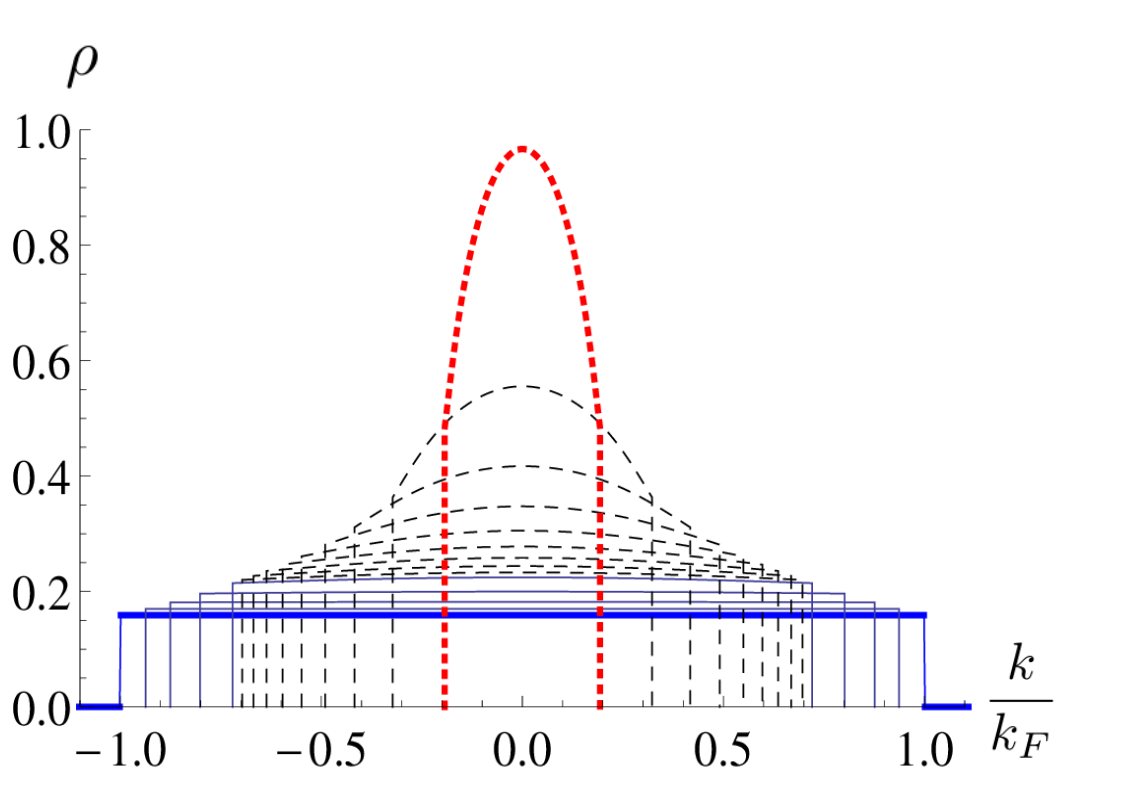}
\caption{Density of pseudo-momenta $\rho$ as a function of dimensionless pseudo-momentum $k/k_F$ for several interaction strengths. Different colors and line styles represent various approximations. From bottom to top, one sees the exact result in the Tonks-Girardeau regime (thick, blue), then four curves corresponding to dimensionless parameters $\alpha\!=\!10$, $5$, $3$ and $2$ respectively (solid, blue) obtained from analytical methods and a Monte-Carlo algorithm to solve the Lieb equation Eq.~(\ref{Fredholm}) (indistinguishable from each other). Above, another set of curves corresponds to interaction strengths from $\alpha\!=\!1.8$ to $\alpha\!=\!0.4$ with step $-0.2$ and from bottom to top (dashed, black), obtained from a Monte-Carlo algorithm, where again analytical and numerical results are indistinguishable. Finally, I also plot the result at $\alpha\!=\!0.2$ from the method of Appendix \ref{Fabrikantmethod} (dotted, red).}
\label{Fig6}
\end{figure}

An alternative interpretation is utterly based on the density of quasi-momenta in the original units, using Fig.~\ref{Fig6}. This quantity interpolates between a top-hat function in the Tonks-Girardeau regime, and a Dirac-delta for noninteracting bosons. The density of quasi-momenta is relatively flat in a wide range of large interaction strengths, explaining why the approximation $e(\gamma)\!\simeq\!\tilde{e}_0(\gamma)$ works so well from the Tonks-Girardeau regime to intermediate interaction strengths.

To sum up with, in this section I have explained how coordinate Bethe Ansatz gives access to the exact ground-state energy of the Lieb-Liniger model, both at finite boson number and in the thermodynamic limit, as a set of coupled equations. The latter is easily solved numerically for a given value of the interaction strength, but its analytical, exact solution is still unknown. In recent years, both weak- and strong-coupling regimes have been theoretically addressed in a systematic way, allowing in principle to obtain exact expansions to arbitrary order, but the procedures remain quite complicated when high orders are required.

Our understanding of the exact solution is improving too. I have performed a tentative partial resummation of the strong-coupling series expansion, and the weak-coupling one seems to contain a rich and interesting structure involving the Riemann zeta function at odd arguments. These expressions are known with high enough accuracy to numerically match at intermediate coupling. The relative error is of the order of a few per thousands over the whole range of interaction strengths, and semi-analytical techniques allow to reach machine precision if needed.
Hence, while the Lieb-Liniger model was 'solved' from the very beginning in the sense that its exact ground-state energy was expressed in closed form as the solution of a set of equations, it is now solved in a stronger sense, both analytically and numerically. Strictly speaking, the problem of the ground-state energy is still open.

An even stronger definition of 'solving' a model includes the knowledge of correlation functions, an issue that I tackle from the next section on.

\section{Local correlation functions}

One of the main limitations of the coordinate Bethe Ansatz approach is that it provides only with an implicit knowledge of the wavefunction close to the thermodynamic limit, since it is a superposition of an exponentially large number of terms as a function of the particle number. Direct calculation of correlation functions based on the explicit many-body wavefunction remains a formidable task and, for many practical purposes, is not attainable.

The algebraic Bethe Ansatz approach, on the other hand, provides a more compact expression for the eigenstates. This formulation in turn allows to express many correlation functions as Fredholm determinants. Although very elegant, these expressions still require some work to provide useful results. Actually, thanks to interesting mathematical properties, the explicit knowledge of the many-body wavefunction is not necessary to obtain the \textit{local} correlation functions of the Lieb-Liniger model, and coordinate Bethe Ansatz is sufficient.

The $k$-body, local correlation function in the ground state is defined as
\begin{eqnarray}
\label{defgk}
g_k=\frac{\langle [\hat{\psi}^{\dagger}(0)]^k[\hat{\psi}(0)]^k\rangle}{n_0^k},
\end{eqnarray}
where $\langle . \rangle$ represents the ground-state average. With this choice of normalization, $g_k$ represents the probability of observing $k$ bosons simultaneously at the same place. As a consequence of this definition, the following equality holds trivially:
\begin{eqnarray}
g_1(\gamma)=1.
\end{eqnarray}
Indeed, the numerator of the right-hand side of Eq.~(\ref{defgk}) coincides with the mean linear density since the gas is uniform, and so does its denominator.

It is expected that all higher-order local correlations depend on the interaction strength. The following qualitative properties are quite easily obtained: in the non-interacting gas, i.e. if $\gamma\!=\!0$, $g_k\!=\!1$ to all orders $k$. In the Tonks-Girardeau regime, since interactions mimic the Pauli principle, two bosons can not be observed simultaneously at the same place. A fortiori, larger local clusters are forbidden too, so $g_k^{TG}\!=\!0$ to all orders $k>1$. Inbetween, at finite interaction strength the probability to observe $k\!+\!1$ particles at the same place is strictly lower than the one to observe $k$ particles, thus $0\!<\!g_{k+1}(\gamma)\!<\!g_k(\gamma)$ for any finite $\gamma$. One also expects $|g_{k+1}'(\gamma)|\!>\!|g_k'(\gamma)|\!>\!0$ (every local correlation function, except $g_1$, is a strictly-decreasing function of $\gamma$). These properties imply that high-order correlation functions are difficult to measure by in-situ observations, in particular close to the Tonks-Girardeau regime.

After these general comments, I turn to more specific cases. I will focus on $g_2$ and $g_3$, that are the most experimentally-relevant local correlation functions \cite{Kinoshita2005, Tolra2004, ArmijoJacqminKheruntsyanBouchoule2010, Haller2011}. From the theoretical point of view, the second-order correlation function is easily obtained from the ground-state energy, as the Hellmann-Feynman theorem yields \cite{GangardtShlyapnikov2003}
\begin{eqnarray}
\label{HF}
g_2(\gamma)\!=\!e'(\gamma).
\end{eqnarray}
According to this equation, the fact that $e$ is an increasing function of the Lieb parameter $\gamma$ is a direct consequence of the positivity of $g_2$.

As a general rule, all local correlation functions are expected to be related (possibly in a fairly non-trivial way) to moments of the density of pseudo-momenta, defined as
\begin{eqnarray}
\label{defmoments}
 e_k(\gamma)=\frac{\int_{-1}^1dz\, z^k g[z;\alpha(\gamma)]}{\{\int_{-1}^1dz\,g[z;\alpha(\gamma)]\}^{k+1}}.
\end{eqnarray}
Note that odd-order moments are null, since $g$ is an even function of $z$, $e_0\!=\!1$, and $e_2(\gamma)\!=\!e(\gamma)$. In particular, the third-order local correlation function governs the rates of inelastic processes, such as three-body recombination and photoassociation in pair collisions. It is expressed in terms of the two first non-trivial moments as \cite{Cheianov, Smith}
\begin{eqnarray}
\label{g3formula}
g_3(\gamma)=\frac{3}{2\gamma}\frac{de_4}{d\gamma}-\frac{5e_4}{\gamma^2}+\left(1+\frac{\gamma}{2}\right)\frac{de_2}{d\gamma}-2\frac{e_2}{\gamma}-3\frac{e_2}{\gamma}\frac{de_2}{d\gamma}+9\frac{e_2^2}{\gamma^2}.
\end{eqnarray}
This expression is significantly more complicated than Eq.~(\ref{HF}), and the situation is not likely to improve at higher orders, where similar expressions are still unknown.

In \cite{Cheianov}, the solution to the Bethe Ansatz equations has been found numerically, and useful approximations to the three-body local correlation function have been obtained by fitting this numerical solution, namely:
\begin{eqnarray}
g_3(\gamma)\simeq \frac{1-6\pi^{-1} \gamma^{1/2}+1.2656 \gamma-0.2959 \gamma^{3/2}}{1-0.2262 \gamma-0.1981 \gamma^{3/2}}, 0\leq \gamma \leq 1,
\end{eqnarray}
\begin{eqnarray}
g_3(\gamma)\simeq \frac{0.705-0.107 \gamma+5.08*10^{-3} \gamma^2}{1+3.41 \gamma+0.903 \gamma^2+0.495 \gamma^3}, 1\leq \gamma \leq 7,
\end{eqnarray}
\begin{eqnarray}
g_3(\gamma)\simeq \frac{16\pi^6}{15\gamma^6}\frac{9.43-5.40 \gamma+\gamma^2}{89.32+10.19 \gamma+\gamma^2}, 7\leq \gamma \leq 30
\end{eqnarray}
with a relative error lower than $2\%$ according to the authors. For $\gamma\geq 30$, it is tacitly assumed that the available strong-coupling expansions of $g_3$ are at least as accurate.

\begin{figure}
\includegraphics[width=9cm, keepaspectratio, angle=0]{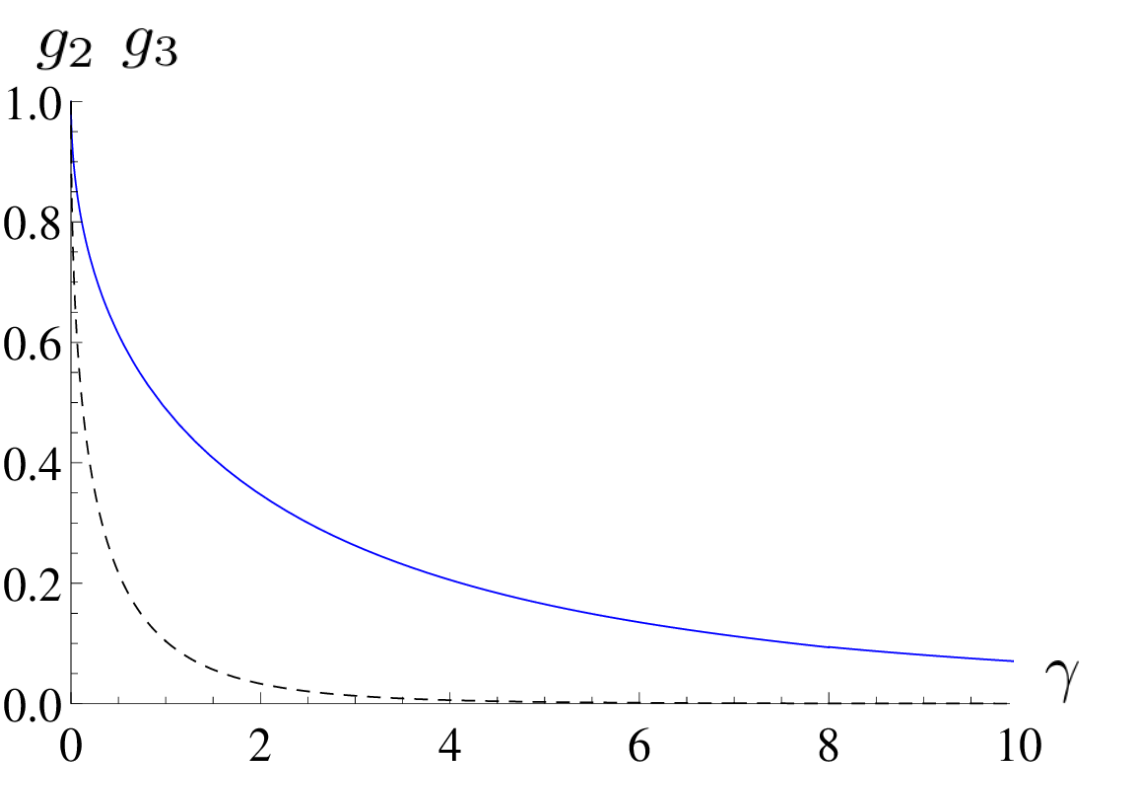}
\caption{Dimensionless local correlation functions $g_2$ (solid, blue) and $g_3$ (dashed, black) as functions of the dimensionless interaction strength $\gamma$. Three-body processes are strongly suppressed at high interaction strength but become of the same order of magnitude as two-body processes in the quasi-condensate regime.}
\label{Fig7}
\end{figure}

Actually, the dominant term of the strong-coupling asymptotic expansion of all local correlation functions is known and reads \cite{GangardtShlyapnikov2003NJP, KormosMussardoTrombettoni}
\begin{eqnarray}
g_k\simeq_{\gamma\gg 1}\frac{k!}{2^k}\left(\frac{\pi}{\gamma}\right)^{k(k-1)}I_k
\end{eqnarray}
where
\begin{eqnarray}
I_k=\int_{-1}^1dk_1\,\dots\int_{-1}^1dk_k\!\prod_{\{i<j\leq k\}}\!(k_i-k_j)^2, 
\end{eqnarray}
and has even been generalized to \cite{NandaniRomerTanGuan}:
\begin{eqnarray}
g_k=_{\gamma\gg 1}\frac{\left(\prod_{j=1}^kj!\right)^2\left(1-\frac{2}{\gamma}\right)^{k^2-1}}{\left[\prod_{j=1}^{k-1}(2j-1)!!\right]^2(2k-1)!!}\left(\frac{\pi}{\gamma}\right)^{k(k-1)}+\dots
\end{eqnarray}
The fourth-order local correlation $g_4(\gamma)$ has been constructed using a different approach in \cite{Pozsgay2011} (see Appendix \ref{gkk}), but this correlation function has not been probed experimentally yet.

\begin{figure}
\includegraphics[width=9cm, keepaspectratio, angle=0]{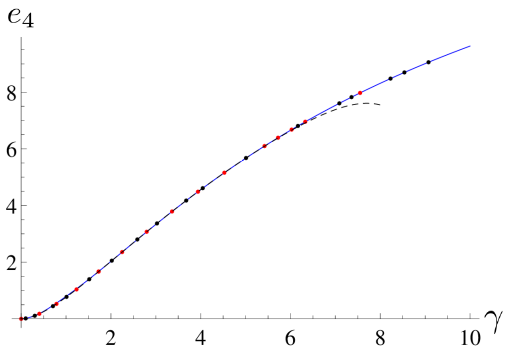}
\caption{Dimensionless fourth moment of the distribution of quasi-momenta, $e_4$, as a function of the dimensionless interaction strength $\gamma$. The analytical result from the conjecture (\ref{e4conj}) (solid, blue), and from the conjecture in the weakly-interacting regime, Eq.~(\ref{conjweak}), with appropriate coefficients (dashed, black), are in excellent agreement with independent accurate numerical evaluations (red and black dots).}
\label{FFig8}
\end{figure}

As one can see on the previous examples, a key ingredient to evaluate a local correlation function $g_k$ by coordinate Bethe Ansatz is to evaluate the moments of the density of pseudo-momenta, given by Eq.~(\ref{defmoments}), to the corresponding order. Thanks to my good knowledge of $g(z;\alpha)$, I have access to their strongly-interacting expansion. In particular, I could evaluate $e_4$, that enters $g_3$ through Eq.~(\ref{g3formula}). Based once more on strong-coupling expansions to order $20$, I have conjectured that
\begin{align}
\label{conjstrong}
e_{2k}(\gamma)=\left(\frac{\gamma}{2+\gamma}\right)^{\!2k}\sum_{i=0}^{+\infty}\frac{\pi^{2(k+i)}}{(2+\gamma)^{3i}}\mathcal{L}_{2k,i}(\gamma),
\end{align}
where $\mathcal{L}_{2k,i}$ are polynomials with rational coefficients, such that 
\begin{eqnarray}
\mathcal{L}_{2k,0}=\frac{1}{2k+1},
\end{eqnarray}
and $\mathcal{L}_{2k,i\geq 1}$ is of degree $i\!-\!1$. This generalizes the corresponding conjecture for $e_2$, Eq.~(\ref{exp}). In particular, I have identified:
\begin{align}
\label{e4conj}
&\mathcal{L}_{4,1}(\gamma)=\frac{32}{35},\nonumber\\
&\mathcal{L}_{4,2}(\gamma)=-\frac{1984}{1575}\gamma+\frac{3424}{1575},\nonumber\\
&\mathcal{L}_{4,3}(\gamma)=\frac{8192}{3465}\gamma^2-\frac{37376}{5775}\gamma+\frac{169728}{45045},\nonumber\\
&\mathcal{L}_{4,4}(\gamma)=-\frac{47104}{9009}\gamma^3+\frac{59337728}{3378375}\gamma^2-\frac{61582336}{3378375}\gamma+\frac{137573632}{23648625},\nonumber\\
&\mathcal{L}_{4,5}(\gamma)=\frac{192512}{15015}\gamma^4-\frac{765952}{15925}\gamma^3+\frac{80326709248}{1206079875}\gamma^2-\frac{594448384}{14189175}\gamma+\frac{295196160000}{38192529375},\nonumber\\
&\mathcal{L}_{4,6}(\gamma)=-\frac{335872}{9945}\gamma^5\!+\!\frac{132872192}{984555}\gamma^4\!-\!\frac{2316542492672}{10416144375}\gamma^3\!+\!\frac{3689660465152}{18091198125}\gamma^2\nonumber\\
&\!-\!\frac{184095784026112}{2406129350625}\gamma\!+\!\frac{12238234443776}{1260353469375}.
\end{align}
It is worth mentionning that the validity range of the strong-coupling approximation in $1/\gamma$ increases towards weaker interactions with $k$, as illustrated in Fig.~\ref{FFig8}.

In the weakly-interacting regime, I have conjectured that the even moments have the following structure:
\begin{align}
\label{conjweak}
e_{2k}(\gamma)=\sum_{i=0}^{+\infty}\frac{\tilde{a}_{2k,i}}{\pi^i}\gamma^{k+i/2},
\end{align}
generalizing my previous conjecture Eq.~(\ref{myconjweake}) for the ground-state energy. Using Eq.~(\ref{gcorr}), I derived
\begin{align}
\tilde{a}_{2k,0}=\binom{2k}{k}-\binom{2k}{k+1}=\frac{1}{k+1}\binom{2k}{k}= C_k,
\end{align}
where $C_k$ denotes the $k$-th Catalan number. This is in agreement with a well-known result in random-matrix theory.

I have also obtained
\begin{align}
\tilde{a}_{2k,1}\!=\!\binom{2k}{k}\!
-
\frac{2^{4k}}{\binom{2k+1}{k}}\frac{1}{k+1}\sum_{i=0}^k\left[\frac{1}{2^{2i}}\binom{2i}{i}\right]^2,
\end{align}
but when both $k$ and $i$ are strictly higher than one, the exact coefficients $\tilde{a}_{2k,i}$ are still unknown. In the end, to lowest order the $k$-body local correlation function reads \cite{GangardtShlyapnikov2003NJP}
\begin{eqnarray}
g_k\simeq_{\gamma\ll 1}1-\frac{k(k-1)}{\pi}\sqrt{\gamma}.
\end{eqnarray}

\section{Non-local correlation functions, notion of connection}

By essence, local correlations are far from providing as much information on a system as non-local ones, i.e. at finite spatial separation. It is usual to investigate the $k$-body density matrices, defined as
\begin{eqnarray}
&&\rho_k(x_1,\dots,x_k;x_1',\dots x_k')\nonumber\\
&&=\int dx_{k+1}\,\dots \int dx_N\,\psi_N^*(x_1',\dots, x_k',x_{k+1},\dots,x_N)\,\psi_N(x_1,\dots,x_N),
\end{eqnarray}
and related to the local correlation functions through the relation
\begin{eqnarray}
g_k=\frac{N!}{(N-k)!}\frac{\rho_k(0,\dots,0;0,\dots,0)}{n_0^k}.
\end{eqnarray}
Traditionally in condensed-matter physics, one is more interested in their large-distance behavior, since it characterizes the type of ordering.
In particular, the one-body correlation function $g_1$ acquires a non-trivial structure in the relative coordinate, that depends on the interaction strength. As an introduction to this topic, I sum up the main results in the Tonks-Girardeau regime.

\subsection{One-body non-local correlation function in the Tonks-Girardeau regime}

The one-body, non-local correlation function of a translation-invariant system reads
\begin{eqnarray}
 g_1(x)=\frac{\langle\hat{\psi}^{\dagger}(x)\hat{\psi}(0)\rangle}{n_0},
\end{eqnarray}
where $x$ denotes the relative coordinate, i.e. the distance between two points. Even in the Tonks-Girardeau regime, its exact closed-form expression is unknown, but it can be studied asymptotically. I use the notation $z\!=\!k_Fx$, where $k_F\!=\!\pi n_0$ is the Fermi wavevector in 1D. I recall the large-distance expansion derived in \cite{Jimbo}(with signs of the coefficients corrected as in \cite{Gangardtbis}):
\begin{eqnarray}
\label{long}
g_1^{TG}(z)\!=\!\frac{G(3/2)^4}{\sqrt{2|z|}}\!\left[1\!-\!\frac{1}{32z^2}\!-\!\frac{\cos(2z)}{8z^2}\!-\!\frac{3}{16}\frac{\sin(2z)}{z^3}\!+\!\frac{33}{2048}\frac{1}{z^4}\!+\!\frac{93}{256}\frac{\cos(2z)}{z^4}\!+\!\dots\right]\!\!,
\end{eqnarray}
where $G$ is the Barnes function, defined by $G(1)\!=\!1$ and the functional relation $G(z+1)\!=\!\Gamma(z)G(z)$, $\Gamma$ being the Euler Gamma function. Since $g_1^{TG}(z)\to_{z\to +\infty} 0$, there is no long-range order. The decay is algebraic, so one speaks of a quasi-long-range order, which is quite slow here.

The general large-distance structure has been identified as \cite{Vaidya}
\begin{eqnarray}
\label{Vaidyastructure}
\!\!\!\!g_1^{TG}(z)\!=\!\frac{G(3/2)^4}{\sqrt{2|z|}}\!\left[1\!+\!\sum_{n=1}^{+\infty}\!\frac{c_{2n}}{z^{2n}}\!+\!\sum_{m=1}^{+\infty}\!\frac{\cos(2mz)}{z^{2m}}\sum_{n=0}^{+\infty}\frac{c_{2n,m}'}{z^{2n}}\!+\!\sum_{m=1}^{+\infty}\!\frac{\sin(2mz)}{z^{2m+1}}\sum_{n=0}^{+\infty}\!\frac{c_{2n,m}''}{z^{2n}}\!\right]\!,
\end{eqnarray}
in agreement with the fact that $g_1(z)$ is an even function in a Galilean-invariant model. Few coefficients have been explicitly identified, however, and Eq.~(\ref{long}) remains the reference to date.

\begin{figure}
\includegraphics[width=8cm, keepaspectratio, angle=0]{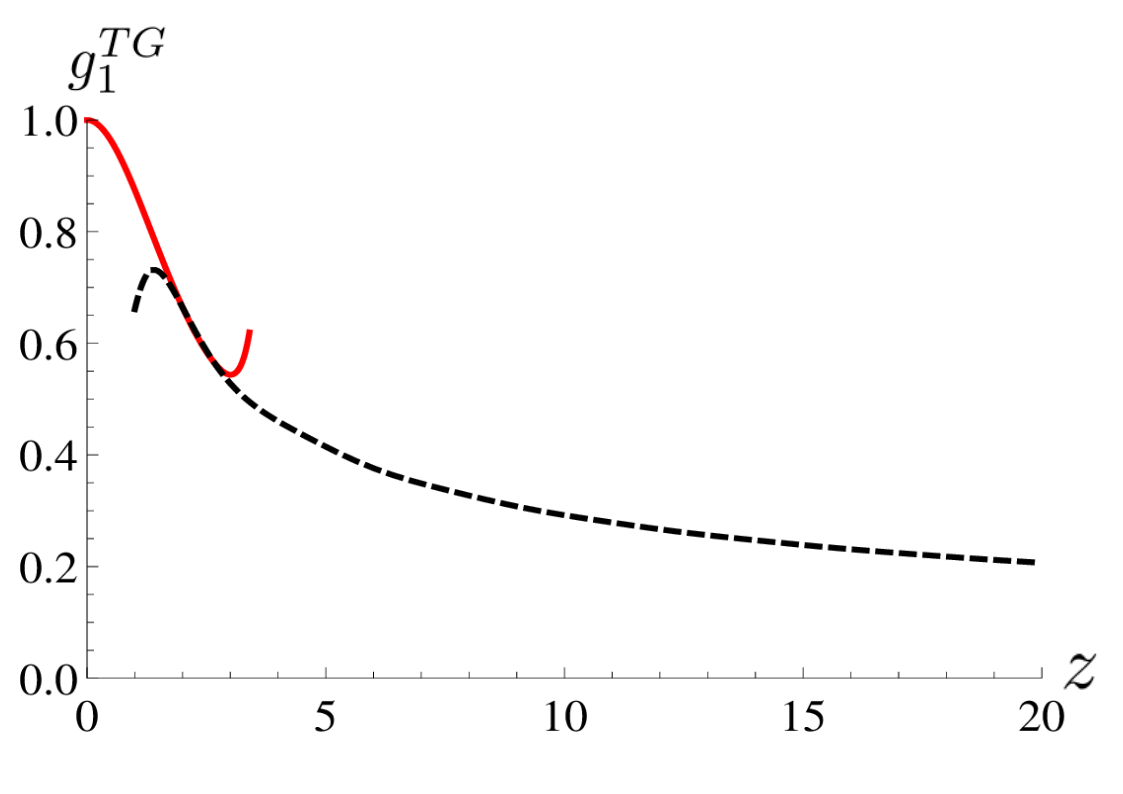}
\caption{Dimensionless one-body correlation function $g_1^{TG}$ in the Tonks-Girardeau regime as a function of the dimensionless distance $z$. Short-distance asymptotics given by Eq.~(\ref{short}) (solid, red) and large-distance asymptotics given by Eq.~(\ref{long}) (dashed, black) overlap at intermediate distances.}
\label{Fig9}
\end{figure}

At short distances, using the same technique as in \cite{Forrester} to solve the sixth Painlev\'e equation, I have obtained the following expansion, where I have added six orders compared to \cite{Jimbo}:
\begin{eqnarray}
\label{short}
&&g_1^{TG}(z)\!=\!\sum_{k=0}^{8}\frac{(-1)^kz^{2k}}{(2k+1)!}\!+\!\frac{|z|^3}{9\pi}\!-\!\frac{11|z|^5}{1350\pi}\!+\!\frac{61|z|^7}{264600\pi}\!+\!\frac{z^8}{24300\pi^2}\!-\!\frac{253|z|^9}{71442000\pi}\!-\!\frac{163z^{10}}{59535000\pi^2}\nonumber\\
&&+\!\frac{7141|z|^{11}}{207467568000\pi}\!+\!\frac{589 z^{12}}{6429780000\pi^2}\!-\!\frac{113623|z|^{13}}{490868265888000\pi}\!-\!\frac{2447503 z^{14}}{1143664968600000\pi^2}\nonumber\\
&&+\!\left(\frac{1}{40186125000\pi^3}\!+\!\frac{33661}{29452095953280000\pi}\right)|z|^{15}\!+\!\frac{5597693z^{16}}{140566821595200000\pi^2}\!+\dots\nonumber\\
&&
\end{eqnarray}
The first sum is the truncated Taylor series associated to the function $\sin(z)/z$, and corresponds to the one-body correlation function of noninteracting fermions,
\begin{eqnarray}
g_1^F(z)=\frac{\sin(z)}{z},
\end{eqnarray}
while the additional terms are specific to bosons with contact interactions. The one-body correlation function of Tonks-Girardeau bosons differs from the one of a Fermi gas due to the fact that it depends on the phase of the wavefunction, in addition to its modulus. The full structure, that would be the short-distance equivalent of Eq.~(\ref{Vaidyastructure}), is still unknown. In the end, expansions at short and large distances are known at high enough orders to overlap at intermediate distances \cite{Jimbo, Vaidya, Lenard}, as can be seen in Fig.~\ref{Fig9}.

I turn to the case of finite interactions, where the large-distance regime is amenable to Luttinger liquid theory and its generalizations.

\subsection{Large-distance, one-body correlation function at finite interaction strength from the Tomonaga-Luttinger liquid formalism}

The Tomonaga-Luttinger liquid theory is a suitable framework to obtain the large-distance, one-body correlation function. The result reads \cite{Haldane1981}
\begin{eqnarray}
g_1^{TL}(z)=\frac{1}{|z|^{\frac{1}{2K}}}\sum_{m=0}^{+\infty}B_m\frac{\cos(2mz)}{z^{2Km^2}}.
\end{eqnarray}
We already know that the case $K\!=\!1$ corresponds to the Tonks-Girardeau regime. By comparison with Eq.~(\ref{Vaidyastructure}) above, it appears that the Tomonaga-Luttinger approach, although it correctly predicts the behaviour of the dominant term, is not able to provide the full structure. This is in contrast with the two-body correlation function, whose equal-time structure is exact in the Tonks-Girardeau regime, as shown in chapter \ref{secII}.

However, Equation (\ref{Vaidyastructure}) has been generalized to finite interaction strengths in \cite{Didier}, by regularization of the Tomonaga-Luttinger liquid formalism, predicting that
\begin{eqnarray}
\label{Didierstructure}
\!\!\!\!g_1^{RTL}(z)\!=\!\frac{G(3/2)^4}{\sqrt{2|z|^{\frac{1}{K}}}}\!\left[1\!+\!\sum_{n=1}^{+\infty}\!\frac{c_{n}(K)}{z^{2n}}\!+\!\sum_{m=1}^{+\infty}\!\frac{\cos(2mz)}{z^{2Km^2}}\sum_{n=0}^{+\infty}\frac{c_{n,m}'}{z^{2n}}\!+\!\sum_{m=1}^{+\infty}\!\frac{\sin(2mz)}{z^{2Km^2+1}}\!\sum_{n=0}^{+\infty}\!\frac{c_{n,m}''}{z^{2n}}\!\right]\!\!.
\end{eqnarray}
The problem of the extraction of the amplitudes $B_m$, or $\{c_n, c_{n,m}', c_{n,m}''\}$ by Bethe Ansatz or alternative techniques is once more the main difficulty. A few of them have been obtained semi-analytically in \cite{Shashi2012}.

\subsection{Short-distance, one-body correlation function from integrability, notion of connection}

At arbitrary interaction strength, due to Galilean invariance, the short-distance series expansion of the one-body density matrix,
\begin{eqnarray}
\rho_1(x,x';\gamma)=\int dx_2\,\dots \int dx_N\,\psi^*_N(x,x_2,\dots,x_N)\,\psi_N(x',x_2,\dots,x_N),
\end{eqnarray}
can be written as
\begin{eqnarray}
\label{densmat}
\rho_{1}(x,\,x';\gamma) =\frac{1}{L}\sum_{l=0}^{+\infty}c_l(\gamma)(n_0|x\!-\!x'|)^l.
\end{eqnarray}
The list of coefficients $\{c_l\}$ can be constructed from integrability at arbitrary interaction strength. The procedure relies on conservation laws. The most common are the number of particles, total momentum and energy, that are eigenvalues of their associated operators: particle number, momentum and Hamiltonian. In an integrable model, these quantities are conserved too, as well as infinitely many others, called higher energies and written $E_n$, that are eigenvalues of peculiar operators called higher Hamiltonians, written $\hat{H}_{n}$, that have the same Bethe eigenvector $\psi_N$ as the Hamiltonian. To obtain the results presented in Ref.~\cite{OlshaniiDunjkoMinguzziLang}, I have used and compared several strategies, sketched in \cite{Gutkin1985, Davies1990, DaviesKorepin2011}. All of them are quite technical, but a systematic procedure and a few general properties have emerged in the course of the derivation.

I have defined the notion of connection for the one-body density matrix as a functional relation $\mathcal{F}$ that connects one of the coefficients $c_l$ of Eq.~(\ref{densmat}) to a local correlation function, via moments of the density of pseudo-momenta and their derivatives, that reads
\begin{eqnarray}
\label{defconnection}
\mathcal{F}\left[c_l(\gamma),g_k(\gamma),\{e_{2n}(\gamma),e_{2n}'(\gamma),\dots\},\gamma\right]=0.
\end{eqnarray}
Connections encompass many relationships scattered throughout the literature in a unified description. Each of them is unambiguously denoted by the pair of indices $(l,k)$, where by convention an index is set to $0$ if the corresponding quantity does not appear in Eq.~(\ref{defconnection}). This compact notation is convenient, as it allows to list and classify the connections.

To illustrate this point, I recall the first few connections, obtained from conservation laws. I find
\begin{eqnarray}
c_0=g_1=e_0=1,
\end{eqnarray}
yielding the connections (0,0) and (0,1), as well as
\begin{eqnarray}
c_1=0,
\end{eqnarray}
denoted by (1,0). The connection (2,2) is
\begin{eqnarray}
-2c_2+\gamma g_2=e_2,
\end{eqnarray}
while (0,2) is obtained by applying the Hellmann-Feynman theorem to the Lieb-Liniger Hamiltonian Eq.~(\ref{HLL}), and is nothing else than Eq.~(\ref{HF}).
Then, combining the connections (2,2) and (0,2) yields (2,0), that reads
\begin{eqnarray}
c_2=\frac{1}{2}(\gamma e_2'-e_2).
\end{eqnarray}
The main result of \cite{OlshaniiDunjkoMinguzziLang} is the derivation of the connection (4,3), that reads
\begin{align}
24 c_{4} - 2 \gamma^2 g_{3} = e_{4}-\gamma e_{4}'.
\label{Maineq}
\end{align}
This derivation involves an operator $\hat{H}_{4}$ that yields, when applied to a Bethe eigenstate $\psi_N$, the fourth integral of motion $E_4$, such that 
\begin{align}
E_{4}= \sum_{i=1}^{N} k_i^4.
\,\,
\end {align}
The higher Hamiltonian $\hat{H}_4$ can be written explicitly as \cite{Davies1990, DaviesKorepin2011, OlshaniiDunjkoMinguzziLang}
\begin{eqnarray}
\label{H_4}
%
&&\hat{H}_{4}=\sum_{i=1}^{N} \frac{\partial^4}{\partial x_{i}^4}\nonumber\\
&&+12c^2\sum_{i=1}^{N-2} \sum_{j=i+1}^{N-1}\sum_{k=j+1}^{N} \delta(x_{i}-x_{j}) \delta(x_{j}-x_{k})\nonumber\\
&&-2c\!\sum_{i=1}^{N-1} \sum_{j=i+1}^{N}\!\left\{\!\left(\frac{\partial^2}{\partial x_{i}^2}\!+\!\frac{\partial^2}{\partial x_{j}^2}\!+\!\frac{\partial^2}{\partial x_{i} \partial x_{j}} \right)\!\delta(x_{i}\!-\!x_{j})\!+\!\delta(x_{i}\!-\!x_{j})\!\left(\frac{\partial^2}{\partial x_{i}^2}\!+\!\frac{\partial^2}{\partial x_{j}^2}\!+\!\frac{\partial^2}{\partial x_{i} \partial x_{j}}\right)\!\right\}\nonumber\\
&&+2c^2 \sum_{i=1}^{N-1} \sum_{j=i+1}^{N} \delta^2(x_{i}-x_{j})\nonumber\\
&&=\hat{h}_4^{(1)}+\hat{h}_4^{(2)}+\hat{h}_4^{(3)}+\hat{h}_4^{(4)}.
\end{eqnarray}
Let me comment on the physical meaning of Eq.~(\ref{Maineq}) in view of Eq.~(\ref{H_4}), from which it is derived. The fact that $g_3$ appears in the connection (4,3) stems from $\hat{h}_4^{(2)}$ in Eq.~(\ref{H_4}), that involves three-body processes provided that $N\!\geq\!3$. The coefficient $c_4$, that stems from $\hat{h}_4^{(1)}$, is related to the higher kinetic energy in that the momentum operator applied to the density matrix generates the coefficients of its Taylor expansion when taken at zero distance.

In the course of the derivation, the requirement that $\hat{H}_4$ is divergence-free, i.e. contains no $\delta(0)$ term in spite of $\hat{h}_4^{(4)}$ being mathematically ill-defined (it contains an operator $\delta^2$), yields the connection (3,2), first obtained in \cite{OlshaniiDunjko2003} from asymptotic properties of Fourier transforms:
\begin{eqnarray}
\label{conn23}
c_3=\frac{\gamma^2}{12}g_2.
\end{eqnarray}
The connection (3,0) follows naturally by combination with (0,2) and reads:
\begin{eqnarray}
c_3=\frac{\gamma^2}{12}e_2'.
\end{eqnarray}
In Ref.~\cite{OlshaniiDunjkoMinguzziLang}, we also proposed an alternative derivation of $(3,2)$, as a corollary of the more general result:
\begin{eqnarray}
\label{rhok3}
\rho_k^{(3)}(0,\dots;0,\dots)=\frac{N-k}{12}c^2\rho_{k+1}(0,\dots;0,\dots),
\end{eqnarray}
where $\rho_k$ is the $k$-body density matrix expanded as
\begin{eqnarray}
\!\!\!\!\rho_k(x_1,\dots,x_k;x_1',\dots,x_k')=\sum_{m=0}^{+\infty}\!\rho_k^{(m)}\!\!\left(\!\frac{x_1\!+\!x_1'}{2},x_2,\dots,x_k;x_2',\dots,x_k'\!\right)\!|x_1\!-\!x_1'|^m,
\end{eqnarray}
a form that naturally emerges from the contact condition. Equation (\ref{rhok3}) can be written as
\begin{eqnarray}
c_3^{(k)}=\frac{\gamma^2}{12}g_{k+1},
\end{eqnarray}
where $c_3^{(k)}$ is the third-order coefficient of the Taylor expansion of the $k$-body density matrix, and provides an example of generalized connection, a notion that remains in limbo.

As a last step, combining the connection (0,3), Eq.~(\ref{g3formula}), with the connection (4,3), Eq.~(\ref{Maineq}), yields the connection (4,0) first published in \cite{Olshaniiter}:
\begin{align}
\label{c4}
c_4(\gamma)\!=\!\frac{\gamma e_4'}{12}\!-\!\frac{3}{8}e_4\!+\!\frac{2\gamma^2\!+\!\gamma^3}{24}e_2'\!-\!\frac{\gamma e_2}{6}\!-\!\frac{\gamma e_2e_2'}{4}\!+\!\frac{3}{4}e_2^2.
\end{align}
More generally, all correlations of the model are encoded in the connections of type $(l,0)$ and $(0,k)$, as a consequence of integrability.

\begin{figure}
\includegraphics[width=8cm, keepaspectratio, angle=0]{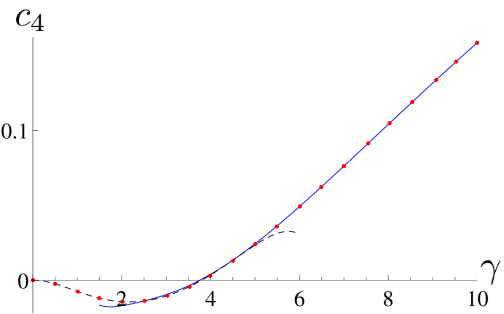}
\caption{Dimensionless coefficient $c_4$ as a function of the dimensionless interaction strength $\gamma$, as predicted from the conjectures (solid, blue) and (dashed, black), compared to accurate numerics (red dots). A sign inversion occurs at $\gamma\simeq 3.8$.}
\label{Fig10}
\end{figure}

\begin{figure}
\includegraphics[width=8cm, keepaspectratio, angle=0]{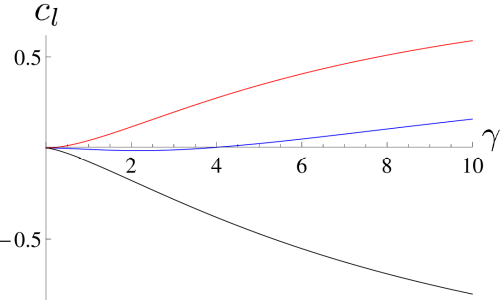}
\caption{Dimensionless coefficients $c_2$, $c_4$  and $c_3$ (black, blue, red and from bottom to top respectively) as predicted from connections combined to conjectures for the moments of the density of pseudo-momenta, as functions of the dimensionless interaction strength $\gamma$.}
\label{Fig11}
\end{figure}

Combining the results given above, I have access to the first few coefficients $\{c_l\}_{l=0,\dots, 4}$ of the Taylor expansion of $g_1$. Contrary to $c_0$ and $c_1$ that are constant, $c_2$ and $c_3$ that are monotonous, $c_4(\gamma)$ changes sign when the interaction strength takes the numerical value
\begin{eqnarray}
\gamma_c=3.8160616255908\dots
\end{eqnarray}
obtained with this accuracy by two independent methods, based on a numerical and a semi-analytical solution of the Bethe Ansatz equations respectively. This is illustrated in Figs.~\ref{Fig10} and \ref{Fig11}. It was previously known that $c_4$ changes sign, as obtained from numerical analysis in \cite{CauxCalabreseSlavnov}, but the only certitude was that $1<\gamma_c<8$. 

\section{Momentum distribution and Tan's contact}

In addition to real-space correlations, through ballistic expansion of the atomic cloud, experimentalists have also access to the Fourier transform of the non-local, static correlation functions. Only the first few orders bear specific names and have been investigated by now. The Fourier transform of the one-body correlation function $g_1$ is the momentum distribution, while the momentum space representation of $g_2$ is known as the static structure factor. The momentum distribution is measured with ever increasing accuracy in various systems, from a 3D Fermi gas over the whole range of interaction strengths \cite{Jin2005, Jin2010} to Bose-Einstein condensates \cite{Wild2012, Clement2016} and the 1D Bose gas \cite{Jacqmin2012}.

The momentum distribution of the Lieb-Liniger model, defined as
\begin{eqnarray}
n(p)=n_0\int_{-\infty}^{+\infty}e^{i\frac{p}{\hbar}x}g_1(x),
\end{eqnarray}
is difficult to access from integrability. For this, reason most studies are based on fully numerical methods so far \cite{XuRigol2015, CauxCalabreseSlavnov, AstrakharchikGiorgini2003}. Analytically, it is quite natural, as usual, to treat the Tonks-Girardeau regime as a warm-up. As illustrated in Fig.~\ref{Fig9}, expressions for the one-body correlation function $g_1^{TG}(z)$ obtained at small and large distances match at intermediate distances, but separately none is appropriate for a direct Fourier transform.

A step forward is made by noticing that the low- and high-momentum expansions of $n(p)$ can be deduced from the large- and short-distance asymptotics of $g_1(z)$ respectively, according to the following theorem \cite{Bleistein}: if a periodic function $f$ is defined on an interval $[-L/2,L/2]$ and has a singularity of the form $f(z)=|z-z_0|^{\alpha}F(z)$, where $F$ is a regular function, $\alpha>-1$ and not an even integer, the leading term of the Fourier transform reads
\begin{eqnarray}
\label{asympfourier}
&&\!\!\int_{-L/2}^{L/2}dz\, e^{-ikz}f(z)=_{|k|\to +\infty}2\cos\!\left[\frac{\pi}{2}(\alpha\!+\!1)\right]\Gamma(\alpha\!+\!1)e^{-ikz_0}F(z_0)\frac{1}{|k|^{\alpha+1}}+O\!\left(\frac{1}{|k|^{\alpha+2}}\right).\nonumber\\
&&
\end{eqnarray}
A legitimate accuracy requirement is that the expansions of $n(p)$ should overlap at intermediate momenta. It is more or less fullfilled in the Tonks-Girardeau regime, but this is not the case yet at finite interaction strengths. It is known, however, that at small momenta the momentum distribution of the Tonks-Girardeau gas, $n^{TG}(p)$, scales like $p^{-1/2}$, in strong contrast with a noninteracting Fermi gas, as usual for correlation functions of odd order, linked to the phase observable. This result can be extended to arbitrary interactions using the Tomonaga-Luttinger liquid theory, and one finds that $n^{TL}(p)$ scales like $p^{\frac{1}{2K}-1}$.

\begin{figure}
\includegraphics[width=10cm, keepaspectratio, angle=0]{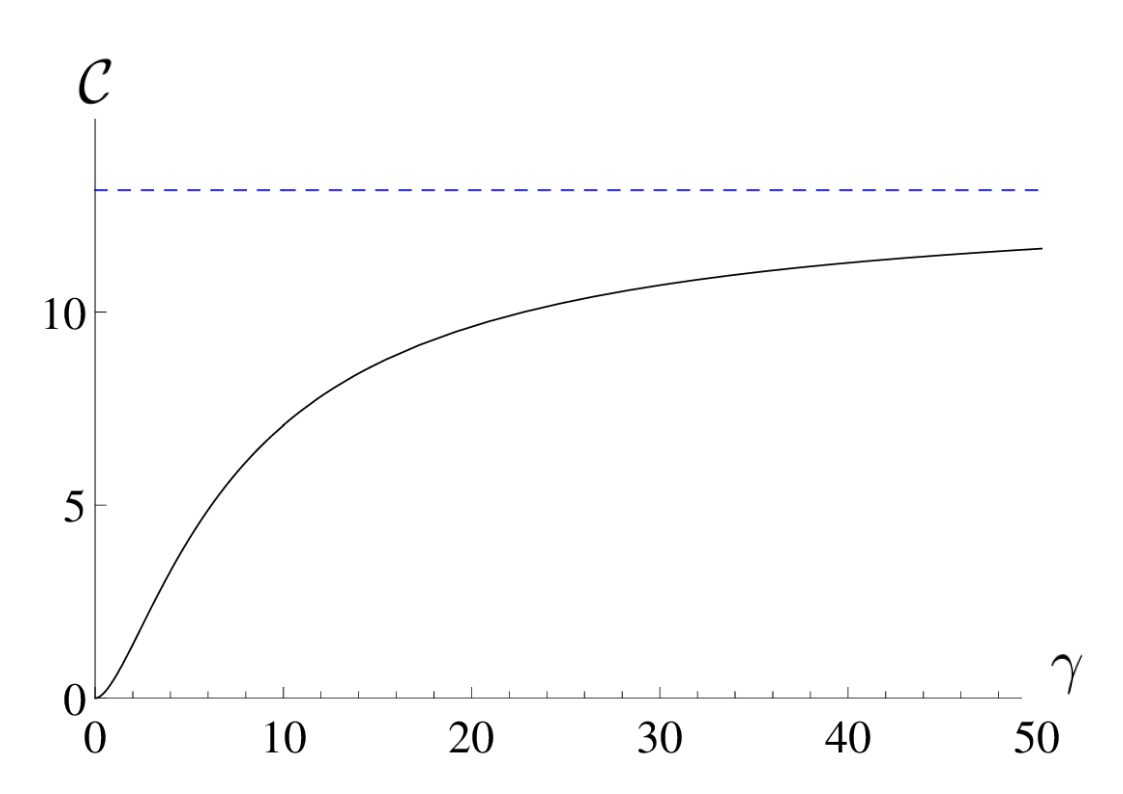}
\caption{Dimensionless Tan's contact $\mathcal{C}\!=\!\frac{2\pi}{L}\frac{1}{n_0^4}C(\gamma)$ as a function of the dimensionless interaction strength $\gamma$ (solid, black) and its value in the Tonks-Girardeau limit, $\mathcal{C}^{TG}\!=\!\frac{4\pi^2}{3}$ (dashed, blue)}
\label{FigTan}
\end{figure}

At large momenta and in the Tonks-Girardeau regime, to leading order the momentum distribution scales like $1/p^4$ \cite{MinguzziVignoloTosi2002}, again in contrast with a noninteracting Fermi gas where such a tail does not exist due to the finite Fermi sea structure. This power law associated with the Tonks-Girardeau gas is not affected by the interaction strength, showing its universality, and stems for the $|z|^3$ non-analyticity in $g_1$ according to Eq.~(\ref{asympfourier}). The coefficient of the $1/p^4$ tail is called Tan's contact, and is a function of the coupling $\gamma$ \cite{OlshaniiDunjko2003}. As such, it yields an experimental means to evaluate the interaction strength, but as Tan has shown in a series of articles in the case of a Fermi gas \cite{Tan1, Tan2, Tan3}, and others in a Bose gas \cite{BraatenKangPlatter}, it also gives much more information about the system. For instance, according to Tan's sweep relation, Tan's contact in 1D is related to the ground-state energy $E_0$ of the gas according to 
\begin{equation}
C=-\frac{m^2}{\pi \hbar^4} \frac{\partial E_0}{\partial (1/g_{1D})},
\label{eq:defC}
\end{equation}
that can be rewriten in dimensionless units as
\begin{equation}
\label{Chom}
C(\gamma)=n_0^4\frac{L}{2\pi}\gamma^2g_2(\gamma),
\end{equation}
and is illustrated in Fig.~\ref{FigTan}. Written in this form, it becomes clear that this quantity is governed by the two-body correlations, which is a priori surprizing for a quantity associated to a one-particle observable.

\section{Breakdown of integrability, BALDA formalism}

\subsection{Effect of a harmonic trap}

In current experimental realizations of 1D gases, external trapping along the longitudinal direction often breaks translational invariance, spoiling integrability. Due to this external potential, real systems are inhomogeneous, and their theoretical description requires modifications. Let us assume that the atoms are further confined in the longitudinal direction by an external potential $ V_{ext}(x)$, describing the optical or magnetic trapping present in ultracold atom experiments. Then, as a generalization of the Lieb-Liniger model, the Hamiltonian of the system reads
\begin{equation}
\label{HLLtrap}
H=\sum_{j=1}^N \left[-\frac{\hbar^2}{2m}\frac{\partial^2}{\partial x_j^2} + V_{ext}(x_j)+\frac{g_{\mathrm{1D}}}{2}\sum_{\{l\neq j\}} \delta(x_j-x_\ell)\right].
\end{equation}
In the case of a harmonic confinement, the only one I will consider here, $V_{ext}(x)\!=\!m \omega_0^2 x^2/2$. Introducing the harmonic-oscillator length $a_{ho}\!=\!\sqrt{\hbar/(m\omega_0)}$ and the one-dimensional scattering length $a_{\mathrm{1D}}\!=\!- 2 \hbar^2/(g_{\mathrm{1D}} m)$, in the inhomogeneous system the dimensionless LDA parameter corresponding to the Lieb parameter $\gamma$ in the homogeneous gas is
\begin{eqnarray}
\alpha_0=\frac{2 a_{ho}}{|a_{\mathrm{1D}}|\sqrt{N}}.
\end{eqnarray}
Due to the additional term compared to the homogeneous case, new tools are needed to derive the dynamics of a system described by Eq.~(\ref{HLLtrap}). The few-particle problem is exactly solvable for $N\!=\!2$ \cite{Busch1998} and $N\!=\!3$ \cite{Brouzos2012} with analytical techniques or using a geometrical ansatz \cite{Wilson2014}, but the thermodynamic limit requires a different approach.

\subsection{Local-density approximation for the density profile in the Tonks-Girardeau regime}

In the Tonks-Girardeau regime, characterized by $\alpha_0\!\to\!+\infty$, a generalized Bose-Fermi mapping allows for an exact solution of the Schr\"odinger equation associated to Eq.~(\ref{HLLtrap}) \cite{Girardeau2001}. However, this exact solution is restricted to infinite interaction strength, and not utterly trivial. It is thus instructing to study an approximate method that would be easier to handle, and generalizable to arbitrary interaction strengths.

The local-density approximation (LDA) provides an approach to this problem. It is expected to be reliable for sufficiently large systems, where finite size corrections and gradient terms in the density profile are negligible. Its interest lies also in its generality, as LDA can be applied to various systems, and does not depend on quantum statistics. 

In the Tonks-Girardeau regime, predictions of LDA can be compared to the exact solution, in particular it has been checked numerically in \cite{VignoloMinguzzi2001} that the Thomas-Fermi density profile predicted by the LDA becomes exact in the thermodynamic limit, as illustrated in Fig.~\ref{FigLDA}. This exact equivalence can be proven rigorously (see Appendix \ref{LDATG}) and reads
\begin{eqnarray}
\label{densTGLDA}
n^{TG}(x;N)\!=\!\sum_{n=0}^{N-1}|\phi_n(x)|^2\sim_{N\to+\infty}n^{TF}(x;N)\!=\!\frac{1}{\pi a_{ho}}\sqrt{2N\!-\!\left(\frac{x}{a_{ho}}\right)^2},
\end{eqnarray}
where the eigenfunctions of the harmonic oscillators are \cite{YukalovGirardeau2005}
\begin{eqnarray}
\phi_n(x)=\frac{e^{-x^2/(2a_{ho}^2)}}{(\sqrt{\pi}2^nn!a_{ho})^{1/2}}H_n\!\left(\frac{x}{a_{ho}}\right),
\end{eqnarray}
and $H_n$ are the Hermite polynomials, defined as $H_n(x)\!=\!(-1)^ne^{x^2}\frac{d^n}{dx^n}e^{-x^2}$.

\begin{figure}
\includegraphics[width=8cm, keepaspectratio, angle=0]{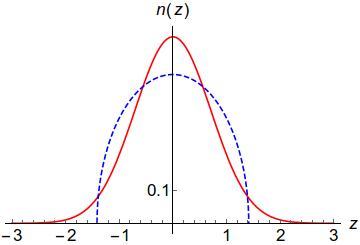}
\includegraphics[width=8cm, keepaspectratio, angle=0]{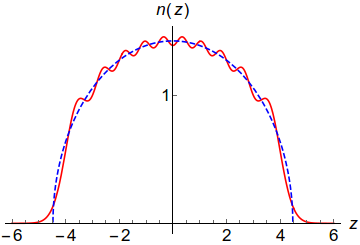}
\includegraphics[width=8cm, keepaspectratio, angle=0]{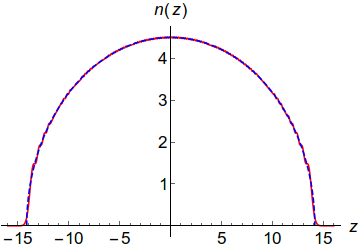}
\caption{Exact density profile of a harmonically trapped Tonks-Girardeau gas (solid, red) and as predicted by LDA (dashed, blue) in units of the inverse harmonic oscillator size $1/a_{ho}$, as a function of the dimensionless distance $z=x/a_{ho}$, for $1$, $10$ and $100$ particles respectively from top to bottom and left to right.} 
\label{FigLDA}
\end{figure}

\subsection{From local-density approximation to Bethe Ansatz LDA}

In order to describe a one-dimensional, harmonically-trapped Bose gas, a possible strategy is to try and combine the local-density approximation, exact for a trapped Tonks-Girardeau gas in the thermodynamic limit, with Bethe Ansatz, exact for the uniform Lieb-Liniger gas at arbitrary interaction strength. This combination leads to the Bethe Ansatz local-density approximation (BALDA) formalism, that predics the thermodynamics of a trapped gas at arbitrary interaction strength.

To do so, in \cite{LangVignoloMinguzzi2017} we employed the density functional approach, previously developed and illustrated in \cite{Volosniev2014, Decamp2016} in the fermionic case. In detail, it consists in defining an energy functional $E_0[n]$ of the local density $n(x)$ which, in the local-density approximation, reads
\begin{equation}
E_0[n]=\int dx \, \left\{\epsilon(n) + [V_{ext}(x)-\mu] n(x)\right\},
\label{eq:LDA}
\end{equation}
where $\epsilon$ is the ground-state energy density of the homogeneous gas and $\mu$ is its chemical potential. Minimizing this functional, i.e. setting $\delta E_0/\delta n\!=\!0$, yields an implicit equation for the density profile,
\begin{equation}
\frac{3}{2} \frac{\hbar^2}{m} n^2 e- \frac{g_{\mathrm{1D}}n}{2}e'(\gamma) =\mu -V_{ext}(x)=\mu\, \left(1-\frac{x^2}{R_{TF}^2}\right),
\label{eq:rho}
\end{equation}
where
\begin{eqnarray}
R_{TF}\!=\!\sqrt{\frac{2\mu}{m\omega_0^2}}
\end{eqnarray}
is the Thomas-Fermi radius in a harmonic trap, and the dimensionless average ground-state energy per particle $e$ defined in Eq.~(\ref{energy}) is such that
\begin{eqnarray}
\epsilon(n)\!=\!\frac{\hbar^2}{2 m} n^3 e(\gamma).
\end{eqnarray}
The chemical potential is fixed by imposing the normalization condition 
\begin{eqnarray}
\label{norm}
N=\int dx\, n(x).
\end{eqnarray}
Note that in the homogeneous case, Eq.~(\ref{eq:LDA}) would yield
\begin{eqnarray}
\mu=\frac{n_0^2\hbar^2}{2m}(3e-\gamma e'),
\end{eqnarray}
as expected from thermodynamics \cite{LiebLiniger1963}.

\subsection{Tan's contact of a trapped Bose gas}

To illustrate the BALDA formalism, in Ref.~\cite{LangVignoloMinguzzi2017} we have investigated the Tan contact of a trapped one-dimensional Bose gas. Before explaining my own contributions to the problem, let me summarize a few important analytical results previously obtained by other authors.

The high-momentum tail of a harmonically-trapped Bose gas in the Tonks-Girardeau regime scales like $p^{-4}$, as in the homogeneous case \cite{MinguzziVignoloTosi2002}, and this power law still holds at arbitrary interaction strength \cite{OlshaniiDunjko2003}. It has even been shown that such a tail is the exact dominant term at finite temperature in the Tonks-Girardeau regime \cite{VignoloMinguzzi2013}.

Motivated by a recent experiment realizing a 1D gas of fermions with $SU(\kappa)$ symmetry with up to $\kappa\!=\!6$ spin components \cite{Pagano2014}, a step forward has been made for interacting spin-balanced harmonically-trapped Fermi gases of arbitrary spin. The two first corrections to the fermionic Tonks-Girardeau regime have been obtained within the local-density approximation in \cite{Decamp2016}. This readily yields the result for the Lieb-Liniger model with an additional harmonic trap, considering a theorem that states the equivalence between a balanced one-dimensional gas of fermions with $SU(\kappa\!=\!+\infty)$ symmetry and a spinless 1D Bose gas \cite{YangZhuang2011}. Another important result of \cite{Decamp2016} relies on comparison of the BALDA result to a numerical exact solution from DMRG, that shows remarkable agreement at large interaction strengths. It was not possible, however, to attain higher orders analytically for this Fermi gas, because the strong-coupling expansion of its ground-state energy is known to third order only \cite{GuanMaWilson2012}.

In \cite{LangVignoloMinguzzi2017}, I have obtained Tan's contact for the Bose gas to $4^{th}$ order in the inverse coupling, and developed a procedure to evaluate this expansion to arbitrary order for a harmonically-trapped gas, from the corresponding asymptotic energy of an homogeneous gas at next order. This procedure can be applied to bosons and fermions alike.
Within the LDA, Tan's contact of the inhomogeneous gas reads
\begin{equation}
\label{CLDA}
C_{LDA}=g_{\mathrm{1D}}^2 \frac{m^2}{2 \pi \hbar^4} \int dx \, n^2(x) \left.\frac{\partial e}{\partial \gamma}\right|_{n_0=n(x)}\!\!\!\!\!\!\!\!\!\!\!\!\!\!\!\!.
\end{equation}
This expression readily generalizes the known result for the homogeneous gas, Eq.~(\ref{Chom}). To perform the calculation explicitly, it is necessary to dispose of a model of equation of state $e(\gamma)$ for the homogeneous gas. For noninteracting spinless fermions or the Tonks-Girardeau gas, the LDA calculation can be performed this way, but requires the knowledge of the first correction in $1/\gamma$. In the case of the Lieb-Liniger model with arbitrary coupling constant, I have relied on the strong- and weak-coupling expansions found in Sec.~\ref{GSenergy}.

First, I derived the strong-coupling expansion of Tan's contact for a harmonically-trapped gas, based on the corresponding expansion of the ground state-energy of the homogeneous system, Eq.~(\ref{strongcouplinge}). To quantify the interaction strength in the trapped gas, I used the dimensionless unit $\alpha_0$, such that
\begin{eqnarray}
g_{\mathrm{1D}}\!=\!\hbar \omega_0 a_{ho}\sqrt{N}\alpha_0.
\end{eqnarray}
I also introduced the rescaled variables
\begin{eqnarray}
&&\overline{n}= \frac{n \, a_{ho}}{\sqrt{N}},\,\,\,\overline{\mu}= \frac{\mu}{N\hbar\omega_0},\,\,\,\overline{x}= \frac{x}{R_{TF}}.
\end{eqnarray}
Combining Eq.~(\ref{eq:rho}), the normalization condition Eq.~(\ref{norm}) and these scalings, I obtained the following set of equations:
\begin{eqnarray}
\label{eq1strong}
\frac{\pi^2}{6}\sum_{k=0}^{+\infty}\frac{(k+3)e_k}{\alpha_0^k}\overline{n}^{k+2}(\overline{x};\alpha_0)=(1-\overline{x}^2)\overline{\mu}(\alpha_0),
\end{eqnarray}
where $e_k$ is defined as in Eq.~(\ref{strongcouplinge}), as well as
\begin{eqnarray}
\label{eq2strong}
1\!=\!\sqrt{2\,\overline{\mu}}\int_{-1}^1\!d\overline{x}\,\,\overline{n}(\overline{x}).
\end{eqnarray}

Then, I developed an efficient procedure, that allows to calculate the strong-coupling expansion of Tan's contact to arbitrary order. This procedure relies on the following expansions:
\begin{eqnarray}
\overline{\mu}=\sum_{k=0}^{+\infty}\frac{c_k}{\alpha_0^k},
\end{eqnarray}
and 
\begin{eqnarray}
\overline{n}(\overline{x})=\sum_{j=0}^{+\infty}\frac{b_j}{\alpha_0^j}f_j(\overline{x}),
\end{eqnarray}
where $\{c_k\}_{k\geq 0}$ and $\{b_j\}_{j\geq 0}$ are numerical coefficients, and $\{f_j\}_{j\geq 0}$ is a set of unknown functions. Injected into Eqs.~(\ref{eq1strong}) and (\ref{eq2strong}), they yield a consistency condition:
\begin{eqnarray}
b_jf_j(\overline{x})=\sum_{m=0}^jb_{mj}(1-\overline{x}^2)^{(m+1)/2},
\end{eqnarray}
where $\{b_{mj}\}$ are unkwown coefficients of an upper triangular matrix, so that the previous equations become
\begin{eqnarray}
\label{eq1strongfin}
\frac{\pi^2}{6}\sum_{k=0}^{+\infty}\frac{(k+3)e_k}{\alpha_0^k}\left(\sum_{j=0}^{+\infty}\frac{1}{\alpha_0^j}\sum_{m=0}^jb_{mj}(1-\overline{x}^2)^{(m+1)/2}\right)^{k+2}\!\!\!\!=\!\left(1-\overline{x}^2\right)\sum_{k=0}^{+\infty}\frac{c_k}{\alpha_0^k}
\end{eqnarray}
and
\begin{eqnarray}
\label{eq2strongfin}
1=32\sum_{k=0}^{+\infty}\frac{c_k}{\alpha_0^k}\left[\sum_{j=0}^{+\infty}\frac{1}{\alpha_0^j}\sum_{m=0}^jb_{mj}2^mB\left(\frac{m+3}{2},\frac{m+3}{2}\right)\right]^2,
\end{eqnarray}
where $B$ is the Euler Beta function.

Equations (\ref{eq1strongfin}) and (\ref{eq2strongfin}) are the final set of equations. Solving them when truncated to order $n$ requires the solution at all lower orders. Moreover, at each step Eq.~(\ref{eq1strongfin}) splits into $n\!+\!1$ independent equations, obtained by equating the coefficients of $(1-\overline{x}^2)^{(1+m)/2}_{m=0,\dots,n}$ in the left- and right-hand sides. One thus needs to solve a set of $n\!+\!2$ equations to obtain $c_n$ and $\{b_{mn}\}_{m=0,\dots, n}$. Fortunately, $n$ of them, giving $b_{mj}$, $m\geq 1$, are fully decoupled.

As a final step, Eq.~(\ref{CLDA}) yields Tan's contact. In natural units imposed by the scaling, i.e. taking \cite{Matveeva2016}
\begin{eqnarray}
\overline{C}_{LDA}=C_{LDA} \frac{a_{ho}^3}{N^{5/2}},
\end{eqnarray}
the final equation reads:
\begin{eqnarray}
\!\!\!\!\!\!\!\overline{C}_{LDA}\!=\!-\frac{\pi}{3\sqrt{2}}\sqrt{\sum_{k'=0}^{+\infty}\frac{c_{k'}}{\alpha_0^{k'}}}\sum_{k=0}^{+\infty}\frac{(k\!+\!1)e_{k+1}}{\alpha_0^k}\!\int_{-1}^1\!d\overline{x}\left(\sum_{j=0}^{+\infty}\frac{1}{\alpha_0^j}\sum_{m=0}^jb_{mj}(1\!-\!\overline{x}^2)^{(m+1)/2}\!\right)^{k+4}\!\!\!\!\!\!\!\!\!\!.
\end{eqnarray}
In spite of the global minus sign, Tan's contact is a non-negative quantity because $e_1\!<\!0$ and corrections decrease quickly enough. At order $n$, the condition $k'+k+j'=n$, where $j'$ is the power of $\alpha_0$ in the integrand, shows that the coefficient of order $n$ is a sum of $\binom{n+2}{n}$ integrals. One of them involves $e_{n+1}$, so $e(\gamma)$ must be known to order $n\!+\!1$ in $1/\gamma$ to obtain the expansion of Tan's contact to order $n$ in $1/\alpha_0$.

Following this approach, the strong-coupling expansion reads:
\begin{eqnarray}
\label{eq:strongc}
&&\overline{C}_{LDA}=\frac{128\sqrt{2}}{45 \pi^3}+\frac{1}{\alpha_0}\left(-\frac{8192}{81\pi^5}+\frac{70}{9\pi^3}\right)+\frac{\sqrt{2}}{\alpha_0^2}\left(\frac{131072}{81\pi^7}-\frac{30656 }{189\pi^5}-\frac{4096}{525 \pi^3}\right)\nonumber\\
&&+\frac{1}{\alpha_0^3}\left(-\frac{335544320}{6561  \pi^9}+\frac{4407296}{729\pi^7}+\frac{872701}{2025\pi^5}-\frac{112}{3\pi^3}\right)\nonumber\\
&&+\frac{\sqrt{2}}{\alpha_0^4}\left(\frac{47982837760 }{59049  \pi^{11}}-\frac{717291520 }{6561 \pi^9}-\frac{108494512 }{10935 \pi^7}+\frac{2112512}{1701 \pi^5}+\frac{65536}{2205\pi^3}\right)\nonumber\\
&&+\dots
\end{eqnarray}
This expression agrees with the zero order one obtained in \cite{OlshaniiDunjko2003}, and with the one derived for a $\kappa$-component balanced spinful Fermi gas to order two \cite{Decamp2016} in the infinite spin limit.

In the weak-coupling regime, I also derived an expression for Tan's contact by combining the weak-coupling expansion of the homogeneous gas to the local-density approximation. Using the same notations as above, I obtained
\begin{eqnarray}
\label{Tanbof}
\sum_{k=0}^{+\infty}\frac{a_k}{4}(4-k)\overline{n}^{\frac{2-k}{2}}(\overline x;\alpha_0)\alpha_0^{\frac{k+2}{2}}=\left(1-\overline{x}^2\right)\overline{\mu}(\alpha_0).
\end{eqnarray}
In this regime, it is not obvious to what order truncation should be performed to obtain a consistent expansion at a given order, nor to find the variable in which to expand, as can be seen by evaluating the first orders.

Considering only the $k\!=\!0$ term in Eq.~(\ref{Tanbof}) yields
\begin{eqnarray}
\overline{n}\,(\overline{x})=\left(\frac{9}{32\alpha_0}\right)^{1/3}(1-\overline{x}^2)
\end{eqnarray}
and
\begin{eqnarray}
\overline{\mu}\,(\alpha_0)=\left(\frac{9}{32}\right)^{1/3}\alpha_0^{2/3}.
\end{eqnarray}
The expansion to next order is problematic. If one retains terms up to $k\!=\!1$, corresponding to the Bogoliubov approximation, since the coefficient $a_1$ is negative the equation of state becomes negative at sufficiently large density. Then, it is not possible to use it to perform the local-density approximation. One may also recall that the LDA breaks down at very weak interactions, where it is not accurate to neglect the quantum tails in the density profile \cite{Petrov2004, Astrakharchik2014, Olshanii2015}.

In the end, the weak-coupling expansion to lowest order reads
\begin{equation}
\label{eq:weakc}
\overline{C}_{LDA}=\frac{1}{5\pi}\left(\frac{3}{2}\right)^{2/3}\alpha_0^{5/3},
\end{equation}
in agreement with \cite{OlshaniiDunjko2003}.

\begin{figure}
\includegraphics[width=12cm, keepaspectratio, angle=0]{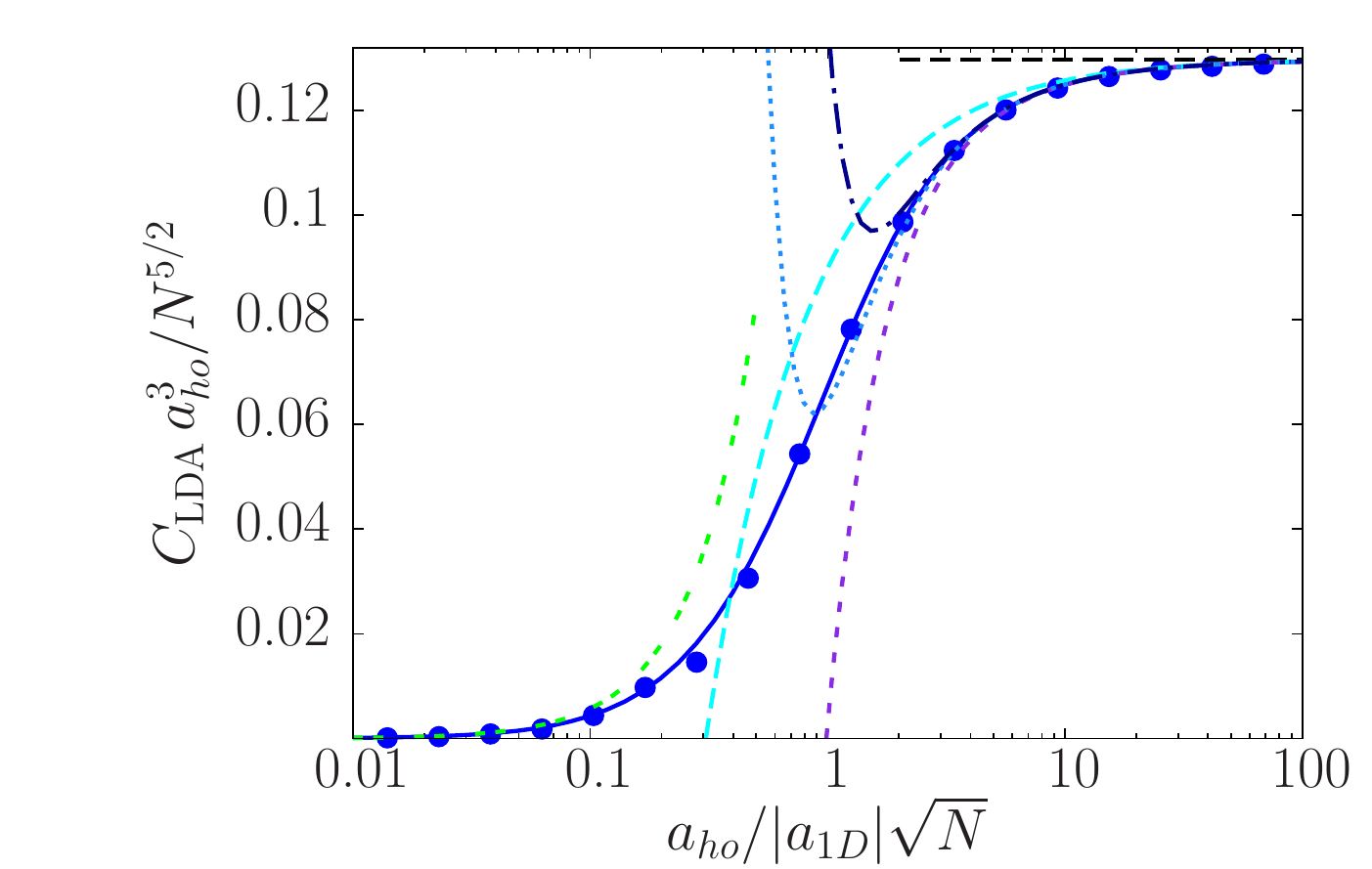}
\caption{Scaled Tan's contact for a 1D Bose gas (in units of $N/a_{ho}$) as a function of the
dimensionless interaction strength $\alpha_0/2=a_{ho} /(|a_{\mathrm{1D}}|\sqrt{N})$. Results from the strong-coupling
expansion: Tonks-Girardeau (horizontal long-dashed line, black), 1$^{st}$ order correction
(long-dashed, cyan), 2$^{nd}$ order correction (short-dashed, purple), 3$^{rd}$ order correction (dotted,
light blue), 4$^{th}$ order correction (dot-dashed, dark blue). Results at arbitrary interactions:
conjecture (blue dots), exact equation of state (data from Ref.~\cite{OlshaniiDunjko2003}, continuous,
blue). The weak-coupling expansion is also shown (double-dashed, green).}
\label{Fig12}
\end{figure}

Figure \ref{Fig12} summarizes our results for Tan's contact. Notice that, although the contact is scaled by the overall factor $N^{5/2}/a_{ho}^3$, it still depends on the number of bosons through the factor $\alpha_0/2=a_{ho}/|a_{1D}|\sqrt{N}$.
We have also applied the LDA numerically to the strong-coupling conjecture, Eq.~(\ref{exp}). The result is extremely close to the one obtained from the numerical solution of the Bethe Ansatz equation of state in \cite{OlshaniiDunjko2003}. By comparing the strong-coupling expansion with the results of the full calculation, we notice that the expansion (\ref{eq:strongc}) is valid down to $\alpha_0\simeq 6$, and provides a useful analytical expression for Tan's contact in an harmonic trap. In order to accurately describe the regime of lower interactions, a considerable number of terms would be needed in the strong-coupling expansion of the equation of state. The use of the conjecture (\ref{exp}, \ref{Lpol}) is thus a valuable alternative with respect to solving the Bethe Ansatz integral equations, the weak coupling expansion being applicable only for very weak interactions $\alpha_0\lesssim 0.1$.

\section{Summary of this chapter/r\'esum\'e du chapitre}

This chapter was devoted to the ground-state energy and static correlations of the Lieb-Liniger model. It began with a relatively detailed account of Lieb and Liniger's beautiful derivation of the integral equations that encode all the ground-state properties of their model in the thermodynamic limit. This procedure, based on Bethe Ansatz, does not rely on any approximation whatsoever. The Lieb-Liniger model is thus 'exactly solved', in the sense that its ground-state energy is expressed in closed form as the solution of relatively few equations, but this is not the end of the story, as this solution is not explicit at that stage.

To make quantitative predictions, the integral equation that governs the exact ground-state energy can be solved numerically, but this is not in the spirit of integrable models, that are supposed to be exactly solvable by fully-analytical techniques. Sophisticated mathematical methods give access to weak- and strong-coupling expansions of the ground-state energy, and are valid at arbitrary order. I have improved on a powerful method based on orthogonal polynomials, designed to study the strongly-interacting regime. I have studied it in detail to put into light its advantages and drawbacks, and proposed an alternative method explained in appendix, that is even more powerful.

The Lieb-Liniger model is thus exactly solved in a broader sense, that does not exhaust the problem however. Indeed, evaluating the numerical coefficients of these expansions is quite tedious, past the very first orders. My contributions to the problem are twofold. I have obtained the exact analytical coefficients of the strong-coupling expansion up to order twenty, while former studies stopped at order eight. In the weak-coupling regime, from numerical data available in the literature, I guessed the exact value of the third-order coefficient of the weak-coupling expansion. I also refined Tracy and Widom's conjecture on the structure of the weak-coupling expansion. 

As far as the ground-state energy is concerned, the next and last step towards the exact analytical solution would be to identify the generating function of either of these expansions, and sum the series explicitly. I took the first step by identifying patterns in the strong-coupling expansion, that enabled me to conjecture a partially-resummed form, whose validity range and accuracy are considerably enhanced compared to bare expansions. I even identified one type of coefficients involved. Most of my results are conjectural, which should not be surprizing as the mathematical field of analytical number theory is full of conjectures.

This solution, in turn, allowed me to obtain the local correlation functions of the Lieb-Liniger model, that give information on its degree of fermionization of the system and on its stability. Generalizing my previous conjectures on the energy to higher moments of the density of pseudo-momenta, I improved on the analytical evaluations of these quantities.

Then, I turned to non-local correlation functions. While long-range correlations are amenable to the Tomonaga-Luttinger liquid framework, their short-range expansion can be obtained systematically by Bethe Ansatz techniques. Focusing on the one-body correlation function, I constructed higher Hamiltonians and conserved quantities up to order four. I introduced the notion of connection to denote equations that relate local correlation functions, coefficients of their short-range series expansion, and moments of the density of pseudo-momenta. I derived most of them up to order four, simplifying some of the existing derivations, and identified them with most of the celebrated results in the literature, now unified in a single formalism. As a new result, I evaluated the fourth-order coefficient of the one-body correlation function semi-analytically, and found the interaction strength at which it changes sign with extraordinary precision.

The Fourier transform of the one-body correlation function is known as the momentum distribution, and is also amenable to perturbative expansions. The coefficient of its high-momentum universal $p^{-4}$ tail is known as Tan's contact. I used this observable to illustrate an extension of the Bethe Ansatz formalism to an harmonically trapped gas, in a non-integrable regime. This extension is realized by combination with the local-density approximation. I have developed a procedure that yields the expansion of Tan's contact for the trapped Bose gas to arbitrary order in the strongly-interacting regime, and used it to evaluate the corrections to the Tonks-Girardeau regime up to fourth order.
\\
\\
\\
\\
\\
Ce chapitre \'etait exclusivement consacr\'e \`a l'\'energie de l'\'etat fondamental du mod\`ele de Lieb et Liniger, ainsi qu'\`a ses fonctions de corr\'elation statiques \`a l'\'equilibre thermodynamique. J'y ai d\'etaill\'e dans un premier temps les principales \'etapes du raisonnement de Lieb et Liniger, fond\'e sur l'Ansatz de Bethe, pour ramener la solution exacte de leur mod\`ele \`a celle d'un ensemble restreint d'\'equations int\'egrales. Le mod\`ele est d\`es lors consi\'er\'e comme r\'esolu, mais dans une acception restreinte car l'\'energie de l'\'etat fondamental n'est pas exprim\'ee de fa\c con explicite \`a ce stade.

Les \'equations qui codent cette derni\`ere peuvent \^etre r\'esolues assez facilement par int\'egration num\'erique, d\`es lors qu'on fixe une valeur de l'intensit\'e des interactions. Toutefois, ce n'est pas vraiment l'esprit de la physique des mod\`eles int\'egrables, dont on vante souvent le m\'erite qu'ils ont de pouvoir \^etre r\'esolus analytiquement et de fa\c con exacte. Des m\'ethodes math\'ematiques relativement sophistiqu\'ees permettent de construire le d\'eveloppement limit\'e et asymptotique de l'\'energie de l'\'etat fondamental en fonction de l'intensit\'e des interactions. Je me suis concentr\'e sur la m\'ethode d\'evelopp\'ee par Zoran Ristivojevic, l'ai \'etudi\'ee en d\'etail pour en d\'evoiler les qualit\'es et d\'efauts, puis me suis attel\'e \`a la t\^ache d'y apporter des am\'eliorations et en proposer une alternative plus efficace.

Le mod\`ele de Lieb et Liniger est d\`es lors r\'esolu dans un sens plus vaste encore, qui ne suffit n\'eanmoins pas pour clore le probl\`eme. En effet, il s'av\`ere difficile, en pratique, de calculer les coefficients des diff\'erents d\'eveloppements en s\'erie analytiquement au-del\`a des premiers ordres. J'ai pouss\'e le d\'eveloppement \`a l'ordre vingt dans le r\'egime des fortes interactions, soit douze ordres plus loin que ce qui avait \'et\'e fait jusque l\`a, et j'ai identifi\'e le coefficient exact du troisi\`eme ordre dans le d\'eveloppement \`a faible interaction en m'appuyant sur des donn\'ees num\'eriques existantes.

L'ultime \'etape pour obtenir l'\'energie exacte de l'\'etat fondamental serait d'identifier la fonction g\'en\'eratrice de l'un ou l'autre des d\'eveloppements en s\'erie, et de sommer cette derni\`ere. J'ai entam\'e cette ascension finale en identifiant certains sch\'emas dans le d\'eveloppement \`a forte interaction, qui m'ont permis d'effectuer une resommation partielle de la s\'erie et, de ce fait, d'accro\^itre consid\'erablement le domaine de validit\'e de l'approximation par troncation \`a un ordre donn\'e. Ces r\'esultats sont pour la plupart conjecturaux, ce qui n'est pas particuli\`erement \'etonnant au vu de leur appartenance \`a la th\'eorie analytique des nombres.

Ces solutions approch\'ees extr\^emement pr\'ecises m'ont permis d'obtenir avec une pr\'ecision \'equivalente les corr\'elations locales du deuxi\`eme et troisi\`eme ordre, qui renseignent sur le degr\'e de fermionisation et la stabilit\'e du gaz, apr\`es avoir propos\'e quelques conjectures suppl\'ementaires concernant les moments d'ordre sup\'erieur de la densit\'e de pseudo-impulsions.

J'ai ensuite tourn\'e mon attention vers les corr\'elations non-locales. Ces derni\`eres sont bien d\'ecrites \`a grande distance par la th\'eorie des liquides de Tomonaga-Luttinger, tandis que leurs corr\'elations \`a courte distance sont accessibles, une fois encore, par Ansatz de Bethe. La construction des Hamiltoniens d'ordre sup\'erieur et des quantit\'es conserv\'ees associ\'ees m'a mis sur la voie du concept de connexion, que j'ai d\'efini comme \'etant une \'equation liant une fonction de corr\'elation locale, un coefficient du d\'eveloppement en s\'erie \`a courte distance d'une fonction de corr\'elation non-locale (la fonction de corr\'elation \`a un corps dans les cas que j'ai trait\'es), et divers moments de la distribution des quasi-impulsions. J'ai obtenu explicitement la majeure partie des connexions d'ordre inf\'erieur ou \'egal \`a quatre, et ai reconnu en ces derni\`eres bon nombre de r\'esultats consid\'er\'es comme importants dans la litt\'erature, qui n'avaient pas encore \'et\'e envisag\'es sous l'angle d'un formalisme unique. J'ai notamment pu \'evaluer l'intensit\'e des interactions pour laquelle le quatri\`eme coefficient de la fonction de corr\'elation \`a un corps s'annule et change de signe, avec une pr\'ecision in\'egal\'ee.

La transform\'ee de Fourier de la fonction de corr\'elation \`a un corps est plus connue sous le nom de distribution des impulsions. Un d\'eveloppement asymptotique de cette derni\`ere donne acc\`es au coefficient de sa d\'ecroissance en $p^{-4}$ \`a haute impulsion, connu sous le nom de contact de Tan. Je me suis appuy\'e sur l'exemple de cette observable pour illustrer une extension de l'Ansatz de Bethe au gaz de Lieb-Liniger plac\'e dans un pi\`ege harmonique, dans un r\'egime o\`u il n'est pas int\'egrable, fond\'ee sur une combinaison avec l'approximation de la densit\'e locale. J'ai d\'evelopp\'e une proc\'edure qui donne le contact de Tan \`a un ordre arbitraire, et l'ai utilis\'ee pour le calculer jusqu'\`a l'ordre quatre.

\section{Outlook of this chapter}

This chapter shows that, even at a basic level, the Lieb-Liniger model has not revealed all its secrets yet.

The exact, analytical expression of the ground-state energy is, more than ever before, a thriving open problem. I am not able to quantify the efforts still required to understand one of the asymptotic structures fairly enough to sum the series, but my feeling is that the problem is easier at strong coupling. In particular, a special role seems to be played by the quantity $1\!+\!2/\gamma$, associated to a ratio of Fredholm determinants \cite{NardisPanfil2016}. However, the new terms identified at weak coupling are far more interesting, as they involve the zeta function, the most celebrated one in the field of analytical number theory.

Several other examples of structures involving zeta functions had already been reported on, in the statistical physics, quantum field theory and string theory literature. This function appears in the Feynman diagrams of quantum electrodynamics, the $\phi^4$ model and correlation functions of the Heisenberg spin chain \cite{BoosKorepin2001}, but its presence in as simple a model as the Lieb-Liniger Bose gas came as a surprise, so I hope that it will attract attention.

Knowledge of the exact ground-state energy might help ingenuous mathematicians to prove new theorems on the zeta function. Examples I have in mind are the irrationality of $\zeta(3)/\pi^3$, that explicitly appears in the weak-coupling expansion of $e(\gamma)$, and of $\zeta(2n+1)$ for $n\geq 2$ (it is only known that an infinity of them is irrational \cite{Rivoal2000}, and a few improvements thereof). I conjecture that multiple zeta functions are also involved in this expansion at higher orders, $\zeta(3)^2$ to begin with.

The techniques used in the case of the Lieb-Liniger model could also be useful when applied to closely related models such as the Yang-Gaudin model, or for an extension to the super Tonks-Girardeau regime.
In particular, again from numerical data of Ref.~\cite{Prolhac}, I have conjectured that the weak-coupling expansion of the ground-state energy of the attractive spin-$1/2$ $\delta$-Fermi gas reads
\begin{eqnarray}
e(\gamma)=\sum_{k=0}^{+\infty}\tilde{b}_k\frac{\gamma^k}{\pi^{2k-2}}.
\end{eqnarray}
The known exact coefficients are \cite{TracyWidomII}
\begin{eqnarray}
\tilde{b}_0=\frac{1}{12}, \,\tilde{b}_1=-\frac{1}{2}, \,\tilde{b}_2=\frac{1}{6},
\end{eqnarray}
Prolhac guessed \cite{Prolhac}
\begin{eqnarray}
\tilde{b}_3\!=\!-\zeta(3), 
\end{eqnarray}
and from his numerical data I inferred
\begin{eqnarray}
\tilde{b}_4\!=\!-\frac{3}{2}\zeta(3),\,\, \tilde{b}_5\!=\!-3\zeta(3).
\end{eqnarray}
Numerical data may also allow to gain insight in the weak-coupling expansion of the higher moments of the density of pseudo-momenta, i.e. in the coefficients $a_{2k,i}$ in Eq.~(\ref{conjweak}).

Another purely theoretical issue that may allow to gain insight in the model and some mathematical aspects is the equivalence between the approach followed in the main text, and the alternative point of view based on a peculiar nonrelativistic limit of the sinh-Gordon model, investigated in the references associated to Appendix \ref{gkk}, that involves other integral equations. The equivalence of their formulations of the third-order local correlation functions has not been rigorously verified yet. It is not clear either whether the notion of connection can be adapted to this context.

As far as the BALDA formalism is concerned, the exact thermodynamics of a harmonically trapped gas is not explicitly known, but approximations could be improved by guesses and summations as in the homogeneous case. It could also be extended to other types of trapping, as an alternative to the techniques used in Ref.~\cite{BrunDubail2017}. In particular, the term beyond Tan's tail of the momentum distribution is still widely unexplored, both in the homogeneous and trapped case. For a homogeneous gas, it should be derived from higher-order connections.

In an experimental perspective, it is also important to investigate finite temperature thermodynamics of the Lieb-Liniger model \cite{Wang2013, LangHekkingMinguzzi2015, DeRosi2017}, that can be exactly obtained from the thermodynamic Bethe Ansatz approach introduced by Yang and Yang \cite{YangYang1969}. Needless to say, analytical approximate solutions are even more difficult to obtain in this case, but the interplay of statistics in $k$-space and interactions should be more tangible at finite temperature. A series of theoretical works have already tackled thermal correlation functions of the Lieb-Liniger model. Analytical approximate expressions have been obtained for the non-local $g_2$ correlation function in various regimes, and compared to numerical simulations in \cite{Kheruntsyan2003, Sykes2008, Deuar2009}, while $g_3$ has been studied in \cite{KormosMussardoTrombettoni}.

\newpage

\chapter{Dynamical structure factor of the Lieb-Liniger model and drag force due to a potential barrier}
\label{secIV}

\section{Introduction}

In this chapter, whose original results are mostly based on Refs.~\cite{LangHekkingMinguzzi2015, LangHekkingMinguzzi2017}, I take the next step towards the full characterization of a 1D Bose gas through its correlation functions. Going beyond static correlation functions, dynamical ones in energy-momentum space provide another possible way to understand a system, but their richer structure makes them harder to evaluate, and their theoretical study may involve fairly advanced techniques. Two observables attract peculiar attention: the Fourier transform of Green's function, a.k.a. the spectral function, and of the density-density correlations, known as the dynamical structure factor. The latter is quite sensitive to both interactions and dimensionality, providing an ideal observable to probe their joint effect.

Another strong motivation lies in the fact that equilibrium dynamical correlation functions yield valuable information about the response of a fluid to a weak external excitation. This response is the central object of linear response theory, that gives insight into slightly out of equilibrium situations. In this perspective, the dynamical structure factor governs the response of a fluid to a weak external potential locally coupled to its density. More precisely, it is linked to the drag force experienced by a single impurity, that characterizes the viscosity of its flow. I take this opportunity to dwell on the issue of superfluidity, a concept associated to the dramatic phenomenon of frictionless flow observed in quantum fluids below a critical velocity.

This chapter is organized as follows: first, I recall a few experimental facts related to superfluidity and their historical interpretation, then I present Landau's criterion for superfluidity, and the drag force criterion as a generalization thereof.

Following chapter \ref{secIII}, I still consider the Lieb-Liniger model. The Tonks-Girardeau regime is amenable to exact calculations, and at finite interaction strength I use the Tomonaga-Luttinger liquid framework, keeping in mind that its validity range is limited to low energies or small flow velocities. Refining the analysis to get closer to experimental situations, I also investigate finite temperature, as well as the effect of the barrier width on the drag force, putting into light a quasi-superfluid supersonic regime.

To finish with, as a first step towards beyond Luttinger liquid quantitative treatment, I examine the exact excitation spectrum of the Lieb-Liniger model using coordinate Bethe Ansatz techniques, and give quantitative bounds for the validity range of the Tomonaga-Luttinger liquid framework in terms of interaction strength.
\\
\\
\\
\\
\\
Dans ce chapitre, je franchis un pas de plus vers la caract\'erisation compl\`ete d'un gaz de Bose unidimensionnel par ses fonctions de corr\'elation, en consid\'erant une facette suppl\'ementaire de ces derni\`eres, \`a savoir leur dynamique dans l'espace des \'energies et impulsions. Leur structure s'av\`ere plus complexe, et par cons\'equent plus difficile \`a obtenir que dans le cas statique. Deux observables attirent particuli\`erement l'attention: la transform\'ee de Fourier de la fonction de Green, connue sous le nom de fonction spectrale, et celle des corr\'elations spatio-temporelles en densit\'e, appel\'ee facteur de structure dynamique. Ce dernier est particuli\`erement sensible \`a la fois \`a la dimension du syst\`eme et aux interactions, ce qui en fait une observable de choix pour sonder leurs effets conjoints.

Une autre motivation, et non des moindres, vient du fait que les fonctions de corr\'elation dynamiques \`a l'\'equilibre renseignent sur la r\'eponse du syst\`eme \`a d'\'eventuelles perturbations de nature externe. La th\'eorie de la r\'eponse lin\'eaire permet en effet d'en d\'eduire la dynamique de situations l\'eg\`erement hors \'equilibre. Dans cette perspective, le facteur de structure dynamique gouverne la r\'eponse d'un fluide \`a une barri\`ere de potentiel de faible amplitude coupl\'ee localement \`a la densit\'e. Plus pr\'ecis\'ement, la donn\'ee du facteur de structure dynamique et de la forme pr\'ecise de la barri\`ere de potentiel, qui mod\'elise un faisceau laser ou une impuret\'e, permet d'obtenir la force de tra\^in\'ee, qui caract\'erise la viscosit\'e de l'\'ecoulement. J'en profite pour m'attarder quelque peu sur la notion de superfluidit\'e, traditionnellement associ\'ee \`a un \'ecoulement parfait en-dessous d'une vitesse critique.

Le chapitre s'organise de la mani\`ere suivante: dans un premier temps, je rappelle certains r\'esultats exp\'erimentaux historiques et leur interpr\'etation th\'eorique. Un pas d\'ecisif a notamment \'et\'e franchi par Landau lorsqu'il a \'enonc\'e son crit\`ere de superfluidit\'e, dans la filiation duquel s'inscrit le crit\`ere de la force de tra\^in\'ee, sur lequel se fonde mon \'etude.

Dans la veine du Chapitre \ref{secIII}, je consid\`ere encore une fois un gaz de Bose unidimensionnel, d\'ecrit par le mod\`ele de Lieb et Liniger. Moyennant l'hypoth\`ese de faible barri\`ere d'\'epaisseur nulle, je traite le r\'egime de Tonks-Girardeau sans autre approximation. Dans le cas d'une interaction finie, je m'appuie sur le mod\`ele de Tomonaga-Luttinger, tout en gardant en m\'emoire que son domaine de validit\'e est limit\'e \`a une zone restreinte de basse \'energie ou de faible vitesse.

Afin de gagner en r\'ealisme, je m'int\'eresse aussi aux effets thermiques et \`a une barri\`ere de potentiel d'\'epaisseur non-nulle, ce qui me permet de mettre en \'evidence un r\'egime supersonique quasi-superfluide, caract\'eris\'e par l'\'evanescence de la force de tra\^in\'ee. L'\'etape suivante pour raffiner l'analyse serait d'envisager un mod\`ele au-del\`a du liquide de Luttinger. Afin de faire des pr\'evisions quantitatives dans ce cadre, dans un premier temps il s'av\`ere n\'ecessaire d'\'evaluer le spectre d'excitation du mod\`ele de Lieb et Liniger, que j'obtiens par Ansatz de Bethe. La comparaison de cette solution exacte et des pr\'edictions du mod\`ele de Tomonaga-Luttinger dans sa version standard met en lumi\`ere ses limites, et me permet de donner pour la premi\`ere fois une borne sup\'erieure quantitative \`a son domaine de validit\'e.

\section{Conceptual problems raised by superfluidity, lack of universal criterion}

Attaining a full understanding of the microscopic mechanisms behind superfluidity is among the major challenges of modern physics. An historical perspective shows that experiments constantly challenge theoretical understanding \cite{Leggett1999, Balibar2007, Griffin2009, Albert2009} and that, although four Nobel Prizes have already been awarded for seminal contributions to this complicated topic (to Landau in 1962, Kapitza in 1978, Lee, Osheroff and Richardson in 1996 and to Abrikosov, Ginzburg and Leggett in 2003), interest in the latter shows no sign of exhaustion whatsoever. 

\subsection{Experimental facts, properties of superfluids}

Superfluids are one of the most appealing manifestations of quantum physics at the macroscopic scale. They seem to defy gravity and surface tension by their ability to flow up and out of a container, or through narrow slits and nanopores at relatively high velocity. Another famous property is the fountain effect \cite{Allen1938}: when heat is applied to a superfluid on one side of a porous plug, pressure increases proportionally to the heat current so that the level of the free surface goes up, and a liquid jet can even occur if pressure is high enough. If the liquid were described by classical hydrodynamics, the vapor pressure would be higher on the warm side so that, in order to maintain hydrostatic equilibrium, the liquid level would have to go down.

Superfluidity is also characterized by a sharp drop of viscosity and thermal resistivity at the transition temperature, as shown by the early historical experiments involving liquid ${^4}$He. In this system, the superfluid transition is observed at a temperature $T_{\lambda}\!\simeq$ 2.2K, the lambda point, separating its phases called He I (above) and He II (below) \cite{Keesom1936, Kapitza1938, AllenMisener1938}.

These facts remind of the sudden fall of resistivity previously witnessed in superconductors, hinting at an analogy, or even a deep connection between both phenomena. Superconductivity is traditionally explained by the formation of Cooper pairs of electrons in a metal, as prescribed by the Bardeen-Cooper-Schrieffer (BCS) theory \cite{BCSa, BCSb}. For this picture to emerge in the context of superfluids, it took the unexpected observation of superfluidity in ${^3}$He \cite{Osheroff1972}, at temperatures lower than 2.6mK. This historical step bridged the superfluidity of helium and the phenomenon of superconductivity, as the underlying mechanism was identified as the formation of pairs of fermionic ${^3}$He atoms \cite{Leggett1975}.

The superfluidity of ${^4}$He, however, seemed associated to the BEC of these bosonic atoms \cite{Tisza1947}. London was the first to relate superfluidity to Bose-Einstein condensation, through the heuristic observation that the experimental value of the superfluid critical temperature measured in ${^4}$He is close to the theoretical condensation temperature of an ideal Bose gas at the same density (an intuition sometimes referred to as the 'London conjecture') \cite{London1938}. This picture has to be nuanced as helium is a strongly-interacting liquid, the ideal Bose gas is actually not superfluid, and the superfluid fraction $n_s/n$ is equal to one at $T\!=\!0$, while only ten percent of the atoms are Bose condensed in ${^4}$He.

This notion of superfluid fraction (in analogy to the condensate fraction in BEC) stems from the Tisza-Landau two-fluid model \cite{Tisza1938, Landau1941}, that pictures quantum fluids as containing two impenetrable parts, a normal (associated to the index $n$) and a fully superfluid one (indexed by $s$), such that the total density reads $n\!=\!n_n\!+\!n_s$. The normal part behaves like a Newtonian, classical fluid, while the superfluid component does not carry entropy and has zero viscosity. In particular, the two-fluid hydrodynamic second sound velocity is associated to superfluid density. This collective mode is an entropy wave, with constant pressure, where superfluid and normal densities oscillate with opposite phases.

For decades, the two isotopes of helium have been the only known examples of quantum fluids, as Helium is the only element that is naturally liquid at the very low temperatures where quantum effects arise. Much later, starting during the very last years of the $20^{th}$ century, superfluidity has also been observed in ultracold gases, at temperatures of the order of a few dozens of nK. Through their high degree of tunability, such systems provide a versatile tool to study superfluidity in simplified situations. A paradigmatic example is the weakly-interacting Bose gas \cite{Raman1999, Onofrio2000}, which is far less complex than helium as its interactions have a simpler structure, are much weaker, and its excitation spectrum features phonons but no complicated roton excitation. Superfluidity has then been studied along the BEC-BCS crossover \cite{Miller2007, Weimer2015}, and in a Bose-Fermi counterflow, where a Bose-Einstein condensate plays the role of an impurity in a degenerate Fermi fluid \cite{Sushi2015}. Ultracold atoms also allow to study superfluidity on a lattice \cite{Greiner2002}, where it is opposed to a Mott insulating phase. In the superfluid phase, the atoms can hop from one site to another and each of them spreads out over the whole lattice with a long-range phase coherence. More recently, polaritons in microcavities have provided a new kind of system to explore the very rich physics of non-equilibrium quantum fluids \cite{Amo2009, Boulier2016, Gallemi2017}, up to room temperature \cite{Lerario2017}.

Even in view of this collection of experimental results, theoretical characterization of superfluidity remains quite challenging. Frictionless flow is the historical criterion; the existence of quantized vortices, quantized circulation, persistent currents (i.e. metastability of superflow states) or absence of a drag force are also commonly invoked, but to what extent these manifestations are equivalent or complementary to each others is far from being settled. This puzzling situation is even more problematic in cases that have not been considered at the beginning. For instance, superfluidity could exist in 1D, where BEC does not, and out-of-equilibrium gases complexify this picture, leading to scenarios where a few possible definitions are satisfied, and others not \cite{KillingBerloff}. In our current understanding, superfluidity is rather an accumulation of phenomena, preventing, so far, a straighforward and universal description to emerge.

Equilibrium superfluidity can be probed experimentally through the Hess-Fairbank effect: when the walls of a toroidal container or a bucket are set into rotation adiabatically with small tangential velocity, a superfluid inside stays at rest while a normal fluid would follow the container. This leads to a nonclassical rotational inertia, that can be used to determine the superfluid fraction, providing an indirect validation of the two-fluid description \cite{HessFairbank1967}. A superfluid is also described by a macroscopic wave function $\psi(\vec{r})$ \cite{London1938}, as are Bose-Einstein condensates and superconductors, which implies phase coherence. The superfluid wave function can be expressed as
$\psi(\vec{r})\!=\!|\psi( \vec{r} )|e^{i\phi(\vec{r})}$ in modulus-phase representation, and the superfluid velocity $\vec{v}_s$ is characterized by the gradient of the phase $\phi$ through the relation
\begin{eqnarray}
\label{irrot}
\vec{v}_s=\frac{\hbar}{m}\vec{\nabla} \phi(\vec{r}),
\end{eqnarray}
where $\vec{\nabla}$ is the nabla operator. A consequence of Eq.~(\ref{irrot}), is that the flow is always irrotational ($\mathrm{curl}(\vec{v}_s)\!=\!\vec{0}$), a characteristic shared by Bose-Eintein condensates. The phase $\phi$ is single-valued, leading to the existence of quantized vortices (as long as the fluid is not confined to 1D), as first predicted in helium \cite{Onsager1949} and experimentally observed in the same system, long before the ultracold gases experiments already evoked in chapter~\ref{secII}.

Among all possible criteria for superfluidity, I will delve deeper into the so-called drag force criterion, which is one of the most recent. Due to its historical filiation, I will first introduce the most celebrated and famous criterion for superfluidity: Landau's criterion.

\subsection{Landau's criterion for superfluidity}

Why should superfluids flow without friction, while normal fluids experience viscosity? The first relevant answer to this crucial question was provided by Landau, who proposed a mechanism to explain why dissipation occurs in a normal fluid, and under what condition it is prevented \cite{Landau1941}. His phenomenological argument, of which I give a simplified account, is based on the following picture: in a narrow tube, fluid particles experience random scattering from the walls, that are rough at the atomic level. This mechanism transfers momentum from the fluid to the walls, leading to friction in a normal fluid.

Formally, in the reference frame moving with the fluid, let us denote by $E_0$ the energy of the fluid and by $P_0$ its momentum. If it starts moving with the walls, its motion must begin through a progressive excitation of internal moves, therefore by the appearance of elementary excitations. If $p$ denotes the momentum of an elementary excitation and $\epsilon$ its energy, then $\epsilon(p)$ is the dispersion relation of the fluid. Thus, $E_0\!=\!\epsilon(p)$ and $P_0\!=\!p$.

Then, going back to the rest frame of the capillary, where the fluid flows with velocity $\vec{v}$, the energy $E$ of the fluid in this frame of reference is obtained by means of a Galilean transformation and reads
\begin{eqnarray}
E=\epsilon+\vec{p}\cdot\vec{v}+\frac{Mv^2}{2},
\end{eqnarray}
where $M\!=\!Nm$ is the total mass of the fluid and $Mv^2/2$ its kinetic energy. The energy variation caused by dissipation through an elementary excitation is $\epsilon(p) + \vec{p}\cdot\vec{v}$, and is necessarily negative. It is minimal when $\vec{v}$ and $\vec{p}$ are anti-parallel, imposing $\epsilon-pv\leq 0$, and as a consequence the flow should be dissipationless at velocities lower than
\begin{eqnarray}
\label{Landcrit}
v_c=\mathrm{min}_p\left[\frac{\epsilon(p)}{p}\right].
\end{eqnarray}
Equation (\ref{Landcrit}) links the microscopic observable $\epsilon$ to a macroscopic one, the critical velocity $v_c$. A direct consequence of this equation is that systems such that $\mathrm{min}_p[\epsilon(p)/p]=0$ can not be superfluid. In particular, the minimum of $\epsilon(p)/p$ is solution to $\partial_p(\epsilon/p)=0$, hence
\begin{eqnarray}
\label{condvc}
\left(\frac{\partial \epsilon}{\partial p}\right)_{\!v=v_c}\!\!\!\!=\frac{\epsilon}{p}.
\end{eqnarray}
Equation (\ref{condvc}) means that, at $v_c$, the group and phase velocities of the fluid coincide. Graphically, the tangent to the excitation spectrum coincides with the line between this point and the origin. In particular, for a system to be superfluid, its dispersion relation should not be tangent to the $p$-axis at the origin. In an ideal Bose gas, where $\epsilon(p)\!=\!\frac{p^2}{2m}$, the rough walls can always impart momentum to the fluid, leading to viscous friction. More generally, gapless systems with zero slope at the origin in energy-momentum space are not superfluid. On the contrary, helium is superfluid according to Eq.~(\ref{Landcrit}) in view of its dispersion relation, and so is the weakly-interacting Bose gas, as shown by Bogoliubov who derived the approximate spectrum \cite{Bogoliubov1947}
\begin{eqnarray}
\epsilon_{Bog}(p)=\sqrt{p^2c^2+\left(\frac{p^2}{2m}\right)^2}.
\end{eqnarray}

Predicting the existence of a critical velocity is one of the major contributions of Landau's to the theory of superfluids. It has been observed in Bose gases \cite{Raman1999, Onofrio2000}, where $v_c$ is of the order of a few mm/s, showing that Landau's criterion is qualitatively correct, studied in 2D, both experimentally \cite{Desbuquois2012} and theoretically \cite{Singh2017}, and then along the BEC-BCS transition \cite{Miller2007}. In the BCS regime, pair-breaking excitations are expected to limit $v_c$, while for a Bose-Fermi counterflow, Landau's criterion can be adapted and becomes \cite{Castin2015}
\begin{eqnarray}
v_c^{BF}=\mathrm{min}_p\left[\frac{\epsilon^F(p)+\epsilon^B(p)}{p}\right].
\end{eqnarray}

Upon closer inspection, the critical velocity as predicted by Landau's criterion is overevaluated compared to most experiments, sometimes by one order of magnitude. A first explanation is that nonlinear processes are neglected in Landau's approach, such as vortices in 2D, and vortex rings in 3D \cite{Feynman1957, Zwerger2000}. In 1D also, it would certainly yield a too high value as well, as it neglects solitons \cite{Hakim1997, Pavloff2002}. It is also important to keep in mind that Landau's argument is purely kinematical and classical in its historical formulation. There is no guarantee that one can apply it to understand dynamical and quantum aspects of superfluidity. Another criticism is that Galilean invariance is a crucial assumption thus the criterion does not apply to inhomogeneous systems \cite{Fedichov2001}.

More generally, correlations, fluctuations and interactions should be addressed correctly to quantitatively understand the mechanisms behind superfluidity. Coming back to Eq.~(\ref{Landcrit}), it is possible that even when the line of slope $v$ intersects the spectrum, the transition probability to this state is strongly suppressed due to interactions or to the specific kind of external perturbing potential. These issues can be tackled using a more involved formalism, that I will use hereafter.

\section{Drag force as a generalized Landau criterion}

In a seminal article \cite{AstrakharchikPitaevskii2004}, Astrakharchik and Pitaevskii developed a quantitative approach to the problem of superfluidity of a generic fluid, whose basic idea relies on an analogy with classical physics, where an object moving in a fluid experiences viscosity. At the classical level, viscous friction is described phenomenologically by a force opposed to the direction of motion, and that depends on its velocity. In first approximation, this force scales linearly with velocity for a laminar flow, and quadratically in the turbulent regime. Its prefactor is usually considered as a phenomenological quantity, that depends on the viscosity of the fluid and on the shape of the object.

This drag force is not a fundamental ingredient of the theory as directly defined in its principles, but arises due to collective, extremely complicated phenomena. The quantum framework, however, is appropriate to describe the motion of a single impurity, immerged into a fluid and coupled to its density at the atomic scale. Prior to any calculation, based on classical fluid dynamics and Landau's criterion, one can expect the following behavior: if an impurity moves slowly enough inside a superfluid, then its motion does not lead to friction, and as a consequence, in its frame of reference the velocity of the flow remains constant.

In a setup with periodic boundary conditions (to avoid revivals due to rebounces on walls, that complexify the analysis), a persistent flow should be observed, which is one manifestation of superfluidity. According to Newton's laws of motion, if perchance they hold in this context, the drag force experienced by the impurity must be strictly zero. Above a critical velocity, however, superfluidity can not be sustained anymore. Then, it is expected that the impurity experiences a drag force from the fluid, and slows down.

Defining this drag force at the quantum statistical level was the first challenge to change this intuition into a quantitative theory, since the very notion of force is usually absent from the formalisms of quantum physics. A first possible definition, already proposed in \cite{Pavloff2002}, is
\begin{eqnarray}
\label{defF}
\vec{F}=-\int d^dr|\psi(\vec{r})|^2\vec{\nabla}U,
\end{eqnarray}
where $\psi$ is the macroscopic wavefunction that describes a Bose-condensed fluid, and $U$ the potential that models to the perturbation due to the impurity. Equation (\ref{defF}) can be seen as the semi-classical analog to the classical definition of a force in terms of the gradient of a potential. Actually, the drag force corresponds to the opposite of the force one should exert on the impurity to keep its velocity constant.

The impurity adds a perturbation term to the Hamiltonian, written as
\begin{eqnarray}
H_{pert}=\int d^dr|\psi(\vec{r})|^2U(\vec{r}-\vec{v}t),
\end{eqnarray}
where $\vec{v}$ is the constant velocity at which the impurity moves inside the fluid. In the case of a point-like impurity, the potential reads
\begin{eqnarray}
U(\vec{r})=g_i\delta(\vec{r}),
\end{eqnarray}
where $g_i$ is the impurity-fluid interaction strength. This picture of a point-like polaron is quite realistic if one has in mind an experience involving neutrons as impurities, for instance. In \cite{AstrakharchikPitaevskii2004}, it is also assumed that the impurity is heavy, so that it does not add a significant kinetic energy term to the fluid-impurity Hamiltonian.

\begin{figure}
\includegraphics[width=7cm, keepaspectratio, angle=0]{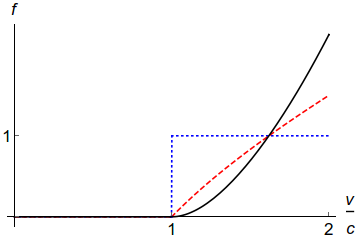}
\caption{Dimensionless drag force  $f\!=\!F_d/\!\left(\frac{s_{d-1}}{(2\pi)^{d-1}}\frac{m^dg_i^2n_{d}}{\hbar^{d+1}}\right)$ in a Bose gas due to a heavy, point-like impurity, as a function of the Mach number $v/c$, as predicted by the Bogolibov theory in dimension 1 (dotted, blue), 2 (dashed, red) and 3 (solid, black)}
\label{FF}
\end{figure}

Still in \cite{AstrakharchikPitaevskii2004}, this drag force formalism has been applied to the weakly-interacting Bose gas, as described by the Gross-Pitaevskii equation, and the norm of the drag force has been evaluated in dimension one (where the result was already known \cite{Pavloff2002}), two and three, using the Born approximation that supposes sufficiently low values of $g_i$. For compactness, I merge the results into a single expression, namely
\begin{equation}
F_{d}(v)=\frac{s_{d-1}}{(2\pi)^{d-1}}\frac{m^dn_{d}g_i^2}{\hbar^{d+1}}\left(\frac{v^2-c_d^2}{v}\right)^{d-1}\Theta(v-c_d),
\end{equation}
where $n_d=N/L^d$ is the $d$-dimensional density, $s_{d-1}$ the area of the unit sphere, $c_d$ is the sound velocity in the mean field approach and plays the role of a critical velocity due to the Heaviside $\Theta$ function, as illustrated in Fig.~\ref{FF}.

The dimensionless drag force profile is strongly dimension-dependent, except at a special point where all dimensionless drag forces are equal, irrespective of the dimension, when the Mach number is equal to the golden ratio, i.e. $v/c_d\!=\!(1\!+\!\sqrt{5})/2$. More importantly, at the mean field level a subsonic flow is superfluid. In this sense, the drag force formalism can be seen as a quantitative extension of Landau's criterion \cite{AstrakharchikPitaevskii2004}. Note that, for an ideal Bose gas, the sound velocity vanishes and thus it is not superfluid, confirming that interactions are a crucial ingredient of superfluidity.

We are now equipped with a qualitative criterion for superfluidity \cite{Pavloff2002, AstrakharchikPitaevskii2004}: \textit{if there exists a non-zero flow velocity such that the drag force is strictly zero, then the superfluid fraction is equal to one}. As an addendum to this criterion \cite{AstrakharchikPitaevskii2004, LangHekkingMinguzzi2015}: \textit{the higher the drag force, the farther the flow is from being superfluid}.

An advantage of the drag force criterion is that it accomodates perfectly well with the historical definition of superfluidity as a flow without viscosity. A major drawback, already noticed by the authors of Ref.~\cite{AstrakharchikPitaevskii2004} themselves, is the lack of obvious way to define a superfluid fraction from the drag force, that would coincide with the one predicted by the two-fluid approach.

There is also a second definition of the drag force \cite{AstrakharchikPitaevskii2004}, that has been the most popular in the subsequent literature:
\begin{eqnarray}
\label{defFbis}
\dot{E}=-\vec{F}\cdot\vec{v},
\end{eqnarray}
where $\dot{E}$ is the statistical average energy dissipation per unit time. This definition is in analogy with the classical mechanics formula that links energy dissipation per unit time, i.e. power, to the force responsible for energy transfer to the environment.

Within this approach, the energy variation per unit time due to the impurity is calculated first, and the drag force is deduced from the latter. Definition (\ref{defFbis}) is quite convenient for experimentalists, in the sense that energy dissipation is related to the heating rate, which is a measurable quantity \cite{Desbuquois2012, Weimer2015}. It is quite complicated to probe such tiny drag forces (of the order of a few nN) in a direct way, although a recent proposal to study superfluidity of light based on an optomechanical, cantilever beam device, seems quite promising \cite{Larre2015}.

From the theoretical point of view, Eq.~(\ref{defFbis}) does not provide a simple means to evaluate the drag force analytically in full generality. To go further, a useful approximation was developped in \cite{AstrakharchikPitaevskii2004}. If a weak potential barrier or impurity is stirred along the fluid, putting it slightly out of equilibrium, then linear response theory holds, and the average energy dissipation per unit time is linked to the dynamical structure factor (see Appendix \ref{linrespDSFF} for more details on this observable)
\begin{equation}
S_{d}(\vec{q},\omega)\!=\!\int_{-\infty}^{+\infty}\!\!\!\!dt\!\int\! d^dr\ \!e^{i(\omega t-\vec{q}\cdot \vec{r})} \langle \delta n(\vec{r},t)\delta n(\vec{0},0)\rangle,
\end{equation}
by the relation, valid in arbitrary dimension \cite{LangHekkingMinguzzi2015}:
\begin{eqnarray}
\label{linF}
&&\dot{E}\!=\!-\frac{1}{2\pi\hbar V_d}\!\int_0^{+\infty}\!\!\!d\omega\!\! \int \!\!\frac{d^dq}{(2\pi)^d} S_{d}(\vec{q},\omega)|U_{d}(\vec{q},\omega)|^2\omega,
\end{eqnarray}
where 
\begin{eqnarray}
U_{d}(\vec{q},\omega)=\int_{-\infty}^{+\infty}dt\,\int d^{d}r\,e^{i(\omega t-\vec{q}\cdot \vec{r})}\,U_d(\vec{r},t)
\end{eqnarray}
is the Fourier transform of the potential barrier $U_d(\vec{r},t)$, that defines the perturbation part of the Hamiltonian as 
\begin{eqnarray}
H_{pert}=\int d^dr\, U_d(\vec{r},t)n_d(\vec{r}).
\end{eqnarray}
In 1D, equation (\ref{linF}) was obtained in \cite{AstrakharchikPitaevskii2004}, and recovered from the Fermi Golden rule in \cite{ChernyCauxBrand2012}. The assumption of a weak fluid-barrier coupling is not well controlled in the derivation, but I will assume that it is fulfilled all the same. Note that two quantities are involved in the integrand of the right-hand side of Eq.~(\ref{linF}): the Fourier transform of the potential barrier, and the dynamical structure factor of the gas.

This quantity is worth studying for itself and has been measured by Bragg spectroscopy since the early days of ultracold gases experiments \cite{Stenger1999, Stamper1999}. For this reason, I will devote next paragraph to the dynamical structure factor, starting as usual by considering the Tonks-Girardeau regime to gain some insight, before turning to the more complicated issue of finite interaction strengths. From now on in this chapter, I will focus on the 1D Bose gas as described by the Lieb-Liniger model, postponing the issue of higher dimensions to chapter \ref{secV}.

\section{Dynamical structure factor and drag force for a Tonks-Girardeau gas}

\subsection{Dynamical structure factor}

\begin{figure}
\includegraphics[width=7cm, keepaspectratio, angle=0]{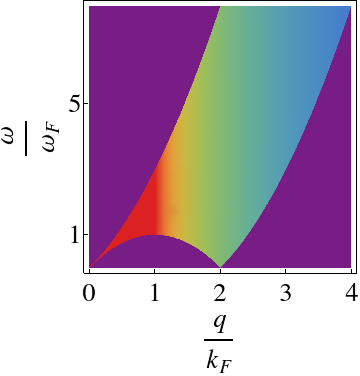}
\caption{Dynamical structure factor of a Tonks-Girardeau gas at zero temperature in energy-momentum space, where warmer colors correspond to higher values. Excitations occur only inside the area bounded by the excitation spectra of the gas, and are prevented in a low-energy region due to kinematical constraints in 1D. Note that the spectrum is quasi-linear at the origin and around the umklapp point ($q\!=\!2k_F, \omega\!=\!0$), which is a signature of well-defined phonon-like excitations.}
\label{STGplot}
\end{figure}

As a first step towards the drag force, I evaluate the dynamical structure factor of a one-dimensional Bose gas,
\begin{equation}
\label{DSF}
S(q,\omega)=\int_{-\infty}^{+\infty}\!\!\!\int_{-\infty}^{+\infty}\!\!\!dx\,dt\,e^{i(\omega t-qx)} \langle \delta n(x,t) \delta n(0,0) \rangle,
\end{equation}
in the Tonks-Girardeau regime where the Lieb-Liniger model is equivalent to a gas of noninteracting fermions for this observable, due to the Bose-Fermi mapping. Calculating fermionic density-density correlations using Wick's theorem yields after Fourier transform, as detailed in Appendix \ref{derDSFTG}, the well-known result \cite{VignoloMinguzziTosi2001, ChernyBrand2005}:
\begin{equation}
\label{TonksDSF}
S^{TG}(q,\omega)=\frac{m}{\hbar |q|}\Theta\!\left[\omega_+(q)-\omega\right]\Theta\!\left[\omega-\omega_-(q)\right],
\end{equation}
where
\begin{eqnarray}
\omega_+(q)=\frac{\hbar}{2m}(q^2+2qk_F)
\end{eqnarray}
and
\begin{eqnarray}
\omega_-(q)=\frac{\hbar}{2m}|q^2-2qk_F|
\end{eqnarray}
are the limiting dispersion relations. They represent the boundaries of the energy-momentum sector where particle-hole excitations can occur according to energy conservation in the thermodynamic limit, known as the particle-hole continuum and illustrated in Fig.~\ref{STGplot}. They also correspond to the type I (+) and type II (-) excitation spectra of the Lieb-Liniger model at infinite interaction strength, already evoked in chapter \ref{secII}.

At zero temperature, the dynamical structure factor features jumps from a strictly zero to a finite value at these thresholds, but is regularized by smooth smearing at finite temperature. A natural question is up to what energies and temperatures the phonon-like excitations, characterized by the nearly-linear spectrum at low energy, are well-defined when thermal effects come into play.

The dynamical structure factor can also be obtained from the fluctuation-dissipation theorem as
\begin{eqnarray}
\label{flucdiss}
S(q,\omega)=\frac{2\hbar}{1\!-\!e^{-\beta \hbar \omega}}\Im\!\left[\chi_{nn}(q,\omega)\right],
\end{eqnarray}
relating it to the imaginary part of the linear density-density response function $\chi_{nn}$. Lindhard's expression \cite{Ashcroft} is valid for noninteracting fermions and, consequently, also for the Tonks-Girardeau gas:
\begin{equation}
\label{LinR}
\chi_{nn}(q,\omega)=\frac{1}{L}\sum_k\frac{n_F(k)-\!n_F(k+q)}{\hbar\omega+\epsilon(k)-\epsilon(k+q)+i0^+},
\end{equation}
where
\begin{eqnarray}
n_F(k)=\frac{1}{e^{\beta[\epsilon(k)-\mu]}+1}
\end{eqnarray}
is the Fermi-Dirac distribution and $\epsilon(k)\!=\!\frac{\hbar^2 k^2}{2m}$ is the dispersion relation of noninteracting spinless fermions, $\mu$ the chemical potential, and the infinitesimal imaginary part $i0^+$ ensures causality. From Eqs.~(\ref{flucdiss}) and (\ref{LinR}), using the property
\begin{eqnarray}
\frac{1}{X+i0^+}=\mathrm{P.P.}\!\left(\frac{1}{X}\right)-i\pi \delta(X),
\end{eqnarray}
where P.P. is the principal part distribution. I deduce a practical expression of the finite-temperature dynamical structure factor in the thermodynamic limit:
\begin{eqnarray}
\!\!\!\!\!\!\small{S^{TG}_{T>0}(q,\omega)=\int_{-\infty}^{+\infty}\!dk\, \frac{n_F(k)\!-\!n_F(k+q)}{1-e^{-\beta\hbar\omega}}\,\delta\!\left[\omega-\omega_q(k)\right]},
\end{eqnarray}
where $\omega_q(k)=\frac{1}{\hbar}[\epsilon(k+q)-\epsilon(k)]$. This is equivalent to
\begin{equation}
\small{S^{TG}_{T>0}(q,\omega)\!=\!\int_{-\infty}^{+\infty}\! \! dk \, n_F(k)\left[1-n_F(k+q)\right]\delta\left[\omega-\omega_q(k)\right]}.
\end{equation}

\begin{figure}
\includegraphics[width=7cm, keepaspectratio, angle=0]{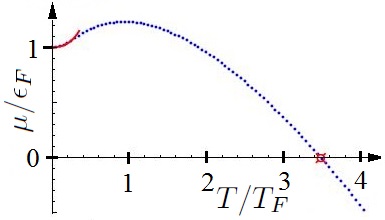}
\caption{Dimensionless chemical potential of a noninteracting one-dimensional Fermi gas as a function of the dimensionless temperature. Blue dots represent numerical data, the red curve corresponds to the Sommerfeld approximation at low temperature, and the red symbol to the analytical result for the annulation of the chemical potential. The chemical potential starts increasing with $T$, reaches a maximum and then decreases monotonically, contrary to the 3D case where it is a strictly decreasing function of temperature.}
\label{tonksmuT}
\end{figure}

Either of them can be used to obtain, after a few algebraic manipulations, the final expression
\begin{equation}
\label{DSFTG}
S^{TG}_{T>0}(q,\omega)=\frac{m}{\hbar|q|} \frac{n_F\!\left[\frac{\hbar\omega-\epsilon(q)}{\hbar^2 q/m}\right]-n_F\!\left[\frac{\hbar\omega+\epsilon(q)}{\hbar^2 q/m}\right]}{1-e^{-\beta\hbar\omega}}.
\end{equation}
To complete the calculation and make quantitative predictions, it is first necessary to determine the temperature dependence of the chemical potential, that appears in the Fermi-Dirac distribution. It is obtained by numerical inversion of the equation fixing the density $n_0$,
\begin{equation}
\label{invmu}
\frac{1}{2\pi}\int_{-\infty}^{+\infty}\mathrm{d}k \ n_F(k) =n_0.
\end{equation}
Equation (\ref{invmu}) can not be solved analytically in full generality. At low temperature, Sommerfeld's expansion \cite{Ashcroft} yields the approximate result \cite{LangHekkingMinguzzi2015}:
\begin{eqnarray}
\frac{\mu(T)}{\epsilon_F}\simeq_{T \ll T_F} 1\!+\!\frac{\pi^2}{12}\left(\frac{T}{T_F}\right)^2,
\end{eqnarray}
where $\epsilon_F\!=\!\epsilon(k_F)\!=\!k_BT_F$ is the Fermi energy, $k_B$ is Boltzmann's constant, and $T_F$ the Fermi temperature of the gas. Note that the first correction to the ground-state chemical potential is exactly opposite to the 3D case. At high temperature, classical expansion yields
\begin{eqnarray}
\label{muhighT}
\frac{\mu(T)}{\epsilon_F} \simeq_{T\gg T_F} -\frac{T}{2T_F}\ln\left(\frac{T}{T_F}\right).
\end{eqnarray}
The chemical potential being negative at high temperature according to Eq.~(\ref{muhighT}), by continuity there must be a temperature $T_0$ at which it changes sign. The latter is evaluated analytically as \cite{LangHekkingMinguzzi2015}
\begin{equation}
\frac{T_0}{T_F}=\frac{4}{\pi}\frac{1}{[(\sqrt{2}-1)\zeta(1/2)]^2}\simeq 3.48,
\end{equation}
where $\zeta$ is the Riemann zeta function. These results are illustrated in Fig.~\ref{tonksmuT}.

\begin{figure}
\includegraphics[width=8cm, keepaspectratio, angle=0]{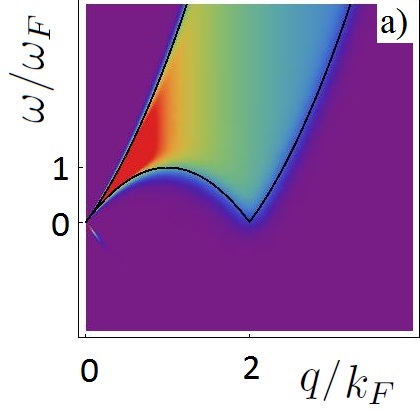}
\includegraphics[width=8cm, keepaspectratio, angle=0]{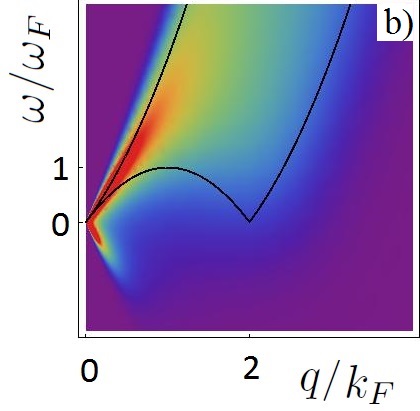}
\includegraphics[width=8cm, keepaspectratio, angle=0]{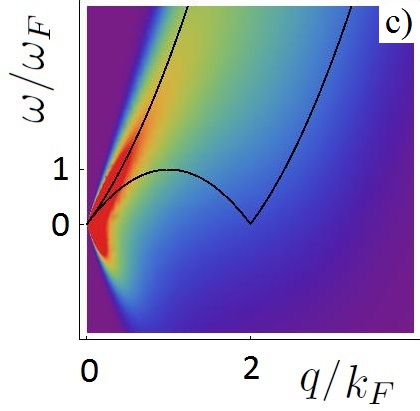}
\includegraphics[width=8cm, keepaspectratio, angle=0]{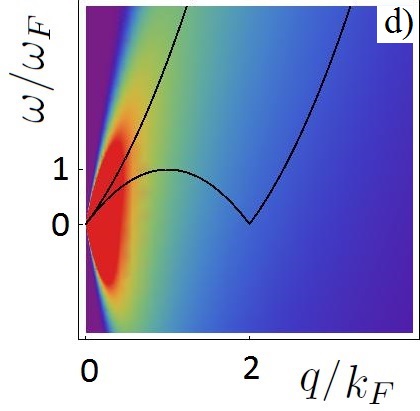}
\caption{Dynamical structure factor $S^{TG}(q,\omega)$ of the Tonks-Girardeau gas in the thermodynamic limit for several dimensionless temperatures, $T/T_F=0.1,0.5,1$ and $4$, in panels (a), (b), (c), and (d), respectively. Warmer colors are associated to higher values. Solid black lines correspond to the limiting dispersion relations $\omega_+$ and $\omega_-$, defining the excitation domain at $T\!=\!0$.}
\label{tonks}
\end{figure}

Then, I use the numerical solution of Eq.~(\ref{invmu}) to evaluate the dynamical structure factor of the Tonks-Girardeau gas at finite temperature from Eq.~(\ref{DSFTG}). As shown in Fig.~\ref{tonks}, the dynamical structure factor of the Tonks-Girardeau gas, $S^{TG}(q,\omega)$, is quite sensitive to temperature. At finite temperature, the range of allowed excitations spreads beyond the type I and type II spectra, since the dynamical structure factor takes into account thermally-activated excitations. The latter can even occur at $\omega\!<\!0$, meaning that energy can be emitted, but such a case has not been reported on yet in ultracold atom experiments. The quasi-linear shape of the spectrum near the origin and the umklapp point $(2k_F,0)$ fades out at temperatures larger than the order of $0.2\,T_F$. When temperature is of the order of or higher than $T_F$, this theoretical analysis is not quite relevant since the gas is very likely not to be one-dimensional anymore in experiments using current trapping techniques.

\begin{figure}
\includegraphics[width=9cm, keepaspectratio, angle=0]{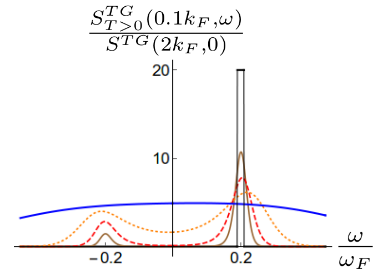}
\caption{Sections of the dynamical structure factor of the Tonks-Girardeau gas at $q\!=\!0.1k_F$, in units of the dynamical structure at the umklapp point at zero temperature, as a function of the dimensionless energy $\omega/\omega_F$, at various temperatures $T/T_F\!=\!0, 0.1, 0.2, 0.5$ and $5$ (thick, black), (brown), (dashed, red), (dotted, orange) and (thick, blue) respectively.}
\label{tonksbis}
\end{figure}

In Fig.~\ref{tonksbis}, for various finite temperatures, I represent sections of the dynamical structure factor at a momentum $q\!=\!0.1k_F$, near the origin. The divergence of the dynamical structure factor at $T\!=\!0$ and $q\!=\!0$, and the high values that it takes close to the origin dramatically decrease once temperature is taken into account. An emission peak, whose position is symmetric to the absorption one already present at $T\!=\!0$ with respect to $\omega\!=\!0$, but whose amplitude is lower, appears at finite temperature. The ratio of their heights is given by the detailed balanced relation
\begin{eqnarray}
S(q,-\omega)=e^{-\beta\hbar\omega}S(q,\omega).
\end{eqnarray}
Both peaks form quite well-defined phonon dispersions at very low temperature, but start overlapping if $T\!\gtrsim\!0.2\,T_F$. At higher temperatures, of the order of a few $T_F$, one can not distinguish them anymore and the dynamical structure factor becomes symmetric with respect to $\omega\!=\!0$.

\subsection{Drag force due to a mobile, point-like impurity in the Tonks-Girardeau regime}

In 1D and at arbitrary temperature, the drag force reads
\begin{equation}
\label{drag}
F_{T>0}\!=\!\frac{1}{2\pi\hbar}\int_0^{+\infty}\!\!dq\,|U(q)|^2 q\,S_{T>0}(q,qv)(1-e^{-\beta\hbar qv}).
\end{equation}
The graphical interpretation of Eq.~(\ref{drag}) is that the drag force, that measures dissipation, is obtained by integration of the dynamical structure factor weighted by other quantities, along a line of slope $v$ in energy-momentum space. At zero temperature, the lower bound of the dynamical structure factor coincides with the lower excitation spectrum, and the link with the graphical interpretation of Landau's criterion becomes obvious. However, new ingredients are taken into account in the drag force formalism: if the weight of excitations is zero, i.e. for a vanishing dynamical structure factor, excitations do not occur even if the integration line crosses the excitation spectrum. Moreover, the precise shape of the potential barrier is also taken into account here, and plays a major role as will be seen below.

In the Tonks-Girardeau regime, with a potential barrier $U(x,t)\!=\!U_b\,\delta(x-vt)$, combining Eq.~(\ref{drag}) and Eq.~(\ref{TonksDSF}), a simple expression is obtained at $T\!=\!0$ \cite{AstrakharchikPitaevskii2004, ChernyCauxBrand2012}, namely
\begin{equation}
\label{basicFTG}
F^{TG}(v)\!=\!\frac{2 U_b^2 n_0 m}{\hbar^2}\left[\Theta(v-v_F)+\frac{v}{v_F}\Theta(v_F-v)\right]\!.
\end{equation}
Thus, within linear response theory, drag force is a linear function of the barrier velocity if $v\!<\!v_F$, and saturates when $v\!>\!v_F$. As I will show below, this saturation to a constant finite value is actually an artifact, due to the idealized Dirac-$\delta$ shape of the potential barrier. Equation~(\ref{basicFTG}) shows that the drag force is non-zero if the velocity of the perturbing potential is finite, meaning that energy dissipation occurs as long as the barrier is driven along the fluid. Thus, according to the drag force criterion, the Tonks-Girardeau gas is not superfluid even at zero temperature.

Equation (\ref{drag}) also allows to discuss thermal effects on the drag force. At finite temperature, it reads \cite{LangHekkingMinguzzi2015}
\begin{equation}
\frac{F_{T>0}^{TG}}{F^{TG}(v_F)}\!=\!\frac{1}{2}\sqrt{\frac{T}{T_F}}\int_0^{\beta mv^2/2}\!\!\!\!\frac{d\epsilon}{\sqrt{\epsilon}(e^{\epsilon-\beta\mu(T)}+1)}.
\end{equation}
The integral can easily be evaluated numerically. As a main result, thermal effects cause a depletion of the drag force close to the Fermi velocity, while at low velocity the drag force profile remains linear. The intriguing fact that, at fixed velocity, the drag force decreases when temperature increases, might be due to the fact that I do not take any barrier renormalization into account.

\subsection{Effect of a finite barrier width on the drag force}

In Ref.~\cite{LangHekkingMinguzzi2015}, I have also investigated the effect of a finite barrier width on the drag force in the Tonks-Girardeau regime. My main theoretical motivation was that Eq.~(\ref{basicFTG}) predicts a saturation of the drag force at high velocities, which does not seem realistic. Among the infinitely many possible barrier shapes, experimental considerations suggest to consider a gaussian barrier. This profile models a laser beam, often used as a stirrer in experiments. I have focused on the case of a blue-detuned, repulsive laser beam. Then, the perturbation part of the Hamiltonian reads
\begin{equation}
\label{Pot}
 H_{pert}\!=\!\int_0^L\!dx\,\sqrt{\frac{2}{\pi}}\frac{U_b}{w}e^{-\frac{2(x-vt)^2}{w^2}}\psi^\dagger(x)\psi(x),
\end{equation}
where $U_b$ is the height of the barrier, and $w$ its waist. Prefactors have been chosen to recover a $\delta$-potential in the limit $w\!\to\!0$. The Fourier transform of the potential in Eq.~(\ref{Pot}) reads
\begin{equation}
\label{TFPot}
U(q)=U_b\, e^{-\frac{q^2 w^2}{8}},
\end{equation}
and the drag force at $T\!=\!0$ is readily obtained as
\begin{equation}
\!\!F_{w>0}(v)=\frac{U_b^2}{2\pi\hbar}\int_0^{+\infty}\!\!dq\,e^{-\frac{q^2 w^2}{4}} q\,S(q,qv).
\end{equation}
In the Tonks-Girardeau gas case, I have obtained an explicit expression at $T\!=\!0$,
\begin{eqnarray}
\label{FwTG}
\!\!\!\!\!\!\!\!\small{\frac{F_{w>0}^{TG}(v)}{F^{TG}(v_F\!)}\!=\!\frac{\sqrt{\pi}}{4}\frac{1}{w k_F}\!\left\{\mathrm{erf}\!\left[w k_F\!\left(\!1\!+\!\frac{v}{v_F}\!\right)\!\right]\!-\!\mathrm{erf}\!\left[w k_F\!\left|1\!-\!\frac{v}{v_F}\right|\right]\right\}},
\end{eqnarray}
where $\mathrm{erf}(x)\!=\!\frac{2}{\sqrt{\pi}}\int_0^x du \,e^{-u^2}$ is the error function. The more general case where both waist and temperature are finite is obtained by inserting Eqs.~(\ref{DSFTG}) and (\ref{TFPot}) into Eq.~(\ref{drag}), and the integral is then evaluated numerically.

\begin{figure}
\includegraphics[width=9cm, keepaspectratio, angle=0]{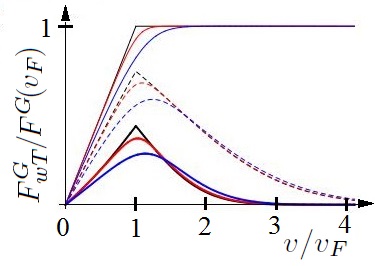}
\caption{Drag force at finite barrier waist and temperature in the Tonks-Girardeau gas, $F_{w>0, T>0}^{TG}(v)$, in units of $F^{TG}(v_F)$, as a function of the dimensionless barrier velocity $v/v_F$. Solid lines stand for a dimensionless waist $w k_F\!=\!0$, dashed lines for $w k_F\!=\!0.5$ and thick lines for $w k_F\!=\!1$. For a given set of curves, temperature increases from $0$ to $0.1\,T_F$ and $0.5\,T_F$ from top to bottom, in black, red and blue respectively.}
\label{fT}
\end{figure}

All these results are illustrated in Fig.~\ref{fT}. While for a delta potential the drag force saturates at supersonic flow velocities, for any finite barrier width the frag force vanishes at sufficiently large velocities. According to the drag force criterion, this means that the flow is close to being superfluid in this regime.

This important result deserves being put in perspective. The first theoretical consideration of the drag force due to a Gaussian laser beam dates back to Ref.~\cite{Pavloff2002}, that focuses on the weakly-interacting regime of the Lieb-Liniger model, treated through the Gross-Pitaevskii equation. At the time I wrote \cite{LangHekkingMinguzzi2015}, I was not aware of this paper. My contribution is still seemingly pioneering in the strongly-interacting regime.

\begin{figure}
\includegraphics[width=7cm, keepaspectratio, angle=0]{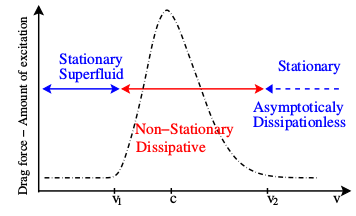}
\caption{Picture taken from Ref.~\cite{Albert2009}, showing the three regimes in the drag force profile, put to light in a Bose gas and conjectured in a general quantum fluid. The sound velocity $c$ corresponds to the critical velocity in the Landau mean-field picture. Due to various effects, the flow is really superfluid only below a lower threshold velocity, the true critical velocity $v_1$, generically lower than $c$. This velocity would be null in the Tonks-Girardeau regime. A dissipative regime occurs above $v_1$, and dissipation takes finite values up to an upper critical velocity $v_2$ above which it can be neglected, and the flow can be considered as quasi-superfluid. }
\label{AlbertSF}
\end{figure}

It can can inferred that a strong suppression of the drag force at supersonic barrier velocities is common to all interaction regimes. The range of velocities where this occurs corresponds to a 'quasi-superfluid' regime. More generally, a typical damping profile, sketched in Fig.~\ref{AlbertSF}, can decently be expected. Actually, it had already been observed in several experimental situations upon close inspection \cite{Engels2007, Dries2010}.

After our work \cite{LangHekkingMinguzzi2015}, several works have investigated similar points more in detail. The authors of Ref.~\cite{Singh2016}, using both numerical and analytical methods, have identified nonzero temperature, circular motion of the stirrer, and the density profile of the atomic cloud as additional key factors influencing the magnitude of the critical velocity in a weakly-interacting Bose gas. According to the terminology introduced above, a quasi-superfluid regime has been predicted in higher dimensions too. In \cite{Pinsker2017}, the definition Eq.~(\ref{defF}) of the drag force has been used to consider the effect of a Gaussian barrier on a weakly-interacting Bose gas, using the Bogoliubov formalism in 2D and 3D. Within this approach, the critical velocity still coincides with the sound velocity, and after reaching a peak around twice the sound velocity, in 3D the drag force decreases monotonically. Note, however, that the predicted drag force profile is smooth even at zero temperature, which is not the case in the Tonks-Girardeau gas. This may be due to the fact that a few corrections to linear response are included as well within this approach.

In the next section, I shall investigate the whole region between the Tonks-Girardeau and Gross-Pitaevskii regimes, at intermediate interaction strengths between the bosons of the fluid.

\section{Dynamical structure factor and drag force for a 1D Bose gas at finite interaction strength}

\subsection{State of the art}

Evaluating the dynamical structure factor of the Lieb-Liniger model at finite interaction strength is challenging, and several approaches have been undertaken along the years.

As was the case for thermodynamic quantities, perturbation theory allows to evaluate dynamical ones in the strongly-interacting regime as corrections to the Tonks-Girardeau regime. Such a perturbative approach has been developed to first order in $1/\gamma$ at $T\!=\!0$ in \cite{ChernyBrand2005}, and extended to finite temperature in \cite{ChernyBrand2006}. By qualitative comparison of the results obtained in these references, and in the Tonks-Girardeau regime studied above, I conclude that the dynamical structure factor at $\gamma\!\gtrsim\!10$ still looks pretty much like in the TG regime. In particular, the low-temperature phonon-like tail starting from the origin at $\omega\!<\!0$ can be observed both in Ref.~\cite{ChernyBrand2006} and in panel a) of Fig.~\ref{tonks}. A notable difference to the limit $\gamma\!=\!+\infty$ is that excitations are progressively suppressed close to the umklapp point $(q\!=\!2k_F,\omega\!=\!0)$ when the interaction strength is decreased, and a crude extrapolation suggests that it tends towards a superfluid behavior. However, one should keep in mind that first-order corrections to the Tonks-Girardeau regime are expected to be reliable only as far as $\gamma \gtrsim 10$.

In this respect, the Tomonaga-Luttinger liquid formalism, whose use in this context was first suggested in Ref.~\cite{AstrakharchikPitaevskii2004}, is more versatile as it can be used at arbitrary interaction strengths. However, it is also expected to be accurate only inside a small, low-energy sector. This is not necessarily redhibitory to study superfluidity, since the latter is defined through the low-velocity behavior of the drag force, that is dominated by the low-energy sector of the dynamical structure factor, close to the umklapp point. However, the quasi-superfluid, supersonic regime, is definitely out of reach with this method.

Finding the exact dynamical structure factor at arbitrary interaction strength and energies actually required the development of fairly involved algebraic Bethe Ansatz techniques. The numerical evaluation of form factors at finite $N$ has been implemented in the Algebraic Bethe Ansatz-based Computation of Universal Structure factors (ABACUS) algorithm \cite{CauxABACUS}, first at zero \cite{CauxCalabrese2006}, then even at finite temperature \cite{PanfilCaux2014}. This was a major breakthrough in the long-standing issue of dynamical correlation functions of integrable models, and is one of the most important theoretical achievements in this field in the early 2000's.

This exact solution tends to validate the main qualitative predictions of the Imambekov-Glazman (IG) liquid theory, developed in parallel as an extension of the standard Tomonaga-Luttinger liquid theory to a wider range of energies \cite{Pustilnik2006, Khodas2007, Khodas2007bis, ImambekovGlazman2008, ImambekovGlazman2009}. The dynamical structure factor of a 1D Bose gas features power-law behaviors along the type I and type II excitation branches at $T\!=\!0$, with a sharp response at the upper threshold in the case of repulsive interactions, and at the lower one if they are attractive \cite{CalabreseCaux2007}.

Based on numerical data produced with the ABACUS code, a phenomenological expression has been proposed for the dynamical structure factor, that incorporates the TL and IG liquid predictions as special cases \cite{ChernyBrand2008}. Later on, the dynamical structure factor of ${^4}$He has also been obtained numerically, this time with Quantum Monte Carlo techniques, and has also shown beyond-Luttinger liquid behavior \cite{Bertaina2016}.

Measures of the dynamical structure factor of an array of 1D gases have shown remarkable agreement with the algebraic Bethe Ansatz predictions for the Lieb-Liniger model over a wide range of interaction strengths. They have definitely confirmed the need of a beyond-Luttinger liquid theory approach to the problem at high energies \cite{Clement2015, Meinert2015}.

As far as the drag force is concerned, its first evaluation at arbitrary interaction strength was reported on in \cite{AstrakharchikPitaevskii2004}, and relied on the Tomonaga-Luttinger liquid framework. Then, the dynamical structure factor as predicted by the ABACUS algorithm, once combined with Eq.~(\ref{drag}) and numerically integrated, yielded the drag force due to a point-like impurity at arbitrary interaction strength \cite{ChernyCauxBrand2009, ChernyBrand2009}. The conclusion of this study is that, in this configuration, the Lieb-Liniger model is never strictly superfluid in the thermodynamic limit according to the drag force criterion.

I shall study the dynamical structure factor and drag force within the linear Tomonaga-Luttinger liquid theory despite its shortcomings, for the following reasons: first of all, it is currently the simplest approach allowing to make fully analytical, quantitative predictions at finite interaction strength. Moreover, its validity range is still not known quantitatively, and asks for additional studies. It is also the first step towards more accurate predictions, e.g. from the Imambekov-Glazman liquid theory, and towards generalizations of the Tomonaga-Luttinger framework to multi-component gases.

\subsection{Dynamical structure factor from the Tomonaga-Luttinger liquid theory}

Starting from the real-space density-density correlations of a Tomonaga-Luttinger liquid at $T\!=\!0$ and in the thermodynamic limit, Eq.~(\ref{nn}), Fourier transform with respect to time and space yields the dominant terms of the dynamical structure factor of gapless 1D models. Since this quantity is symmetric with respect to $q\leftrightarrow -q$, I shall write the result for $q>0$ \cite{AstrakharchikPitaevskii2004, LangHekkingMinguzzi2015} (I refer to Appendix \ref{DSFTLLder} for a detailed derivation):
\begin{eqnarray}
\label{DSFLL}
S^{TL}(q,\omega)\!\!\!\!\!\!\!\!&&\simeq K|q|\delta[\omega-\omega(q)]+B_1(K)\left[\omega^2-(q-2k_F)^2v_s^2\right]^{K-1}\Theta[\omega-|q-2k_F|v_s] \nonumber \\
&&=S_0^{TL}(q,\omega)+S_1^{TL}(q,\omega)
\end{eqnarray}
when read in the same order, where 
\begin{eqnarray}
\label{B1}
B_1(K)\!=\!\frac{A_1(K)}{(2k_Fv_s)^{2\lbrace K-1\rbrace}}\frac{1}{\Gamma(K)^2}\frac{1}{v_s}
\end{eqnarray}
is a non-universal coefficient. It is the single mode form factor of the dynamical structure factor, and is related to the phonic form factor of the density-density correlation function, $A_1(K)$, already defined in Eq.~(\ref{nnTLL}).

In Eq.~(\ref{DSFLL}), $S_0$ displays a sharp peak, in exact correspondence to the linear phonon-like dispersion $\omega(q)\!=\!qv_s$. Its divergence and zero widths are artifacts due to the spectrum linearization. If $v\leq v_s$, it does not contribute to the drag force in this framework, and if $v\ll v_s$, it is also true according to more accurate descriptions, so I will not devote much attention to $S_0$ anymore, but rather focus on the second contribution to the dynamical structure factor, denoted by $S_1$.

There are two linear limiting dispersion relations described by $S_1$. They are symmetric with respect to $q\!=\!2k_F$, and form a triangular shape above the umklapp point $(2k_F,0)$. Actually, these excitation spectra correspond to the linearization of $\omega_-$, so one can write $\omega_-^{TL}\!=\!|q-2k_F|v_s$, and
\begin{eqnarray}
S_1^{TL}(q,\omega)=B_1(K)[\omega^2-(\omega_-^{TL})^2]^{K-1}\Theta(\omega-\omega_-^{TL}).
\end{eqnarray}
The slopes of the limiting dispersions in $S_0$ and $S_1$ depend on the interaction strength via $v_s$. Hence, measuring the excitation spectrum of a 1D Bose gas at low energy provides an indirect way to determine the sound velocity.

\begin{figure}
\includegraphics[width=7cm, keepaspectratio, angle=0]{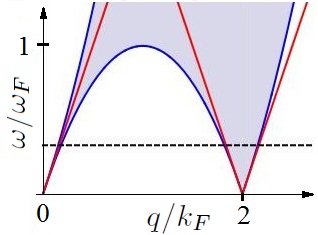}
\caption{Definition domain of the dynamical structure factor of the Tonks-Girardeau gas at $T\!=\!0$, in the plane $(q,\omega)$, in units of $(k_F,\omega_F)$. I superimposed the exact result (shaded gray area delimited by the blue curves) to the result in the Tomonaga-Luttinger liquid framework for dimensionless parameters $K\!=\!1$ and $v_s/v_F\!=\!1$ (red). In the latter, the domain consists in a line starting from the origin, and the area included in the infinite triangle starting from the umklapp point $(0, 2k_F)$. The upper energy limit of potential validity of the Tomonaga-Luttinger liquid theory is approximately given by the dashed line.}
\label{compTLLTGepsilonII}
\end{figure}

To make quantitative predictions, the first requirement is to evaluate $v_s$, or equivalently $K$, as well as the form factor $A_1$. Both have already been obtained in the Tonks-Girardeau regime in chapter \ref{secII}, readily allowing to make quantitative predictions in this case. If they had not been obtained yet, the Luttinger parameter $K$ could be determined so as to reproduce the phonon-like dispersion relation at the origin, and $A_1$ so as to reproduce the exact dynamical structure factor at the umklapp point. Comparison between the exact and linearized spectra in the Tonks-Girardeau regime is made in Fig.~\ref{compTLLTGepsilonII}, confirming that the Tomonaga-Luttinger liquid formalism is intrinsically limited to low energies.

In particular, within this formalism it is impossible to make quantitative predictions around the top of the type II excitation spectrum, where curvature effects are important. This is not the only problem, actually. In the Tonks-Girardeau regime, the exact dynamical structure factor scales like $1/q$ inside its definition domain, whereas the Tomonaga-Luttinger liquid prediction of Eq.~(\ref{DSFLL}) is constant in the triangular domain of the umklapp region when $K\!=\!1$. Both results coincide only along a vertical line starting from the umklapp point, and this line is finite since the TLL formalism utterly ignores the upper excitation spectrum.

Although the TLL result is fairly disappointing at first when compared to the Tonks-Girardeau exact one, I recall that, contrary to the Bogoliubov formalism, it predicts the existence of excitations near the umklapp point, which is not obvious. Moreover, at this stage it is not excluded that the agreement with the exact dynamical structure factor of the Lieb-Liniger model may be better at finite interaction strength. More generally, it is interesting to evaluate the validity range of the standard Tomonaga-Luttinger liquid theory as precisely as possible, but this requires a comparison point. A possible generalization, where it is possible to compare the result with an exact prediction, concerns thermal effects, already investigated in the Tonks-Girardeau regime through Eq.~(\ref{DSFTG}). In the Tomonaga-Luttinger liquid formalism, the dynamical structure factor at finite temperature is obtained by Fourier transform of Eq.~(\ref{nnfftT}).

On the one hand, I obtain (I refer to Appendix \ref{DSFTLLTder} for details) \cite{LangHekkingMinguzzi2015}
\begin{eqnarray}
\label{S0LLT}
S_{0,T>0}^{TL}(q,\omega)=\frac{K|q|}{1\!-\!e^{-\beta\hbar\omega(q)}}\!\left\lbrace\delta[\omega-\omega(q)]+e^{-\beta\hbar\omega(q)}\delta[\omega+\omega(q)]\right\rbrace,
\end{eqnarray}
and on the other hand \cite{LangHekkingMinguzzi2015},
\begin{eqnarray}
\label{S1LLT}
&&S_{1,T>0}^{TL}(q,\omega)=C(K,T)\frac{1}{v_s}\left(L_Tk_F\right)^{2(1-K)} e^{\frac{\beta\hbar\omega}{2}} \nonumber \\
&&B\left\{\frac{K}{2}+i\frac{\beta\hbar}{4\pi}[\omega+(q-2k_F)v_s],\frac{K}{2}-i\frac{\beta\hbar}{4\pi}[\omega+(q-2k_F)v_s]\right\}\nonumber \\
&&B\left\{\frac{K}{2}+i\frac{\beta\hbar}{4\pi}[\omega-(q-2k_F)v_s],\frac{K}{2}-i\frac{\beta\hbar}{4\pi}[\omega-(q-2k_F)v_s]\right\},
\end{eqnarray}
where $C(K,T)$ is a dimensionless prefactor and $B(x,y)\!=\!\frac{\Gamma(x)\Gamma(y)}{\Gamma(x+y)}$ is the Euler Beta function.

While the value of the prefactor $C(K,T)$ in Eq.~(\ref{S1LLT}) is fixed by the exact result at the umklapp, $v_s(T)$ should be evaluated independently. This can be done at very low temperatures, by identification with the phonon modes, whose slope is $v_s$. One finds that below $T\!\simeq\!0.2\,T_F$, which is approximately the highest temperature where these phonons are well defined, $v_s$ does not significantly vary with $T$.

\begin{figure}
\includegraphics[width=6cm, keepaspectratio, angle=0]{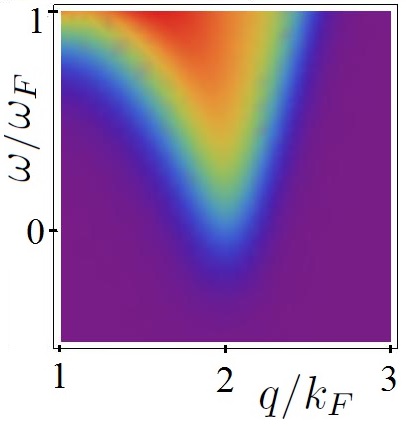}
\includegraphics[width=6cm, keepaspectratio, angle=0]{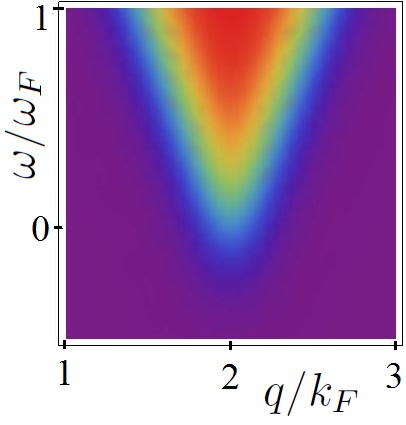}
\caption{Dynamical structure factor of the Tonks-Girardeau gas at $T\!=\!0.1\,T_F$ in the plane $(q,\omega)$ in units of $(k_F,\omega_F)$ in the vicinity of the umklapp point, as predicted from the Bose-Fermi mapping (left panel), and for a Tomonaga-Luttinger liquid (right panel). The exact temperature dependence is quite well reproduced in the Tomonaga-Luttinger liquid framework. Differences stem mostly from nonlinearities, which are not taken into account in this case.}
\label{compTLLTGT}
\end{figure}

Comparison of the approximation Eq.~(\ref{S1LLT}) and the exact Tonks-Girardeau result is shown in Fig.~\ref{compTLLTGT}. Their agreement is quite remarkable, and the validity range of the TLL framework is even increased compared to the $T\!=\!0$ case. Surprising at first, this fact can be understood at the real-space level. Correlations decay exponentially at large distances according to Eq.~(\ref{nnfftT}), thus the neglected part is not as important. However, at slightly higher temperatures the agreement would break down abruptly. The conclusion is that the Tomonaga-Luttinger liquid framework is valid at very low temperatures only.

Coming back to $T\!=\!0$, at finite interaction strength the Tomonaga-Luttinger liquid formalism makes predictions that have not been investigated quantitatively so far. Here, I try and fill this gap, to allow for a subsequent comparison with more powerful techniques. A first necessary condition is to evaluate $v_s(\gamma)$. From its thermodynamic definition comes
\begin{eqnarray}
\label{soundgamma}
v_s(\gamma)\!=\!\frac{v_F}{\pi}\left[3e(\gamma)-2\gamma \frac{de}{d\gamma}(\gamma)+\frac{1}{2}\gamma^2 \frac{d^2e}{d\gamma^2}(\gamma)\right]^{1/2}.
\end{eqnarray}
Analytical expansions of the sound velocity at large and small interaction strength can be found in the literature. The first- and second-order corrections to the Tonks-Girardeau regime in $1/\gamma$ are given in \cite{Cazalilla2004}, they are calculated to fourth order in \cite{Zvonarev} and up to eighth order in \cite{Ristivojevic}. In the weakly-interacting regime, expansions are found in \cite{Cazalilla2004} and \cite{Wadati}. Eqs.~(\ref{conjWidom}, \ref{exp}, \ref{Lpol}) for the dimensionless energy per particle obtained in chapter \ref{secIII} considerably increase the accuracy compared to these works, after straightforward algebra.

Interestingly, it is not necessary to evaluate the ground-state energy per particle $e(\gamma)$ to obtain $v_s(\gamma)$, but actually sufficient to know the density of pseudo-momenta $g$ at the edge of the Fermi sea, i.e. at $z\!=\!1$, due to the useful equality \cite{KorepinBogoliubovIzergin}
\begin{eqnarray}
\label{gatone}
\frac{v_s(\gamma)}{v_F}=\frac{1}{\{2\pi g[1;\alpha(\gamma)]\}^2}.
\end{eqnarray}
Reciprocally, if $v_s$ is already known with high accuracy from Eq.~(\ref{soundgamma}) applied to a reliable equation of state $e(\gamma)$, then Eq.~(\ref{gatone}) provides an excellent accuracy test for a proposed solution $g$ to the Lieb equation (\ref{Fredholm}), since it allows to check its value at the edge of the interval $[-1,1]$, where it is the most difficult to obtain with most known methods.

\begin{figure}
\includegraphics[width=8.3cm, keepaspectratio, angle=0]{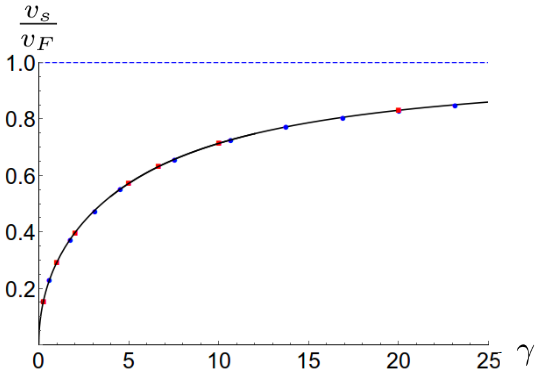}
\caption{Dimensionless sound velocity $v_s/v_F$, where $v_F$ is the Fermi velocity, as a function of the dimensionless Lieb parameter $\gamma$, from numerical solution of the Lieb equation (blue dots), compared to values found in the literature \cite{Cazalilla2004, CauxCalabrese2006} (red squares), and analytical result from Eqs.~(\ref{exp}, \ref{Lpol}, \ref{soundgamma})(solid, black). The Tonks-Girardeau limit is indicated by the dashed blue line.}
\label{FigIV6}
\end{figure}

I have used both approaches to evaluate the sound velocity over a wide range of strong to intermediate interaction strengths with excellent accuracy, as illustrated in Fig.~\ref{FigIV6}. In particular, the fact that $v_s\!\to_{\gamma\to 0} 0$ implies that $g(z;\alpha)\!\to_{z\to 1,\alpha\to 0}+\infty$, hinting at the fact that polynomial expansion methods are not appropriate at very low interaction strength, as expected due to the vicinity of the singularity.

As far as the dynamical structure factor is concerned, there are two possible points of view at this stage. Either one is interested in qualitative properties, i.e. in its global shape as a function of $\gamma$, and one can divide the result by the unknown coefficient $A_1(K)$, or one is rather motivated by quantitative evaluations, e.g. for a comparison with experiments. Then, it becomes necessary to evaluate $A_1(K)$. The solution to this tough problem is provided in \cite{Shashi2011, Shashi2012}. The form factor is extracted numerically as the solution of a complicated set of coupled integral equations, whose analytical solution stays out of reach.

My philosophy in this thesis is to rely on analytical expressions as often as possible, so additional efforts are required. Instead of trying and solve the set of equations of Ref.~\cite{Shashi2012}, I have extracted data points from the figure provided in this very reference. It turns out that it is especially well approached by a very simple fit function, of the form
\begin{eqnarray}
\label{A1fit}
\frac{A_1(K)}{\pi^{2(K-1)}}\!=\!\frac{1}{2}e^{-\alpha (K-1)},
\end{eqnarray}
where $\alpha\!\simeq\!3.8$ up to data extraction errors. This expression is approximately valid for $K\!\in\![1,2]$, for $K\!\gtrsim\!2$ data can not be read on the graph because $A_1$ is too small then.

A few comments are in order: first, Eq.~(\ref{A1fit}) is certainly not the exact solution, in view of the extreme complexity of the equations it is derived from. This could be verified by evaluating $A_1(K)$ at $K\!<\!1$ (i.e. in the super Tonks-Girardeau regime), where a discrepency with the extrapolated fit function is very likely to become obvious. However, Eq.~(\ref{A1fit}) may be equivalent, or at least close to being so, to the exact solution at $K\!\gtrsim\!1$, in view of the remarkable agreement with numerical data in this range.

\begin{figure}
\includegraphics[width=8cm, keepaspectratio, angle=0]{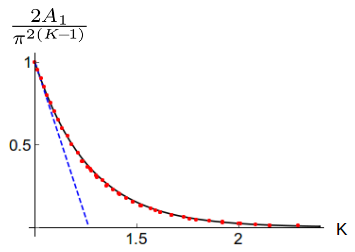}
\caption{Reduced form factor of a Tomonaga-Luttinger liquid describing the Lieb-Liniger model, as a function of the dimensionless Luttinger parameter $K$. The fit function Eq.~(\ref{A1fit}) with $\alpha\!=\!1\!+\!4\ln(2)$ (solid, black) is quite accurate over a wider range than the first order Taylor expansion, Eq.~(\ref{A1ChernyBrand}) (dashed, blue) compared to the graphical data from Ref.~\cite{Shashi2012} (red dots).}
\label{FigPirate}
\end{figure}

If equation (\ref{A1fit}) were exact, to infer the value of $\alpha$, I could rely on the exact expansion close to $K\!=\!1$ \cite{ChernyBrand2009},
\begin{eqnarray}
\label{A1ChernyBrand}
\frac{A_1(K)}{\pi^{2(K-1)}}\!=_{K\gtrsim 1}\frac{1}{2}\{1\!-\![1\!+\!4\ln(2)](K\!-\!1)\}+O[(K\!-\!1)^2],
\end{eqnarray}
and by identification of both Taylor expansions at first order in the variable $K\!-\!1$, deduce that $\alpha\!=\!1\!+\!4\ln(2)\!\simeq\!3.77$, which is actually quite close to the value obtained by fitting on data points. The agreement is not perfect at higher values of $K$, as can be seen in Fig.~\ref{FigPirate}. Another clue, if needs be, that Eq.~(\ref{A1fit}) is not exact is that it does not agree with the high-$K$ expansion obtained in \cite{AstrakharchikPitaevskii2004}.

\begin{figure}
\includegraphics[width=8cm, keepaspectratio, angle=0]{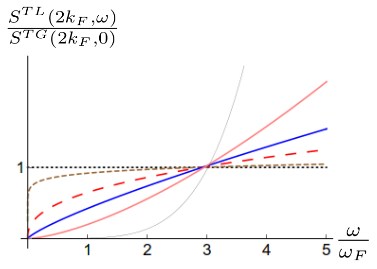}
\caption{Dynamical structure factor of the Lieb-Liniger gas as predicted by the Tomonaga-Luttinger liquid theory, Eq.~(\ref{DSFLL}), with the expression of the form factor Eq.~(\ref{A1fit}), along the umklapp line and in units of the dynamical structure factor of the Tonks-Girardeau gas at the umklapp point, as a function of the dimensionless energy. The various curves correspond respectively to values of $K$ that correspond to Lieb parameters $\gamma\!=\!+\infty$ (dotted, black), $\gamma\!=\!100$ (dashed, brown), $\gamma\!=\!20$ (long-dashed, red), $\gamma\!=\!10$ (thick, blue), $\gamma\!=\!5$ (pink) and $\gamma\!=\!1$ (thin, gray), to be compared to the corresponding figure in Ref.~\cite{CauxCalabrese2006}.}
\label{FigPirateS}
\end{figure}

All together, these results allow to predict quantitatively the dynamical structure factor near the umklapp point in the Tomonaga-Luttinger liquid framework, represented in Fig.~\ref{FigPirateS}. The values chosen for the interaction strength are in exact correspondence to those of Ref.~\cite{CauxCalabrese2006}, allowing for a direct comparison with the exact result from the ABACUS. The agreement is excellent for most values, except at too high energies because the Tomonaga-Luttinger liquid does not predict the upper threshold, and for $\gamma\!=\!1$, where Eq.~(\ref{A1fit}) is likely to be used outside its range of validity.

\subsection{Drag force from the Tomonaga-Luttinger liquid formalism}

The dynamical structure factor gives access to the drag force through Eq.~(\ref{drag}). First, I have addressed the simplest case, $T\!=\!0$ and $w\!=\!0$, that yields \cite{LangHekkingMinguzzi2015} (see Appendix \ref{FTLder} for details)
\begin{eqnarray}
\label{FPit}
\!\!F^{TL}(v)\!&=&\!\frac{U_b^2}{2\pi\hbar}\int_0^{+\infty}\!\!dq\!\!\! \quad\!\!q\,S^{TL}(q,qv) \nonumber \\
\!\!&=&\!\!\frac{U_b^2}{2\pi\hbar}\frac{B_1(K)}{v_s^2}\frac{\sqrt{\pi}\Gamma(K)(2k_Fv_s)^{2K}}{\Gamma(K\!+\!1/2)}\frac{\left(\frac{v}{v_s}\right)^{2K-1}}{\left[1\!-\!\left(\frac{v}{v_s}\right)^2\right]^{\!K+1}},
\end{eqnarray}
in agreement with \cite{AstrakharchikPitaevskii2004} in the limit $v/v_s\!\ll\!1$. At low velocities, the drag force scales as a power law $v^{2K-1}$, that depends on the interaction strength in a non-trivial way.  A comparison with the Tonks-Girardeau result at $K\!=\!1$ leads to the determination of the exact form factor,
\begin{eqnarray}
B_1(K\!=\!1)\!=\!\frac{1}{2v_F}.
\end{eqnarray}

\begin{figure}
\includegraphics[width=7cm, keepaspectratio, angle=0]{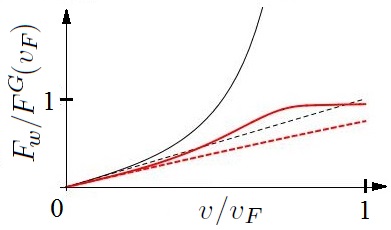}
\caption{Drag force in units of $F^{TG}(v_F)$ as a function of the velocity $v$ (in units of $v_F$), as predicted for a Tonks-Girardeau gas (dashed lines) and a Tomonaga-Luttinger liquid at dimensionless parameter $K\!=\!1$ (solid lines), at $T\!=\!0$. Thin black curves correspond to a dimensionless waist $wk_F\!=\!0$ and thick red curves to a finite waist $wk_F\!=\!0.5$. The TLL prediction is valid provided that $v\!\ll\!v_F$.}
\label{ForceVS}
\end{figure}

Then, I have generalized the expression of the drag force to finite laser waist $w$. In the Tonks-Girardeau regime, I obtained the analytical, simple expression \cite{LangHekkingMinguzzi2015}
\begin{equation}
\label{FLLwK1}
F_{w>0,K=1}^{TL}(v)\!=\!\frac{2U_b^2n_0 m}{\hbar^2}\frac{1}{(2wk_F)^2}\left[e^{\!-\frac{ w^2k_F^2}{(1+v/v_F)^2}}\!-\!e^{\!-\frac{w^2k_F^2}{(1-v/v_F)^2}}\right]\!,
\end{equation}
allowing for a quantitative comparison with the exact result, Eq.~(\ref{FwTG}). I have used Eqs.~(\ref{FwTG}), (\ref{FPit}) and (\ref{FLLwK1}) to plot the curves in Fig.~\ref{ForceVS}, showing that the Tomonaga-Luttinger liquid model predictions in the Tonks-Girardeau regime are valid for velocities $v\!\ll\!v_F$, as expected since the dynamical structure factor is well approximated by the TLL model at low energy only. The exact drag force is all the better approached as the potential is wide. It is always linear near the origin, but its slope depends on the barrier waist $w$.

At arbitrary interaction strength, the expression of the drag force in the case of a finite-width potential is given by \cite{LangHekkingMinguzzi2015}
\begin{eqnarray}
\label{FLw}
&&\frac{F_{w>0}^{TL}(v)}{F^{TL}(v)}=\frac{1}{w k_F}\!\sum_{k=0}^{+\infty}\!\frac{(-1)^k}{k!}\!\left(\frac{w k_F}{1\!+\!\frac{v}{v_s}}\right)^{2k+1}\!\!\!\!\!\!\!{_2F_1}\!\left(-1\!-\!2k,\!K;2K;-\frac{2v}{v_s\!-\!v}\right),
\end{eqnarray}
where ${_2F_1}$ is the hypergeometric function. I have verified numerically that for $wk_F\!\lesssim\!1$, truncating the series to low orders is a very good approximation.

\begin{figure}
\includegraphics[width=7cm, keepaspectratio, angle=0]{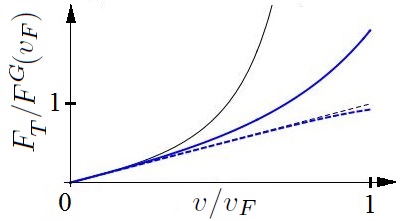}
\caption{Drag force in units of $F^{TG}(v_F)$ as a function of the velocity $v$ (in units of $v_F$), as predicted for a Tonks-Girardeau gas (dashed lines) and a Tomonaga-Luttinger liquid (solid lines), at $w\!=\!0$. Thin black curves correspond to $T\!=\!0$ and thick blue curves to $T\!=\!0.1\,T_F$.}
\label{ForceVT}
\end{figure}

The effect of temperature on the drag force is obtained by integrating numerically Eq.~(\ref{drag}) with the input of Eqs.~(\ref{S0LLT}) and (\ref{S1LLT}). In Fig.~\ref{ForceVT}, I plot the drag force at $T\!=\!0$ and finite temperature as a function of the velocity for a Tonks-Girardeau gas, as obtained from the exact solution and the Tomonaga-Luttinger liquid approach.

As a main result, I have shown that in the strongly-interacting regime $K\!\gtrsim\!1$, the Tomonaga-Luttinger liquid theory reproduces quite well the exact dynamical structure factor of the Tonks-Girardeau gas around the umklapp point, and its drag force at low velocities, even for a finite-width potential barrier. This allows to use the Tomonaga-Luttinger liquid theory to predict the generic low-velocity behavior of the drag force at large to intermediate interactions, as a complementary approach to the Bogoliubov treatment at weak interactions.

\section{Exact excitation spectra from integrability}

To go beyond the standard Tomonaga-Luttinger liquid using analytical methods, three types of quantities have to be evaluated with the highest possible accuracy. They are the form factors, that give the weights of the different contributions to the dynamical structure factor, the edge exponents, that describe power laws at the thresholds, and the excitation spectra of the Lieb-Liniger model, that fix their locations.

This section is more specifically devoted to the excitation spectra. Lieb studied them in \cite{Lieb1963}, and much to his surprise, found that the excitation spectrum of his model was two-fold. The Bogoliubov spectrum corresponds to the type-I spectrum for weak interactions, but the nature of the type-II spectrum in this regime was elucidated later on, when a new solution to the non-linear Schr\"odinger equation was found \cite{Kulish}. This spectrum is most probably linked to solitons, as suggested by a fair number of works \cite{Khodas2008, Karpiuk2012, Syrwid2015, Karpiuk2015, Sato2016}.

At arbitrary interaction strength, coordinate Bethe Ansatz yields the exact excitation spectrum, both at finite particle number and in the thermodynamic limit. It is technically harder to obtain than the energy, however. The set of Bethe Ansatz equations (\ref{logBAE}) for the finite-$N$ many-body problem can not be solved analytically in full generality, but the solution is easily obtained numerically for a few bosons. To reach the thermodynamic limit with several digits accuracy, the Tonks-Girardeau case suggests that $N$ should be of the order of a hundred. Although the interplay between interactions and finite $N$ may slow down the convergence at finite $\gamma$ \cite{CauxCalabrese2006}, a numerical treatment is still possible.

In \cite{LangHekkingMinguzzi2017}, I have addressed the problem directly in the thermodynamic limit, where it reduces to two equations \cite{Pustilnik2014, Ristivojevic}:
\begin{eqnarray}
\label{eqP}
 p(k;\gamma)=2\pi\hbar\,Q(\gamma)\left|\int_1^{k/Q(\gamma)}\!\!dy\,g[y;\alpha(\gamma)]\right|
\end{eqnarray}
and
\begin{eqnarray}
\label{EqEE}
\epsilon(k;\gamma)=\frac{\hbar^2Q^2(\gamma)}{m}\left|\int_1^{k/Q(\gamma)}\!\!dy\,f[y;\alpha(\gamma)]\right|,
\end{eqnarray}
where
\begin{eqnarray}
Q(\gamma)= \frac{n_0}{\int_{-1}^1dy\,g[y;\alpha(\gamma)]}
\end{eqnarray}
is a non-negative quantity known as the Fermi rapidity. It represents the radius of the quasi-Fermi sphere, and equals $k_F$ in the Tonks-Girardeau regime. The function $f$ that appears in Eq.~(\ref{EqEE}) satisfies the integral equation
\begin{eqnarray}
\label{secondlieb}
f(z;\alpha)-\frac{1}{\pi}\int_{-1}^1dy\,\frac{\alpha}{\alpha^2+(y-z)^2}f(y;\alpha)=z,
\end{eqnarray}
referred to as the second Lieb equation in what follows.

For a given interaction strength $\gamma$, the excitation spectrum is obtained in a parametric way as $\epsilon(k;\gamma)[p(k;\gamma)], k\!\in\![0,+\infty[$.  Why is it so, then, that Lieb predicted two excitation spectra, and not just one? The answer was rather clear at finite $N$ from general considerations on particle-hole excitations. In the thermodynamic limit considered presently, the type I and type II spectra could be interpreted as a single parametric curve, but the type I part corresponds to $|k|/Q\geq 1$ and thus to quasi-particle excitations, while the type II dispersion is obtained for $|k|/Q \leq 1$. Thus, the latter is associated to processes taking place inside the quasi-Fermi sphere, which confirms that they correspond to quasi-hole excitations, in agreement with the finite $N$ picture.

Using basic algebraic manipulations on Eqs.~(\ref{eqP}) and (\ref{EqEE}), I have obtained a few general properties:

 (a) The ground state $(p\!=\!0, \epsilon\!=\!0$) trivially corresponds to $k\!=\!Q(\gamma)$, confirming that $Q$ represents the edge of the Fermi surface.
 
 (b) The quasi-momentum $k\!=\!- Q(\gamma)$ corresponds to the umklapp point $(p\!=\!2p_F,\epsilon\!=\!0)$, always reached by the type II spectrum in the thermodynamic limit, regardless of the value of $\gamma$.
 
 (c) The maximal excitation energy associated to the type II curve lies at $k\!=\!0$ and corresponds to $p\!=\!p_F$.
 
 (d) If $k\!\leq\!Q(\gamma)$, $p(-k)\!=\!2p_F\!-\!p(k)$ and $\epsilon(-k)\!=\!\epsilon(k)$, hence $\epsilon^{II}(p)\!=\!\epsilon^{II}(2p_F\!-\!p)$, generalizing to finite interaction strength the symmetry $p\leftrightarrow 2p_F\!-\!p$ already put into light in the Tonks-Girardeau regime.
 
 (e) The type I curve $\epsilon^I(p)$ repeats itself, starting from the umklapp point, shifted by $2p_F$ in $p$. Thus, what is usually considered as a continuation of the type II branch can also be thought as a shifted replica of the type I branch.
 
 (f) Close to the ground state, $\epsilon^I(p)=-\epsilon^{II}(-p)$. This can be proven using the following sequence of equalities: $\epsilon^I(p)\!=\!\epsilon^I(p\!+\!2p_F)\!=\!-\epsilon^{II}(p\!+\!2p_F)\!=\!-\epsilon^{II}(2p_F\!-\!(-p))\!=\!-\epsilon^{II}(-p)$.
 
These symmetry properties are useful in the analysis of the spectra, and provide stringent tests for numerical solutions. Before calculating the excitation spectra, let me make a few technical comments. The momentum $p$ is relatively easy to obtain if $k/Q\!\leq\!1$, since the Lieb equation (\ref{Fredholm}) has been solved with high accuracy in chapter \ref{secIII}. Otherwise, if $k/Q\!>\!1$, the Lieb equation is solved numerically at $z\!>\!1$ from the solution at $z\!\leq\!1$. Thus, the type II spectrum between $p\!=\!0$ and $p\!=\!2p_F$ is a priori easier to obtain than the exact type I spectrum.

A new technical difficulty whose solution is not readily provided by the evaluation the ground-state energy comes from the second Lieb equation (\ref{secondlieb}), which is another type of integral equation, whose exact solution at arbitrary interaction strength is also unknown. A possible tactics to solve Eq.~(\ref{secondlieb}) is to adapt the orthogonal polynomial method used to solve the first Lieb equation (\ref{Fredholm}), yielding an approximate solution for $\alpha\!>\!2$ \cite{Ristivojevic, LangHekkingMinguzzi2017}. In the weakly-interacting regime, no systematic method has been developed so far, but since $f$ is well-behaved at low $\alpha$, numerical solutions are rather easily obtained. Moreover, at $\alpha\!>\!2$ the strong-coupling expansion converges faster to the exact solution and generates far fewer terms than was the case for the density of pseudo-momenta $g$ \cite{LangHekkingMinguzzi2017}.

I also noticed that Eq.~(\ref{secondlieb}) appears in the alternative approach to the Lieb-Liniger model, based on a limit case of the sinh-Gordon model, where this equation must be solved to obtain the ground-state energy and correlation functions, cf Appendix \ref{gkk}. It also appears in other contexts, such as the problem of two circular disks rotating slowly in a viscous fluid with equal angular velocities \cite{Cooke1956}, or of the radiation of water waves due to a thin rigid circular disk in three dimensions \cite{Farina2010}.

In the end, I have access to accurate analytical estimates of the type II branch, provided that $\alpha\!>\!2$. As far as the type I spectrum is concerned, an additional limitation comes from the fact that the approximate expressions for $g(z;\alpha)$ and $f(z;\alpha)$ are valid only if $|z-y|\!\leq\!\alpha$. This adds the restriction $|k|/Q(\alpha)\!\leq\!\alpha\!-\!1$, which is not constraining as long as $\alpha\!\gg\!1$, but for $\alpha\!\gtrsim\!2$, the validity range is very narrow around $p\!=\!0$.

To bypass this problem, one can use an iteration method to evaluate $g$ and $f$. In practice, this method is analytically tractable at large interactions only, as it allows to recover at best the first few terms of the exact $1/\alpha$ expansion of $\epsilon(k;\alpha)$ and $p(k;\alpha)$ (to order $2$ in \cite{Ristivojevic}). Another difficulty is that these approximate expressions are not of polynomial type, and it is then a huge challenge to substitute the parameter $k$ and express $\epsilon(p)$ explicitly, forcing to resort on approximations at high and small momenta.

\begin{figure}
\includegraphics[width=9cm, keepaspectratio, angle=0]{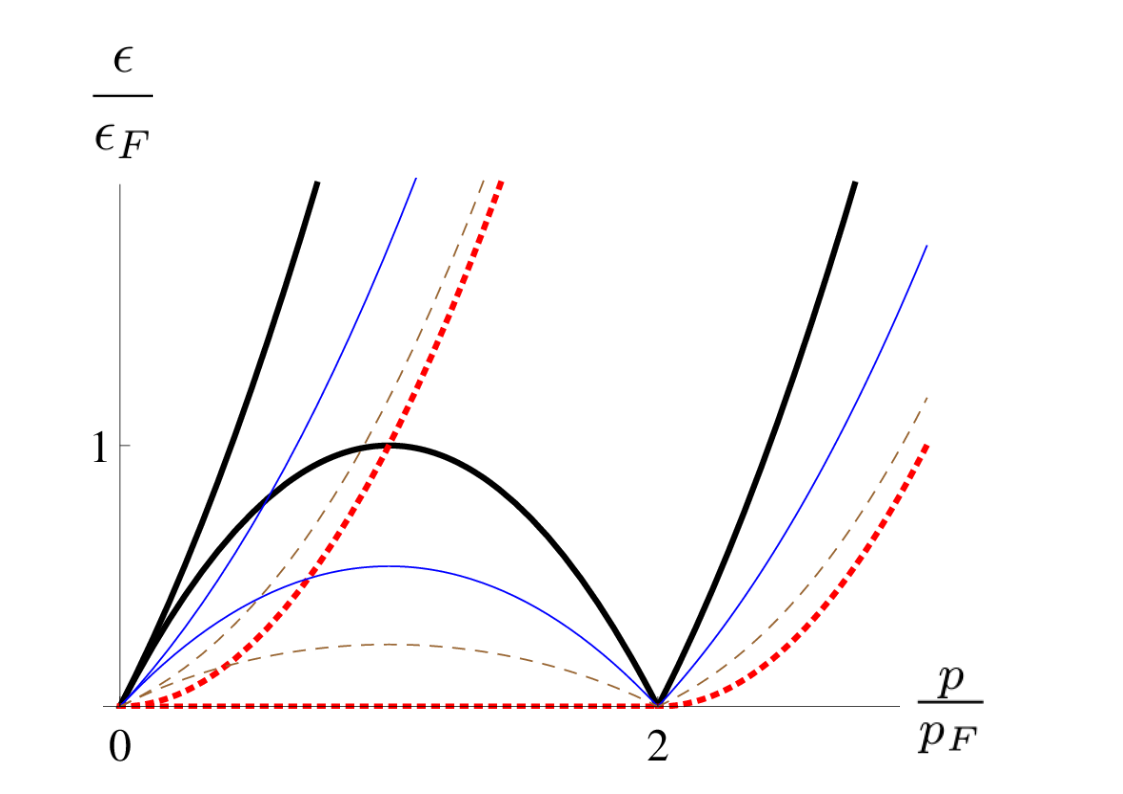}
\caption{Type I and type II excitation spectra of the Lieb-Liniger model for several values of the interaction strength, from the noninteracting Bose gas (dashed, red) to the Tonks-Girardeau regime (thick, black) with intermediate values $\alpha\!=\!0.6$ (dashed, brown) and $\alpha\!=\!2$ (solid, blue).}
\label{Figepsilon}
\end{figure}

Both excitation spectra are shown in Fig.~\ref{Figepsilon} for several values of the interaction strength, as obtained from the most appropriate method in each case. Note that the area below the type II spectrum, as well as the maximal excitation energy at $p_F$, are both increasing functions of the Lieb parameter $\gamma$ and vanish for a noninteracting Bose gas.

\begin{figure}
\includegraphics[width=8cm, keepaspectratio, angle=0]{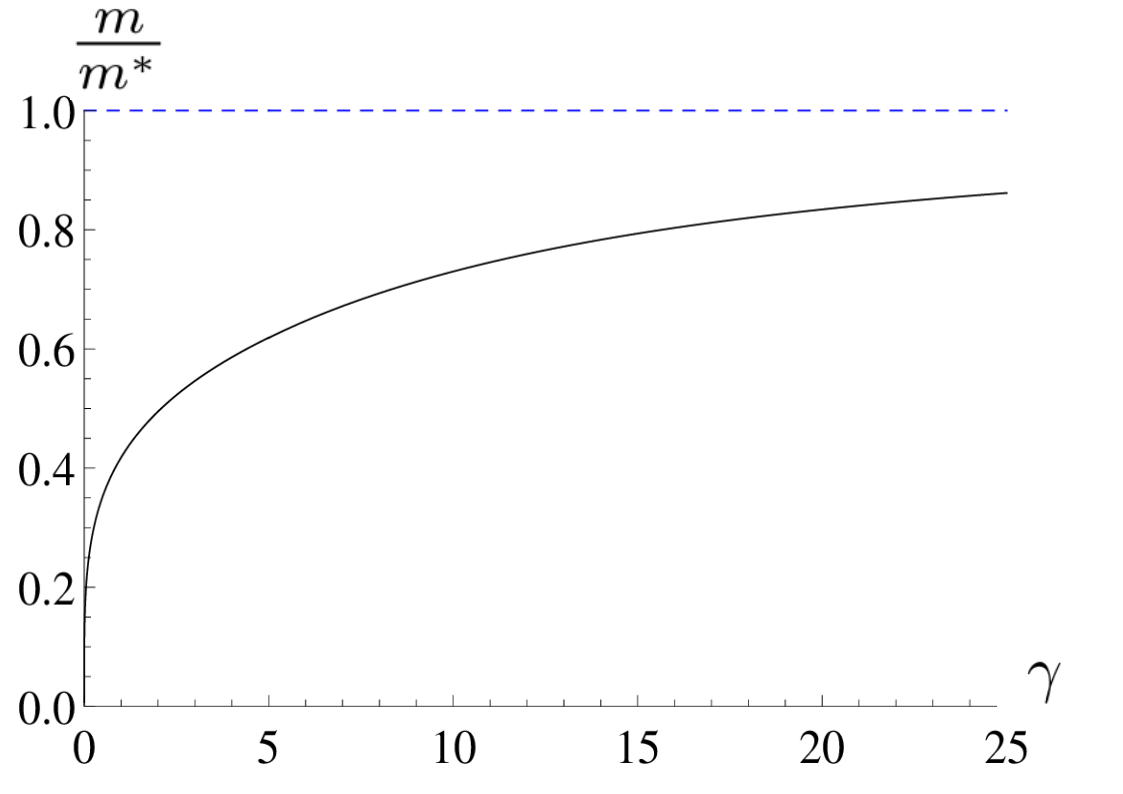}
\caption{Dimensionless inverse renormalized mass $m/m^*$ obtained with Eqs.~(\ref{minveq}), (\ref{soundgamma}), (\ref{exp}) and (\ref{conjWidom}), as a function of the dimensionless Lieb parameter $\gamma$ (black). The exact value in the Tonks-Girardeau regime is plotted in dashed blue as a comparison.}
\label{Figlbis}
\end{figure}

At small momenta, the type I spectrum can be expressed by its series expansion in $p$ \cite{MatveevFurusaki2013, RistivojevicMatveev2016}, that reads
\begin{eqnarray}
\label{expI}
\epsilon^I(p;\gamma)=_{p\simeq 0}v_s(\gamma)p+\frac{p^2}{2m^*(\gamma)}+\frac{\lambda^*(\gamma)}{6}p^3+\dots.
\end{eqnarray}
By comparison, in the Tonks-Girardeau regime, as follows from Eq.~(\ref{EETG}), the coefficients in Eq.~(\ref{expI}) are $v_s\!=\!v_F$, $m^*\!=\!m$, $\lambda^*=0$, and all higher-order coefficients are null as well. At finite interaction strength, $v_s$ can be seen as a renormalized Fermi velocity, and $m^*$ is interpreted as an effective mass, whose general expression is \cite{Ristivojevic, MatveevPustilnik2016}
\begin{eqnarray}
\label{minveq}
\frac{m}{m^*}=\left(1-\gamma \frac{d}{d\gamma}\right)\sqrt{\frac{v_s}{v_F}},
\end{eqnarray}
shown in Fig.~\ref{Figlbis}. Note that for a noninteracting Bose gas, $m^*\!=\!m$, causing a discontinuity at $\gamma\!=\!0$. This means that the ideal Bose gas is not adiabatically connected to the weakly-interacting regime in 1D.

As far as the type II spectrum is concerned, the properties (a)-(f) detailed above suggest another type of expansion, whose truncation to first order has been anticipated in \cite{Shamailov2016}:
\begin{eqnarray}
\label{expII}
\frac{\epsilon^{II}(p;\gamma)}{\epsilon_F}=\sum_{n=1}^{+\infty}\epsilon_n(\gamma)\left(\frac{p}{p_F}\right)^{2n}\left(2-\frac{p}{p_F}\right)^{2n},
\end{eqnarray}
where $\{\epsilon_n(\gamma)\}_{n\geq 1}$ are dimensionless functions. The property (f) enables me to write
\begin{eqnarray}
\label{expIIbis}
\epsilon^{II}(p;\gamma)=v_s(\gamma)p-\frac{p^2}{2m^*(\gamma)}+\dots,
\end{eqnarray}
and equating both expressions to order $p^2$, I find in particular that
\begin{eqnarray}
\label{f1}
\epsilon_1(\gamma)=\frac{v_s(\gamma)}{v_F}.
\end{eqnarray}
A similar result was recently inferred from a Monte-Carlo simulation of one-dimensional $^4$He in \cite{Bertaina2016}, and proved by Bethe Ansatz applied to the hard-rods model in \cite{Motta2016}. Using the same approach to next order, I also obtained
\begin{eqnarray}
\label{f2}
\epsilon_2(\gamma)=\frac{1}{4}\left(\frac{v_s(\gamma)}{v_F}-\frac{m}{m^*(\gamma)}\right).
\end{eqnarray}

\begin{figure}
\includegraphics[width=8cm, keepaspectratio, angle=0]{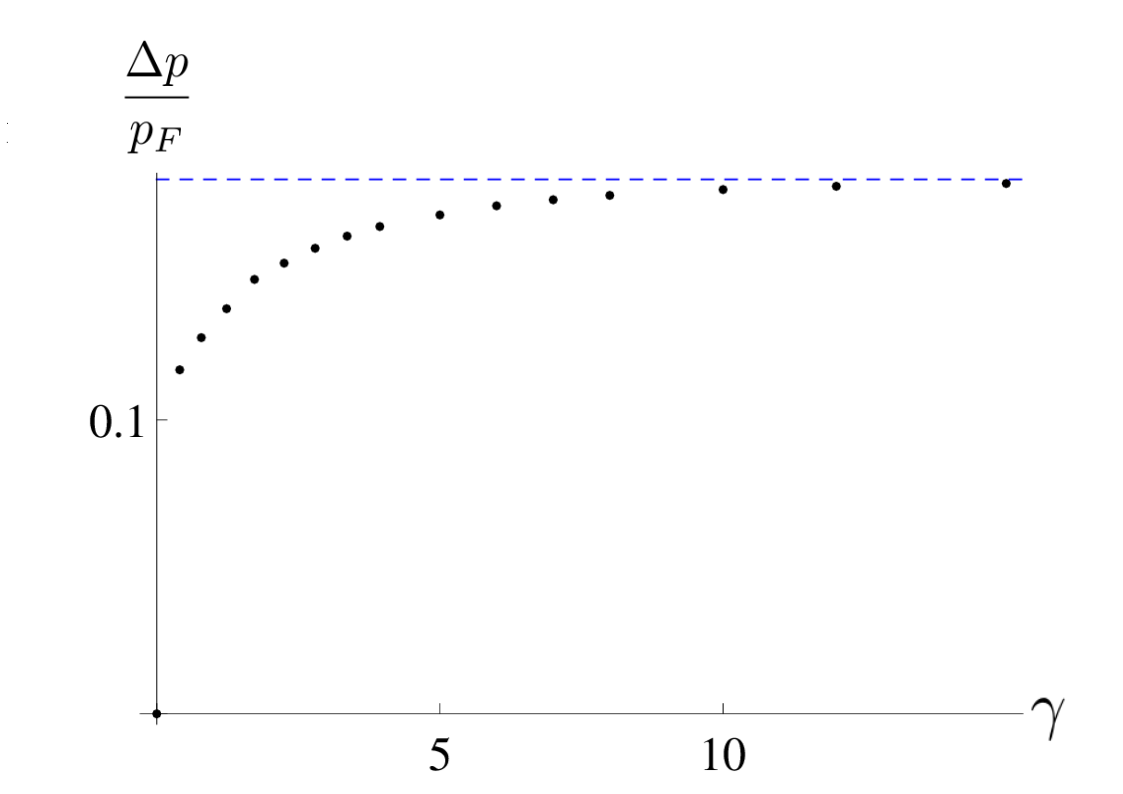}
\includegraphics[width=8cm, keepaspectratio, angle=0]{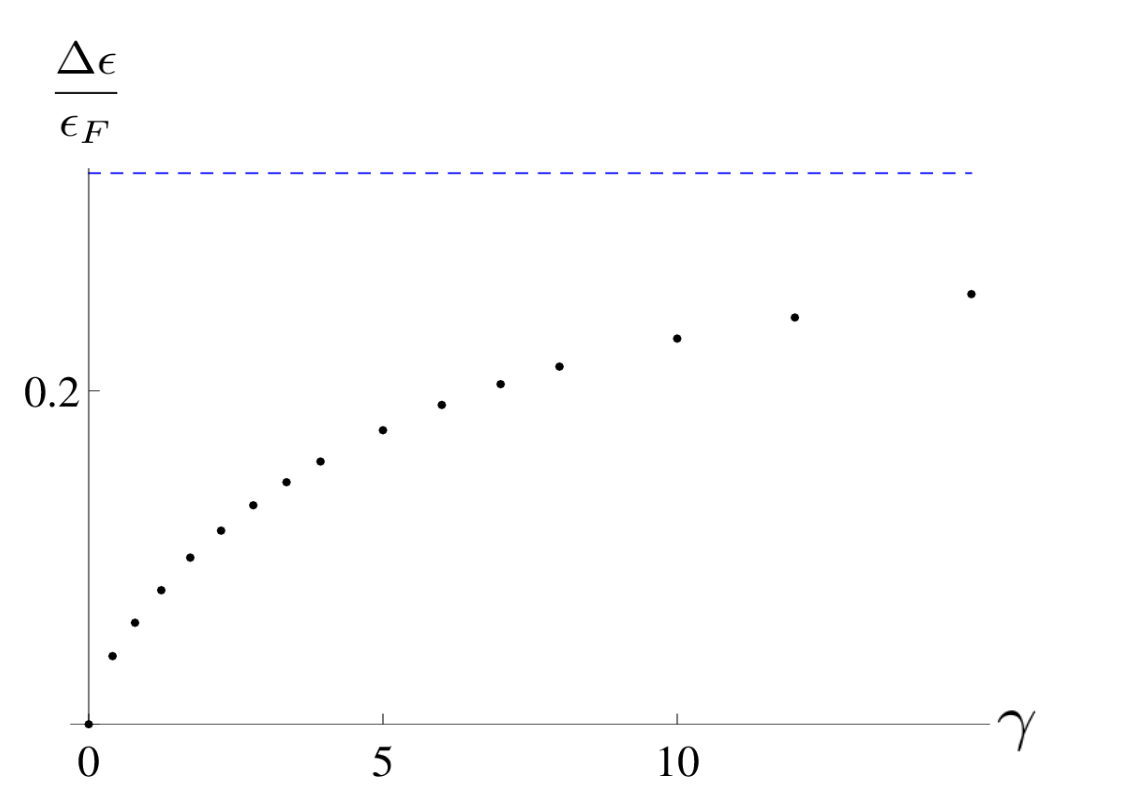}
\caption{Upper bounds for the validity range of the Tomonaga-Luttinger liquid framework in dimensionless momentum (left panel) and dimensionless energy (right panel) around the umklapp point $(p\!=\!2p_F, \epsilon\!=\!0)$ of the of the Lieb-Liniger model for the excitation spectrum $\epsilon^{II}$, as functions of the dimensionless interaction strength $\gamma$. Dots represent the numerical estimate at finite interaction strength, the dashed blue curve corresponds to the Tonks-Girardeau regime, where $\Delta \epsilon^{TG}/\epsilon_F\simeq 0.3$ and $\Delta p^{TG}/p_F\simeq 0.2$}
\label{FigLutt}
\end{figure}

It is now possible to compare the exact spectrum to the truncated series obtained from Eq.~(\ref{expII}), to investigate the validity range of various approximate expressions. I denote by $\Delta p$ and $\Delta \epsilon$ respectively the half-width of momentum around the umklapp point, and the maximum energy, such that the linearized spectrum $\epsilon^{TL}\!=\!v_s|p\!-\!2p_F|$ is exact up to ten percent. These quantities, shown in Fig.~\ref{FigLutt}, should be considered as upper bounds of validity for dynamical observables such as the dynamical structure factor. The Tomonaga-Luttinger liquid theory works better at large interaction strength.

\begin{figure}
\includegraphics[width=8cm, keepaspectratio, angle=0]{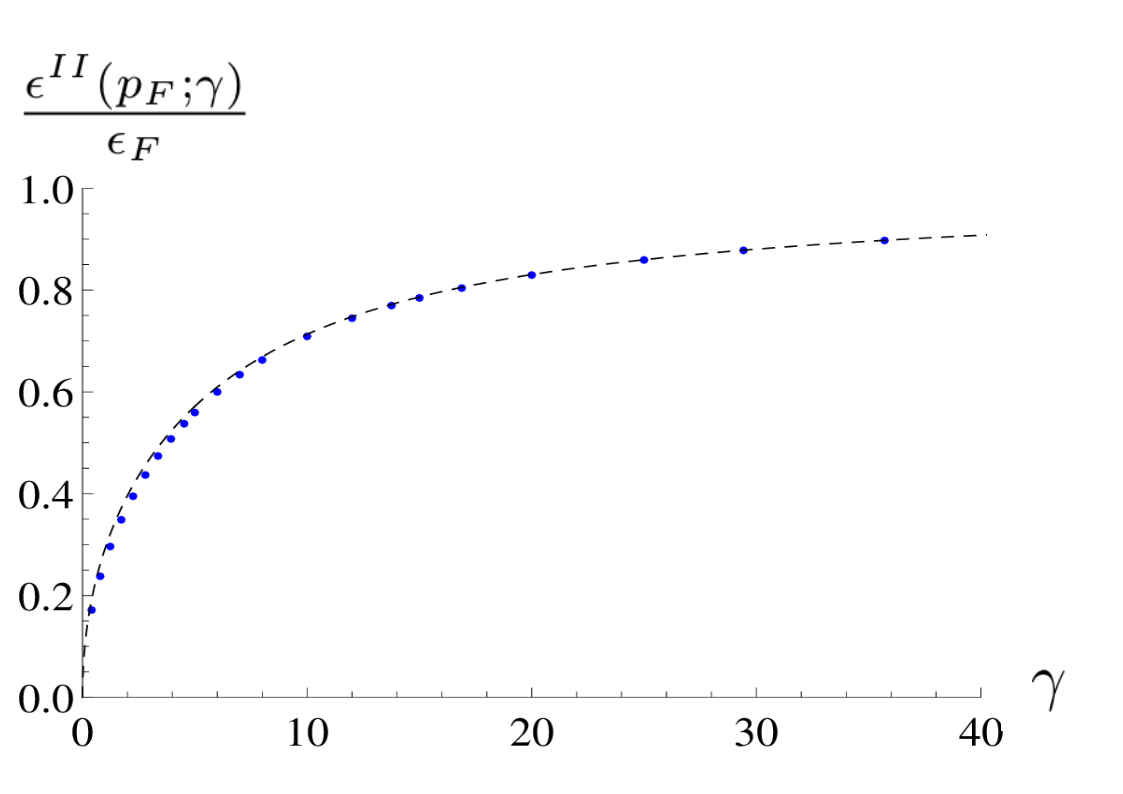}
\includegraphics[width=8cm, keepaspectratio, angle=0]{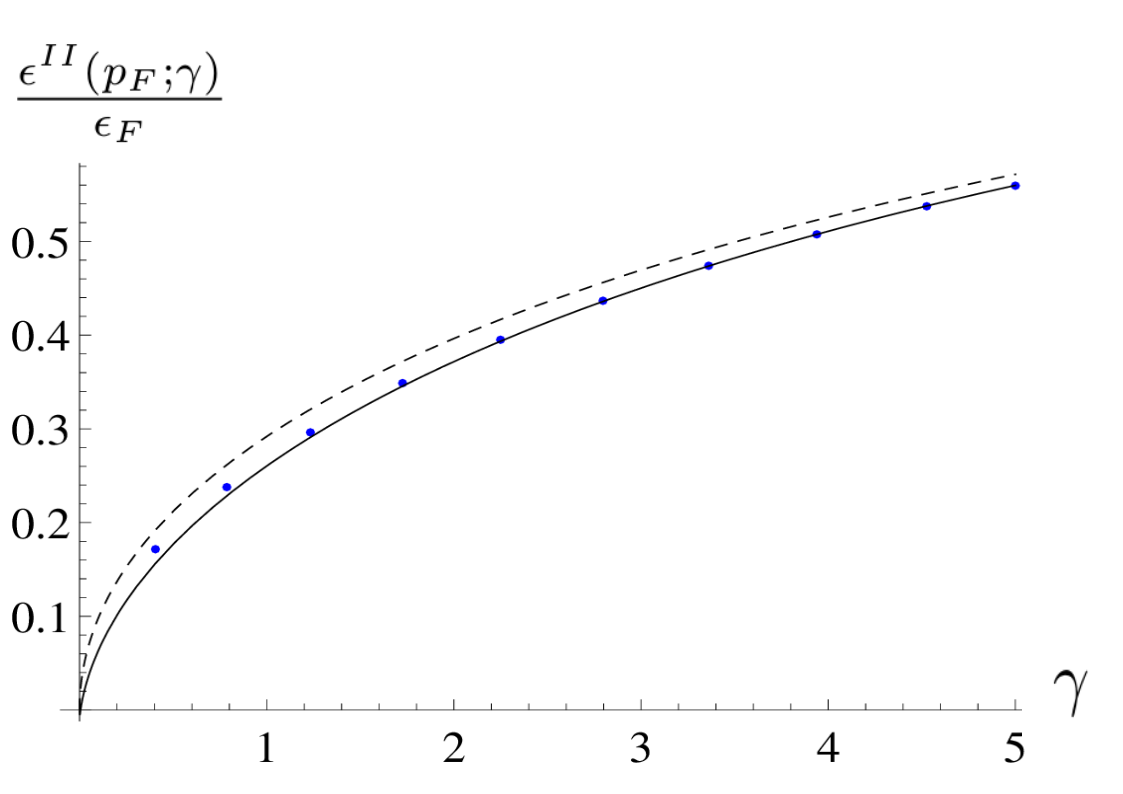}
\caption{Maximum of the type II spectrum, $\epsilon^{II}(p_F;\gamma)$, in units of the Fermi energy $\epsilon_F$, as a function of the dimensionless interaction strength $\gamma$. The left panel shows the first order approximation in Eq.~(\ref{expII}) taking $\epsilon_{n\geq2}=0$ (dashed, black), compared to exact numerical data (blue dots). The right panel shows a zoom close to the origin and the second order approximation (solid, black). Agreement with the exact result is significantly improved when using this correction.}
\label{Fig8}
\end{figure}

Including the quadratic term in Eq.~(\ref{expIIbis}) and neglecting higher-order ones is actually a complete change of paradigm, from massless bosonic to massive fermionic excitations at low energy, at the basis of the Imambekov-Glazman theory of beyond-Luttinger liquids. Systematic substitution of the variable $k$ in Eqs.~(\ref{eqP}) and (\ref{EqEE}), and higher-order expansions suggest that higher-order terms in expansion (\ref{expII}) can be neglected in a wide range of strong to intermediate interaction strengths. Figure \ref{Fig8} shows the local maximum value $\epsilon^{II}(p_F)$ of the Lieb-II excitation spectrum, as obtained from a numerical calculation as well as the expansion (\ref{expII}) truncated to orders one and two. I find that the result to order one is satisfying at large $\gamma$, but the second order correction significantly improves the result at intermediate values of the Lieb parameter. Numerical calculations show that third- and higher-order corrections are negligible in a wide range of strong interactions.

\section{Summary of this chapter/r\'esum\'e du chapitre}

This chapter starts with an historical account of a few experiments that allowed major breakthroughs in the understanding of superfluidity. Discovered in ${^4}$He, dramatic suppression of viscosity was then witnessed in ${^3}$He, in ultracold atom experiments and in polaritons. The conceptual difficulties raised by the notion of superfluidity are such that no universal definition or characterization means has been found so far.

The most celebrated criterion for superfluidity is due to Landau, who predicted a thorough absence of viscosity at zero temperature below a critical velocity, that coincides with the sound velocity at the mean field level. In the filiation of this criterion, the concept of drag force due to an impurity or a laser beam stirred in the fluid allows to study friction in a quantitative way. It generalizes Landau's arguments, taking into account transition probabilities to excited states and the precise shape of the potential barrier. Linear response theory allows to express the drag force due to a weak barrier given the profile of the latter and another observable, the dynamical structure factor.

The Tonks-Girardeau regime gives an opportunity to find the drag force in a strongly-interacting Bose gas in the linear response framework, without additional approximation. In the thermodynamic limit, the energy-momentum profile of possible excitations shows that a low-energy region is forbidden in 1D, and the low-energy excitations are dominated by processes that occur close to the umklapp point. To obtain the dynamical structure factor at finite temperature, I first had to derive the temperature profile of the chemical potential. Then, thermal effects on the dynamical structure factor consist essentially in a broadening of the momentum-energy sector where excitations can occur, and above $T\!\gtrsim\!0.2\,T_F$, well-defined phonon excitations disappear progressively.

As far as the drag force is concerned, at zero temperature, linear response theory predicts that it saturates at supersonic flow velocities, which seems quite unrealistic, and at finite temperature the drag force is lower than it would at $T\!=\!0$ close to the Fermi velocity, which is also disturbing. The first point is solved by taking into account the barrier width, assuming a Gaussian profile. Then, instead of saturating, at high velocities the drag force vanishes. In other words, I have predicted the existence of a quasi-superfluid, supersonic regime.

At finite interaction strengths, several approaches exist to evaluate the dynamical structure factor, ranging from perturbation theory to the Tomonaga-Luttinger liquid framework, its extension to higher energies through the Imambekov-Glazman liquid formalism, and numerical solution based on algebraic Bethe Ansatz methods. Although experiments have demonstrated the predictive power of the more advanced techniques, I have used the standard Tomonaga-Luttinger liquid formalism all the same. I have calculated the dynamical structure factor and generalized it to finite temperature, where a comparison to the exact Tonks-Girardeau result has allowed me to quantitatively show that the effective description is limited to low temperatures, of the order of $T\!\lesssim\!0.2\,T_F$.

At zero temperature and finite interaction strength, making quantitative predictions within the Tomonaga-Luttinger liquid framework requires the knowlege of the sound velocity. The latter is obtained exactly by Bethe Ansatz. It is also important to evaluate the first form factor. I have found an accurate fitting function that allows for a comparison with the exact result from algebraic Bethe Ansatz. Then, I have evaluated the drag force at low velocities, in the simplest case and also at finite barrier width and temperature, where I have compared the predictions with the exact result in the Tonks-Girardeau regime. This shows that the drag force evaluated in the Tomonaga-Luttinger liquid framework is correct up to $v\!\lesssim\!0.2\,v_F$.

To go beyond the standard Luttinger liquid treatment, it is important to accurately evaluate the excitation spectra. This is done exactly by Bethe Ansatz, through a procedure that I explain in detail. I have obtained several symmetry properties of the Lieb II excitation spectrum, found several approximations in terms of the sound velocity and effective mass, and evaluated their range of validity.
\\
\\
\\
\\
\\
Ce chapitre d\'ebute par un bref rappel des principales exp\'eriences historiques qui ont mis en \'evidence la superfluidit\'e et ont conduit \`a ses premi\`eres interpr\'etations th\'eoriques. D\'ecouverte gr\^ace \`a l'isotope \`a quatre nucl\'eons de l'h\'elium sous forme liquide, la spectaculaire disparition de la viscosit\'e en-dessous d'une temp\'erature critique a ensuite \'et\'e observ\'ee pour l'isotope \`a trois nucl\'eons du m\^eme \'el\'ement, puis dans des syst\`emes d'atomes froids et plus r\'ecemment dans les polaritons. Les difficult\'es conceptuelles soulev\'ees par la superfluidit\'e sont telles qu'aucun crit\`ere universellement valide n'a pu \^etre d\'efini \`a ce jour pour la d\'efinir ou la caract\'eriser.

Parmi les multiples crit\`eres propos\'es, celui d\^u \`a Landau, qui pr\'evoit notamment l'existence d'une vitesse critique en-dessous de laquelle un \'ecoulement devient superfluide, a sans doute eu le plus fort impact. Dans son \'etroite filiation, le concept de force de tra\^in\'ee dans le r\'egime quantique, due \`a une impuret\'e mobile ou \`a un faisceau laser qui parcourt le fluide, permet d'\'evaluer quantitativement l'effet de la viscosit\'e. Le crit\`ere fond\'e sur l'absence de force de tra\^in\'ee dans le r\'egime superfluide g\'en\'eralise celui de Landau, en prenant en compte les probabilit\'es de transition vers des \'etats excit\'es et le profil spatial de la barri\`ere de potentiel. Le formalisme de la r\'eponse lin\'eaire permet d'\'evaluer la force de tra\^in\'ee en fonction de la vitesse d'\'ecoulement, si on se donne un profil de potentiel et le facteur de structure dynamique du fluide.

Dans le r\'egime de Tonks-Girardeau, j'ai pu \'evaluer le facteur de structure dynamique, puis la force de tra\^in\'ee dans le cadre de la r\'eponse lin\'eaire. \`A la limite thermodynamique, la r\'egion de l'espace des \'energies et impulsions dans laquelle les excitations peuvent avoir lieu forme un continuum. Ce dernier est coup\'e des \'energies nulles par une r\'egion vide d'excitations, \`a part \`a l'origine et au point dit d'umklapp, dont le voisinage domine les processus de dissipation. Pour obtenir le facteur de structure dynamique \`a temp\'erature finie, il m'a fallu dans un premier temps calculer le profil en temp\'erature du potentiel chimique. L'effet de la temp\'erature sur le facteur de structure dynamique consiste essentiellement en un \'elargissement du domaine d'excitations possibles, jusqu'\`a des \'energies n\'egatives qui correspondent \`a une \'emission vers le milieu ext\'erieur, et en une r\'eduction de la probabilit\'e des excitations \`a faible vecteur d'onde. On constate \'egalement que les phonons sont de moins en moins bien d\'efinis \`a mesure que la temp\'erature augmente, jusqu'\`a ne plus \^etre identifiables.

En ce qui concerne la force de tra\^in\'ee, \`a temp\'erature nulle, le formalisme de la r\'eponse lin\'eaire pr\'evoit un profil lin\'eaire en fonction de la vitesse si celle-ci est plus faible que la vitesse de Fermi, qui est \'egalement celle du son, puis une saturation dans le r\'egime supersonique, qui para\^it peu r\'ealiste. Ce probl\`eme est r\'esolu par la prise en compte de l'\'epaisseur de la barri\`ere de potentiel, assimil\'ee \`a une Gaussienne. D\`es lors, \`a grande vitesse la force de tra\^in\'ee s'effondre, ce qui conduit \`a un r\'egime quasi-superfluide inclus dans le domaine supersonique.

Plusieurs strat\'egies sont possibles pour prendre en compte les interactions dans le cas g\'en\'eral. Le formalisme des liquides de Tomonaga-Luttinger s'applique approximativement \`a basse \'energie et faible vitesse, celui des liquides d'Imambekov-Glazman a un domaine de validit\'e plus cons\'equent, et enfin des m\'ethodes num\'eriques pour r\'esoudre l'Ansatz de Bethe alg\'ebrique donnent un r\'esultat quasi-exact. Bien que les exp\'eriences mettent en \'evidence le pouvoir pr\'edictif des m\'ethodes les plus avanc\'ees, dans un premier temps je me contente d'utiliser le formalisme des liquides de Tomonaga-Luttinger, que j'ai g\'en\'eralis\'e \`a temp\'erature non-nulle et compar\'e autant que possible au traitement exact du r\'egime de Tonks-Girardeau, qu'il reproduit plut\^ot bien \`a basse \'energie et tr\`es basse temp\'erature.

\`A temp\'erature nulle et intensit\'e des interactions arbitraire, afin de faire des pr\'edictions quantitatives dans le formalisme des liquides de Tomonaga-Luttinger, il faut au pr\'ealable \'evaluer la vitesse du son, obtenue par Ansatz de Bethe, et le premier facteur de forme. Ce dernier est tr\`es difficile \`a calculer, et requiert des techniques avanc\'ees fond\'ees sur l'Ansatz de Bethe alg\'ebrique. Par ajustement d'une expression approch\'ee sur les donn\'ees num\'eriques disponibles dans la litt\'erature, j'ai pu obtenir une bonne approximation de cette quantit\'e. Ceci m'a permis en particulier de comparer les pr\'edictions du formalisme de Tomonaga-Luttinger au facteur de structure dynamique exact le long d'une ligne \`a la verticale du point d'umklapp, qui se traduit par un accord surprenant dans une large gamme d'intensit\'e des interactions, \`a \'energie suffisamment faible. J'ai ensuite \'evalu\'e la force de tra\^in\'ee, et me suis appuy\'e une fois de plus sur la solution exacte dans le r\'egime de Tonks-Girardeau pour montrer que la solution effective s'applique bien aux vitesses faibles.

Pour aller plus loin que le formalisme des liquides de Tomonaga-Luttinger, il est essentiel d'\'evaluer avec pr\'ecision le spectre d'excitation du mod\`ele de Lieb et Liniger. J'en ai \'etudi\'e les propri\'et\'es de sym\'etrie et ai trouv\'e un d\'eveloppement en s\'erie appropri\'e pour l'exprimer dans le cas g\'en\'eral, dont les premiers termes d\'ependent de la vitesse du son et de la masse effective, obtenues par Ansatz de Bethe. J'ai compar\'e diverses approximations au r\'esultat exact et \'evalu\'e leur domaine de validit\'e.

\section{Outlook of this chapter}

Although it has already led to a fair number of new results, the project associated to this chapter is not at an end. My analytical analysis stays at a basic level compared to my initial ambitions, as its accuracy does not compete yet with the numerical results from the ABACUS algorithm. I am only one step away, however, from making quantitative predictions based on the Imambekov-Glazman theory, as I only miss the edge exponents, that should be at reach in principle. My analytical estimates of the excitation spectrum of the Lieb-Liniger model are already fairly accurate. They could be further improved at small and intermediate interaction strength, by including next order in the truncated series. In particular, the function $\epsilon_3(\gamma)$ in Eq.~(\ref{expII}), depends on $\lambda^*(\gamma)$ in Eq.~(\ref{expI}), that could be explicitly evaluated by straightforward algebra from results of Refs.~\cite{MatveevFurusaki2013, RistivojevicMatveev2016}, see Ref.~\cite{Ristivojevic2017} for recent improvements in this direction.

Whatever level of sophistication should be employed to evaluate the dynamical structure factor, I already foresee a lot of exciting open problems related to the drag force, that could be answered at a basic level within current means. For instance, the anisotropic drag force due to spin-orbit coupling constitutes a new thread of development \cite{He2014, Liao2016}. In the context of Anderson localization, a random potential may break down superfluidity at high disorder. This issue was pioneered in \cite{Albert2010}, and it has been recently shown that a random potential with finite correlation length gives rise to the kind of drag force pattern that I have put into light in this chapter \cite{ChernyCauxBrand2017}. Other types of barrier may also be considered, for instance a shallow lattice \cite{ChernyCauxBrand2009}. Those having an appropriate profile, whose Fourier transform does not overlap with the energy-momentum area where the dynamical structure factor is zero, lead to exact superfluidity in the framework of linear response theory, according to the drag force criterion. However, even if linear response theory predicts superfluidity, to be certain that it is the case, drag force should be accounted for at higher orders in perturbation theory, or even better, non-perturbatively. This was investigated in \cite{ChernyCauxBrand2012} for the Tonks-Girardeau gas.

It has been argued that when quantum fluctuations are properly taken into account, they impose a zero critical velocity as there always exists a Casimir type force \cite{RobertsPomeau2005, Sykes2009}. However, the latter is of a far lesser amplitude, and another key ingredient, neglected in this approach and mine, is finite mass impurity. It can be taken into account in the drag force formalism \cite{ChernyCauxBrand2012}, or in the Bose polaron framework, where the full excitation spectrum of the gas and impurity is considered non-perturbatively \cite{Schecter2012}. The situation is radically different then, as strict superfluidity seems possible if $m_i\!<\!+\infty$ \cite{Lychkovskiy2014}. More generally, even in the drag force context, at finite $N$ a small gap appears at the umklapp point \cite{ChernyCauxBrand2009}, allowing superfluidity at very low velocities \cite{Schenke2012}. It would also be interesting to relate the drag force to supercurrent decay, which has emerged as the standard observable to study superfluidity of mesoscopic systems. These works show that superfluidity is rather expected at the mesoscopic scale than the macroscopic one.

The drag force formalism, as it has been applied so far, is not fully satisfying, as it should be used in Newton's equations to predict the equation of motion of the impurity, rather than just checking whether the drag force is zero or not. The only example thereof that I know is \cite{ChernyCauxBrand2009}, although studies of long-time velocity as a function of the initial one are flourishing in the literature based on the Bose polaron formalism \cite{Lychkovskiy2014, Gamayun2017}. Moreover, at the moment the drag force formalism does not take into account the inhomogeneities, that are essential \cite{Fedichov2001}. They have been taken into account in the Tomonaga-Luttinger liquid framework in \cite{Orignac2012}, but to fully understand the back action of the drag on the local density profile, it should be taken into account dynamically, so that numerical support is still needed to follow the center of mass position correctly \cite{Caux2016, Robinson2016}. To finish with, experimentally, there are some contexts where the drag force should still be measured, for instance in polaritons \cite{Berceanu2012}, or in a 1D gas.

To enlarge the scope of this study to other fields of physics, let me mention that the concept of superfluidity is also studied in astrophysics since neutron stars may have a superfluid behavior \cite{Migdal1960, Page2011}, and in cosmology \cite{Volovik2001}. In a longer run perspective, as far as technological applications of superfluids beyond mere cooling are concerned, stability against thermal fluctuations or external perturbations is crucial. The key parameters, critical temperature and critical velocity, are typically highest in the strongly-correlated regime, where the interactions stabilizing the many-body state are peculiarly strong. 

\newpage

\chapter{Dimensional crossovers in a gas of noninteracting spinless fermions}
\label{secV}

\section{Introduction}

In order to describe ultracold atom experiments with high accuracy, in addition to selecting a model for the interactions, several other aspects have to be taken into account, such as finite temperature, system size and number of particles, or inhomogeneities of the atomic cloud. I have illustrated such refinements on the example of the Lieb-Liniger model in the previous chapters. Taking several effects into account simultaneously is technically challenging, but it is not easy either to rank them by relative importance, to decide which of them could be neglected or should be incorporated first.

However, among all assumptions that could be questioned, there is one on which I have not insisted on yet. Most of the low-dimensional gases created so far are not strictly one-dimensional, but involve a multi-component structure. For instance, when an array of wires is created, the gases confined in two different ones may interact through their phase or density. Even in a tightly-confined single gas, the conditions requested to reach a strictly one-dimensional regime are not ideally fulfilled, and the gas is actually quasi-1D, with several modes in momentum-energy space. This slight nuance may be more important to take into account than any of the effects listed before. Although considering even a few modes is a theoretical challenge of its own, it did not refrain pioneers to investigate the dimensional crossover to higher dimensions already mentionned in chapter \ref{secII}, whose technical difficulty lies at an even higher level, but is accessible to experiments.

To have a chance of treating the dimensional crossover problem analytically and exactly, I have considered the most conceptually simple system, i.e. a noninteracting gas. In this case, dimensional crossover is obviously not realized by interactions, but rather by a transverse deconfinement, through a progressive release of trapping. This scenario is not the most commonly considered in the literature, so Ref.~\cite{LangHekkingMinguzzi2016}, on which this chapter is based, is quite pioneering and original compared to my other works presented above. The simplifications made at the beginning enable me to study dynamical observables such as the dynamical structure factor and drag force.

The chapter is organized as follows: in a first time, I calculate the dynamical structure factor of a noninteracting Fermi gas in higher dimensions. Then, I develop a formalism that describes the multi-mode structure, and use it to recover the previous results by adding more modes up to observing a dimensional crossover. The same work is done on the drag force, then.

In a second time, I consider the effect of a harmonic trap in the longitudinal direction within the local-density approximation, and show that the trap increases the effective dimension of the system, allowing to simulate a gas in a box up to dimension $d\!=\!6$.

To conclude the chapter, I develop a multimode Tomonaga-Luttinger model to describe the noninteracting Fermi gas, and apply it throughout the dimensional crossover from 1D to 2D.
\\
\\
\\
\\
\\
Pour d\'ecrire avec pr\'ecision une exp\'erience d'atomes froids, choisir un mod\`ele appropri\'e est un \'etape oblig\'ee, mais encore faut-il prendre en compte divers aspects du probl\`eme, comme la temp\'erature, la taille du syst\`eme et le nombre d'atomes, ainsi que les inhomog\'en\'eit\'es du profil de densit\'e du gaz, autant de param\`etres qui peuvent avoir leur importance. J'ai d\'ej\`a illustr\'e ces diff\'erents points dans les chapitres pr\'ec\'edents, en m'appuyant sur le mod\`ele de Lieb et Liniger. La prise en compte de plusieurs param\`etres en m\^eme temps s'av\`ere rapidement probl\'ematique pour des raisons techniques, mais il est a priori difficile de savoir lesquels n\'egliger en toute s\'ecurit\'e, ou prendre en compte en priorit\'e, sachant qu'ils peuvent avoir des effets inverses qui se compensent partiellement et n'apportent pas grand chose, de ce fait, en comparaison d'une analyse moins fine.

Toutefois, dans la liste \'etablie plus haut, j'ai pass\'e sous silence un aspect important de la mod\'elisation, \`a savoir le choix d'affecter au syst\`eme une dimension donn\'ee. La plupart des gaz ultrafroids de basse dimension r\'ealis\'es exp\'erimentalement ne s'accomodent pas \`a la perfection d'une description unidimensionnelle, mais poss\`edent une structure multimode, qui leur donne le statut de gaz quasi-1D. Par exemple, un gaz dans un r\'eseau optique peut \^etre vu comme un ensemble de gaz unidimensionnels s\'epar\'es spatialement, mais dans la pratique des couplages peuvent avoir lieu entre ces diff\'erents gaz. M\^eme dans un gaz unique fortement confin\'e, les conditions th\'eoriques pour le rendre strictement unidimensionnel ne sont pas toujours remplies, et le gaz est alors quasi-1D dans la pratique. Ces consid\'erations de dimension peuvent s'av\'erer plus importantes que tous les autres effets r\'eunis, de par leurs importantes cons\'equences en basse dimension. Bien que la prise en compte ne serait-ce que de deux ou trois modes puisse s'av\'erer tr\`es technique sur le plan th\'eorique, cela n'a pas emp\^ech\'e certains pionniers d'ouvrir le champ de recherche associ\'e au changement de dimension, extr\^emement compliqu\'e dans la th\'eorie, mais r\'ealisable en pratique.

Afin de maximiser mes chances de r\'eussir \`a traiter un exemple analytiquement et de mani\`ere exacte, j'ai consid\'er\'e le mod\`ele le plus simple possible, un gaz id\'eal. Dans ce cas, les effets dimensionnels ne sont bien \'evidemment pas dus aux interactions, mais au rel\^achement progressif d'une contrainte de confinement lat\'eral. La situation s'av\`ere suffisamment simple pour me permettre de consid\'erer des observables \`a la structure riche, comme le facteur de structure dynamique et la force de tra\^in\'ee.

Le chapitre s'organise de la sorte: dans un premier temps, je calcule le facteur de structure dynamique d'un gaz de fermions en dimensions deux et trois, apr\`es quoi je d\'eveloppe le formalisme ad\'equat pour \'etudier des structure multimodes, et en guise d'illustration, je m'en sers pour retrouver ces r\'esultats \`a travers le passage vers la dimension sup\'erieure. J'adapte ensuite tout cela \`a la force de tra\^in\'ee.

Dans un second temps, j'envisage un confinement harmonique dans la direction longitudinale, que je traite dans le cadre de l'approximation de la densit\'e locale. Je montre que ce pi\`ege a pour effet d'augmenter la dimension effective du syst\`eme, ce qui permet de simuler jusqu'\`a six dimensions dans une bo\^ite. Enfin, j'\'etends le formalisme des liquides de Tomonaga-Luttinger \`a des situations multi-modes, et l'applique jusqu'\`a la limite de dimension deux afin d'illustrer ses capacit\'es pr\'edictives.

\section{Energy-momentum space dimensional crossover in a box trap}

\begin{figure}
\includegraphics[width=12cm, keepaspectratio, angle=0]{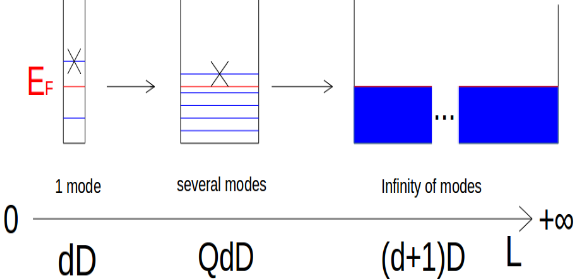}
\caption{Illustration of the dimensional crossover concept in energy-momentum space. Consider a $d$-dimensional gas of noninteracting spinless fermions. In the transverse direction, the Fermi energy is low enough, so that only one mode is selected. When a transverse dimension of the box increases, new modes are available below the Fermi energy. In the limit of an infinite transverse direction, they form a continuum, and the system is $(d\!+\!1)$-dimensional.}
\label{Figcross}
\end{figure}

In this section, I consider $N$ non-interacting spinless fermions of mass $m$ in an anisotropic paralellepipedic box confinement at zero temperature. I assume that the length $L_x$ of the box trap is much larger than its width $L_y$ and height $L_z$, giving it the shape of a beam, and that the gas confined in the latter is uniform. Thanks to recent developments, this seemingly much-idealized situation can be approached experimentally, in an optical box trap \cite{VanEs2010, Gaunt2013, Mukherjee2017, Hueck2017}. If at least one of the transverse sizes is small enough, i.e. such that the energy level spacing is larger than all characteristic energy scales of the problem (given by temperature, or chemical potential), then the gas is confined to 2D or even to 1D, since the occupation of higher transverse modes is suppressed. In the following, I study the behavior of the system as transverse sizes are gradually increased and transverse modes occupied. This yields a dimensional crossover from 1D to 2D, and eventually 3D, whose principle is sketched in Fig.~\ref{Figcross}.

An interesting observable in this context is the dynamical structure factor, already considered in the previous chapter. To begin with, I study the effect of space dimension by direct calculation. In arbitrary dimension $d$ in a box-trap, the dynamical structure factor can be calculated as
\begin{equation}
\label{defS}
S_{d}(\vec{q},\omega)\!=\!V_d\!\int_{-\infty}^{+\infty}\!\!\!\!dt\!\int\! d^dr\ \!e^{i(\omega t-\vec{q}\cdot \vec{r})} \langle \delta n_d(\vec{r},t)\delta n_d(\vec{0},0)\rangle,
\end{equation}
where $V_d$ is the volume of the system, $\hbar \vec{q}$ and $\hbar\omega$ are the transferred momentum and energy in the Bragg spectroscopy process, $\delta n_d(\vec{r},t)\!=\!n_d(\vec{r},t)\!-\!N/V_d$ is the local fluctuation of the density operator around its average value, and $\langle \dots \rangle$ denotes the equilibrium quantum statistical average. Note that I have slightly modified the expression of the drag force compared to the previous chapter, so that its unit does not depend on $d$. Putting the prefactor $V_d$ before would only have lead to heavier notations.

If the gas is probed in the longitudinal direction, i.e. specializing to $\vec{q}\!=\!q\vec{e}_x$, where $\vec{e}_x$ is the unit vector along the $x$-axis, also denoted by $x_1$, the first coordinate in dimension $d$, then
\begin{eqnarray}
\label{dirS}
S_d(q\vec{e}_x,\omega)\!=\!V_d \int \frac{d^dk}{(2\pi)^{d-1}}\Theta\!\left(\!\epsilon_F\!-\!\sum_{i=1}^d\epsilon_{k_{x_i}}\!\right)\!\Theta\!\left(\sum_{i=1}^{d}\epsilon_{k_{x_i}\!+q\delta_{i,1}}\!-\!\epsilon_F\!\right)\!\delta[\omega\!-\!(\omega_{k_{x_1}\!+q}\!-\!\omega_{k_{x_1}}\!)],
\end{eqnarray}
which is more convenient than Eq.~(\ref{defS}) to actually perform the calculations. From Eq.~(\ref{dirS}), I have computed the dynamical structure factor of a $d$-dimensional Fermi gas in the thermodynamic limit, for $d\!=\!1,2,3$, and in particular, I recovered the known result in 1D, given by Eq.~(\ref{DSFTG}), and 3D \cite{Nozieres, Fetter}. As far as the two-dimensional case is concerned, I have given details of calculations in Ref.~\cite{LangHekkingMinguzzi2016}.

Then, looking for a general expression that would depend explicitly on $d$, I have realized that these results can be written in a compact form as
\begin{eqnarray}
\label{Sd}
 &&S_{d}(q \vec{e}_x,\omega)=V_ds_{d}\left(\frac{m}{2\pi\hbar q}\right)^d \nonumber\\
 &&\left[\Theta(\omega_{+}\!-\!\omega)\Theta(\omega\!-\!\omega_{-})(\omega_{+}\!-\!\omega)^{\frac{d\!-\!1}{2}}[\omega\!-\!\mathrm{sign}(q\!-\!2k_{F\!})\omega_{-}]^{\frac{d\!-\!1}{2}}\right.\nonumber\\
 &&+ \Theta(2k_{F}\!-\!q)\Theta(\omega_{-}\!-\!\omega)\left.\left\{[(\omega_{-\!}\!+\!\omega)(\omega_{+}\!-\!\omega)]^{\frac{d\!-\!1}{2}}\!\!\!-\![(\omega_{+}\!+\!\omega)(\omega_{-\!}\!-\!\omega)]^{\frac{d\!-\!1}{2}}\right\}\right]\!\!,
\end{eqnarray}
where
\begin{eqnarray}
k_{F}=\left[\frac{N}{V_d}\frac{(2\pi)^d}{\Omega_d}\right]^{1/d}
\end{eqnarray}
is the modulus of the $d$-dimensional Fermi wavevector of the gas,
\begin{eqnarray}
\Omega_d\!=\!\frac{\pi^{\frac{d}{2}}}{\Gamma\!\left(\frac{d+2}{2}\right)}
\end{eqnarray}
is the volume of the unit $d$-dimensional ball,
\begin{eqnarray}
s_d=\frac{2\pi^{\frac{d+1}{2}}}{\Gamma\!\left(\frac{d+1}{2}\right)}
\end{eqnarray}
is the area of the unit $d$-sphere, and keeping the same notation as in the previous chapter,
\begin{eqnarray}
\omega_{\pm}\!=\!\left|\frac{\hbar q^2}{2m}\!\pm\!\frac{\hbar k_{F}q}{m}\right|.
\end{eqnarray}

Equation (\ref{Sd}) clearly shows that the case $d\!=\!1$ is special, in the sense that the second term vanishes, then. The dynamically-forbidden, low-energy region seen in the previous chapter is one more specificity of a 1D gas. Another interesting aspect of Eq.~(\ref{Sd}) is that it explicitly depends on an integer parameter, that I have denoted by $d$ for obvious reasons. Let us call it $n$, forget about physics for a while and look at (\ref{Sd}) with the eyes of a pure mathematician. When thinking about a strategy to prove that the property $P(n)$, that means '(\ref{Sd}) holds if $d\!=\!n$', is true for any natural integer $n$, the first one that comes to mind is induction on this index. For a mathematician, this ought to be a reflex, but a physicist would argue that there is actually no need for such a proof, since $P(n)$ is already proven for all physically-relevant dimensions, $1, 2$ and $3$.

An interesting issue has been raised, though: a tool would be needed to perform the induction step $P(n)\!\to\!P(n\!+\!1)$, and multimode structures are natural candidates to play this role. The very possibility of performing any of the peculiar induction steps is in itself an appreciable opportunity, as it yields an alternative to direct calculation and, as such, a way to cross-check long and tedious derivations. One can even hope that this step would yield new insights into dimensional crossovers, allowing to revisit dimensional-dependent phenomena.

As an illustration, I consider the dimensional crossover of the dynamical structure factor from 1D to 2D, obtained by populating higher transverse modes of the atomic waveguide in a quasi-1D (Q1D) structure. I write the two-dimensional fermionic field operator as
\begin{eqnarray}
\hat{\psi}(x,y)=\sum_{k_x}\sum_{k_y}\frac{e^{ik_xx}}{\sqrt{L_x}}\frac{e^{ik_yy}}{\sqrt{L}_y}\hat{a}_{k_xk_y},
\end{eqnarray}
where $k_{x,y}\!=\!\frac{2\pi}{L_{x,y}}j_{x,y}$, with $j_{x,y}$ an integer, and $\hat{a}_{k_xk_y}\!=\!\hat{a}_{\vec{k}}$ is the fermionic annihilation operator, such that $\{\hat{a}_{\vec{k}},\hat{a}^{\dagger}_{\vec{k}'}\}\!=\!\delta_{\vec{k},\vec{k}'}$, and $\langle \hat{a}_{\vec{k}}^{\dagger}\hat{a}_{\vec{k}'}\rangle=\delta_{\vec{k},\vec{k'}}n_F(\epsilon_k)$. Then, applying Wick's theorem, I find that
\begin{eqnarray}
\label{Wick}
\langle\delta n(\vec{r},t)\delta n(\vec{0},0)\rangle&=&\!\frac{1}{(L_xL_y)^2}\sum_{\vec{k},\vec{k}'}\!e^{-i[(\vec{k}-\vec{k}')\cdot \vec{r}-(\omega_{k_x}+\omega_{k_y}-\omega_{k_x'}-\omega_{k_y'})t]}\nonumber\\
&&n_F(\epsilon_{k_x}\!\!+\!\epsilon_{k_y})[1\!-\!n_F(\epsilon_{k_x'}\!+\!\epsilon_{k_y'})].
\end{eqnarray}
Substituting Eq.~(\ref{Wick}) into (\ref{defS}), the dynamical structure factor reads
\begin{eqnarray}
 S_{Q1}(q\vec{e}_x,\omega)\!&=&\!\frac{L_x}{L_y}\int_{-\infty}^{+\infty}\!dt\!\int_{-L_x/2}^{L_x/2}\!dx\int_{-L_y/2}^{L_y/2}\!dy e^{i(\omega t-q_x x)}\nonumber\\
 &&\!\!\frac{1}{(2\pi)^2}\int_{-\infty}^{+\infty}\!dk_x\!\!\int_{-\infty}^{+\infty}\!\!dk_x'\!\sum_{k_y,k_y'}\!\!\!e^{-i[(k_x-k_x')x+(k_y-k_y')y-(\omega_{k_x}+\omega_{k_y}-\omega_{k_x'}-\omega_{k_y'})t]}\nonumber\\
 &&n_F(\epsilon_{k_x}\!\!+\!\epsilon_{k_y})[1\!-\!n_F(\epsilon_{k_x'}\!+\!\epsilon_{k_y'})].
 \end{eqnarray}
 A few additional algebraic manipulations and specialization to $T\!=\!0$ yield
 \begin{eqnarray}
 \label{Equa5}
 S_{Q1}(q\vec{e}_x,\omega)\!\!\!&=&\!\!\sum_{k_y}2\pi L_x\!\!\int_{-\infty}^{+\infty}\!dk_x\Theta[\epsilon_F\!-\!(\epsilon_{k_x}\!+\!\epsilon_{k_y})]\Theta[\epsilon_{k_x\!+\!q_x}\!+\!\epsilon_{k_y}\!-\!\epsilon_F]\delta[\omega\!-\!(\omega_{k_x+q_x}\!-\!\omega_{k_x})]\nonumber\\
 \!\!\!&=&\!\!\!\!\!\sum_{j_y=-M}^MS_1(q\vec{e}_x,\omega;\tilde{k}_{F}[j_y/M]),
\end{eqnarray}
where $S_1(q\vec{e}_x,\omega;\tilde{k}_F[j_y/M])$ is the 1D dynamical structure factor where the chemical potential has been replaced by $\epsilon_F-\epsilon_{k_y}$, or equivalently, where the Fermi wavevector $k_{F,1}$ is replaced by 
\begin{eqnarray}
\tilde{k}_{F}[j_y/M]=k_{F}\sqrt{1\!-\!\frac{j_y^2}{\tilde{M}^2}}.
\end{eqnarray}

This defines the number of transverse modes, $2M+1$, through
\begin{eqnarray}
M\!=\!I\left[\frac{k_FL_y}{2\pi}\right]\!=\!I(\tilde{M}),
\end{eqnarray}
where $I$ is the floor function. In the large-$M$ limit, the Riemann sum in Eq.~(\ref{Equa5}) becomes an integral, and then
\begin{eqnarray}
\label{crossdim}
\frac{k_{F}L_y}{\pi}\!\int_0^1\!dx\,S_1\left(q\vec{e_x},\omega;k_F\sqrt{1-x^2}\right)=S_2(q\vec{e}_x,\omega),
\end{eqnarray}
providing the dimensional crossover from 1D to 2D. More generally, one can start from any dimension $d$ and find, after relaxation of a transverse confinement,
\begin{eqnarray}
\label{induc}
&&S_{Qd}(q\vec{e}_x,\omega)=\!\sum_{j=-M}^M\!S_d(q\vec{e}_x,\omega;\tilde{k}_{F}[j/M])_{\stackrel{\longrightarrow}{_{M\to +\infty}}}\nonumber\\
&&\frac{k_{F}L_{d+1}}{\pi}\!\!\int_0^1\!\!dx\,S_d(q\vec{e_x},\omega;\tilde{k}_{F}[x])=S_{d+1}(q\vec{e}_x,\omega).
\end{eqnarray}
If used repeatedly, Eq.~(\ref{induc}) allows to evaluate the dynamical structure factor up to any dimension, and is the key to the induction step.

\begin{figure}
\includegraphics[width=9cm, keepaspectratio, angle=0]{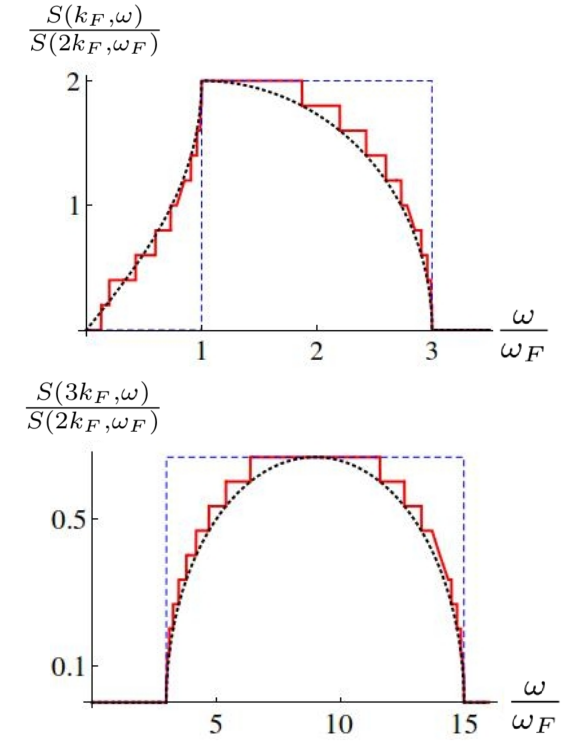}
\caption{Dynamical structure factor of a noninteracting Fermi gas $S(q,\omega)$, in units of $S(q\!=\!2k_F,\omega\!=\!\omega_F)$, for dimensionless wavevectors $q/k_F\!=\!1$ (upper panel) and $q/k_F\!=\!3$ (lower panel), as a function of frequency $\omega$ in units of the Fermi frequency, in 1D (dashed, blue) and Q1D for $2M\!+\!1\!=\!21$ modes (solid, red) compared to 2D (dotted, black). Few modes are needed for the Q1D system to display a similar behavior as the 2D one, such as the shark fin shape in the upper panel.}
\label{Sq13}
\end{figure}

I illustrate numerically the dimensional crossover from 1D to 2D using Eq.~(\ref{induc}). Figure~\ref{Sq13} shows the dynamical structure factor as a function of the frequency $\omega$ for two choices of wavevector $q$. Sections are made at fixed $q$, rather than $\omega$, because such curves are accessible to experiments. In each panel, this observable is given for a 1D gas, a Q1D gas where $M\!=\!10$ and 2D for a comparison. Notice that only a few modes are needed to recover higher-dimensional physics within a very good approximation, since in this example, the staircase shape taken by the dynamical structure factor of the Q1D gas mimics already quite well the 2D one.

I proceed, following the spirit of chapter \ref{secIV} where this analysis was done for a weakly-interacting Bose gas, by studying the effect of dimension on the drag force. I recall that if a weak potential barrier or impurity is stirred along the fluid, putting it slightly out of equilibrium, then in linear response theory the average energy dissipation per unit time is linked to the dynamical structure factor by the relation
\begin{eqnarray}
&&\langle\overline{\dot{E}}\rangle\!=\!-\frac{1}{2\pi\hbar V_d}\!\int_0^{+\infty}\!\!\!\!d\omega\!\! \int \!\!\frac{d^dq}{(2\pi)^d} S_{d}(\vec{q},\omega)|U_{d}(\vec{q},\omega)|^2\omega,
\end{eqnarray}
where $U_{d}(\vec{q},\omega)$ is the Fourier transform of the potential barrier $U_d(\vec{r},t)$ defining the perturbation part of the Hamiltonian $H_{pert}\!=\!\int d^dr\,U_d(\vec{r},t)n_d(\vec{r})$.

\begin{figure}
\includegraphics[width=8cm, keepaspectratio, angle=0]{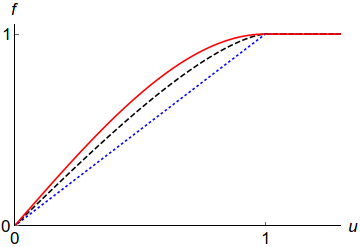}
\caption{Dimensionless drag force due to an infinitely thin potential barrier, $f\!=\!F/F(v_F)$, as a function of the dimensionless flow velocity $u\!=\!v/v_F$, in dimensions 1 (dotted, blue), 2 (dashed, black) and 3 (solid, red). All of them experience a saturation at supersonic velocity flows.}
\label{F123}
\end{figure}

With a delta-potential barrier $U_d(\vec{r},t)\!=\!U_d \delta(x\!-\!vt)$ in the direction $x$, covering the whole waveguide in the transverse directions, the drag force reads
\begin{eqnarray}
\label{explF}
 F_{d}(v)=\frac{U_d^2}{2\pi\hbar V_d}\int_0^{+\infty}\!\!\!dq\,qS_{d}(q\vec{e_x},qv)
\end{eqnarray}
in arbitrary dimension. Using the general notation $v_{F,d}\!=\!\frac{\hbar k_{F,d}}{m}$ to denote the Fermi velocity in dimension $d$, from Eqs.~(\ref{explF}) and (\ref{Sd}), I find that for $v\leq v_{F,d}$,
\begin{equation}
\label{F1}
F_{1}(v)\!=\!\frac{2U_1^2mn_1}{\hbar^2}\frac{v}{v_{F,1}},
\end{equation}
\begin{eqnarray}
\label{F2}
\!\!\!\!\!\!\!\!F_{2}(v)\!=\!\frac{2U_2^2mn_{2}}{\hbar^2}\frac{2}{\pi}\!\left[\frac{v}{v_{F,2}}\sqrt{1\!-\!\left(\!\frac{v}{v_{F,2}}\!\right)^2}+\arcsin\!\left(\frac{v}{v_{F,2}}\right)\!\right]
\end{eqnarray}
and
\begin{eqnarray}
\label{F3}
F_{3}(v)=\frac{2U_3^2mn_{3}}{\hbar^2}\frac{3}{2}\frac{v}{v_{F,3}}\left[1-\frac{1}{3}\left(\frac{v}{v_{F,3}}\right)^2\right].
\end{eqnarray}
If $v\!>\!v_{F,d}$, for the potential barrier considered, the drag force saturates and takes a universal value
\begin{eqnarray}
F_{d}(v_{F,d})=\frac{2U_d^2mn_{d}}{\hbar^2}.
\end{eqnarray}
Then, I have recovered these results by applying the cross-dimensional approach from dimension $d$ to dimension $(d\!+\!1)$, which validates this technique once more. The drag force profiles are plotted simultaneously in Fig.~\ref{F123}. The effect of dimension on the drag force is less impressive than on the dynamical structure factor.

From Equations (\ref{F1}), (\ref{F2}) and (\ref{F3}), it is difficult to guess a general formula, valid for any integer dimension $d$. This is in stark contrast to the dynamical structure factor, where a close inspection was sufficient to infer Eq.~(\ref{Sd}). In particular, dimension two looks fairly weird, as it involves an arcsin function. In absence of any intuition, I carried out the calculation from Eqs.~(\ref{Sd}) and (\ref{explF}), and found that the general expression actually reads
\begin{eqnarray}
\label{Fdd}
F_d(u_d\leq 1)\!\!\!&=&\!\!\!\frac{2U_d^2mn_d}{\hbar^2}\frac{2}{\sqrt{\pi}(d+1)}\frac{\Gamma(\frac{d+2}{2})}{\Gamma(\frac{d+1}{2})}(1-u_d^2)^{\frac{d-1}{2}}\nonumber\\
&&\!\!\!\left[(1\!+\!u_d)\,{_2}F_1\!\!\left(\!1,\!\frac{1\!-\!d}{2};\!\frac{d\!+\!3}{2};\!-\frac{1\!+\!u_d}{1\!-\!u_d}\right)\!-\!(u_d\!\rightarrow\!-u_d)\right]\!\!,
\end{eqnarray}
where ${_2}F_1$ is the hypergeometric function, and I used the notation $u_d\!=\!v/v_{F,d}$. Actually, this expression can be simplified in even and odd dimensions separately, but the final expression remains rather heavy all the same, see Ref.~\cite{LangHekkingMinguzzi2016} for details. According to the drag force criterion, the non-interacting Fermi gas is not superfluid, as expected since superfluidity is a collective phenomenon.

\section{Dimensional crossovers in a harmonic trap}
\label{trap}
After discussing the dimensional crossover in energy space in a box trap, I focus here on dimensional crossovers in the experimentally relevant case of a harmonically trapped gas. To begin with, I consider a 1D Fermi gas, longitudinally confined by a harmonic trap described by the potential $V(x)\!=\!\frac{1}{2}m\omega_{0}^2x^2$, where $\omega_{0}$ is the frequency of the trap. Assuming a slow spatial variation of the density profile of the gas along $x$, the local-density approximation accurately describes the density profile of the gas, as shown in chapter \ref{secIII}.

Within the same approximation of a slowly-varying spatial confinement, for wavevectors $q$ larger than the inverse size of the spatial confinement $1/R_{TF}$, where $R_{TF}$ is the Thomas-Fermi radius, the dynamical structure factor $S_{1,h.o.}(q,\omega)$ of the harmonically trapped gas is given by the spatial average
\begin{eqnarray}
\label{SLDA}
S_{1,h.o.}(q,\omega)\!\!\!\!\!\!\!&&=\!\frac{1}{2R_{TF}}\!\!\int_{-R_{TF}}^{R_{TF}}\!\!dx\,S_{1,hom}[q,\omega;n_1(x)]\nonumber\\
&&=\!\int_0^1dz\,S_{1,hom}(q,\omega;n_1\sqrt{1-z^2})
\end{eqnarray}
after the change of variable $z\!=\!x/R_{TF}$, where $S_{1,hom}[q,\omega;n]$ is the dynamical structure factor of a 1D homogeneous gas, and the linear density $n_1\!=\!\frac{2N}{\pi R_{TF}}$. In other words, the local-density approximation assumes that portions of the size of the confinement length scale $a_{h.o.}\!=\!\sqrt{\frac{\hbar}{m\omega_0}}$ can be considered as homogeneous, and that their responses are independent from each other \cite{Golovach2009}. The validity of this approximation for the dynamical structure factor has been verified in \cite{VignoloMinguzzi2001}, by comparison with the exact result.

\begin{figure}
\includegraphics[width=9cm, keepaspectratio, angle=0]{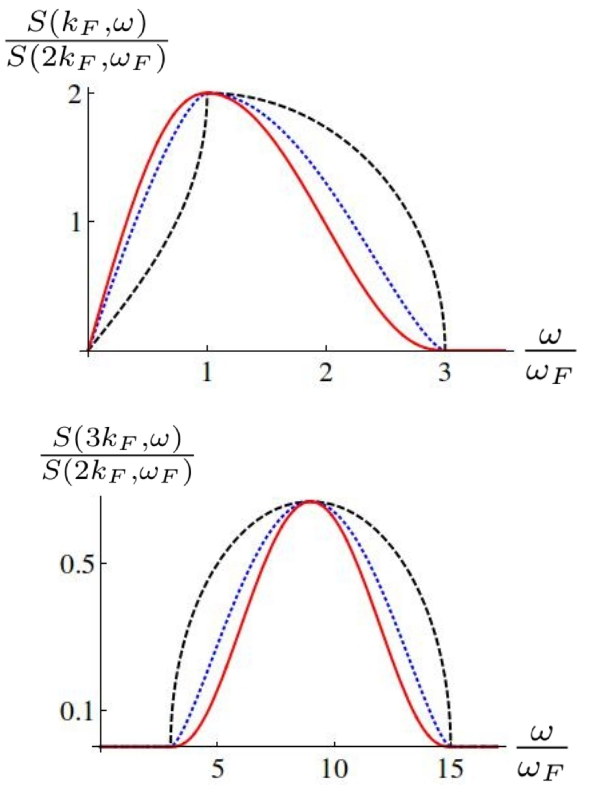}
\caption{Dynamical structure factor of a noninteracting Fermi gas, $S(q,\omega)$, in units of $S(q\!=\!2k_F,\omega\!=\!\omega_F)$, for dimensionless wavevectors $q/k_F\!=\!1$ (upper panel) and $q/k_F\!=\!3$ (lower panel), as a function of frequency $\omega$ in units of the Fermi frequency $\omega_F$, in 2D (dashed, black), 4D (dotted, blue) and 6D (solid, red), as obtained from a harmonically trapped gas, in a quantum simulator perspective}
\label{S36D}
\end{figure}

Interestingly, equation (\ref{SLDA}) has the same structure as Eq.~(\ref{crossdim}), thus establishing the equivalence, for the dynamical structure factor, of a 1D harmonic trapped gas and a 2D gas in a box. More generally, a similar procedure yields the following property: \textit{In reduced units, the dynamical structure factor of a harmonically trapped ideal gas in $d$ dimensions as predicted by the LDA is the same as in a box trap in $2d$ dimensions.} In Figure~\ref{S36D}, I illustrate the latter on the dynamical structure factor of an ideal Fermi gas in a box in dimensions $d\!=\!2,4,6$, as can be simulated by a harmonically-confined gas in dimension $d\!=\!1,2,3$ respectively.

This correspondence between a 2dD box trap and a dD harmonic trap can be inferred directly from the Hamiltonian of the system: for a box trap there are d quadratic contributions stemming from the kinetic energy, whereas for a harmonic confinement there are 2d quadratic terms originating from both kinetic and potential energy. Since, in a semiclassical treatment, each term contributes in a similar manner, harmonic confinement leads to a doubling of the effective dimensionality of the system in this noninteracting case. This is expected not only for the dynamical structure factor, but is also witnessed in other quantities such as the density of states, the condensate fraction of a Bose gas below the critical temperature, and virial coefficients \cite{Bahauddin} for instance. 

The relevance of the dimensional crossover as a tool to prove dimensional-dependent properties by induction should be revised in view of this correspondence. In its light, the induction tool suddenly becomes far more interesting, in particular in the context of a 3D harmonically-trapped gas, as it corresponds to a uniform gas in 6D, whose dynamical structure factor is difficult to evaluate by direct calculation.

Actually, harmonic trapping in a longitudinal dimension does not necessarily increase the effective dimension of the system. To illustrate this point, I analyze the dynamical structure factor of a harmonically-confined gas in the experimentally relevant case where only the central part of the cloud is probed, over a radius $r\!<\!R_{TF}$. Assuming that $r$ is larger than the characteristic variation length of the external confinement, and using again the local density approximation, Eq.~(\ref{SLDA}) transforms into
\begin{eqnarray}
 S_{1,h.o.}(q,\omega;r)=\int_{0}^{r/R_{TF}}\!\!\!dx\,S_{1}\left(q,\omega;n_1\sqrt{1-x^2}\right).
\end{eqnarray}
An explicit expression is obtained by evaluating the integral
\begin{eqnarray}
S\!=\!\int_0^{r/R_{TF}}\!\!dx\,\Theta\left(q^2\!+\!2q \sqrt{1\!-\!x^2}\!-\!\omega\right)\Theta\left(\omega\!-\!|q^2\!-\!2q \sqrt{1\!-\!x^2}|\right),
\end{eqnarray}
where $\omega$ and $q$ are expressed in reduced units such that $k_F\!=\!1$ and $\omega_F\!=\!1$. The final expression reads
\begin{eqnarray}
\label{I}
 &&S=\Theta(\omega_+\!-\!\omega)\Theta(\omega\!-\!\omega_-)\min\!\left[\!\frac{r}{R_{TF}},\sqrt{1-\left(\frac{\omega-q^2}{2q}\right)^2}\right]\nonumber\\
 &&+\Theta(2\!-\!q)\Theta(\omega_-\!-\!\omega)\min\!\left[\!\frac{r}{R_{TF}},\sqrt{1-\left(\frac{\omega-q^2}{2q}\right)^2}\right]\nonumber\\
 &&-\Theta(2\!-\!q)\Theta(\omega_-\!-\!\omega)\min\!\left[\!\frac{r}{R_{TF}},\sqrt{1-\left(\frac{\omega+q^2}{2q}\right)^2}\right]\!\!.
\end{eqnarray}

Equation (\ref{I}) displays another kind of crossover between the dynamical structure factor of a 1D gas in a box and the one of a 2D gas in a box. In order to obtain the 1D behavior, $r/R_{TF}$ must be the minimal argument in Eq.~(\ref{I}) above, while a 2D behavior is obtained when it is the maximal one.

\begin{figure}
\includegraphics[width=8.5cm, keepaspectratio, angle=0]{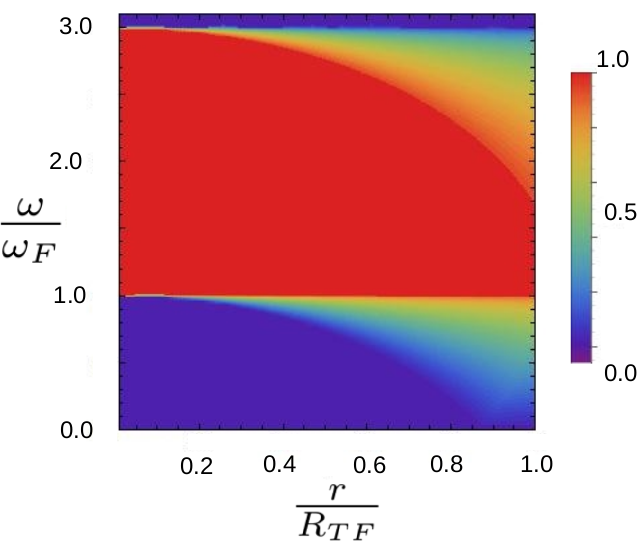}
\caption{Reduced dynamical structure factor $S(q\!=\!k_F,\omega;r)/r$ in units of $S_1(q\!=\!k_F,\omega)$ in the plane $(r,\omega)$, where $r$ is the probed length of the gas in units of the Thomas-Fermi radius $R_{TF}$, and $\omega$ the energy in units of $\omega_F$. If $r\!\ll\!R_{TF}$, the 1D box result is recovered, while $r\to R_{TF}$ yields the 2D box result. Excitations below the lower excitation branch $\omega_-$ appear progressively as the dimensionless ratio $r/R_{TF}$ is increased.}
\label{r}
\end{figure}

Figure \ref{r} shows the section of the dynamical structure factor $S_{1,h.o.}(q\!=\!k_F,\omega;r)$, as a function of energy, at varying the size $r$ of the probed region, at $q\!=\!k_F$. In essence, to get close to 1D behavior in spite of the longitudinal trapping potential, one should take the smallest $r$ compatible with the condition $r\!\gtrsim\!1/q$, that ensures the validity of the LDA, and with $1\!-\!r/R_{TF}\!\ll\!1$ in order to detect enough signal.

After this study of the effect of a trap, that was inspired by experiments, I investigate another point, motivated by theoretical issues, anticipating a possible generalization to interacting systems. In 1D, I have widely used the Tomonaga-Luttinger liquid approach to tackle the dynamical correlations, in chapters \ref{secII} and \ref{secIV}. Here, I proceed to consider the noninterating Fermi gas as a testbed to develop a generalized Tomonaga-Luttinger liquid framework in higher dimensions.

\section{Low-energy approach for fermions in a box trap, cross-dimensional Luttinger liquid}
\label{Luttinger}

\begin{figure}
\includegraphics[width=9cm, keepaspectratio, angle=0]{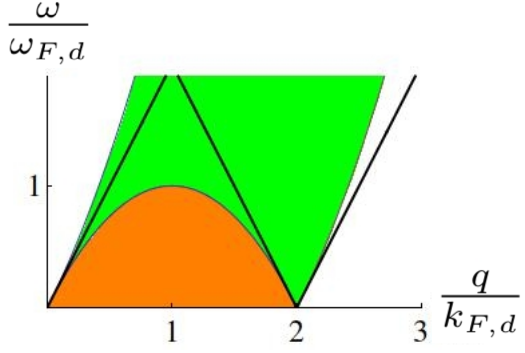}
\caption{Definition domain of the dynamical structure factor of a Fermi gas in the plane $(q,\omega)$ in units of $(q_{F,d},\omega_{F,d})$. Colored areas represent the domain where single particle-hole excitations can occur. The light green one is found in any integer dimension $d\!\in\!\{1,2,3\}$, while the dark orange one is specific to $d\!>\!1$. Black straight lines correspond to linearization of the lower excitation spectrum in the Tomonaga-Luttinger formalism in 1D.}
\label{SdD}
\end{figure}

In chapter \ref{secII}, I have pointed out that the Tomonaga-Luttinger liquid approach breaks down when $d\!>\!1$. Explanations of this fact often rely of fairly non-trivial arguments. However, I shall show that the dynamical structure factor provides a quite simple and pictural illustration and explanation thereof. As can be seen in Fig.~\ref{SdD} and in Eq.~(\ref{Sd}), in 2D and 3D, since excitations are possible at energies lower than $\omega_{-}$ and down to $\omega\!=\!0$ for any $q\!<\!2k_F$, no linearization of the lower branch of the excitation spectrum is possible. This, in turn, can be interpreted as a dramatic manifestation of the standard Tomonaga-Luttinger liquid theory breakdown.

Many attempts have been made to generalize the Tomonaga-Luttinger model to higher dimensions, as an alternative to Fermi liquids to describe interacting systems. An intermediate issue is whether or not the TLL applies to Q1D systems. As an answer to both questions at once, I have tried and constructed a Tomonaga-Luttinger model in higher dimension, defining a multimode Tomonaga-Luttinger model (M-TLM). Indeed, if $d\!>\!1$, the emergence of contributions to the dynamical structure factor at energies lower than $\omega_-$ can be interpreted as contributions of transverse modes of a 1D gas.

\begin{figure}
\includegraphics[width=9cm, keepaspectratio, angle=0]{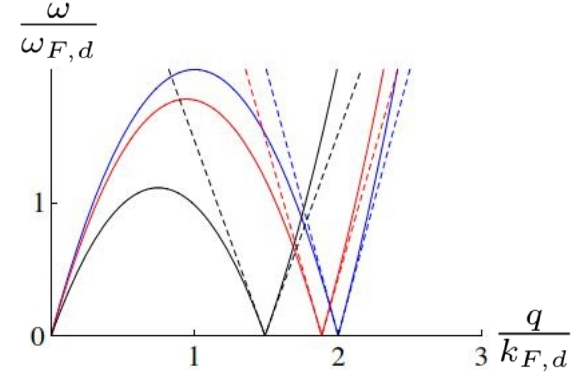}
\caption{Lower boundary of the definition domain of the dynamical structure factor for a Q1D gas with three modes, in the plane $(q,\omega)$  in units of $(k_F,\omega_F)$, as found in the Tomonaga-Luttinger formalism (dashed), compared to the exact solution (solid)}
\label{multilin}
\end{figure}

As an appetizer, note that all these modes, taken separately, display a linear structure in their excitation spectrum at low energy, as illustrated in Fig.\ref{multilin}. This means that each mode, taken separately, can be described by a Tomonaga-Luttinger liquid. Thus, applying Eq.~(\ref{crossdim}) to the Tomonaga-Luttinger model, in Q1D the dynamical structure factor reads
\begin{eqnarray}
\label{multiLutt}
S^{TL}_{Q1}(q,\omega)\!=\!\sum_{j=-M}^MS_{1}^{TL}(q,\omega;\tilde{k}_F[j/M]).
\end{eqnarray}

The question is, up to what point the small errors for each mode in the framework of the effective theory amplify or cancel when adding more modes, especially in the limit $M\!\to\!+\!\infty$, that corresponds to the crossover to 2D. To address this question, I have carried out the procedure explicitly on the example of the 1D to 2D crossover and compared the prediction of the cross-dimensional Tomonaga-Luttinger liquid theory to the exact solution. Combining Eq.~(\ref{multiLutt}) to the dynamical structure factor of a 1D gas, I have found
\begin{eqnarray}
\!\!\!\!\!\!\!\!\!\!\!&&S^{TL}_{Q1}(q,\omega)=L_x\frac{m}{4\pi\hbar}\frac{1}{\tilde{M}}\sum_{j=-M}^M\frac{1}{\sqrt{1\!-\!\frac{j^2}{\tilde{M}^2}}}\Theta\!\left(\omega\!-\!\left|q\!-\!2k_F\sqrt{1\!-\!\frac{j^2}{\tilde{M}^2}}\right|\!v_F\sqrt{1\!-\!\frac{j^2}{\tilde{M}^2}}\right)\nonumber\\
\!\!\!\!\!\!\!\!\!\!\!&&\to_{M\to +\infty}\!L_x\frac{m}{2\pi\hbar}\int_0^1\!dx\,\frac{1}{\sqrt{1\!-\!x^2}}\Theta\!\left(\omega\!-\!\left|q\!-\!2k_F\sqrt{1\!-\!x^2}\right|\!v_F\sqrt{1\!-\!x^2}\right)\!=\!S_2^{TL}(q,\omega).
\end{eqnarray}
Evaluating the integral yields
\begin{eqnarray}
\!\!\!\!\!\!\!\!\!S^{TL}_2(q,\omega)\!=\!\frac{mL_x}{2\pi\hbar}\!\left[\Theta(q\!-\!2k_F)S_>(q,\omega)\!+\!\Theta(2k_F\!-\!q)S_<(q,\omega)\right]\!
\end{eqnarray}
with
\begin{eqnarray}
\label{Sup}
S_>(q,\omega)\!\!&=&\!\!\!\Theta\left(\frac{\hbar q^2}{8m}-\omega\right)\Theta(\tilde{q}v_F-\omega)\arcsin\left[\frac{q}{4k_F}\left(1-\sqrt{1-\frac{8m\omega}{\hbar q^2}}\right)\right]\nonumber\\
\!\!&+&\!\!\!\Theta(4k_F-q)\Theta(\omega-\tilde{q}v_F)\Theta\left(\frac{\hbar q^2}{8m}-\omega\right)\arcsin\!\left[\!\frac{q}{4k_F}\left(1-\sqrt{1-\frac{8m\omega}{\hbar q^2}}\right)\right]\!\nonumber\\
\!\!&+&\!\!\!\arccos\!\left[\frac{q}{4k_F}\!\!\left(1\!+\!\!\sqrt{1\!-\!\frac{8m\omega}{\hbar q^2}}\right)\right]+\Theta\left(\omega-\frac{\hbar q^2}{8m}\right)\Theta(\omega-\tilde{q}v_F)\frac{\pi}{2}
\end{eqnarray}
and 
\begin{eqnarray}
\label{Sinf}
S_<(q,\omega)\!\!&=&\!\!\!\Theta(|\tilde{q}|v_F-\omega)\arcsin\left[\frac{q}{4k_F}\left(1+\sqrt{1+\frac{8m\omega}{\hbar q^2}}\right)\right]\nonumber\\
\!\!&+&\!\!\!\Theta\left(\frac{\hbar q^2}{8m}-\omega\right)\arcsin\!\left[\!\frac{q}{4k_F}\!\!\left(\!1\!-\!\sqrt{1\!-\!\frac{8m\omega}{\hbar q^2}}\!\right)\!\right]\!\!-\!\arcsin\!\left[\frac{q}{4k_F}\!\!\left(1\!+\!\sqrt{1\!-\!\frac{8m\omega}{\hbar q^2}}\right)\!\right]\nonumber\\
\!\!&+&\!\!\!\Theta(\omega-|\tilde{q}|v_F)\frac{\pi}{2},
\end{eqnarray}
where $\tilde{q}\!=\!q\!-\!2k_F$.

\begin{figure}
\includegraphics[width=8.5cm, keepaspectratio, angle=0]{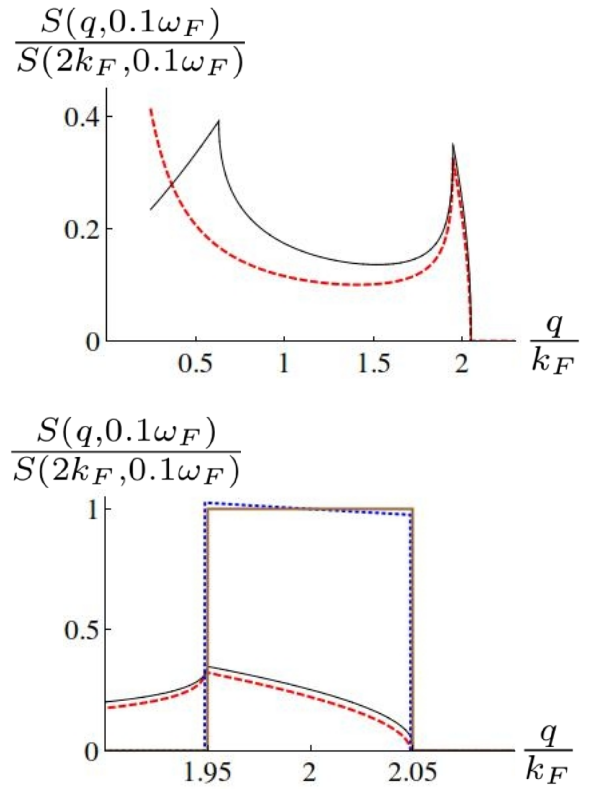}
\caption{Section of the dynamical structure factor $S(q,\omega\!=\!0.1\,\omega_F)$ in units of $S(q\!=\!2k_F,\omega\!=\!0.1\,\omega_F)$ as a function of $q$ in units of $k_F$, at fixed energy $\omega\!=\!0.1\,\omega_F$. The exact result in 2D (dashed, red) is compared to the M-TLM prediction (solid, black) in the upper panel. The lower panel shows a zoom into the backscattering region near $q\!=\!2k_F$. It compares the 2D exact (dashed, red) and the M-TLM model (solid, black) to the exact (thick, brown) and TLM (dotted, blue) results in 1D.}
\label{Vs4}
\end{figure}

I illustrate Eqs.~(\ref{Sup}) and (\ref{Sinf}) in Fig.\ref{Vs4}, that compares sections of the dynamical structure factor as predicted by the multi-mode Tomonaga-Luttinger model and by exact calculation, as a function of $q$ at $\omega\!=\!0.1\,\omega_F$. This low-energy value has been chosen in view of the known validity range of the TLL already studied in 1D. Around the umklapp point $(q\!=\!2k_F,\omega\!=\!0)$ lies a sector where the effective model is in a rather good quantitative agreement with the exact result in 2D. Discrepencies between the Tomonaga-Luttinger model and the exact solution of the original one at low $q$ are due to the fact that for a given point, the TLM slightly overestimates the value of the dynamical structure factor for larger $q$ and underestimates it at lower $q$, as can be seen in the 1D case. Combined with the fact that the curvature of the dispersion relation is neglected, and that the density of modes is lower at low $q$, this explains the anomalous cusp predicted by the M-TLM at low $q$. Note however that this result is by far closer to the 2D exact result than the 1D one in the large $M$ case, showing that there is a true multi-mode effect.

I find that the limiting prediction of the multimode Tomonaga-Luttinger model for a noninteracting gas is in quantitative agreement with the exact 2D result for $\omega\!\ll\!\omega_F$ and $|q\!-\!2k_F|\!\ll\!2k_F$. Similar conditions have to be met in 1D in order to ensure the validity of the Tomonaga-Luttinger model, therefore my heuristic construction is quite satisfactory from this point of view. It is not fully satisfying, however, in the sense that one needs to start from 1D, and the 2D model that would directly yield this result is unknown.

\section{Summary of this chapter/r\'esum\'e du chapitre}
In this chapter, I have investigated the dynamical structure factor and drag force of a noninteracting Fermi gas as functions of the dimension of space. It turns out that dimension has a dramatic effect on the dynamical structure factor, whose strongest manifestation is a low-energy forbidden region in momentum-energy space, that becomes completely filled in higher dimensions. This effect on the dynamical structure factor allows to forecast, by adiabatic continuation, the transition from Luttinger to Fermi liquid behavior in interacting systems. In comparison, the effect on the drag force is not so huge, as it is dominated by excitations close to the umklapp point. 

Then, I have investigated multi-mode systems obtained by releasing a transverse trapping and demonstrated the dimensional crossover allowed by this structure. Actually, the mathematical property hidden behing dimensional crossover is as simple as the crossover from Riemann sums to integrals.  I have studied the effect of a longitudinal harmonic trap and shown, within the local-density approximation, that each degree of trapping is equivalent, for the noninteracting gas, to an additional effective dimension. This property allows to simulate up to six dimensions. The multimode structure allows to prove general results by induction on space dimension, which is quite useful in this context. I have also shown that dimension enhancement does not occur if the dynamical structure factor of a 1D trap in a longitudinally trapped gas is probed only close to the center of the trap.

To finish with, I have turned to the issue of extensions of the TLL formalism to higher dimensions, in view of future applications to interacting systems. I have proposed a model of multimode Tomonaga-Luttinger liquid, whose dimensional crossover to 2D reproduces the exact result at sufficiently low energies with satisying accuracy, close to the umklapp point.
\\
\\
\\
\\
\\
Dans ce chapitre, j'ai \'etudi\'e le facteur de structure dynamique et la force de tra\^in\'ee en fonction de la dimension du syst\`eme, dans le cas d'un gaz de fermions sans interaction. Il s'av\`ere que la dimension a un effet important sur le facteur de structure dynamique, qui s'annule dans une zone de basse \'energie uniquement en dimension un. La possibilit\'e d'observer des excitations dans cette derni\`ere en dimension sup\'erieure peut s'interpr\'eter, par extrapolation adiabatique \`a un gaz pr\'esentant des interactions, comme une manifestation de la transition entre un liquide de Luttinger et un liquide de Fermi.

Afin de mieux comprendre la transition dimensionnelle, j'ai \'etudi\'e l'apparition d'une structure multimode par relaxation d'un degr\'e de confinement transverse. La dimension du syst\`eme augmente alors progressivement, et on peut observer comment la zone dynamiquement interdite en dimension un s'emplit progressivement du fait de l'augmentation du nombre de modes, jusqu'\`a \^etre compl\`etement combl\'ee en dimension deux, qui correspond \`a un nombre infini de modes. J'ai aussi mis en \'evidence une autre mani\`ere d'augmenter la dimension d'un syst\`eme, cette fois de mani\`ere effective, par confinement harmonique selon une direction longitudinale. Chaque degr\'e de libert\'e entrav\'e augmente la dimension effective d'une unit\'e d'apr\`es l'approximation de la densit\'e locale, ce qui permet de simuler un gaz de dimension six. Toutefois, cette augmentation de dimension n'a pas lieu si on sonde uniquement la r\'egion centrale du pi\`ege, o\`u le gaz est relativement uniforme. Pour finir, je me suis appuy\'e sur ce m\^eme exemple d'un gaz sans interaction pour d\'evelopper un formalisme de liquide de Tomonaga-Luttinger multimode, dont j'ai test\'e la validit\'e jusqu'au passage \`a la dimension deux, o\`u ses pr\'edictions restent correctes \`a basse \'energie au voisinage du point de r\'etrodiffusion.

\section{Outlook of this chapter}

A few issues dealt with in this chapter could be investigated further as was done in chapter \ref{secIV} for a Bose gas, such as the effect of finite temperature on the dimensional structure factor of a Fermi gas in dimensions two and three. It is not obvious how dimensional crossovers would manifest themselves at finite temperature, and the issue deserves attention. The effect of the barrier width on the drag force profile is also unknown yet in higher dimensions, but I expect that a quasi-superfluid regime exists in this configuration too. The multimode structure leading to dimensional crossovers may be investigated for virtually any observable and for other sufficiently simple systems, offering a wide landscape of perspectives.

The most thriving issue, however, is by far the adaptation of the multicomponent approach to interacting systems. Some results, such as the $d\leftrightarrow 2d$ correspondence in a harmonic trap, are likely not to be robust. The multimode Tomonaga-Luttinger liquid formalism, however, can definitely be adapted to multicomponent interacting system, by choosing the type of interactions considered. There are essentially two types of terms that emerge \cite{CazalillaHo2003}, leading respectively to density-coupled gases, or to couplings of a cosine type, that correspond to the Sine-Gordon model. They should be a low-dimensional description of multicomponent Lieb-Liniger or Yang-Gaudin type gases.

The simpler case is assuredly the density-coupled multicomponent gas, whose Hamiltonian,
\begin{eqnarray}
\label{HLLcoupled}
\!\!\!\!H^{TL}_M\!=\!\sum_{i=0}^{M\!-\!1} \frac{\hbar v_i}{2\pi}\!\int_0^L\!\!dx\!\left[K_i(\partial_x\Phi_i)^2\!+\!\frac{1}{K_i}(\partial_x\theta_i)^2\right]\!+\!\frac{1}{2}\sum_{i=0}^{M\!-\!1}\sum_{j=0}^{M\!-\!1} (1\!-\!\delta_{ij})g_{ij}\!\int_0^L\!dx\, \frac{\partial_x\theta_i}{\pi}\frac{\partial_x\theta_j}{\pi},
\end{eqnarray}
is quadratic in the fields and can be diagonalized explicitly \cite{MatveevGlazman1993}. This case will be the subject of a later publication. A few results concerning dynamical correlations are even already available for a few components \cite{Iucci2007, Orignac2011}, and a general formalism based on generalized hypergeometric series has been developped to deal with an arbitrary number of components \cite{OrignacCitro2012}.

Next step would be the extension of the Imambekov-Glazman formalism to multimode gases, and the development of Bethe Ansatz techniques to the dynamics of multi-component Lieb-Liniger and Yang-Gaudin models.

In view of the technical difficulty of the dimensional crossover problem even for a few modes, it might be that quantum simulation will be needed to solve it. However, if only a few modes are needed to recover higher-dimensional physics as was the case for a noninteracting Fermi gas, then there is hope that some cases are at reach. Whether or not such solutions could help to better understand or solve higher-dimensional models is not obvious, nor the way a model transforms along the dimensional crossover. As an example of this problematics, the Tonks-Girardeau gas is equivalent to a gas of noninteracting fermions for a few observables in 1D, but no such correspondence is known in higher dimension. To what conditions would an indifferentiate 1D gas become a noninteracting Fermi gas or a unitary Bose gas in 2D is far from obvious.

\newpage

\chapter{General conclusion/le mot de la fin}

In conclusion, in this thesis, I have studied the effects of interactions, quantum and thermal fluctuations on a one-dimensional Bose gas.

In the introductory chapter \ref{secII}, I have recaped a few known hallmarks of one-dimensional quantum systems, such as collectivization of motion and excitations, that prevents the existence of well-defined quasi-particles and seals the breakdown of Fermi liquid theory. Fermionization of interacting bosons manifests itself through the appearance of a Fermi sea structure in quasi-momentum space, and in real space, through a fictitious Pauli principle that is not due to statistics but to interactions. For systems with spin, the charge and spin sectors of the Hilbert space decouple, and their excitations split in real space too, challenging the notion of elementary particle. All of these effects are consequences of the crossing topological constraint, that enhances the role of fluctuations.
Another striking consequence of dimensional reduction is the Mermin-Wagner theorem, that states the impossibility of spontaneous symmetry breaking in many models. The latter do not undergo phase transitions but rather smooth crossovers, withdrawing interest to their phase diagramms. An alternative paradigm consists in characterizing systems through their correlation functions, either local or non-local, in real or in energy-momentum space, at or out of equilibrium. These correlation functions are probed on a daily basis in ultracold atom setups.

Low-dimensional quantum gases are obtained in experiments by strong confinement along transverse directions, allowed by trapping and cooling. They can be created in an optical lattice, leading to an ensemble of wires, or on a microchip that provides a single gas, both situations corresponding to open boundary conditions. Ring geometries, that realize periodic boundary conditions, are also available thanks to magnetic trapping, radio-frequency fields or a combination, using time-average adiabatic potentials.

A fair number of simple models describing these gases are integrable, meaning that their scattering matrix verifies the Yang-Baxter equation. This situation is more likely to appear in low dimension, and allows to obtain the exact ground-state energy, the excitation spectrum and even, at the price of huge efforts, correlation functions, using Bethe Ansatz techniques. Other theoretical tools allow for a quite detailed analytical study of low-dimensional models, such as the exact Bose-Fermi mapping, that states the formal equivalence, for many observables, between the Tonks-Girardeau gas of strongly-interacting bosons, and a fictitious gas of noninteracting spinless fermions. Many models belong to the universality class of Tomonaga-Luttinger liquids, which is completely solved by bosonization, and yields the structure of the short-time, large-distance, and low-energy correlation functions of these models. These correlations are critical at $T\!=\!0$ in the thermodynamic limit, as they decay algebraicly in space, which is one more hallmark of one-dimensional physics. At finite temperature, their decay becomes exponential. Conformal field theory can be used as an alternative formalism to obtain these correlation functions, and requires less calculation efforts. The validity range of both approaches is investigated by comparison with exact results in the Tonks-Girardeau regime, or from Bethe Ansatz when they are available.

In chapter \ref{secIII}, I have studied one of the most famous models of 1D gases, where bosonic atoms interact through a local potential, a.k.a. the Lieb-Liniger model. I have recalled the Bethe Ansatz procedure to obtain its ground-state energy in closed form in the thermodynamic limit, as the solution of a set of coupled integral equations. Approximate solutions of these equations can be constructed systematically with arbitrary accuracy in the weakly-interacting and strongly-interacting regimes, by identification of the corresponding series expansions. In the weak-coupling regime, I have identified the general pattern of this series, and guessed the exact value of the third-order coefficient. In the strongly-interacting regime, I have pushed the expansion to an unprecedented order and inferred a partially-resummed structure. I have also developed a semi-analytical technique that works in all interaction regimes. In the end, these methods give access to the whole range of interaction strengths with excellent accuracy.

Then, I have turned to the more intricate issue of local correlation functions. The one-body local correlation is trivial, the second- and third-order ones can be expressed in terms of moments of the density of pseudo-momenta, already studied to obtain the energy. I have found new expressions in the weak-coupling regime and conjectured the global structure in the strongly-interacting regime, improving analytical estimates.

The one-body correlation function acquires a non-trivial structure at finite space separation. Tomonaga-Luttinger liquid theory predicts that it vanishes at infinity, meaning the absence of off-diagonal long-range order in 1D, but its decay is algebraic, which is a signature of quasi-long-range order. While the large-distance behavior is universal, at short distance it depends on the microscopic details of the model. Here, the coefficients of the short-distance series expansion can be obtained by Bethe Ansatz, through relations that I have called 'connections', that link them to local correlation functions and moments of the density of pseudo-momenta. I have derived the first few connections, and noticed that they correspond to well-known results, that are gathered and classified for a first time in a single formalism. I have also given new and shorter derivations of a few of them. Then, by Bethe Ansatz, I have evaluated the first few coefficients of the short-distance series expansion of the one-body correlation function explicitly, and found that the first that changes sign when the interaction strength varies is the forth one, at a value that I have evaluated with very high accuracy.

The Fourier transform of the one-body correlation function is the momentum distribution, whose large-momentum tail scales like an inverse quartic law. Its coefficient depends on the interaction strength and is known as Tan's contact, it contains much information on the microscopic details of the model. I have chosen this observable to illustrate a method to solve the Lieb-Liniger model in the case of an additional harmonic trap, that breaks its integrability. The technique relies on a combination of Bethe Ansatz and the local-density approximation, whose acronym is BALDA. Within the latter, I have found a procedure, valid in the strongly-interacting regime, to obtain Tan's contact to arbitrary order in the inverse coupling.

In chapter \ref{secIV}, I have considered correlation functions in energy-momentum space. To be more specific, I have focused on the dynamical structure factor, i.e. the absorption spectrum of the gas, probed by Bragg scattering. When an impurity or a laser beam is stirred along the fluid and couples locally to its density sufficiently weakly, then linear response theory applies and allows to evaluate energy dissipation, or equivalently the drag force, once the dynamical structure factor and the shape of the potential barrier are known. This allows to study superfluidity, as characterized by the absence of viscosity, i.e. of a drag force, below a critical velocity, as a generalization of Landau's criterion. After an introduction to experiments on superfluids, an exposition of Landau's criterion and of the drag force concept in the quantum regime, the dynamical structure factor of the Tonks-Girardeau gas is obtained by Bose-Fermi mapping. It features a low-energy region where excitations are forbidden, whose upper bound corresponds to the lower excitation spectrum of the model. The drag force due to an infinitely thin potential barrier is linear below the Fermi velocity, then it saturates to a finite value.

At finite temperature, I have found that the dynamical structure factor spreads beyond the zero-temperature excitation spectra and acquires excitations at negative energy, that correspond to emission. When temperature increases too much, phonons are not well-defined anymore. I have also studied the effect of a finite barrier width on the drag force. It turns out that in this more realistic picture, the drag force is strongly suppressed at large velocities, putting into light the existence of a quasi-superfluid, supersonic regime.

Several techniques are available to study dynamical correlation functions at arbitrary interaction strength. I have focused on the Tomonaga-Luttinger formalism. By comparison of the effective theory with the exact Tonks-Girardeau predictions, I have studied the validity range of the effective theory and found that it is limited to low energy, low temperature, and low velocity for the drag force. At finite interaction strength, to make quantitative predictions, two parameters are needed: the Luttinger parameter and the first form factor. The former is obtained with high accuracy by coordinate Bethe Ansatz, whereas the form factor requires more advanced techniques. I have guessed an approximate expression that allows to reproduce with satisfying accuracy the exact dynamical structure factor at the vertical of the umklapp point, in a wide range of strong to intermediate couplings.

In view of more sophisticated treatments, for instance within the Imambekov-Glazman liquid formalism, I have obtained another key ingredient to evaluate the dynamical structure factor, i.e. the excitation spectrum of the Lieb-Liniger model. I have identified an exact series expansion of the Lieb-II type spectrum, and expressed the two first coefficients as functions of the sound velocity and the effective mass, found by Bethe Ansatz. Comparison with the exact solution shows that truncating the series to second order is a rather good approximation over a wide range of interactions strengths.

In chapter \ref{secV}, I have turned to the issue of the dimensional crossover from 1D to higher dimensions. There are several ways to address a dimensional crossover, for example coupling 1D gases, or using internal degrees of freedom to create a synthetic dimension, or releasing a transverse trapping. Here, I have focused on the last one, considering a gas of noninteracting fermions in a box trap of tunable size. In a first time, I have obtained the dynamical structure factor of the gas as a function of dimension. A general expression shows that the forbidden low-energy region in 1D is filled with excitations in any higher dimension, providing another example of dramatic dimensional effect. The crossover from 1D to 2D is especially interesting. I have followed it all along and observed the progressive appearance of transverse energy modes by increasing a transverse size of the box. These modes fill the low-energy region progressively, up to a point where no gap remains and dimension two is recovered. Then, I have done the same study for the drag force, on which the effect of dimension is far less spectacular.

Experiments often involve longitudinal harmonic trapping, that can be taken into account in the LDA framework. It turns out that each degree of confinement is equivalent, for the dynamical structure factor, to adding an effective space dimension. This effect is not observed, however, if only the central region of the trap is probed, where the gas is practically homogeneous.

To finish with, in view of future generalizations to interacting systems, I have developed a multimode Tomonaga-Luttinger liquid framework, and tested it along the dimensional crossover from 1D to 2D. Its predictions for a Fermi gas are accurate in the vicinity of the umklapp point of each single mode, and the global one in 2D.
\\
\\
\\
As detailed at the end of each chapter, various research directions open up from my work. Explicit identification of the weak- or strong-coupling series expansion of the ground-state energy of the Lieb-Liniger model may lead to the possibility of a full resummation, that would yield the exact ground-state energy, an achievement whose importance would be comparable to the celebrated solution of the 2D Ising model by Onsager. In view of the weak-coupling expansion, that seems to involve the Riemann zeta function at odd arguments, it might be that this solution could help proving difficult theorems in analytic number theory.

A deeper study of the concept of connexion and their calculation to higher orders may suggest general formulae and solve the Lieb-Liniger model in a stronger sense, or at least, allow to investigate the high-momentum tail of the momentum distribution to next order, beyond Tan's contact, as well as higher-order local correlation functions. Bethe Ansatz, coupled to the local-density approximation, allows to study trapped gases in non-integrable regimes, possibly in an exact manner, and could be tested in other cases than harmonic trapping.

Now that the standard Tomonaga-Luttinger model has been pushed to its limits, next step would be to make quantitative predictions for dynamical observables from the Imambekov-Glazman formalism. The excitation spectrum can be evaluated with excellent accuracy, the form factor is known from algebraic Bethe Ansatz, and the edge exponents are at hand. This approach, or the ABACUS algorithm, could serve for a detailed study of the shape and width of the potential barrier on the drag force profile, in particular to investigate the required conditions to observe a quasi-superfluid supersonic regime.

Dimensional crossover by confinement release could be investigated in other systems and on other observables, to gain insight in the crossover mechanism. In particular, the role that finite temperature could play is not obvious. Multimode Tomonaga-Luttinger liquids coupled through their densities or a cosine term are a first step towards an accurate dynamical description of multi-component models, that would seemingly require a generalized Imambekov-Glazman formalism or a Bethe-Ansatz based treatment.
\\
\\
\\
\\
\\
Dans le Chapitre \ref{secII}, j'ai commenc\'e par rappeler quelques traits sp\'ecifiques aux syst\`emes quantiques unidimensionnels. En guise de premier exemple, je rappelle que tout mouvement y est n\'ecessairement collectif du fait des collisions, ce qui emp\^eche l'existence d'excitation individuelle et de quasi-particule, et enl\`eve tout sens au concept de liquide de Fermi, si utile en dimension sup\'erieure. Les interactions dans les syst\`emes de bosons peuvent conduire \`a l'annulation de la fonction d'onde en cas de contact, comme le principe de Pauli l'imposerait pour des fermions libres, et dans l'espace des quasi-impulsions, on observe la structure caract\'eristique d'une mer de Fermi, ce qui diminue fortement la pertinence de la notion de statistique quantique. Enfin, pour les syst\`emes non polaris\'es en spin, on observe la s\'eparation effective des excitations de charge et de spin, ce qui am\`ene \`a revoir la notion de particule \'el\'ementaire. Tous ces effets sont en fait directement li\'ees \`a la contrainte topologique impos\'ee par la nature unidimensionnelle du syst\`eme, les particules ne pouvant pas se croiser sans entrer en collision.

Une autre cons\'equence frappante de la dimension de dimension est la validit\'e du th\'eor\`eme de Mermin-Wagner dans bon nombre de situations, ce qui emp\^eche toute brisure spontan\'ee de sym\'etrie et conduit \`a des diagrammes de phase moins riches qu'en dimension sup\'erieure. Une alternative possible est d'\'etudier les fonctions de corr\'elation pour soi, qu'elles soient locales ou non-locales, dans l'espace r\'eel ou celui des \'energies, \`a l'\'equilibre ou hors \'equilibre. Elles permettent en particulier de caract\'eriser les gaz d'atomes ultrafroids de basse dimension. Ces derniers sont obtenus par un syst\`eme \`a base de pi\`eges fortement anisotropes. Sur un r\'eseau optique, on peut cr\'eer un ensemble de gaz de g\'eom\'etrie filaire, mais on peut aussi obtenir un gaz unique, et m\^eme imposer les conditions aux limites p\'eriodiques ch\`eres aux th\'eoriciens, en choisissant une g\'eom\'etrie annulaire.

Un certain nombre de mod\`eles simples introduits en physique math\'ematique dans les ann\'ees 1960 s'av\`erent d\'ecrire ces gaz avec une bonne pr\'ecision, et poss\'eder la propri\'et\'e d'int\'egrabilit\'e, ce qui permet d'en faire une \'etude analytique exacte qui couvre l'\'energie et la thermodynamique, le spectre d'excitations et m\^eme les fonctions de corr\'elations, gr\^ace \`a l'Ansatz de Bethe. D'autres outils sp\'ecifiques aux mod\`eles de basse dimension viennent enrichir la panoplie, par exemple la correspondance bosons-fermions, qui permet de traiter de fa\c con exacte le gaz de Tonks-Girardeau, fortement corr\'el\'e, en remarquant qu'il se comporte souvent comme un gaz de fermions id\'eal. Un bon nombre de mod\`eles appartient \`a la classe d'universalit\'e des liquides de Tomonaga-Luttinger, qui est compl\`etement r\'esolu par bosonisation. En particulier, on conna\^it ses fonctions de corr\'elations, qui d\'ecrivent celles de mod\`eles plus compliqu\'es \`a longue distance, aux temps courts et \`a basse \'energie. \`A temp\'erature nulle, les corr\'elations spatiales d\'ecroissent comme des lois de puissance, tandis que leur effondrement est exponentiel \`a longue distance \`a temp\'erature finie, comme le confirme le formalisme de la th\'eorie conforme des champs, tout aussi valide pour \'etudier ce probl\`eme.

Dans le Chapitre \ref{secIII}, j'ai \'etudi\'e le mod\`ele de Lieb et Liniger, qui d\'ecrit un gaz de bosons avec interactions de contact en dimension un. L'Ansatz de Bethe permet d'exprimer son \'energie dans l'\'etat fondamental en tant que solution d'un syst\`eme d'\'equations int\'egrales coupl\'ees, dont des expressions approch\'ees de pr\'ecision arbitraire peuvent \^etre obtenues dans les r\'egimes de faible ou forte interaction. Dans le premier, j'ai r\'eussi \`a identifier un coefficient auparavant inconnu du d\'eveloppement en s\'erie, et dans le r\'egime de forte interaction, j'ai atteint des ordres \'elev\'es et r\'eussi \`a sommer partiellement la s\'erie pour proposer une expression conjecturale plus simple. En parall\`ele, j'ai d\'evelopp\'e une autre m\'ethode, applicable quelle que soit l'intensit\'e des interactions. La combinaison de ces r\'esultats me donne acc\`es \`a l'\'energie avec une pr\'ecision diabolique.

Je me suis ensuite tourn\'e vers les fonctions de corr\'elation locales. Celle \`a un corps est trivial, celles d'ordre deux et trois sont accessibles par Ansatz de Bethe moyennant quelques calculs suppl\'ementaires. Il m'a notamment fallu \'evaluer les premiers moments de la densit\'e de pseudo-impulsions, dont j'ai obtenu des estimations plus pr\'ecises que celles connues jusqu'alors.

La fonction de corr\'elation \`a un corps acquiert une structure int\'eressante si on la consid\`ere d'un point de vue non-local. \`A longue distance, son comportement est celui d'un liquide de Tomonaga-Luttinger. Elle d\'ecro\^it de fa\c con alg\'ebrique, ce qui est caract\'eristique d'un quasi-ordre \`a longue distance. \`A courte distance, le comportement d\'epend fortement des caract\'eristiques microscopiques du mod\`ele. Dans le cas du mod\`ele de Lieb et Liniger, j'ai obtenu les premiers coefficients du d\'eveloppement en s\'erie par Ansatz de Bethe, en \'etablissant des liens entre ceux-ci, les fonctions de corr\'elation locales et les moments de la densit\'e de pseudo-impulsions. J'ai baptis\'ees ces relations 'connexions', et me suis rendu compte qu'elles correspondaient souvent \`a des relations bien connues mais qui n'avaient jamais \'et\'e envisag\'ees comme aussi \'etroitement li\'ees les unes aux autres ni englob\'ees dans un seul et unique cadre interpr\'etatif. Parmi ces coefficients, le premier \`a avoir un comportement non-monotone en fonction de l'intensit\'e des interactions est celui d'ordre quatre, qui change de signe pour une certaine valeur de l'intensit\'e des interactions, que j'ai \'evalu\'ee avec une grande pr\'ecision.

La transform\'ee de Fourier de la fonction de corr\'elation \`a un corps n'est autre que la distribution des impulsions, qui se comporte \`a grande impulsion comme l'inverse de la puissance quatri\`eme de cette derni\`ere. Son coefficient est le contact de Tan, qui d\'epend des interactions et renseigne sur les caract\'eristiques microscopiques du mod\`ele. J'ai choisi cette observable pour illustrer un formalisme hybride combinant l'Ansatz de Bethe et l'approximation de la densit\'e locale, afin d'\'etudier un gaz de Bose confin\'e longitudinalement par un pi\`ege harmonique, qui retire au mod\`ele son int\'egrabilit\'e. J'ai pu notamment d\'evelopper une proc\'edure donnant le contact de Tan \`a un ordre arbitraire dans l'inverse de l'intensit\'e des interactions.

Dans le Chapitre \ref{secIV}, je me suis tourn\'e vers les fonctions de corr\'elation dans l'espace des \'energies, en particulier le facteur de structure dynamique du mod\`ele de Lieb et Liniger, qui repr\'esente le spectre d'absorption de ce gaz. Quand une impuret\'e ou un faisceau laser traverse le fluide et modifie localement sa densit\'e, si l'effet est suffisamment faible, il peut \^etre trait\'e \`a travers le formalisme de la r\'eponse lin\'eaire. Ce dernier permet d'en d\'eduire la force de tra\^in\'ee, li\'ee \`a la dissipation d'\'energie vers le milieu ext\'erieur sous forme d'\'echauffement du gaz. J'ai ainsi pu \'etudier la superfluidit\'e, caract\'eris\'ee par l'absence de viscosit\'e ou de force de tra\^in\'ee en-dessous d'une vitesse d'\'ecoulement critique.

Apr\`es une introduction qui retrace le cheminement historique, un expos\'e des id\'ees qui ont conduit au crit\`ere de Landau et au formalisme de la force de tra\^in\'ee dans le r\'egime quantique, j'ai obtenu le facteur de structure du gaz de Tonks-Girardeau en m'appuyant sur la correspondance bosons-fermions. Une certaine r\'egion de basse \'energie situ\'ee en-dessous du spectre d'excitation inf\'erieur est interdite \`a toute excitation. La force de tra\^in\'ee due \`a une barri\`ere de potentielle infiniment fine est d'abord une fonction lin\'eaire de la vitesse, avant de saturer \`a partir de la vitesse de Fermi. Rien de nouveau sous le soleil, ces r\'esultats \'etant d\'ej\`a connus. En revanche, j'ai ensuite \'etudi\'e l'effet de la temp\'erature sur le facteur de structure dynamique. Cette derni\`ere a pour principal effet d'augmenter la taille de la zone o\`u les excitations peuvent avoir lieu, jusqu'aux \'energies n\'egatives qui correspondent \`a une \'emission. Les phonons sont de plus en plus mal d\'efinis \`a mesure que la temp\'erature augmente, et finissent par \^etre noy\'es dans le continuum. Quant \`a la force de tra\^in\'ee, si on ne suppose plus une barri\`ere de potentiel infiniment fine mais qu'on prend en compte son \'epaisseur, la saturation dispara\^it et la force de tra\^in\'ee finit pratiquement par dispara\^itre dans le r\'egime de grande vitesse, signe d'un comportement quasi-superfluide.

Plusieurs techniques ont d\'ej\`a \'et\'e \'eprouv\'ees pour traiter les interactions quelconques. Je me suis concentr\'e sur le formalisme des liquides de Tomonaga-Luttinger, que j'ai pouss\'e dans ses ultimes retranchements. Par comparaison avec les r\'esultats exacts dans le r\'egime de Tonks-Girardeau, \`a temp\'erature nulle et finie, j'ai pu en \'etudier le domaine de validit\'e, qui est restreint aux basses \'energies, basses temp\'eratures, et faibles vitesses en ce qui concerne la force de tra\^in\'ee. \`A interaction finie, toute pr\'ediction quantitative requiert la connaissance du param\`etre de Luttinger, que j'ai obtenu par Ansatz de Bethe, et du premier facteur de forme, accessible uniquement par des outils avanc\'es. J'en ai obtenu une expression effective qui s'est av\'er\'ee assez efficace pour reproduire le r\'esultat exact \`a la verticale du point de r\'etrodiffusion, \`a basse \'energie et dans un large domaine d'interactions. Un traitement plus sophistiqu\'e requiert en particulier la connaissance pr\'ecise des spectres d'excitation, ce qui est d\'esormais chose faite. J'ai obtenu un d\'eveloppement en s\'erie du spectre de type II, dont les deux premiers ordres s'expriment en fonction de la vitesse du son et de la masse effective, \'evalu\'es avec une pr\'ecision arbitraire par Ansatz de Bethe, et qui suffisent \`a reproduire la solution num\'erique exacte avec une bonne pr\'ecision.

Dans le Chapitre \ref{secV}, je me suis int\'eress\'e au changement progressif de dimension, depuis un gaz unidimensionnel, vers les dimensions sup\'erieures. Cette transition lisse peut \^etre r\'ealis\'ee de diverses mani\`eres, par exemple en couplant entre eux un grand nombre de gaz unidimensionnels, en utilisant des degr\'es de libert\'e internes, ou encore en rel\^achant un confinement transverse. C'est sur ce dernier cas que je me suis concentr\'e, l'illustrant \`a travers l'exemple d'un gaz de fermions id\'eal dans une bo\^ite de taille variable.

Dans un premier temps, j'ai obtenu le facteur de structure dynamique en fonction de la dimension. Son expression g\'en\'erale montre que la r\'egion interdite dans l'espace des \'energies en dimension un devient accessible d\`es que celle-ci augmente. La transition d'une \`a deux dimensions est tout particuli\`erement int\'eressante de ce point de vue. Je l'ai suivie de bout en bout \`a travers l'\'etude de la structure multi-mode qui appara\^it quand on augmente progressivement la taille d'un c\^ot\'e de la bo\^ite. Ces modes emplissent petit \`a petit tout l'espace disponible en-dessous du spectre d'excitation inf\'erieur, jusqu'\`a le combler enti\`erement lorsqu'on atteint la dimension deux. J'ai fait la m\^eme \'etude pour la force de tra\^in\'ee, qui se r\'ev\`ele moins instructive d'un point de vue physique.

Les exp\'eriences mettent souvent en jeu un pi\`ege harmonique longitudinal, qui peut \^etre \'etudi\'e th\'eoriquement dans le cadre de l'approximation de la densit\'e locale. Il s'av\`ere alors que chaque degr\'e de confinement ajout\'e est \'equivalent, pour le facteur de structure dynamique, \`a une augmentation de la dimension de l'espace d'une unit\'e. Ce n'est pas le cas, en revanche, si on se contente de sonder la r\'egion centrale du pi\`ege, o\`u le gaz est relativement homog\`ene.

Pour finir, en vue de g\'en\'eralisations \`a des syst\`emes en interaction, j'ai d\'evelopp\'e un formalisme de liquide de Tomonaga-Luttinger multimode, que j'ai test\'e le long de la transition d'une \`a deux dimensions. Ses pr\'edictions s'av\`erent correctes pour chaque mode au voisinage de son point de r\'etrodiffusion, et \`a la limite de dimension deux, au voisinage de ce point au niveau global.

\newpage

\appendix
\chapter{Complements to chapter \ref{secII}}

\section{Density correlations of a Tomonaga-Luttinger in the thermodynamic limit at zero temperature}
\label{dernn}

The aim of this section is to derive Eq.~(\ref{nnTLL}) in formalism of Tomonaga-Luttinger liquids. Elements of this derivation can be found in various references, see e.g. \cite{GiamarchiBook}.

As a first step, one needs to construct a convenient representation for the density in the continuum. In first quantization, the granular density operator reads
\begin{equation}
\label{densityfirstquant}
 n(x)\!=\!\sum_{i=1}^N \delta(x\!-\!x_i),
\end{equation}
where $\{x_i\}_{i=1,\dots,N}$ label the positions of the point-like particles. This expression is not practical to handle, and needs coarse-graining in the thermodynamic limit. To simplify it, one defines a function $\zeta$ such that $\zeta(x_i)\!=\!i\pi$ at the position of the $i^{th}$ particle and zero otherwise, and applies the property of the $\delta$ composed with a function $f$:
\begin{eqnarray}
 \delta[f(x)]=\sum_{x_i/f(x_i)=0}\frac{1}{|f'(x_i)|}\delta(x-x_i),
\end{eqnarray}
to $f(x)\!=\!\zeta(x)\!-\!i\pi$, so that $f'(x)\!=\!\partial_x\zeta(x)$, and 
\begin{eqnarray}
\delta[\zeta(x)-i\pi]=\sum_{x_i}\frac{1}{|\partial_x\zeta(x_i)|}\delta(x-x_i).
\end{eqnarray}
Summing on $i$ yields
\begin{equation}
 \sum_{i=1}^N \delta[\zeta(x)-i\pi]=\sum_{i=1}^N \sum_{x_i}\frac{1}{|\partial_x\zeta(x_i)|}\delta(x-x_i)=\sum_{i=1}^N\frac{1}{|\partial_x\zeta(x)|}\delta(x-x_i).
\end{equation}
Then, Eq.~(\ref{densityfirstquant}) reads
\begin{equation}
 n(x)\!=\!|\partial_x\zeta(x)|\sum_i \delta[\zeta(x)-i\pi]=\partial_x\zeta(x)\sum_i \delta[\zeta(x)-i\pi].
\end{equation}
One transforms this expression by applying Poisson's formula, namely
\begin{equation}
\sum_{m=-\infty}^{+\infty}g(m)=\sum_{m=-\infty}^{+\infty}\int_{-\infty}^{+\infty}\!dz\,g(z)e^{2im\pi z},
\end{equation}
to the function defined as $g(z)\!=\!\delta[\zeta(x)-\pi z]$, yielding
\begin{eqnarray}
 n(x)\!=\!\frac{\partial_x\zeta(x)}{\pi}\!\sum_{m=-\infty}^{+\infty}e^{2im\zeta(x)}.
\end{eqnarray}
I finally define $\theta(x)\!=\!k_F x\!-\!\zeta(x)$, and rewrite the density-density correlator as

\begin{eqnarray}
\label{sum}
&&\!\!\!\!\!\!\!\!\!\langle n(x,t) n(x',t')\rangle\nonumber\\
&&\!\!\!\!\!\!\!\!\!\!=\!\frac{1}{\pi^2}\!\left\langle\![k_F\!+\!\partial_x \theta (x,t)][k_F\!+\!\partial_{x'} \theta(x',t')]\!\!\sum_{m=-\infty}^{+\infty}\sum_{m'=-\infty}^{+\infty}\!\!\!e^{2im[\theta(x,t)+k_Fx]} e^{2im'[\theta(x',t')+k_Fx']}\!\right\rangle\!\!.
\end{eqnarray}
 
Due to Galilean-invariance of the system, only relative coordinates are important in the thermodynamic limit. Therefore, I set $x'$ and $t'$ to zero in the following. To compute the correlations using Eq.~(\ref{sum}), I split the summations over $m$ and $m'$ into several parts.

First, I compute the leading term, obtained for $m\!=\!m'\!=\!0$. Keeping only the latter is called 'harmonic approximation'. This term corresponds to $\frac{1}{\pi^2}\langle\partial_x \zeta(x,t)\partial_x\zeta(0,0)\rangle$. Using the diagonal form of the Hamiltonian Eq.~(\ref{diagH}), the field expansion over the bosonic basis Eq.~(\ref{thetastatic}), and the equation of motion or the Baker-Campbell-Haussdorff formula for exponentials of operators yields
\begin{equation}
\label{zetaT0}
\partial_x \zeta(x,t)=\pi n_0+\frac{1}{2}\sum_{q\neq 0}\left|\frac{2\pi K}{qL}\right|^{1/2}\!iq\left\{e^{i[qx-\omega(q)t]}b_q-e^{-i[qx-\omega(q)t]}b_q^{\dagger}\right\},
\end{equation}
with $\langle b_q b_{q'}^{\dagger}\rangle\!=\!\delta_{q,q'}$ and $\langle b_q^{\dagger} b_{q'}\rangle\!=\!0$ since $q, q'\!\neq\!0$ and $T\!=\!0$. Also, $\langle b_q^{\dagger}b_{q'}^{\dagger}\rangle\!=\!\langle b_q b_{q'}\rangle\!=\!0$, thus
\begin{eqnarray}
\langle \partial_x \zeta(x,t)\partial_x\zeta(0,0)\rangle-(\pi n_0)^2=\frac{\pi K}{2}\frac{1}{L}\sum_{q\neq 0}|q| e^{iq[x-\mathrm{sign}(q)v_s t]},
\end{eqnarray}
and after a few lines of algebra,
\begin{equation}
\label{redo}
\frac{1}{\pi^2}\langle\partial_x \zeta(x,t)\partial_x\zeta(0,0)\rangle=n_0^2\left(1-\frac{K}{4k_F^2}\left[\frac{1}{(x-v_s t+i\epsilon)^2}+\frac{1}{(x+v_s t-i\epsilon)^2}\right]\right),
\end{equation}
where a short-distance regulator $i\epsilon$ has been added.

Another type of contribution is given by $n_0^2\sum_{m,m'=-\infty,\neq(0,0)}^{+\infty}\langle e^{2im\zeta(x,t)}e^{2im'\zeta(x',t')}\rangle$. I introduce a generating function:
\begin{eqnarray}
G_{m,m'}(x,t;x',t')=e^{2im\zeta(x,t)}e^{2im'\zeta(x',t')},
\end{eqnarray}
and use the identity, valid for two operators A and B both commuting with their commutator:
\begin{equation}
e^{A+B}=e^{A}e^{B}e^{\frac{1}{2}\left[A,B\right]},
\end{equation}
hence
\begin{equation}
G_{m,m'}(x,t;0,0)=e^{2i[m\zeta(x,t)+m'\zeta(0,0)]}e^{\frac{1}{2}\left[2im\zeta(x,t),2im'\zeta(0,0)\right]}.
\end{equation}
Using Eq.~(\ref{thet}),
\begin{eqnarray}
\left[\zeta(x,t),\zeta(0,0)\right]&=&\frac{1}{4}\sum_{q,q^{'}\neq 0}\left|\frac{2\pi K}{qL}\right|^{1/2}\left|\frac{2\pi K}{q^{'}L}\right|^{1/2}\left(e^{i[qx-\omega(q)t]}\left[b_q,b_{q'}^\dagger\right]+e^{-i[qx-\omega(q)t]}\left[b_q^{\dagger},b_{q^{'}}\right]\right)\nonumber \\
&=&i\sum_{q\neq 0}\left|\frac{\pi K}{qL}\right| \sin\left[qx-\omega(q)t\right].
\end{eqnarray}
since $[b_q,b_{q'}^{\dagger}]=\delta_{q,q'}$. Also,
\begin{equation}
m\zeta(x,t)+m'\zeta(0,0)=mk_F x+\frac{1}{2}\sum_{q\neq 0}\left|\frac{2\pi K}{qL}\right|^{1/2}\left[(m e^{i[qx-\omega(q) t]}+m')b_q+h.c.\right],
\end{equation}
where $h.c.$ means 'hermitian conjugate'.
Thus, setting $\alpha_{m,m'}\!=\!m e^{i[qx-\omega(q) t]}+m'$ for concision,
\begin{equation}
G_{m,m'}(x,t;0,0)=e^{imk_Fx+i\sum_{q\neq 0}\left|\frac{2\pi K}{qL}\right|^{1/2}(\alpha_{m,m'}b_q+h.c.)}e^{-2mm' i\frac{1}{L}\sum_{q\neq 0}\left|\frac{\pi K}{q}\right| \sin[qx-\omega(q)t]}.
\end{equation}
We are interested in its statistical average. I use the identity
\begin{eqnarray}
\langle e^A\rangle=e^{\frac{1}{2}\langle A^2\rangle},
\end{eqnarray}
valid for any linear operator $A$, to show that
\begin{eqnarray}
\label{eqA20}
\left\langle e^{i\sum_{q\neq 0}\left|\frac{2\pi K}{qL}\right|^{1/2}(\alpha_{m,m'}b_q+h.c.)}\right\rangle&&=e^{-\sum_{q,q'\neq 0}\left|\frac{\pi K}{qL}\right|^{1/2}\left|\frac{\pi K}{q'L}\right|^{1/2}\left|\alpha_{m,m'}\right|^2\left(\langle b_q{b_{q'}}^{\dagger}\rangle+\langle b_q^{\dagger}b_{q'}\rangle\right)} \nonumber \\
&&=e^{-\sum_{q\neq 0} \frac{\pi K}{|q|L}(m^2+{m'}^2+2mm' \cos\left[qx-\omega(q)t\right])},
\end{eqnarray}
since
\begin{equation}
|\alpha_{m,m'}|^2=(me^{i[qx-\omega(q)t]}+m')(me^{-i[qx-\omega(q)t]}+m')\!=\!m^2\!+\!{m'}^2\!+\!2mm'\cos[qx-\omega(q)t].
\end{equation}
Thus,
\begin{equation}
\langle G_{m,m'}(x,t;0,0)\rangle=e^{2imk_Fx}e^{-\sum_{q\neq 0} \frac{\pi K}{|q|L}(m^2+{m'}^2+2mm' e^{i[qx-\omega(q)t]})}.
\end{equation}
To go further, I have to evaluate $\frac{1}{L}\sum_{q\neq 0} \frac{1}{|q|}(m^2+{m'}^2+2 mm' e^{iq[x-v_s \mathrm{sign}(q)t]})$. First, note that if this series diverges, then $\langle G_{mm^{'}}(x,t;0,0)\rangle\!=\!0$. Rewriting 
\begin{eqnarray}
m^2+{m'}^2+2mm' e^{iq[x-v_s \mathrm{sign}(q)t]}=(m+m')^2- 2mm' \left(1-e^{iq[x-v_s \mathrm{sign}(q)t]}\right),
\end{eqnarray}
one sees that $m\!=\!-m'$ is a necessary condition in the thermodynamic limit to result in a non-vanishing contribution.

In the thermodynamic limit, introducing a regularizing cut-off $e^{-\epsilon q}$ and using the property
\begin{eqnarray}
\frac{1}{q}=\int_0^{+\infty}dy\,e^{-qy},
\end{eqnarray}
one finds
\begin{eqnarray}
&&\int_{0^{+}}^{+\infty}dq\,\frac{e^{-a q}}{q}\left[1-e^{-iqv_st}\cos(qx)\right]\nonumber\\
&&=\lim_{A\to +\infty}\int_0^A dy\int_0^{+\infty}dq\left[e^{-(\epsilon+y)q}-\frac{1}{2}e^{-(\epsilon+y+iv_st-ix)q}-\frac{1}{2}e^{-(\epsilon+y+iv_st+ix)q}\right] \nonumber \\
&&=\frac{1}{2}\ln\!\left[\frac{(\epsilon+iv_st)^2+x^2}{\epsilon^2}\right]=\frac{1}{2}\ln\!\left[\frac{(x-v_st+i\epsilon)(x+v_st-i\epsilon)}{\epsilon^2}\right].
\end{eqnarray}
In the end,
\begin{equation}
\left\langle G_{m,m'}(x,t;0,0)\right\rangle=e^{2imk_Fx}\delta_{m,-m'}\left[\frac{(x-v_s t+i\epsilon)(x+v_s t-i\epsilon)}{\epsilon^2}\right]^{-Km^2}
\end{equation}
yielding the other contributions to Eq.~(\ref{nnTLL}) after replacing the various powers of the regularizing term $\epsilon$ by the non-universal form factors. In particular, I obtain
\begin{eqnarray}
\label{Amcutoff}
A_m=2(\epsilon k_F)^{2Km^2}
\end{eqnarray}
in terms of the small-distance cut-off.

\section{Density correlations of a Tomonaga-Luttinger\\
liquid at finite temperature by bosonization}
\label{dernnT}
This appendix provides an alternative approach to CFT to derive Eq.~(\ref{nnfftT}), based on the Tomonaga-Luttinger liquid formalism. Calculations are far longer, but provide an independent way to check the result and are more elementary in the mathematical sense. I split the calculations into two parts.

\subsection{First contribution to the density correlation}

The beginning of the derivation is essentially the same as in the zero temperature case, already treated in Appendix \ref{dernn}. Using the same notations, I start back from Eq.~(\ref{zetaT0}), up to which the derivation is the same. Now, since $T\!>\!0$ the mean values in the bosonic basis are re-evaluated as $\langle b_q^{\dagger}b_{q'} \rangle\!=\!\delta_{q,q'}n_B(q)$, and $\langle b_qb_{q'}^{\dagger}\rangle\!=\!\delta_{q,q'}[1\!+\!n_B(q)]$, where
\begin{eqnarray}
n_B(q)=\frac{1}{e^{\beta\hbar\omega(q)}-1}
\end{eqnarray}
is the Bose-Einstein distribution of the bosonic modes. Thus,
\begin{eqnarray}
&&[\left\langle\partial_x\zeta(x,t)\partial_x\zeta(0,0)\right\rangle-(\pi n_0)^2]_{T>0} \nonumber \\
&&=\frac{1}{4}\sum_{q\neq 0}\frac{2\pi K}{|q|L}q^2 \left\{e^{i(qx-\omega(q)t)}[1+n_B(q)]+e^{-i(qx-\omega(q)t)}n_B(q)\right\} \nonumber \\
&&=[\left\langle\partial_x\zeta(x,t)\partial_x\zeta(0,0)\right\rangle-(\pi n_0)^2]_{T=0}\nonumber\\
&&+\frac{1}{4}\sum_{q\neq 0}\frac{2\pi K |q|}{L}\left(e^{i[qx-\omega(q)t]}+e^{-i[qx-\omega(q)t]}\right)n_B(q),
\end{eqnarray}
where I have isolated the result at $T\!=\!0$ already evaluated in Appendix \ref{dernn}, and a purely thermal part.

To evaluate this thermal part, I make a few algebraic transformations, take the thermodynamic limit and use the change of variable $\beta\hbar v_s q\to \tilde{q}$ to obtain
\begin{eqnarray}
&&\frac{2\pi}{L}\sum_{q\neq 0}|q|\left(e^{i[qx-\omega(q)t]}+e^{-i[qx-\omega(q)t]}\right)n_B(q)\nonumber\\
&&=\int_{0}^{+\infty}dq\,q\left[e^{iq(x-v_st)}+e^{-iq(x-v_st)}+e^{iq(x+v_st)}+e^{-iq(x+v_st)}\right]n_B(q)\nonumber\\
&&=\frac{1}{L_T^2}\int_0^{+\infty}\!dq\,\frac{q}{e^q-1}\left[e^{iq\frac{(x-v_st)}{L_T}}+e^{iq\frac{(x+v_st)}{L_T}}+e^{-iq\frac{(x-v_st)}{L_T}}+e^{-iq\frac{(x+v_st)}{L_T}}\right],
\end{eqnarray}
where $L_T\!=\!\beta\hbar v_s$ plays the role of a thermal length. I define
\begin{equation}
K(b)=\int_0^{\infty}dy \,y\frac{e^{iby}}{e^y-1},
\end{equation}
and find that
\begin{eqnarray}
K(b)+K(-b)&=&\int_0^{+\infty}dy\,y (e^{iby}+e^{-iby})e^{-y}\sum_{n=0}^{+\infty}e^{-yn}\nonumber \\
&=&-\int_0^{+\infty}dy \sum_{n=0}^{+\infty}\left[\frac{e^{iby}e^{-y}e^{-yn}}{ib-(n+1)}+\frac{e^{-iby}e^{-y}e^{-yn}}{-ib-(n+1)}\right] \nonumber \\
&=&\sum_{n=1}^{+\infty}\int_0^{+\infty}dy\left[\frac{e^{iby}e^{-yn}}{-ib+n}+\frac{e^{-iby}e^{-yn}}{ib+n}\right]\nonumber \\
&=&\sum_{n=1}^{+\infty}\left[\frac{1}{(-ib+n)^2}+\frac{1}{(ib+n)^2}\right] \nonumber \\
&=&2\sum_{n=1}^{+\infty}\frac{n^2-b^2}{(n^2+b^2)^2}.
\end{eqnarray}
Then, I use the property \cite{Gradshteyn}
\begin{equation}
\frac{1}{\sin^2(\pi x)}=\frac{1}{\pi^2 x^2}+\frac{2}{\pi^2}\sum_{n=1}^{+\infty}\frac{x^2+k^2}{(x^2-k^2)^2}
\end{equation}
combined to $\sin(ix)\!=\!i\sinh(x)$, yielding
\begin{equation}
2\sum_{k=1}^{+\infty}\frac{k^2-x^2}{(k^2+x^2)^2}=\pi^2\left[\frac{1}{\pi^2 x^2}-\frac{1}{\sinh^2(\pi x)}\right],
\end{equation}
put back the prefactors and add the known result at $T\!=\!0$ to obtain
\begin{equation}
\frac{\langle n(x,t) n(0,0) \rangle_{m=0}}{n_0^2} = -\frac{K}{4k_F^2}\frac{\pi^2}{L_T^2}\left\{\frac{1}{\sinh^2\left[\frac{\pi(x+v_s t)}{L_T}\right]}+\frac{1}{\sinh^2\left[\frac{\pi(x-v_s t)}{L_T}\right]}\right\}.
\end{equation}

\subsection{Second contribution to the density correlation function}
This time again, the derivation is at first strictly similar to the one at zero temperature. The point where they start to differ is Eq.~(\ref{eqA20}), so I come back to this exact point and find 
\begin{eqnarray}
&&\left\langle e^{i\sum_{q\neq 0}\left|\frac{2\pi K}{qL}\right|^{1/2}(\alpha_{mm'}b_q+h.c.)}\right\rangle \nonumber \\
&&=e^{-\sum_{q,q'\neq 0}\left|\frac{\pi K}{qL}\right|^{1/2}\left|\frac{\pi K}{q'L}\right|^{1/2}|\alpha_{mm'}|^2(\langle b_q b_{q'}^{\dagger}\rangle+\langle b_q^{\dagger} b_{q'}\rangle)} \nonumber \\
&&=e^{-\sum_{q,q'\neq 0}\left|\frac{\pi K}{qL}\right|^{1/2}\left|\frac{\pi K}{q'L}\right|^{1/2}|\alpha_{mm'}|^2\delta_{q,q'}[1+2n_B(q)]},
\end{eqnarray}
thus the generating function at finite temperature reads
\begin{eqnarray}
\!\!\!\!\!\!\!\!\!\!\!\!\!\!\!&&\langle G_{mm'}(x,t;0,0)\rangle_{T>0} \nonumber \\
\!\!\!\!\!\!\!\!\!\!\!\!\!\!\!&&=e^{2imk_F x}e^{-2mm'i\frac{1}{L}\sum_{q\neq 0}\left|\frac{\pi K}{q}\right|\sin[qx-\omega(q)t]}e^{-\sum_{q\neq 0}\left|\frac{\pi K}{qL}\right|(m^2+m'^2+2mm'\cos[qx-\omega(q)t])[1+2n_B(q)]} \nonumber \\
\!\!\!\!\!\!\!\!\!\!\!\!\!\!\!&&=e^{2imk_F x}e^{-\sum_{q\neq 0}\left|\frac{\pi K}{qL}\right|\{m^2+{m'}^2+2mm'e^{i[qx-\omega(q)t]}+2(m^2+m'^2+2mm'\cos[qx-\omega(q)t])n_B(q)\}}\nonumber\\
\!\!\!\!\!\!\!\!\!\!\!\!\!\!\!&&=e^{2imk_F x}e^{-\sum_{q\neq 0}\left|\frac{\pi K}{qL}\right|\{(m+m')^2+2mm'(e^{i[qx-\omega(q)t]}-1)+[(m+m')^2+2mm'(\cos[qx-\omega(q)t]-1)]n_B(q)\}}
\end{eqnarray}
and the condition $m\!=\!-m'$ in the thermodynamic limit yields
\begin{eqnarray}
&&\langle G_{mm'}(x,t;0,0)\rangle_{T>0} \nonumber \\
&&=\delta_{m,-m'}e^{2imk_F x}e^{-m^2 K \int_{q\neq 0}\frac{dq}{|q|}\{1-e^{i[qx-\omega(q)t]}+2\left(1-\cos[qx-\omega(q)t]\right)n_B(q)\}}.
\end{eqnarray}
It involves the integral
\begin{eqnarray}
&&\int_{q\neq 0} \frac{dq}{|q|}\left\{1-e^{i[qx-\omega(q)t]}+2[1-\cos(qx\!-\!\omega(q)t)]n_B(q)\right\} \nonumber \\
&&=2\int_{0}^{+\infty}e^{-\epsilon q}\frac{dq}{q}\left\{1-e^{-iqv_st}\cos(qx)+2[1\!-\!\cos(qx)\cos(qv_st)]n_B(q)\right\},
\end{eqnarray}
where I have restored the regulator neglected up to here to simplify the notations. This integral has been evaluated in imaginary time, defined through $\tau\!=\!it$, in Ref.~\cite{GiamarchiBook}. The result should be
\begin{eqnarray}
\label{F1Gia}
&&F_1(r)=\int_{0}^{+\infty}e^{-\epsilon q}\frac{dq}{q}\left\{1-e^{-qv_s\tau}\cos(qx)+2[1-\cos(qx)\cosh(qv_s\tau)]n_B(q)\right\}\nonumber\\
&&\simeq_{x,v_s\tau\gg \epsilon}\frac{1}{2}\ln\!\left\{\frac{L_T^2}{\pi^2\epsilon^2}\left[\sinh^2\left(\frac{\pi x}{L_T}\right)+\sin^2\left(\frac{\pi v_s\tau}{L_T}\right)\right]\right\}.
\end{eqnarray}
I propose my own derivation of the latter, that relies on the following steps:
\begin{eqnarray}
\!\!\!\!\!\!\!&&\int_{0}^{+\infty} \frac{dq}{q}e^{-\epsilon q}\left\{1-e^{-iqv_s t}\cos(qx)+2[1-\cos(qx)\cos(qv_s t)]n_B(q)\right\} \nonumber \\
\!\!\!\!\!\!\!&&=\!\!\int_{0}^{+\infty}\!\!dq\,e^{-\epsilon q}\!\!\int_0^{+\infty}\!\!dy\,e^{-yq}\!\left[1\!-\!e^{-iqv_s t}\frac{e^{iqx}\!+\!e^{-iqx}}{2}\!+\!2\!\left(\!1\!-\!\frac{e^{iqx}\!+\!e^{-iqx}}{2}\frac{e^{iqv_s t}\!+\!e^{-iqv_s t}}{2}\right)\!\frac{1}{e^{L_Tq}\!-\!1}\right] \nonumber \\
\!\!\!\!\!\!\!&&=\frac{1}{2}\int_0^{+\infty}\!\!dy\int_0^{+\infty}\!\!dq\,e^{-(\epsilon+y)q} \nonumber \\
\!\!\!\!\!\!\!&&\left(\!2\!-\!e^{iq(x-v_st)}\!-\!e^{-iq(x+v_st)}\!+\!\!\sum_{n=0}^{+\infty}\!e^{-L_T q(n+1)}\!\!\left\{4\!-\!\!\left[e^{iq(x+v_st)}\!\!+\!e^{iq(x-v_st)}\!\!+\!e^{-iq(x+v_st)}\!\!+\!e^{-iq(x-v_st)}\right]\right\}\!\right)\nonumber \\
\!\!\!\!\!\!\!&&=\frac{1}{2}\int_0^{A\to +\infty}\!\!dy \,\frac{2}{\epsilon\!+\!y}-\frac{1}{\epsilon\!+\!y\!-\!i(x-v_s t)}-\frac{1}{\epsilon\!+\!y\!+\!i(x+v_s t)} \nonumber \\
\!\!\!\!\!\!\!&&+\sum_{n=0}^{+\infty}\frac{4}{\epsilon\!+\!y\!+L_T(n+1)}-\frac{1}{\epsilon\!+\!y\!+\!L_T(n+1)\!-\!i(x+v_s t)}-\frac{1}{\epsilon\!+\!y\!+\!L_T(n+1)\!+\!i(x-v_s t)} \nonumber \\
\!\!\!\!\!\!\!&&-\frac{1}{\epsilon\!+\!y\!+\!L_T(n+1)\!+\!i(x+v_s t)}-\frac{1}{\epsilon\!+\!y\!+\!L_T(n+1)\!-\!i(x-v_s t)} \nonumber \\
\!\!\!\!\!\!\!&&=\!\frac{1}{2}\ln\!\left[\frac{(x\!+\!v_st\!-\!i\epsilon)(x\!-\!v_st\!+\!i\epsilon)}{\epsilon^2}\prod_{n=0}^{+\infty}\!\left\{\!1\!+\!\frac{(x+v_st)^2}{[\epsilon\!+\!L_T(n\!+\!1)]^2}\!\right\}\!\prod_{n'=0}^{+\infty}\!\left\{\!1\!+\!\frac{(x-v_st)^2}{[\epsilon\!+\!L_T(n'\!+\!1)]^2}\right\}\right]\!\!.
\end{eqnarray}
Then, in the limit $L_T, x,v_st \gg \epsilon$ and using the infinite product expansion of $\sinh$,
\begin{equation}
\sinh(x)=x\prod_{n=1}^{+\infty}\left(1+\frac{x^2}{k^2\pi^2}\right),
\end{equation}
as well as $\sinh^2(x)-\sinh^2(y)=\sinh(x+y)\sinh(x-y)$ and $\sin(ix)=i\sinh(x)$, I recover Eq.~(\ref{F1Gia}) as expected, and Eq.~(\ref{nnfftT}) follows.

Actually, it is even possible to evaluate the integral $F_1(r)$ exactly, providing another derivation of Eq.~(\ref{F1Gia}). To do so, I start from
\begin{eqnarray}
F_1(r)\!=\!\frac{1}{2}\ln\left\{\frac{(x\!+\!v_s t\!-\!i\epsilon)(x\!-\!v_s t\!+\!i\epsilon)}{\epsilon^2}\prod_{n=1}^{+\infty}\left[1\!+\!\frac{(x\!+\!v_s t)^2}{(\epsilon\!+\!nL_T)^2}\right]\prod_{n'=1}^{+\infty}\left[1\!+\!\frac{(x\!-\!v_s t)^2}{(\epsilon\!+\!n'L_T)^2}\right] \right\}
\end{eqnarray}
and rewrite
\begin{eqnarray}
\prod_{n=1}^{+\infty}\left[1+\frac{(x\pm v_s t)^2}{(\epsilon+nL_T)^2}\right]=\prod_{n=0}^{+\infty}\left[1+\frac{\left(\frac{x \pm v_s t}{l_T}\right)^2}{\left(\frac{\epsilon}{L_T}+n\right)^2}\right]\frac{\epsilon^2}{\epsilon^2+(x \pm v_s t)^2}.
\end{eqnarray}
Then, I use the property \cite{Gradshteyn}
\begin{eqnarray}
\left|\frac{\Gamma(x)}{\Gamma(x-iy)}\right|^2=\prod_{k=0}^{+\infty}\left(1+\frac{y^2}{(x+k)^2}\right),\,x\neq 0,-1,-2 \dots
\end{eqnarray}
to obtain
\begin{eqnarray}
F_1(r)\!=\!\frac{1}{2}\ln\!\left[\frac{x^2\!-\!(v_s t\!-\!i\epsilon)^2}{\epsilon^2}\frac{1}{1\!+\!\frac{(x+v_s t)^2}{\epsilon^2}}\frac{1}{1\!+\!\frac{(x-v_s t)^2}{\epsilon^2}}\left|\frac{\Gamma\left(\frac{\epsilon}{L_T}\right)}{\Gamma\!\!\left[\frac{\epsilon-i(x+v_s t)}{L_T}\right]}\right|^2\left|\frac{\Gamma\left(\frac{\epsilon}{L_T}\right)}{\Gamma\!\!\left[\frac{\epsilon-i(x-v_s t)}{L_T}\right]}\right|^2\right]\!.
\end{eqnarray}
To check consistency with Eq.~(\ref{F1Gia}), I take the large thermal length limit combined to the properties $\Gamma(x)\simeq_{x\to 0}1/x$ and $|\Gamma(iy)|^2=\pi/[y\sinh(\pi y)]$ \cite{Gradshteyn}.

\section{Density correlations of a Tomonaga-Luttinger at finite size and temperature by bosonization}
\label{Demo2}
In this appendix, I provide elements of derivation of Eq.~(\ref{nnfftTL}). To do so, I generalize Eq.~(\ref{redo}) to finite size and temperature.
\begin{eqnarray}
&&\langle\partial_x\zeta(x,t)\partial_x\zeta(0,0)\rangle_{L<+\infty, T>0}-(\pi n_0)^2\nonumber\\
&&=\frac{1}{4}\sum_{q\neq 0}\frac{2\pi K}{|q|L}q^2\{e^{i[qx-\omega(q)t]}[1+n_B(q)]+e^{-i[qx-\omega(q)t]}n_B(q)\}\nonumber\\
&&=\langle\partial_x\zeta(x,t)\partial_x\zeta(0,0)\rangle_{L<+\infty}\!-\!(\pi n_0)^2\nonumber\\
&&+\frac{1}{4}\sum_{q\neq 0}\frac{2\pi K}{|q|L}q^2\{e^{i[qx-\omega(q)t]}+e^{-i[qx-\omega(q)t]}\}n_B(q).
\end{eqnarray}
The result at finite size and zero temperature is easily evaluated and yields Eq.~(\ref{nnfftL}). After a few lines of algebra I find that the second part reads
\begin{eqnarray}
\frac{1}{4}\sum_{q\neq 0}\frac{2\pi K}{|q|L}q^2\{e^{i[qx-\omega(q)t]}+e^{-i[qx-\omega(q)t]}\}n_B(q)=\frac{\pi K}{2L}[F(x-v_st)+F(x+v_st)].
\end{eqnarray}
Function $F$ reads
\begin{eqnarray}
&&F(u)=\sum_{q>0}q(e^{iqu}+e^{-iqu})\sum_{n=1}^{+\infty}e^{-L_T qn}\nonumber\\
&&=-\sum_{n=1}^{+\infty}\frac{1}{L_T}\frac{\partial}{\partial n}\left[\sum_{m=1}^{+\infty}\left(e^{i\frac{2\pi}{L}mu}+e^{-i\frac{2\pi}{L}mu}\right)e^{-L_T \frac{2\pi}{L}mn}\right]\nonumber\\
&&=-\frac{\pi}{2L}\sum_{n=1}^{+\infty}\left\{\frac{1}{\sinh^2\left[\frac{\pi}{L}(iu-L_T n)\right]}+\frac{1}{\sinh^2\left[\frac{\pi}{L}(iu+L_T n)\right]}\right\}\nonumber\\
&&=-\frac{\pi}{L}\sum_{n=1}^{+\infty}\frac{\cos^2\left(\frac{\pi u}{L}\right)\sinh^2\left(\frac{\pi L_T}{L} n\right)-\sin^2\left(\frac{\pi u}{L}\right)\cosh^2\left(\frac{\pi L_T}{L} n\right)}{\left[\cos^2\left(\frac{\pi u}{L}\right)\sinh^2\left(\frac{\pi L_T}{L} n\right)+\sin^2\left(\frac{\pi u}{L}\right)\cosh^2\left(\frac{\pi L_T}{L} n\right)\right]^2}\nonumber\\
&&=-\frac{\pi}{L}\sum_{n=1}^{+\infty}\frac{2[\cos\left(\frac{2\pi u}{L}\right)\cosh\left(\frac{2\pi L_T}{L}n\right)-1]}{\left[\cosh\left(\frac{2\pi L_T}{L}n\right)-\cos\left(\frac{2\pi u}{L}\right)\right]^2}\nonumber\\
&&=-\frac{\pi}{2L}\left\{-\frac{1}{\sin^2\left(\frac{\pi u}{L}\right)}+\left[\frac{\theta_1''\left(\frac{\pi u}{L},e^{-\frac{\pi L_T}{L}}\right)}{\theta_1\left(\frac{\pi u}{L},e^{-\frac{\pi L_T}{L}}\right)}-\frac{\theta_1'\left(\frac{\pi u}{L},e^{-\frac{\pi L_T}{L}}\right)^2}{\theta_1\left(\frac{\pi u}{L},e^{-\frac{\pi L_T}{L}}\right)^2}\right]\right\},
\end{eqnarray}
yielding the first term of Eq.~(\ref{nnfftTL}).

The second contribution is obtained by adapting calculation tricks used to evaluate the autocorrelation function of the wavefunction in \cite{Mattson1997, Mattson1997bis}.
To obtain the generating function, it is no more possible to transform the sums into integrals, so I shall evaluate
\begin{eqnarray}
&&\sum_{q\neq 0}\frac{1}{|q|}[(e^{i[qx-\omega(q)t]}-1)(1+n_B(q))+(e^{-i[qx-\omega(q)t]}-1)n_B(q)]\nonumber\\
&&=F_1(q,u)+F_1(q,-v),
\end{eqnarray}
where
\begin{eqnarray}
F_1(q,u)=\sum_{q>0}\frac{1}{q}\left[(e^{iqu}-1)\left(1+\frac{1}{e^{L_Tq}-1}\right)+(e^{-iqu}-1)\frac{1}{e^{L_T q}-1}\right].
\end{eqnarray}
Denoting $a=e^{-i\frac{2\pi}{L}v}$ and $b=e^{2\pi\frac{L_T}{L}n}$,
\begin{eqnarray}
F_1(q,-v)&&=\frac{L}{2\pi}\sum_{n=1}^{+\infty}\frac{1}{n}\left[\frac{a^n-1}{1-(b^{-1})^n}+\frac{a^{-n}-1}{b^n-1}\right]\nonumber\\
&&=\sum_{n=1}^{+\infty}\sum_{k=0}^{+\infty}(b^{-k})^n\frac{1}{n}\left[(a^n-1)+(a^{-n}-1)b^{-n}\right]\nonumber\\
&&=\sum_{k=0}^{+\infty}\sum_{n=1}^{+\infty}\frac{1}{n}[(ab^{-k})^n-(b^{-k})^n]+\sum_{k=1}^{+\infty}\sum_{n=1}^{+\infty}\frac{1}{n}[(a^{-1}b^{-k})^n-(b^{-k})^n]\nonumber\\
&&=\lim_{c\to 1}\sum_{n=1}^{+\infty}\frac{1}{n}c^n(a^n-1)+\sum_{k=1}^{+\infty}\sum_{n=1}^{+\infty}\frac{1}{n}[(ab^{-k})^n-(b^{-k})^n+(a^{-1}b^{-k})^n-(b^{-k})^n]\nonumber\\
&&=T_1+T_2.
\end{eqnarray}
Using
\begin{eqnarray}
\sum_{n=1}^{+\infty}\frac{z^n}{n}=-\ln(1-z),\, |z|<1,
\end{eqnarray}
and doing a bit of algebra, I find
\begin{eqnarray}
&&-T_2=\sum_{k=1}^{+\infty}\left[\ln(1-ab^{-k})-\ln(1-b^{-k})+\ln(1-a^{-1}b^{-k})-\ln(1-b^{-k})\right]\nonumber\\
&&=\ln\left[\prod_{k=1}^{+\infty}\frac{(1-ab^{-k})(1-a^{-1}b^{-k})}{(1-b^{-k})^2}\right]\nonumber\\
&&=\ln\left\{\prod_{k=1}^{+\infty}\left[1+\frac{\sin^2\left(\frac{\pi v}{L}\right)}{\sinh^2\left(\frac{L_T}{L}\pi k\right)}\right]\right\},
\end{eqnarray}
and the property
\begin{eqnarray}
\prod_{k=1}^{+\infty}\left[1\!+\!\frac{\sin^2(\pi z)}{\sinh^2(k\pi\lambda)}\right]=\frac{\theta_1(\pi z,e^{-\pi\lambda})}{\sin(\pi z)\theta_1'(0,e^{-\pi\lambda})}
\end{eqnarray}
allows to conclude after straightforward algebraic manipulations.

\chapter{Complements to chapter \ref{secIII}}

\section{Exact mapping from the Lieb-Liniger model onto the circular plate capacitor}
\label{capa}

In this appendix I illustrate an exact mapping from the Lieb-Liniger model of delta-interacting bosons in 1D discussed in the main text, onto system of classical physics. Historically, both have beneficiated from each other, and limit cases can be understood in different ways according to the context.

Capacitors are emblematic systems in electrostatics lectures. On the example of the parallel plate ideal capacitor, one can introduce various concepts such as symmetries of fields or Gauss' law, and compute the capacitance in a few lines from basic principles, under the assumption that the plates are infinite. To go beyond this approximation, geometry must be taken into account to include edge effects, as was realized by Clausius, Maxwell and Kirchhoff in pioneering works \cite{Clausius, Maxwell, Kirchhoff}. This problem has a huge historical significance in the history of science, since it stimulated the foundation of conformal analysis by Maxwell. The fact that none of these giants managed to solve the problem in full generality, nor anyone else a century later, hints at its tremendous technical difficulty.

Actually, the exact capacitance of a circular coaxial plate capacitor with a free space gap as dielectrics, as a function of the aspect ratio of the cavity $\alpha=d/R$, where $d$ is the distance between the plates and $R$ their radius, reads \cite{Carlson}
\begin{eqnarray}
\label{Ceq}
C(\alpha,\lambda)=2\epsilon_0 R\int_{-1}^1\!dz\,g(z;\alpha,\lambda),
\end{eqnarray}
where $\epsilon_0$ is the vacuum permittivity, $\lambda\!=\!\pm 1$ in the case of equal (respectively opposite) disc charge or potential and $g$ is the solution of the Love equation \cite{Love, Lovebis}
\begin{eqnarray}
\label{Loveeq}
g(z;\alpha,\lambda)=1+\frac{\lambda\alpha}{\pi}\int_{-1}^1\!dy\,\frac{g(y;\alpha,\lambda)}{\alpha^2+(y-z)^2},\,-1\leq z\leq 1.
\end{eqnarray}
This equation turns out to become the Lieb equation (\ref{Fredholm}) when $\lambda\!=\!1$, as first noticed by Gaudin \cite{Gaudin}, and maps onto the super Tonks-Girardeau regime when $\lambda\!=\!-1$. However, the relevant physical quantities are different in the two problems, and are obtained at different steps of the resolution.

In what follows, I shall only consider the case of equally charged discs. At small $\alpha$, i.e. at small gap, using the semi-circular law Eq.~(\ref{guessLieb}), one finds
\begin{eqnarray}
\label{Contactapprox}
C(\alpha)\simeq_{\alpha\ll 1} \frac{\pi \epsilon_0R}{\alpha}\!=\!\frac{\epsilon_0 A}{d},
\end{eqnarray}
where $A$ is the area of a plate, as directly found in the contact approximation.
On the other hand, if the plates are separated from each other and carried up to infinity (this case would correspond to the Tonks-Girardeau regime in the Lieb-Liniger model), one finds $g(z;+\infty)=1$ and thus
\begin{eqnarray}
\label{Capinfinity}
C(\alpha\!\to\!+\infty)\!=\!4\epsilon_0R.
\end{eqnarray}
This result can be understood as follows: at infinite gap, the two plates do not feel each others anymore, and can be considered as two one-plate capacitors in series. The capacitance of one plate is $8\epsilon_0R$ from an electrostatic treatment, and the additivity of inverse capacitances in series then yields the awaited result.

At intermediate distances, one qualitatively expects that the capacitance is larger than the value found in the contact approximation, due to the effect of the fringing electric field outside the cavity delimited by the two plates. The contact approximation shall thus yield a lower bound for any value of $\alpha$, which is in agreement with the property (v) of the main text.

The main results and conjectures in the small gap regime beyond the contact approximation \cite{Hutson, Ignatowski, Polya, Leppington, Atkinson} are summarized and all encompassed in the most general form \cite{Soibelmann}:
\begin{eqnarray}
\label{CSoibelmann}
\mathcal{C}(\epsilon)=\frac{1}{8\epsilon}+\frac{1}{4\pi}\log\left(\frac{1}{\epsilon}\right)+\frac{\log(8\pi)-1}{4\pi}+\frac{1}{8\pi^2}\epsilon\log^2(\epsilon)+\sum_{i=1}^{+\infty}\epsilon^i\sum_{j=0}^{2i}c_{ij}\log^j(\epsilon),
\end{eqnarray}
where traditionally in this community, $2\epsilon\!=\!\alpha$, and $\mathcal{C}\!=\!C/(4\pi\epsilon_0R)$ represents the geometrical capacitance. It is known that $c_{12}\!=\!0$ \cite{Soibelmann}, but higher-order terms are not explicitly known. It has also been shown that
\begin{eqnarray}
\mathcal{C}_{\leq}(\epsilon)=\frac{1}{8\epsilon}+\frac{1}{4\pi}\log\left(\frac{1}{\epsilon}\right)+\frac{\log(4)-\frac{1}{2}}{4\pi}
\end{eqnarray}
is a sharp lower bound \cite{Polya}.

At large $\alpha$, i.e. for distant plates, many different techniques have been considered. Historically, Love used the iterated kernel method. Injecting the right-hand side of Eq.~(\ref{Loveeq}) into itself and iterating, the solution is expressed as a Neumann series \cite{Love}
\begin{eqnarray}
\label{iter}
g(z;\alpha,\lambda)\!=\!1\!+\!\sum_{n=1}^{+\infty}\lambda^n\!\int_{-1}^1dy\,K_n(y\!-\!z)\!=\!\sum_{n=0}^{+\infty}\lambda^ng_n^{I}(z;\alpha)
\end{eqnarray}
with $K_1(x;\alpha)\!=\!\frac{\alpha}{\pi}\frac{1}{\alpha^2+x^2}$ the kernel of Love's equation, which is a Cauchy law (or a Lorentzian), and $K_{n+1}(y\!-\!z;\alpha)\!=\!\int_{-1}^1dx\,K_1(y\!-\!x;\alpha)K_n(x\!-\!z;\alpha)$ is the $(n\!+\!1)$-th order iterated kernel.

It follows from the positivity of the Lorentzian kernel and linearity of the integral that for repulsive plates, $g(z;\alpha,+1)\!>\!1$, yielding a global lower bound in agreement with the physical discussion above. One also finds that $g(z;\alpha,-1)<1$. Approximate solutions are then obtained by truncation to a given order.
One easily finds that
\begin{eqnarray}
 g_1^{I}(z;\alpha)=\frac{1}{\pi}\left[\arctan\left(\frac{1-z}{\alpha}\right)\!+\!\arctan\left(\frac{1+z}{\alpha}\right)\right],
\end{eqnarray}
yielding
\begin{eqnarray}
C_1^{I}(\alpha)=2\epsilon_0R\!\left[4\arctan\!\left(\frac{2}{\alpha}\right)\!+\!\alpha\!\log\left(\frac{\alpha^2}{\alpha^2\!+\!4}\right)\right],
\end{eqnarray}
where $C(\alpha,\lambda)\!=\!\sum_{n=0}^{+\infty}\lambda^nC_n^I(\alpha)$.

Higher orders are cumbersome to evaluate exactly, which is a strong limitation of this method in view of an analytical treatment. Among alternative ways to tackle the problem, I mention Fourier series expansion \cite{Carlson, Norgren}, and those based on orthogonal polynomials \cite{Milovanovic}, that allowed to find the exact expansion of the capacitance to order $9$ in $1/\alpha$ in \cite{Rao} for identical plates, anticipating \cite{Ristivojevic} in the case of Lieb-Liniger model.

\begin{figure}
\includegraphics[width=10cm, keepaspectratio, angle=0]{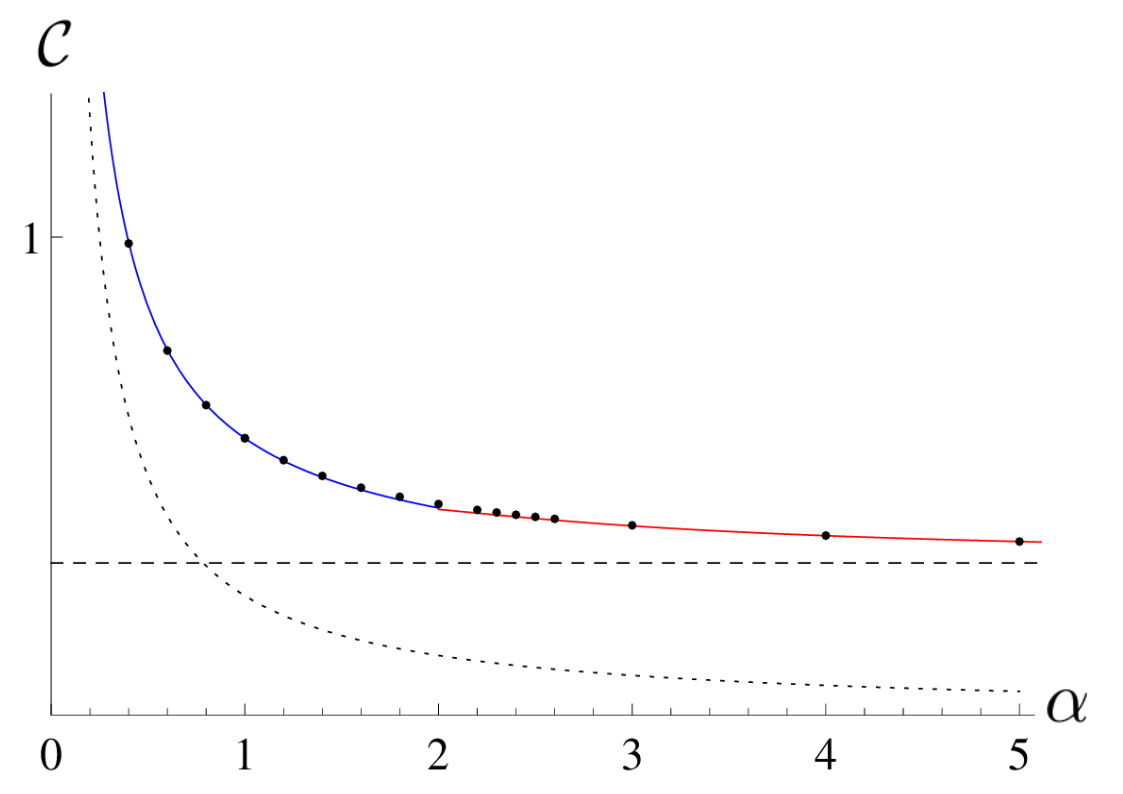}

\caption{Geometrical dimensionless capacitance $\mathcal{C}$ of the parallel plate capacitor as a function of its dimensionless aspect ratio $\alpha$. Results at infinite gap (dashed, black) and in the contact approximation (dotted, black) are rather crude compared to the more sophisticated approximate expressions from Eq.~(\ref{CSoibelmann}) for $\alpha\!<\!2$ and Eq.~(\ref{Cus}) for $\alpha\!>\!2$ (solid, blue and red respectively), when compared to numerical solution of Eqs.~(\ref{Ceq}) and (\ref{Loveeq}) (black dots).} 
\label{FigC}
\end{figure}

In Fig.~(\ref{FigC}), I show several approximations of the geometric capacitance as a function of the aspect ratio $\alpha$. In particular, based on an analytical asymptotic expansion, I have proposed a simple approximation in the large gap regime, namely
\begin{eqnarray}
\label{Cus}
\mathcal{C}(\alpha)\simeq_{\alpha\gg 1}\frac{1}{\pi}\frac{1}{1-2/(\pi\alpha)}-\frac{4}{3\pi^2\alpha^3}\frac{1}{[1-2/(\pi\alpha)]^2}.
\end{eqnarray}

\section{Ristivojevic's method of orthogonal polynomials}
\label{Ristimethods}

In this appendix, I detail Ristivojevic's method, that allows to systematically find approximate solutions to Eq. (\ref{Fredholm}) in the strongly-interacting regime \cite{Ristivojevic, LangHekkingMinguzzi2017}. First, let me recall a few qualitative features of the density of pseudo-momenta, $g(z;\alpha)$. At fixed $\alpha$, $g$ as a function of $z$ is positive, bounded, unique and even. Moreover, it is analytic provided $\alpha\!>\!0$.

Since $g$ is an analytic function of $z$ on the compact $[-1,1]$, at fixed $\alpha$ it can be written as
\begin{eqnarray}
g(z;\alpha)\!=\!\sum_{n=0}^{+\infty}a_n(\alpha)Q_n(z),
\end{eqnarray}
where $\{a_n\}_{n\geq 0}$ are unknown analytic functions and $\{Q_n\}_{n\geq 0}$ are polynomials of degree $n$.

To solve the set of equations (\ref{Fredholm}), (\ref{alphagamma}) and (\ref{energy}), one only needs the values of $g$ for $ z\in [-1,1]$. Thus, a convenient basis for the $Q_n$'s is provided by the Legendre polynomials, defined as
\begin{eqnarray}
P_n(X)\!=\!\frac{(-1)^n}{2^nn!}\left(\frac{d}{dX}\right)^n\![(1\!-\!X^2)^n],
\end{eqnarray}
that form a complete orthogonal set in this range. Furthermore, the Legendre polynomial $P_n$ is of degree $n$ and consists in sums of monomials of the same parity as $n$, so that, since $g$ is an even function of $z$,
\begin{eqnarray}
\label{Legendre}
g(z;\alpha)\!=\!\sum_{n=0}^{+\infty}a_{2n}(\alpha)P_{2n}(z).
\end{eqnarray}
Under the assumption that $\alpha\!>\!2$, since $(y,z)\!\in\![-1,1]^2$, the Lorentzian kernel in Eq. (\ref{Fredholm}) can be expanded as:
\begin{eqnarray}
\label{kernelalphabig}
&&\frac{1}{\pi}\frac{\alpha}{\alpha^2\!+\!(y-z)^2}=\frac{1}{\pi}\sum_{k=0}^{+\infty}\frac{(-1)^k}{\alpha^{2k+1}}\sum_{j=0}^{2k}\binom{2k}{j}y^j(-1)^{j}z^{2k-j}.
\end{eqnarray}
Thus, the combination of Eqs. (\ref{Fredholm}), (\ref{Legendre}) and (\ref{kernelalphabig}) yields
\begin{eqnarray}
 \sum_{n=0}^{+\infty}a_{2n}(\alpha)\!\left[P_{2n}(z)-\frac{1}{\pi}\sum_{k=0}^{+\infty}\frac{(-1)^k}{\alpha^{2k+1}}\sum_{j=0}^{2k}\binom{2k}{j}(-1)^{j}z^{2k-j}\int_{-1}^{1}dy\,y^jP_{2n}(y)\right]\!=\!\frac{1}{2\pi}.
\end{eqnarray}
Following \cite{Ristivojevic}, I introduce the notation
\begin{eqnarray}
F^j_{2n}\!=\!\int_{-1}^1dy\,y^jP_{2n}(y). 
\end{eqnarray}
Due to the parity, $F^j_{2n}\neq 0$ if and only if $j$ is even. An additional condition is that $j\geq n$ \cite{Gradshteyn}. Taking it into account and renaming mute parameters ($k\leftrightarrow n$) yields
\begin{eqnarray}
\label{Interm}
 \sum_{n=0}^{+\infty}\left[a_{2n}(\alpha)P_{2n}(z)-\frac{1}{\pi}\sum_{j=0}^{n}\sum_{k=0}^j\frac{(-1)^n}{\alpha^{2n+1}}a_{2k}(\alpha)\binom{2n}{2j}z^{2(n-j)}F^{2j}_{2k}\right]=\frac{1}{2\pi}.
\end{eqnarray}
The properties of orthogonality and normalization of Legendre polynomials,
\begin{eqnarray}
\int_{-1}^1\!dz\,P_{2i}(z)P_{2j}(z)\!=\!\delta_{i,j}\frac{2}{4j\!+\!1},
\end{eqnarray}
allow to go further. Doing $\int_{-1}^1dz\,P_{2m}(z)\times$ Eq.~(\ref{Interm}) yields:
\begin{eqnarray}
\sum_{n=0}^{+\infty}\left[a_{2n}(\alpha)\delta_{m,n}\frac{2}{4m\!+\!1}-\frac{1}{\pi}\sum_{j=0}^{n}\sum_{k=0}^j\frac{(-1)^n}{\alpha^{2n+1}}a_{2k}(\alpha)\binom{2n}{2j}F^{2j}_{2k}F^{2(n-j)}_{2m}\right]\!=\!\frac{1}{2\pi}F^0_{2m}.
\end{eqnarray}
or after $n-m\to n$:
\begin{eqnarray}
\label{nearlyfinal}
\frac{2a_{2m}(\alpha)}{4m\!+\!1}-\frac{1}{\pi}\sum_{n=0}^{+\infty}\sum_{j=0}^{n}\sum_{k=0}^j\frac{(-1)^{n+m}}{\alpha^{2(n+m)+1}}a_{2k}(\alpha)\binom{2(n+m)}{2j}F^{2j}_{2k}F^{2(n+m-j)}_{2m}\!=\!\frac{1}{2\pi}F^0_0\delta_{m,0}.
\end{eqnarray}
Then, from equation 7.231.1 of \cite{Gradshteyn} and after a few lines of algebra,
\begin{eqnarray}
\label{Fml}
F_{2m}^{2l}\!=\!\frac{2^{2m+1}(2l)!(l+m)!}{(2l+2m+1)!(l-m)!}.
\end{eqnarray}
Inserting Eq.~(\ref{Fml}) into Eq.~(\ref{nearlyfinal}) yields after a few simplifications:
\begin{eqnarray}
\label{finaleq}
\frac{2a_{2m}(\alpha)}{4m+1}-\frac{1}{\pi}\sum_{n=0}^{+\infty}\sum_{j=0}^{n}\sum_{k=0}^j\frac{(-1)^{n+m}}{\alpha^{2(n+m)+1}}C_{m,n,j,k}a_{2k}(\alpha)=\frac{1}{\pi}\delta_{m,0}
\end{eqnarray}
where
\begin{eqnarray}
C_{m,n,j,k}=\frac{2^{2k+1}(j+k)!}{(2j+2k+1)!(j-k)!}\frac{2^{2m+1}(n+2m-j)!(2n+2m)!}{(2n+4m-2j+1)!(n-j)!}.
\end{eqnarray}
To obtain a finite set of equations, I cut off the series in Eq.~(\ref{Legendre}) at an integer value $M\!\geq\!0$. The infinite set of equations (\ref{finaleq}) truncated at order $M$ can then be recast into a matrix form:
\begin{eqnarray}
  \begin{bmatrix}
A
\end{bmatrix}
  \begin{bmatrix}
a_0 \\
a_2 \\
\vdots\\
a_{2M}
\end{bmatrix}
=
  \begin{bmatrix}
\frac{1}{\pi} \\
0 \\
\vdots\\
0
\end{bmatrix},
\end{eqnarray}
where $A$ is a $(M\!+\!1)\!\times\!(M\!+\!1)$ square matrix, inverted to find the set of coefficients $\{a_{2n}(\alpha)\}_{0\leq n\leq M}$. Actually, one only needs to compute $(A^{-1})_{i1}$, $\forall i\in \{1,\dots, M\!+\!1\}$, and combine to Eq. (\ref{Legendre}) to obtain the final result at order $M$. For full consistency with higher orders, one shall expand the result in $1/\alpha$ and truncate it at order $2M\!+\!2$, and to order $2M$ in $z$.

\section{A general method to solve the Lieb equation}
\label{Fabrikantmethod}
In this appendix I explain a method that I have developed to solve the Lieb equation (\ref{Fredholm}). Contrary to Ristivojevic's method presented in the previous appendix, it works at arbitrary coupling, as it does not rely on a strong-coupling expansion of the kernel. However, it also starts with a series expansion of the density of pseudo-momenta,
\begin{eqnarray}
g(z;\alpha)\!=\!\sum_{n=0}^{+\infty}c_{2n}(\alpha)z^{2n},
\end{eqnarray}
injected in the integral equation (\ref{Fredholm}) to transform the latter into an infinite set of algebraic equations for the coefficients $c_{2n}$. To do so, the series expansion of the integral 
\begin{eqnarray}
I_n(\alpha)\!=\!\int_{-1}^1\!dy\,\frac{y^{2n}}{\alpha^2\!+\!(y-z)^2}
\end{eqnarray}
must be known explicitly. $I_n(\alpha)$ is expressed in terms of hypergeometric functions in \cite{Fabrikant1987}. After a series of algebraic transformations detailed in Ref.~\cite{LangHekkingMinguzzi2017}, this integral reads
\begin{eqnarray}
I_n(\alpha)\!\!\!\!&=&\!\!-\frac{1}{\alpha}\sum_{k=0}^{I\left[\frac{2n-1}{2}\right]}\binom{2n}{2k+1}(-1)^k\alpha^{2k+1}z^{2(n-k)-1}\mathrm{argth}\!\left(\frac{2z}{1+z^2+\alpha^2}\right)\nonumber\\
&&\!\!+\frac{1}{\alpha}\left[\arctan\left(\frac{1+z}{\alpha}\right)+\arctan\left(\frac{1-z}{\alpha}\right)\right]\sum_{k=0}^n(-1)^k\binom{2n}{2k}(-1)^k\alpha^{2k}z^{2n-2k}\nonumber\\
&&\!\!+2\sum_{m=0}^{n-1}\sum_{k=0}^m\binom{2m+1}{2k+1}\frac{1}{2n\!-\!2m\!-\!1}(-1)^k\alpha^{2k}z^{2(m-k)},
\end{eqnarray}
then recast into the form
\begin{eqnarray}
I_n(\alpha)\!=\!\sum_{i=0}^{+\infty}d_{2i,n}(\alpha)z^{2i}
\end{eqnarray}
using properties of the Taylor expansions of the functions involved. This tranforms the Lieb equation into
\begin{eqnarray}
\sum_{n=0}^{+\infty}c_{2n}(\alpha)\left[z^{2n}\!-\!\frac{\alpha}{\pi}\sum_{i=0}^{+\infty}d_{2i,n}(\alpha)z^{2i}\right]\!=\!\frac{1}{2\pi}.
\end{eqnarray}
To finish with, the series is truncated to order $M$ in a self-consistent way, to obtain the following set of $M$ linear equations:
\begin{eqnarray}
c_{2n;M}(\alpha)-\frac{\alpha}{\pi}\sum_{i=0}^{M}d_{2n,i}(\alpha)c_{2i;M}(\alpha)=\delta_{n,0}\frac{1}{2\pi},
\end{eqnarray}
where the unknowns are the coefficients $c_{2i;M}(\alpha)$. The solution yields approximate expressions for a truncated polynomial expansion of $g$ in $z$. Since the method is not perturbative in $\alpha$, each coefficient converges faster to its exact value than with the orthogonal polynomial method. I have calculated these coefficients up to order $M\!=\!25$.

\section{Other approaches to local correlation functions}
\label{gkk}

In this appendix, I analyze other approaches to obtain the local correlation functions of a $\delta$-interacting 1D Bose gas. They on mappings from special limits of other models onto the Lieb-Liniger model. The original models are the sinh-Gordon quantum field theory in a non-relativistic limit, the XXZ spin chain in a continuum limit, and $q$-bosons when $q\!\to\!1$.

In the case of the sinh-Gordon model, a special non-relativistic, weak-coupling limit, that keeps the product of the speed of light and coupling constant unchanged, maps its $S$-matrix onto the Lieb-Liniger one, leading to an exact equivalence \cite{KormosMussardoTrombettoni}. In particular, their correlation functions are formally identical \cite{KormosMussardoTrombettoni2010}, with a subtle difference: they refer to the ground state for the Lieb-Liniger model and to the vacuum in the case of the sinh-Gordon model, where the form factors are computed within the Le Clair-Mussardo formalism \cite{LeClairMussardo1999}. Expressions obtained perturbatively in previous works have been resummed non-perturbatively in \cite{KormosChouImambekov2011}, yielding the exact second- and third-order local correlation functions in terms of solutions of integral equations.

A major strength of this approach is that the expressions thereby obtained are also valid at finite temperature and in out-of-equilibrium situations. Another one, though it is rather a matter of taste, is that no derivative is involved, contrary to the direct Bethe Ansatz approach through moments of the density of quasi-momenta, making numerical methods less cumbersome, and more accurate. On the other hand, within this method one needs to solve several integral equations, whose number increases with the order of the correlation function.

The LeClair-Mussardo formalism has also been extended to the non-relativistic limit, that can thus be addressed directly in this formalism, without invoking the original sinh-Gordon model anymore \cite{Pozsgay2011Jstat}. In the same reference, it has also been shown that the LeClair-Mussardo formalism can be derived from algebraic Bethe Ansatz.

Then, additional results have been obtained, based on the exact mapping from a special continuum limit of the XXZ spin chain onto the Lieb-Liniger model \cite{GolzerHolz, Seel2007}. In \cite{Pozsgay2011Jstatmech}, multiple integral formulae for local correlation functions have been obtained, that encompass the previous results from the sinh-Gordon model approach for $g_2$ and $g_3$, and provide the only known expression for $g_4$ to date. This formalism also yields higher-order correlations as well, but no systematic method to construct them has been provided yet.


\section{Density profile of a trapped Tonks-Girardeau gas: LDA and exact result}
\label{LDATG}
The aim of this appendix is to show the equivalence, in the thermodynamic limit, between the density profile predicted by the local-density approximation, and the exact Tonks-Girardeau result obtained by Bose-Fermi mapping. Writing
\begin{eqnarray}
\sigma(x;N)\!=\!\sum_{j=0}^{N-1}\phi_j^2(x),
\end{eqnarray}
where
\begin{eqnarray}
\phi_j(x)=(2^jj!\sqrt{\pi})^{1/2}\exp\left(\frac{x^2}{2}\right)\left(-\frac{d}{dx}\right)^j\exp\left(-x^2\right),
\end{eqnarray}
I shall prove that
\begin{eqnarray}
\sigma(x;N)\sim_{N\to +\infty}\frac{1}{\pi}\sqrt{2N\!-\!x^2}.
\end{eqnarray}
I recall the orthogonality and normalization condition:
\begin{eqnarray}
\int_{-\infty}^{+\infty}\phi_j(x)\phi_k(x)dx=\delta_{j,k}.
\end{eqnarray}
A first derivation based on physical arguments proceeds as follows \cite{Mehta}: $\phi_j$ is the normalized oscillator function, thus $\phi_j^2(x)\,dx$ represents the probability that an oscillator in the $j$-th state is in $[x,x\!+\!dx]$. When $N$ is large, $\sigma$ represents the density of particles at $x$. Since I consider fermions (using the Bose-Fermi mapping) at $T\!=\!0$, there is at most one particle per state and all states are filled up to the Fermi energy. I recall that the differential equation satisfied by $\phi_{N-1}$ is
\begin{eqnarray}
\hbar^2\frac{d^2}{dx^2}\phi_{N-1}(x)+\hbar^2(2N\!-\!1\!-\!x^2)\phi_{N-1}(x)\!=\!0.
\end{eqnarray}
The latter also reads
\begin{eqnarray}
-p_F^2+(2N\!-\!1\!-\!x^2)\hbar^2\!=\!0,
\end{eqnarray}
and in 1D,
\begin{eqnarray}
\sigma(x)\simeq \frac{1}{2\pi\hbar}\int_{-p_F}^{p_F}\!dp,
\end{eqnarray}
yielding the result. A more formal derivation relies on the Christoffel-Darboux formula:
\begin{eqnarray}
\sum_{j=0}^{N-1}\phi_j(x)\phi_j(y)=\left(\frac{N}{2}\right)^{1/2}\left[\frac{\phi_N(x)\phi_{N-1}(y)-\phi_N(y)\phi_{N-1}(x)}{x-y}\right],
\end{eqnarray}
or
\begin{eqnarray}
\sum_{j=0}^{N-1}\phi_j^2(x)=N\phi_N^2(x)-[N(N+1)]^{1/2}\phi_{N-1}(x)\phi_{N+1}(x),
\end{eqnarray}
and the asymptotics of $\phi_{N-1}$, $\phi_N$ and $\phi_{N+1}$.

\chapter{Complements to chapter \ref{secIV}}

\section{Around the notion of dynamical structure factor}
\label{linrespDSFF}
This appendix, mostly based on \cite{Pottier}, gives more details about the notion of dynamical structure factor.




A common method to carry out measurements on a physical system is to submit it to an external force and to observe the way it reacts. For the result of such an experiment to adequately reflect the properties of the system, the perturbation due to the applied force must be sufficiently weak. In this framework, three distinct types of measurement can be carried out: actual response measurements, susceptibility measurements that consist in determining the response of a system to a harmonic force, and relaxation measurements in which, after having removed a force that had been applied for a very long time, one studies the return to equilibrium. The results of these three types of measurements are respectively expressed in terms of response functions, generalized susceptibilities and relaxation functions. In the linear range, these quantities depend solely on the properties of the unperturbed system, and each of them is related to the others.

The object of linear response theory is to allow, for any specific problem, to determine response functions, generalized susceptibilities, and relaxation functions. In the linear range, all these quantities can be expressed through equilibrium correlation funtions of the relevant dynamical variables of the unperturbed system. The corresponding expressions constitute the Kubo formulas.

Let us consider an inelastic scattering process in the course of which, under the effect of an interaction with radiation, a system at equilibrium undergoes a transition from an initial state $|i\rangle$ to a final state $|f\rangle$. The corresponding energy varies from $\epsilon_{i}$ to $\epsilon_{f}$, whereas the radiation energy varies from $E_i$ to $E_f$. Total energy conservation implies that $E_i+\epsilon_{i}=E_f+\epsilon_{f}$. The energy lost by the radiation is denoted by $\hbar \omega\!=\!E_i-E_f=\epsilon_{f}-\epsilon_{i}$, so that absorption corresponds to $\omega\!>\!0$ and emission to $\omega\!<\!0$. I associate an operator $A(r)$ to the system-radiation interaction. For instance, in the case of scattering of light by a fluid in equilibrium, the radiation is scattered by the density fluctuation of the fluid, and therefore the operator $A(r)$ is proportional to the local density fluctuation $\delta n(r)\!=\!n(r)\!-\!\langle n\rangle$.

An incident plane-wave initial state $|k_i\rangle$ is scattered to a final state which, in the framework of the Born approximation for scattering, is considered as a plane-wave state $|k_f\rangle$. The matrix element of this interaction operator between the two states is $\langle k_f|A(r)|k_i\rangle=\int e^{-ik_fr}A(r)e^{ik_ir}dr=A(-q)$, where $q=k_i-k_f$. At lowest perturbation order, the probability per unit time of the process $(|k\rangle,|i\rangle)\to (|k_f\rangle,|f\rangle)$ to occur is given by the Fermi golden rule, that reads
\begin{eqnarray}
W_{(k_f,f),(k_i,i)}=\frac{2\pi}{\hbar}|\langle f|A(-q)|i\rangle|^2\delta[\hbar\omega-(\epsilon_{f}-\epsilon_{i})].
\end{eqnarray}
The total probability per unit time of the process $|k_i\rangle \to |k_f\rangle$ is obtained by weighting $W_{(k_f,f),(k_i,i)}$ by the occupation probability $p_{i}$ of the initial state of the system at equilibrium, and by summing over all initial and final states:
\begin{eqnarray}
W_{(k_f,f),(k_i,i)}=\frac{2\pi}{\hbar}\sum_{i,f}p_{i}|\langle f|A(-q)|i\rangle|^2\delta[\hbar \omega -(\epsilon_{f}-\epsilon_{i})].
\end{eqnarray}
The dynamical structure factor is defined as $S(q,\omega)\!=\!\hbar^2W_{k_i,k_f}$, where $A(r)\!=\!\delta n(r)$. This yields the Lehmann representation of the dynamical structure factor, used for instance in the ABACUS code.

Introducing the Fourier representation of the delta function, $S(q,\omega)$ can be expressed as an autocorrelation function. Since $\delta n(r)$ is hermitian, $\langle f|\delta n(-q)|i\rangle^*=\langle i|\delta n(q)|f\rangle$, so that
\begin{eqnarray}
S(q,\omega)&&=\int_{-\infty}^{+\infty}dt\sum_{i,f}p_{i}\langle i|e^{i\epsilon_{i}t/\hbar}\delta n(q)e^{-i\epsilon_{f}t/\hbar}|f\rangle\langle f|\delta n(-q)|i\rangle e^{i\omega t}\nonumber\\
&&=\int_{-\infty}^{+\infty}dt\,e^{i\omega t}\langle \delta n(q,t)\delta n(-q,0)\rangle\nonumber\\
&&=V \int dr\,e^{-iqr}\!\!\int_{-\infty}^{+\infty}\!\!dt\,e^{i\omega t}\delta n(r,t)\delta n(0,0).
\end{eqnarray}

\section{Dynamical structure factor of the\\
Tonks-Girardeau gas at zero temperature}
\label{derDSFTG}

In this appendix, I shall derive Eq.~(\ref{TonksDSF}). This derivation is quite elementary, but is a good occasion to make a few comments. According to Eq.~(\ref{DSF}), the dynamical structure factor is the Fourier transform of the autocorrelation function of density fluctuations. For the homogeneous gas considered here, the mean value of the density fluctuations is null, thus
\begin{equation}
 \langle \delta n(x,t) \delta n(x',t')\rangle=\langle n(x,t) n(x',t')\rangle-n_0^2.
\end{equation}
I have evaluated the time-dependent density-density correlation of the Tonks-Girardeau gas in chapter \ref{secII}. The strategy based on the brute-force integration of the latter to obtain the dynamical structure factor is quite tedious, I am not even aware of any work where this calculation has been done, nor did I manage to perform it.

In comparison, making a few transformations before doing the integration considerably simplifies the problem. According to the Bose-Fermi mapping, the model can be mapped onto a fictitious one-dimensional gas of noninterating spinless fermions. The fermionic field is expressed in second quantization in terms of fermionic annihilation operators $c_k$ as
\begin{eqnarray}
\psi(x)\!=\!\frac{1}{\sqrt{L}}\sum_ke^{ikx}c_k,
\end{eqnarray}
whose time-dependence in the Schr\"odinger picture is obtained as 
\begin{eqnarray}
 \psi(x,t)\!=\!e^{iHt/\hbar}\psi(x)e^{-iHt/\hbar},
\end{eqnarray}
where the Hamiltonian is
\begin{eqnarray}
H\!=\!\sum_k\frac{\hbar^2k^2}{2m}c_k^{\dagger}c_k=\sum_k \hbar \omega_k c_k^{\dagger}c_k,
\end{eqnarray}
and computed using the equation of motion or the Baker-Campbell-Hausdorff formula, that yields
\begin{eqnarray}
\psi(x,t)\!=\!\frac{1}{\sqrt{L}}\sum_ke^{i(kx-\omega_k t)}c_k.
\end{eqnarray}
Then, since $n(x,t)\!=\!\psi^{\dagger}(x,t)\psi(x,t)$,  Wick's theorem, that applies to noninteracting fermions, yields
\begin{eqnarray}
\langle n(x,t) n(0,0)\rangle&=&\langle \psi^{\dagger}(x,t)\psi(0,0)\rangle \langle \psi(x,t)\psi^{\dagger}(0,0)\rangle\nonumber\\
&+&\langle \psi^{\dagger}(x,t)\psi(x,t)\rangle \langle \psi^{\dagger}(0,0)\psi(0,0)\rangle\nonumber\\
&-&\langle \psi^{\dagger}(x,t)\psi^{\dagger}(0,0)\rangle \langle \psi(0,0)\psi(x,t)\rangle,
\end{eqnarray}
and the last term is null. Then, using the expansion of the fermionic field over fermionic operators, the property $\langle c_k^{\dagger}c_k\rangle\!=\!n_F(k)\delta_{k,k'}$, fermionic commutation relations, $\Theta(-x)\!=\!1\!-\!\Theta(x)$ and specialization to $T\!=\!0$ yield
\begin{eqnarray}
S^{TG}(q,\omega)&=&\int_{-\infty}^{+\infty}\!\!dt\,e^{i\omega t}\!\int_{-\infty}^{+\infty}\!\!dx\,e^{-iqx}\frac{1}{L^2}\sum_{k,k'}e^{-i[(k-k')x-(\omega_k-\omega_{k'})t]}\Theta(k_F\!-\!|k|)\Theta(|k'|\!-\!k_F) \nonumber \\
&=&\frac{1}{L^2}\sum_{k,k'}\Theta(k_F\!-\!|k|)\Theta(|k'|\!-\!k_F)2\pi \delta(k'\!-\!k\!-\!q)2\pi\delta[\omega\!-\!(\omega_{k'}\!-\!\omega_k)].
\end{eqnarray}
Taking the thermodynamic limit, where the sums become integrals,
\begin{eqnarray}
S^{TG}(q,\omega)\!=\!\int dk\,\Theta(k_F\!-\!|k|)\Theta(|k\!+\!q|\!-\!k_F)\delta[\omega-(\omega_{k+q}-\omega_k)].
\end{eqnarray}
In this equation, the delta function is the formal description of energy conservation during the scattering process, which is elastic. The Heaviside distributions mean that the scattering process takes a particle out of the Fermi sea, creating a particle-hole pair.

Actually,
\begin{equation}
\omega_{k+q}-\omega_k=\frac{\hbar}{2m}(q^2+2qk),
\end{equation}
and one can split the problem into two cases.

If $q\!\geq\!2k_F$, then $k\in \left[-k_F,k_F\right]$ and the envelopes are $\frac{\hbar}{2m}(q^2\!-\!2k_F q)\!=\!\frac{\hbar}{2m}|q^2\!-\!2k_F q|\!=\!\omega_-$, and $\frac{\hbar}{2m}(q^2\!+\!2k_F q)\!=\!\omega_+$.

If $0\!\leq\!q\!\leq\!2k_F$, $k\!\in\!\left[k_F\!-\!q,k_F\right]$, and the envelopes are $\frac{\hbar}{2m}[q^2\!+\!2q(k_F-q)]=\frac{\hbar}{2m}(2qk_F-q^2)=\frac{\hbar}{2m}|q^2\!-\!2k_F q|\!=\!\omega_-$ and $\frac{\hbar}{2m}(q^2\!+\!2qk_F)\!=\!\omega_+$.

I evaluate the dynamical structure factor in the case $q\!\geq\!2k_F$:
\begin{equation}
S^{TG}(q,\omega)=\int_{-k_F}^{k_F}dk\,\delta\!\left[\omega-\frac{\hbar}{2m}(q^2\!+\!2qk)\right],
\end{equation}
and using the property of the Dirac distribution,
\begin{eqnarray}
\delta[f(k)]\!=\!\sum_{k_0|f(k_0)=0}\frac{1}{|f'(k_0)|}\delta(k\!-\!k_0),
\end{eqnarray}
by identification here $f'(k)\!=\!-\frac{\hbar q}{m}$ and $k_0\!=\!\frac{1}{2q}(\frac{2m\omega}{\hbar}\!-\!q^2)$.

It implies that $\frac{1}{2q}(\frac{2m\omega}{\hbar}\!-\!q^2)\in \left[-k_F,k_F\right]$, so $\omega\in \left[\omega_-,\omega_+\right]$, $S^{TG}(q,\omega)\!=\!\frac{m}{\hbar \left|q\right|}$ if and only if $\omega\in \left[\omega_-,\omega_+\right]$, $S^{TG}(q,\omega)\!=\!0$ otherwise.
The same conclusion holds if $q\in\left[0,2k_F\right]$, ending the derivation.

\section{Dynamical structure factor of a \\
Tomonaga-Luttinger liquid\\
in the thermodynamic limit}
\label{DSFTLLder}
In this appendix, I give a quite detailed derivation of the dynamical structure factor of a 1D Bose gas, Eq.~(\ref{DSFLL}), obtained from the Tomonaga-Luttinger liquid formalism. My main motivation is that, although the result is well known, details of calculations are scarce in the literature, and some technical aspects are relatively tricky. For clarity, I shall split the derivation into two parts, and evaluate separately the two main contributions, $S_0$ and $S_1$.

\subsection{First contribution: the phonon-like spectrum at the origin in energy-momentum space}
First, I focus on the term denoted by $S_0$ in Eq.~(\ref{DSFLL}), that corresponds to the linearized spectrum at low momentum and energy. According to Eq.~(\ref{nn}), this contribution stems from the terms that diverge on the 'light-cone', whose form factor is known. Their Fourier transform is quite easy, the only subtlety is the need of an additional infinitesimal imaginary part $i\epsilon$ to ensure convergence of the integral. Using the light-cone coordinate $u\!=\!x\!-\!v_st$, I find:
\begin{eqnarray}
F.T.\left[\frac{1}{(x-v_s t+i\epsilon)^2}\right]&&=\int_{-\infty}^{+\infty}dt\,e^{i\omega t}\int_{-\infty}^{+\infty}dx\,e^{-iqx}\frac{1}{(x-v_st+i\epsilon)^2}\nonumber\\
&&=\int_{-\infty}^{+\infty}dt\,e^{i(\omega-qv_s)t}\int_{-\infty}^{+\infty}du\,\frac{e^{-iqu}}{(u+i\epsilon)^2}\nonumber\\
&&=2\pi\delta(\omega-qv_s)\int_{-\infty}^{+\infty}du\,\frac{e^{-iqu}}{(u+i\epsilon)^2}.
\end{eqnarray}
Then, integration by parts and a classical application of the residue theorem on two circular contours in the upper and lower half of the complex plane yield
\begin{eqnarray}
\int_{-\infty}^{+\infty}du\,\frac{e^{-iqu}}{(u+i\epsilon)^2}=-iq\int_{-\infty}^{+\infty}du\,\frac{e^{-iqu}}{u+i\epsilon}
\end{eqnarray}
and
\begin{equation}
\int_{-\infty}^{+\infty}du\,\frac{e^{-iqu}}{(u+i\epsilon)^2}=-2\pi \Theta(q) q.
\end{equation}
In the end,
\begin{eqnarray}
F.T.\left[\frac{1}{(x\pm v_s t+i\epsilon)^2}\right]=-4\pi^2|q|\delta(\omega\!-\!|q|v_s)\Theta(\mp q),
\end{eqnarray}
whence I conclude that
\begin{eqnarray}
S_0(q,\omega)=-\frac{K}{4\pi^2}(-4\pi^2|q|)\delta(\omega-\omega(q))=K|q|\delta[\omega-\omega(q)],
\end{eqnarray}
with $\omega(q)\!=\!|q|v_s$. This result agrees with \cite{Cazalilla2004}
\begin{equation}
\mathcal{I}[\chi_{nn}(q,\omega)]=\frac{v_s K q^2}{2\hbar \omega(q)}\left\{\delta[\omega-\omega(q)]-\delta[\omega+\omega(q)]\right\},
\end{equation}
according to the fluctuation-dissipation theorem at $T\!=\!0$.

\subsection{Second contribution to the dynamical structure factor: the umklapp region}
The most interesting contribution to the dynamical structure factor is denoted by $S_1$ in Eq.~(\ref{DSFLL}), and corresponds to the Fourier transform of the main contribution to the density-density correlation in Eq.~(\ref{nn}), with non-trivial form factor. I shall detail the derivation of an arbitrary contribution, and specialize in the end to this main term.

\begin{figure}
\includegraphics[width=7cm, keepaspectratio, angle=0]{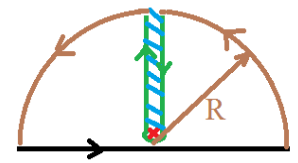}
\includegraphics[width=7cm, keepaspectratio, angle=0]{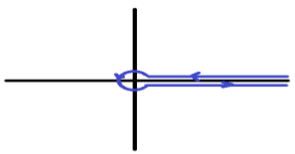}
\caption{The left panel represents the integration contour used in Eq.~(\ref{CC}), where the black line coincide with the x-axis, the branch-cut with the y-axis, and the pole (red) is at $i\epsilon/2$. It contains the H\"ankel contour defined in the right panel, rotated by $\pi/2$ and followed backward.}
\label{Hankel}
\end{figure}

The Fourier transform that should be evaluated here is
\begin{equation}
I(Km^2)\!=\!\int_{-\infty}^{+\infty}\!dt\,e^{i\omega t}\int_{-\infty}^{+\infty}\!dx\,e^{-iqx}\cos(2m k_F x)\left[(x\!-\!v_s t\!+\!i\epsilon)(x\!+\!v_s t\!-\!i\epsilon)\right]^{-Km^2}.
\end{equation}
A natural change of coordinates is given by the light-cone ones, $u=x-v_s t$ and $v=x+v_s t$. The Jacobian of the transformation is
\begin{equation*}
\begin{Vmatrix}
\frac{\partial{x}}{\partial{u}} & \frac{\partial{x}}{\partial{v}} \\
\frac{\partial{t}}{\partial{u}} & \frac{\partial{t}}{\partial{v}}
\end{Vmatrix}
=
\begin{Vmatrix}
\frac{1}{2} & \frac{1}{2} \\
-\frac{1}{2 v_s} & \frac{1}{2 v_s}
\end{Vmatrix}
=\frac{1}{2 v_s},
\end{equation*}
so that, after a rescaling and tranformation of the cosine into complex exponentials,
\begin{eqnarray}
I(Km^2)=\frac{2^{-2Km^2}}{v_s}\int_{-\infty}^{+\infty}du\int_{-\infty}^{+\infty}dv\left(v-\frac{i\epsilon}{2}\right)^{-Km^2}\left(u+\frac{i\epsilon}{2}\right)^{-Km^2} \nonumber \\
\left[e^{iv\left(\frac{\omega}{v_s}-q+2mk_F\right)}e^{-iu\left(\frac{\omega}{v_s}+q-2mk_F\right)}+e^{iv\left(\frac{\omega}{v_s}-q-2mk_F\right)}e^{-iu\left(\frac{\omega}{v_s}+q+2mk_F\right)} \right].
\end{eqnarray}
This in turn can be expressed in terms of
\begin{eqnarray}
J(q;\alpha)\!=\!\int_{-\infty}^{+\infty}dx\,e^{ixq}\left(x-\frac{i\epsilon}{2}\right)^{-\alpha},
\end{eqnarray}
where $\alpha\!>\!0$. Once again, complex analysis is a natural framework to evaluate this integral. The difficulty lies in the fact that, since $\alpha$ is not necessarily integer, the power law represents the exponential of a logarithm, which is multiply defined in the complex plane. To circumvent this problem, a branch cut is introduced, and the integral is not evaluated on a semi-circular contour, but on a more complicated one, sketched in the left panel of Fig.~\ref{Hankel}.

Applying the residue theorem to this contour, I find
\begin{equation}
\label{CC}
0\!=\!\int_{-R}^{R}\!dx\,e^{ixq}\left(x-\frac{i\epsilon}{2}\right)^{-\alpha}\!+\!\int_{brown}\!dz\,e^{izq}\left(z-\frac{i\epsilon}{2}\right)^{-\alpha}\!+\!\int_{green}\!dz\,e^{izq}\left(z-\frac{i\epsilon}{2}\right)^{-\alpha}\,,
\end{equation}
and assuming $q\!>\!0$, $\int_{brown}\!dz\,e^{iqz}\left(z-\frac{i\epsilon}{2}\right)^{-\alpha}\to_{R\to +\infty} 0$, while the first contribution to the right-hand side coincides with the integral $J$.

To finish the calculation, I still need to evaluate the integral over the green contour. To do so, I use the property \cite{Gradshteyn}
\begin{equation}
\label{hank}
\frac{1}{\Gamma(z)}=\frac{i}{2\pi}\int_H\!dt\,(-t)^{-z}e^{-t},
\end{equation}
where $\Gamma(z)$ is the analytic continuation of the Euler Gamma function in the complex plane and $H$ is the H\"ankel contour, sketched in the right panel of Fig.~\ref{Hankel}. A rotation of $-\pi/2$ maps the green contour of the left panel onto the H\"ankel one. Then, it has to be followed backward, and I will then denote it by $BH$. This transformation yields, after a few algebraic manipulations,
\begin{eqnarray}
J(q;\alpha)&=&-i\int_{BH} dz\,e^{qz}\left(-iz-\frac{i\epsilon}{2}\right)^{-\alpha} \nonumber \\
&=&i q^{\alpha-1}(-i)^{-\alpha}e^{-\frac{\epsilon q}{2}}\int_H dz\,e^{-z}(-z)^{-\alpha}\nonumber\\
&=&\frac{i^{\alpha}2\pi q^{\alpha-1}e^{-\frac{\epsilon q}{2}}}{\Gamma(\alpha)}\Theta(q).
\end{eqnarray}
Still according to \cite{Gradshteyn},
\begin{equation}
\int_{-\infty}^{+\infty}dx\,(\beta-ix)^{-\nu}e^{-ipx}=\frac{2\pi p^{\nu-1}e^{-\beta p}}{\Gamma(\nu)}\Theta(p), \mathcal{R}(\nu)>0, \mathcal{R}(\beta)>0,
\end{equation}
that yields the same result even more directly. It is then only a straightforward matter of algebra and combinations to finish the derivation.

\section{Dynamical structure factor of a \\
Tomonaga-Luttinger liquid in the thermodynamic limit at finite temperature}
\label{DSFTLLTder}

In this appendix, I give a quite detailed derivation of Eqs.~(\ref{S0LLT}) and (\ref{S1LLT}). To gain clarity, I split this derivation into several parts, that correspond to the main intermediate results.

\subsection{First contribution to the dynamical structure factor of a Tomonaga-Luttinger liquid at finite temperature}

To obtain the dynamical structure factor at finite temperature, I use the same trick as before, and split it artifically into the zero temperature result and a purely thermal term, which is more convenient to evaluate. Hence,
\begin{eqnarray}
&&S^T(q,\omega)=S_{T>0}(q,\omega)\!-\!S_{T=0}(q,\omega)\nonumber\\
&&=F.T.\left\{\frac{K}{4\pi^2}\int_{q\neq 0}dq |q|\left[ e^{i(qx-\omega(q)t)}+e^{-i(qx-\omega(q)t)}\right]n_B(q) \right\}
\end{eqnarray}
in the thermodynamic limit, and find after straightforward algebra, I find that
\begin{eqnarray}
S_0^T(q,\omega)=\frac{K}{L_T}\left\{\delta\left[\omega+\omega(q)\right]+\delta\left[\omega-\omega(q)\right]\right\}\frac{\beta\hbar \omega(q)}{e^{\beta\hbar \omega(q)}\!-\!1}.
\end{eqnarray}
Eventually,
\begin{eqnarray}
S_{0,T>0}(q,\omega)&=&S_0^T(q,\omega)+S_{0,T=0}(q,\omega)\nonumber\\
&=&\frac{K|q|}{e^{\beta\hbar\omega(q)}-1}\left\{\delta[\omega+\omega(q)]+\delta[\omega-\omega(q)]\right\}+K|q|\delta[\omega-\omega(q)]\nonumber\\
&=&\frac{K|q|}{1-e^{-\beta\hbar\omega(q)}}\left\{\delta[\omega-\omega(q)]+e^{-\beta\hbar\omega}\delta[\omega+\omega(q)]\right\},
\end{eqnarray}
ending the derivation.

\subsection{Second contribution to the dynamical structure factor at finite temperature}
This part of the calculation is, by far, the most difficult. Straightforward algebraic transformations and rescalings show that the integral one needs to evaluate in this situation is
\begin{equation}
 L(a)\!=\!\int_{-\infty}^{+\infty}du\,e^{-iau}\sinh^{-K}(u).
\end{equation}
I have done it stepwise, using the property
\begin{equation}
\Gamma\!\left(-\frac{yz+xi}{2y}\right)\Gamma(1\!+\!z)=(2i)^{z+1}y\,\Gamma\!\left(\!1+\frac{yz-xi}{2y}\right)\int_0^{+\infty}dt\,e^{-tx}\sin^z(ty),
\end{equation}
valid if $\mathcal{R}(yi)\!>\!0$ and $\mathcal{R}(x\!-\!zyi)\!>\!0$, misprinted in \cite{Gradshteyn} so that I first needed to correct it.
A few more algebraic manipulations, such as $\sinh(x)\!=\!i\sin(-ix)$ and splitting the integral, yield the more practical property
\begin{eqnarray}
\int_0^{+\infty}du\,e^{-iau}\sinh^{-K}(u)=\frac{2^{K-1}\Gamma\left(\frac{K+ia}{2}\right)\Gamma(1\!-\!K)}{\Gamma\left(1\!+\!\frac{ia-K}{2}\right)}.
\end{eqnarray}
Another tricky point concerns the branch cut. Prefactors written as $(-1)^{-K}$ are an abuse of notation, and should be interpreted either as $e^{iK\pi}$ or as $e^{-iK\pi}$, to obtain the intermediate expression
\begin{eqnarray}
&&S_{1,T>0}(q,\omega)=\frac{1}{2v_s}(L_Tk_F)^2\left(\frac{\epsilon}{L_T}\right)^{2K}(2\pi)^{2(K-2)}[\Gamma(1\!-\!K)]^2 \nonumber\\
&&\left[\frac{\Gamma\left(\frac{K}{2}+i\frac{\beta \hbar}{4\pi}[\omega+(q-2k_F)v_s]\right)}{\Gamma\left(1-\frac{K}{2}+i\frac{\beta \hbar}{4\pi}[\omega+(q-2k_F)v_s]\right)}+e^{-iK\pi}\frac{\Gamma\left(\frac{K}{2}-i\frac{\beta \hbar}{4\pi}[\omega+(q-2k_F)v_s]\right)}{\Gamma\left(1-\frac{K}{2}-i\frac{\beta \hbar}{4\pi}[\omega+(q-2k_F)v_s]\right)}\right]\nonumber\\
&&\left[\frac{\Gamma\left(\frac{K}{2}-i\frac{\beta \hbar}{4\pi}[\omega-(q-2k_F)v_s]\right)}{\Gamma\left(1-\frac{K}{2}-i\frac{\beta \hbar}{4\pi}[\omega-(q-2k_F)v_s]\right)}+e^{iK\pi}\frac{\Gamma\left(\frac{K}{2}+i\frac{\beta \hbar}{4\pi}[\omega-(q-2k_F)v_s]\right)}{\Gamma\left(1-\frac{K}{2}+i\frac{\beta \hbar}{4\pi}[\omega-(q-2k_F)v_s]\right)}\right]\nonumber\\
\end{eqnarray}
Using the property $\Gamma(z)\Gamma(1\!-\!z)\!=\!\frac{\pi}{\sin(\pi z)}$, after straightforward algebra I finally obtain
\begin{eqnarray}
&&S_{1,T>0}(q,\omega)=\frac{1}{2v_s}\left(\frac{L_T}{2\pi\epsilon}\right)^{2(1-K)}(n_0\epsilon)^2 e^{\frac{\beta\hbar\omega}{2}}\nonumber\\
&&B\left\{\frac{K}{2}+i\frac{\beta\hbar}{4\pi}[\omega+(q-2k_F)v_s],\frac{K}{2}-i\frac{\beta\hbar}{4\pi}[\omega+(q-2k_F)v_s]\right\}\nonumber\\
&&B\left\{\frac{K}{2}+i\frac{\beta\hbar}{4\pi}[\omega-(q-2k_F)v_s],\frac{K}{2}-i\frac{\beta\hbar}{4\pi}[\omega-(q-2k_F)v_s]\right\},
\end{eqnarray}
yielding Eq.~(\ref{S1LLT}), and the expression of the coefficient $C(K,T)$ in terms of the small-distance cut-off.

After this rather long calculation, it is worth checking its consistency with the $T\!=\!0$ case. To do so, I first come back to the property of the Beta function $B(x,y)\!=\!\frac{\Gamma(x)\Gamma(y)}{\Gamma(x+y)}$, as well as the properties of the Gamma function, $\Gamma(\overline{z})\!=\!\overline{\Gamma(z)}$ and \cite{Gradshteyn} $|\Gamma(x\!+\!iy)|\simeq_{y\to +\infty} \sqrt{2\pi}|y|^{x-\frac{1}{2}}e^{-\frac{\pi}{2}|y|}$
to obtain:
\begin{eqnarray}
\label{S1Texp}
 S_{1,T\to 0}(q,\omega)\!=\!B_1(K)\left|\omega^2\!-\!(q\!-\!2k_F)^2 v_s^2\right|^{K-1}e^{-\pi\left|\frac{\beta\hbar}{4\pi}[\omega+(q-2k_F)v_s]\right|}e^{-\pi\left|\frac{\beta\hbar}{4\pi}[\omega-(q-2k_F)v_s]\right|},
\end{eqnarray}
where
\begin{eqnarray}
B_1(K)\!=\!\frac{2(n_0\epsilon)^{2K}}{(2n_0v_s)^{2(K-1)}}\frac{\pi^2}{\Gamma(K)^2}\frac{1}{v_s},
\end{eqnarray}
and combining with Eq.~(\ref{Amcutoff}) I obtain Eq.~(\ref{B1}).
In Eq.~(\ref{S1Texp}), the exponentials vanish except if $\omega\!\geq\!|q-2k_F|v_s$. Then they equal one, and I recover Eq.~(\ref{DSFLL}) as expected.

The special case $K\!=\!1$ is also worth studying on its own as an additional indirect check of Eq.~(\ref{S1LLT}). Assuming $q\!>\!0$,
\begin{eqnarray}
\!\!S_{1,K\!=\!1, T>0}(q,\omega)\!=\!\frac{(k_F\epsilon)^2}{L_T^2}\!\!\int_{-\infty}^{+\infty}\!\!dt \int_{-\infty}^{+\infty}\!\!dx\, e^{i(\omega t-qx)}\frac{e^{2ik_F x}}{\sinh\!\left[\frac{\pi}{L_T}(x\!-\!v_s t)\right]\sinh\!\left[\frac{\pi}{L_T}(x\!+\!v_s t)\right]}.
\end{eqnarray}
so that, after a few algebraic transformations, it boils down to evaluating the integral
\begin{eqnarray}
G(a)\!=\!\int_{-\infty}^{+\infty}du\,\frac{e^{-iau}}{\sinh(u)}.
\end{eqnarray}
This can be done using the property \cite{Gradshteyn}
\begin{equation}
 P.P.\int_{-\infty}^{+\infty}dx\,\frac{e^{-\mu x}}{1-e^{-x}}\!=\!\pi \mathrm{cotan}(\pi\mu),\, 0<\Re(\mu)<1,
\end{equation}
and complex integration, yielding
\begin{eqnarray}
G(a)=P.P.\int_{-\infty}^{+\infty}\!du\,\frac{e^{-\frac{ia+1}{2}u}}{1-e^{-u}}+\frac{1}{2}\mathrm{Residue}(0)=i\pi\left[1\!+\!\tanh\left(\frac{a\pi}{2}\right)\right].
\end{eqnarray}
Rewriting
\begin{eqnarray}
[1+\tanh(a)][1+\tanh(b)]=\frac{e^{a+b}}{\cosh(a)\cosh(b)},
\end{eqnarray}
in the end
\begin{eqnarray}
S_{1,K\!=\!1,T>0}(q,\omega)\!=\!\frac{(k_F\epsilon)^2}{2v_s} \frac{e^{\frac{\beta\hbar\omega}{2}}}{\cosh\left\{\frac{L_T}{4v_s}[\omega\!+\!(q\!-\!2k_F)v_s]\right\}\cosh\left\{\frac{L_T}{4v_s}[\omega\!-\!(q\!-\!2k_F)v_s]\right\}},
\end{eqnarray}
consistent with the case $T\!=\!0$, and with the general case at $K\!=\!1$.

\section{Drag force due to a delta-barrier in the Tomonaga-Luttinger liquid framework}
\label{FTLder}
This appendix gives two derivations of the second line of Eq.~(\ref{FPit}), starting from its first line and Eq.~(\ref{DSFLL}), whose combination yields
\begin{eqnarray}
\label{C71}
F\!\!&=&\!\!\frac{U_b^2}{2\pi\hbar}\int_0^{+\infty}\!\!dq\, q S(q,qv)\nonumber \\
\!\!&=&\!\!\frac{U_b^2 B_1(K)}{2\pi\hbar}\int_0^{+\infty}\!\!dq\, q\left[(qv)^2\!-\!v_s^2(q\!-\!2k_F)^2\right]^{K-1}\Theta\left(qv\!-\!v_s|q\!-\!2k_F|\right)\nonumber\\
\!\!&=&\!\!\frac{U_b^2 B_1(K)}{2\pi\hbar}(v_s^2\!-\!v^2)^{K-1} \int_{q_-}^{q_+}dq\, q\, (q\!-\!q_-)^{K-1}(q_+\!-\!q)^{K-1},
\end{eqnarray}
where I have defined
\begin{eqnarray}
q_{\pm}=\frac{2k_Fv_s}{v_s\mp v}.
\end{eqnarray}
Then, changing of variables as $\tilde{q}=q-q_-$, $\overline{q}=\frac{\tilde{q}}{q_+-q_-}$, up to a global coefficient its boils down to
\begin{eqnarray}
F\propto \int_0^1\!d\overline{q}\,\overline{q}^{K-1}(1-\overline{q})^{K-1}\left(1+\frac{q_+-q_-}{q_-}\,\overline{q}\right)
\end{eqnarray}
with $\frac{q_+-q_-}{q_-}=\frac{2v}{v_s-v}$. This integral can be evaluated using the property \cite{Gradshteyn}
\begin{equation}
B(b,c-b)\, {_2}F_1(a,b;c;z)=\!\int_0^1dx\, x^{b-1}(1-x)^{c-b-1}(1-zx)^{-a} , |z|<1 , \Re(c)>\Re(b),
\end{equation}
(thus for $v/v_s\!<\!1/3$) where $_2F_1$ is the hypergeometric function defined as ${_2}F_1(a,b;c;z)\!=\!\sum_{n=0}^{+\infty}\frac{(a)_n(b)_n}{(c)_n}\frac{z^n}{n!}$ and $(q)_n\!=\!q(q+1)\dots(q\!+\!n\!-\!1)$ is the Pochhammer symbol. I also use the properties of the Beta and Gamma functions, $B(X,Y)\!=\!\frac{\Gamma(X)\Gamma(Y)}{\Gamma(X+Y)}$ and $\Gamma(2z)\!=\!\frac{2^{2z-1}}{\sqrt{\pi}}\Gamma(z)\Gamma(z\!+\!1/2)$,
to rewrite
\begin{eqnarray}
 B(K,K)=\frac{\sqrt{\pi}\Gamma(K)}{2^{2K-1}\Gamma(K\!+\!1/2)},
\end{eqnarray}
and obtain Eq.~(\ref{FPit}) after putting back all prefactors.

Another derivation starts back from the last line of Eq.~(\ref{C71}), which is split into two parts as
\begin{eqnarray}
&&\int_{q_-}^{q_+}dq\, q(q-q_-)^{K-1}(q_+-q)^{K-1} \nonumber \\
&&=\int_{q_-}^{q_+}dq\, (q-q_-)^K(q_+-q)^{K-1}+q_-\int_{q_-}^{q_+}dq\, (q-q_-)^{K-1}(q_+-q)^{K-1}.
\end{eqnarray}
Then, the Beta function naturally appears through the property \cite{Gradshteyn}
\begin{equation}
\int_a^bdx\, (x-a)^{\mu-1}(b-x)^{\nu-1}=(b-a)^{\mu+\nu-1}B(\mu,\nu), b>a, \Re(\mu)>0, \Re(\nu)>0,
\end{equation}
whence
\begin{eqnarray}
&&\frac{F(v)}{(v_s^2-v^2)^{K-1}U_b^2B_1(K)/(2\pi\hbar)}=(q_+-q_-)^{2K}B(K+1,K)+q_-(q_+-q_-)^{2K-1}B(K,K)\nonumber\\
&&=(2k_F)^{2K}\frac{(2vv_s)^{2K}}{(v_s^2-v^2)^{2K}}\frac{\Gamma(K+1)\Gamma(K)}{\Gamma(2K+1)}+\frac{2k_Fv_s (2k_F)^{2K-1}}{v_s+v}\frac{(2vv_s)^{2K-1}}{(v_s^2-v^2)^{2K-1}}\frac{\Gamma(K)^2}{\Gamma(2K)},
\end{eqnarray}
and the property $\Gamma(K+1)\!=\!K\Gamma(K)$ combined to algebraic manipulations yields Eq.~(\ref{FPit}).

\newpage

\end{document}